\documentclass[iop,revtex4,12pt,preprint,longauth]{aa}
\usepackage{graphics,graphicx}
\usepackage{helvet}
\usepackage{subfigure}
\usepackage[utf8]{inputenc}
\usepackage[T1]{fontenc}
\usepackage{soul}
\usepackage[draft]{hyperref}
\usepackage{amsmath, amssymb, gensymb}
\usepackage{morefloats}
\usepackage{float}
\usepackage{ulem}
\usepackage{url}
\usepackage{natbib}
\usepackage{microtype}
\usepackage{multirow}
\usepackage{morefloats}
\usepackage{rotating}
\usepackage{here}
\usepackage{xspace}
\usepackage{color}
\usepackage{lipsum}
\usepackage[a4paper]{geometry}
\usepackage{xspace}
\usepackage{tabularx}
\usepackage{supertabular}
\usepackage{caption}
\newcommand{\NH}{N_{\rm H_2}} 

\def\Pf{\mathcal{P_{\rm{frac}}}}
\def\S{\mathcal{S}}
\def\StimesP{\S\times\Pf}
\newcommand{\plk}{{\it Planck}\xspace}

\def\eg {e.g.,\xspace} 
\def\ie {i.e.,\xspace} 
\def\cmsq  {$\hbox{\textrm{cm}}^{-2}$\xspace}    

\def\slope {0.79}
\def\dslope {0.03}

\def\SPracValue {$0.36^{+0.10}_{-0.17}$}

\begin{document}
\title{A statistical analysis of dust polarization properties in ALMA observations of Class 0 protostellar cores}
\author{V. J. M. Le Gouellec\inst{1,}\inst{2} , A. J. Maury\inst{2,}\inst{3}, V. Guillet\inst{4,}\inst{5}, C. L. H. Hull \inst{6,}\inst{7,}\inst{11,}, J. M. Girart \inst{8,}\inst{9}, A. Verliat\inst{2}, R. Mignon-Risse\inst{2,10}, V. Valdivia\inst{2}, P. Hennebelle \inst{2},  M. González \inst{2,10}, F. Louvet \inst{2}}

\institute{ European Southern Observatory, Alonso de C\'ordova 3107, Vitacura, Casilla 19001, Santiago , Chile
\email{Valentin.LeGouellec@eso.org}
\and
Université Paris-Saclay, CNRS, CEA, Astrophysique, Instrumentation et Modélisation de Paris-Saclay, 91191, Gif-sur-Yvette, France
\and
Harvard-Smithsonian Center for Astrophysics, Cambridge, MA 02138, USA
\and
Université Paris-Saclay, CNRS, Institut d’astrophysique spatiale, 91405, Orsay, France
\and
Laboratoire Univers et Particules de Montpellier, Université de Montpellier, CNRS/IN2P3, CC 72, Place Eugène Bataillon, 34095, Montpellier Cedex 5, France
\and
National Astronomical Observatory of Japan, NAOJ Chile, Alonso de Córdova 3788, Office 61B, 7630422, Vitacura, Santiago, Chile
\and
Joint ALMA Observatory, Alonso de Córdova 3107, Vitacura, Santiago, Chile
\and Institut de Ci\`encies de l’Espai (ICE-CSIC), Campus UAB, Carrer de Can Magrans S/N, E-08193 Cerdanyola del Vallès, Catalonia, Spain
\and
Institut d’Estudis Espacials de Catalunya, E-08030 Barcelona, Catalonia, Spain
\and
Université de Paris, AIM, F-91191, Gif-sur-Yvette, France
\and
NAOJ Fellow
}

\date{Accepted for publication in A\&A on 12 September 2020}

\abstract
{Recent observational progress has challenged the dust grain-alignment theories used to explain the polarized dust emission routinely observed in star-forming cores.}
{In an effort to improve our understanding of the dust grain alignment mechanism(s), we have gathered a dozen ALMA maps of (sub)millimeter-wavelength polarized dust emission from Class 0 protostars, and carried out a comprehensive statistical analysis of dust polarization quantities.}
{We analyze the statistical properties of the polarization fraction $\Pf$ and dispersion of polarization position angles $\S$. More specifically, we investigate the relationship between $\S$ and $\Pf$ as well as the evolution of the product $\StimesP$ as a function of the column density of the gas in the protostellar envelopes. We compare the observed trends with those found in polarization observations of dust in the interstellar medium and in synthetic observations of non-ideal magneto-hydrodynamic (MHD) simulations of protostellar cores.}
{We find a significant $\S \propto \Pf^{-\slope}$ correlation in the polarized dust emission from protostellar envelopes seen with ALMA; the power-law index differs significantly from the one observed by \plk in star-forming clouds.
The product $\StimesP$, which is sensitive to the dust grain alignment efficiency, is approximately constant across three orders of magnitude in envelope column density (from $\NH=10^{22}$\,\cmsq\ to $\NH=10^{25}$\,\cmsq), with a mean value of \SPracValue. This suggests that the grain alignment mechanism producing the bulk of the polarized dust emission in star-forming cores may not depend systematically on the local conditions such as local gas density.
However, in the lowest-luminosity sources in our sample, we find a hint of less efficient dust grain alignment with increasing column density.
Our observations and their comparison with synthetic observations of MHD models suggest that the total intensity versus the polarized dust are distributed at different intrinsic spatial scales, which can affect the statistics from the ALMA observations,
for example by producing artificially high $\Pf$.
Finally, synthetic observations of MHD models implementing radiative alignment torques (RATs) show that the statistical estimator $\StimesP$ is sensitive to the strength of the radiation field in the core. Moreover, we find that the simulations with a uniform perfect alignment (PA) of dust grains yield on average much higher $\StimesP$ values than those implementing RATs; the ALMA values lie among those predicted by PA, and are significantly higher than the ones obtained with RATs, especially at large column densities.}
{Ultimately, our results suggest dust alignment mechanism(s) are efficient at producing dust polarized emission in the various local conditions typical of Class 0 protostars. The grain alignment efficiency found in these objects seems to be higher than the efficiency produced by the standard RAT alignment of paramagnetic grains. Further study will be needed to understand how more efficient grain alignment via, e.g., different irradiation conditions, dust grain characteristics, or additional grain alignment mechanisms can reproduce the observations.}


\titlerunning{Polarization statistics of ALMA observations of Class 0 cores}
\authorrunning{Le Gouellec et al.}

\maketitle

\section{Introduction}
Magnetic fields have been considered to play a key role in the formation of molecular clouds and in the regulation of star formation \citep{Shu1987,McKee1993,McKee2007}. For example, fields are partially responsible for setting the star formation rate \citep{Krumholz&Federrath2019}, as the gas motions tend to follow the orientations of magnetic fields, whose strengths can regulate the gravitational collapse of these structures \citep{Mouschovias1999}. Past observations of molecular clouds have shown that the magnetic field seems to be a key player in the formation of parsec-scale density structures \citep{PlanckXXXV,Soler2019,Seifried2020}, and appears regulate star formation inside these structures \citep{HBLi2017}. One of the main ways to characterize the spatial distribution of magnetic fields is to observe the polarized thermal emission from dust grains. Indeed, since dust grains are not perfectly spherical, they tend to align themselves with the ambient magnetic field under some conditions \citep{Lazarian2007,Andersson2015}, resulting in polarized thermal emission that can be used to infer the magnetic field orientation integrated along the line of sight. This linear polarization emanating from this dust grain population is orthogonal to the magnetic field component projected on the plane of the sky.

Observations of magnetic fields via polarized dust emission are still subject to caveats due to the strong dependence of grain alignment on the local environmental conditions. Understanding the impact of the key factors enabling dust grain alignment via the Radiative Alignment Torques (RATs) theory---such as the degree of anisotropy of the radiation field, the dust grain size distribution, the gas temperature, and the density distribution---has been the focus of numerical works where radiative transfer was performed on magneto-hydrodynamic (MHD) simulations \citep{Padoan2001,Bethell2007,Pelkonen2009,Brauer2016}. One of their main goals was to investigate the widespread phenomenon of depolarization, \ie the drop in the ratio of linearly polarized dust emission to the total intensity emission toward high density zones in molecular clouds and cores; this is the so-called ``polarization hole'' phenomenon.
Single-dish observations of molecular clouds \citep{Poidevin2013,Fissel2016}, single-dish observations of a starless core \citep{Alves2014}, and high-resolution interferometric observations of Class 0 protostellar cores \citep{Hull2014,Galametz2018} found a significant decrease of the polarization fraction with increasing column density, and interpreted this drop as either depolarization caused by disorganized magnetic field lines smeared out in the synthesized beam (i.e., the resolution element of the observations), or as a possible loss of alignment efficiency of the dust particles caused by a lack of irradiation and/or changes in dust grain characteristics toward high column density regions. This depolarization phenomenon was analyzed in the scope of several possible physical explanations: the collisional de-alignment of dust grains due to high gas temperature and density \citep{Reissl2020}; the reddening of the radiation field when reprocessed during its propagation \citep{Lazarian2007}; the change in grain size and shape due to coagulation and formation of icy mantles \citep{Juarez2017}; the lack of the necessary anisotropy in the radiation field as a result of high optical depth \citep[][studied the drop of polarization degree in dense regions of Bok globules]{Brauer2016}; and the level of disorder of the magnetic field lines caused by turbulence \citep{FalcetaGoncalves2008}. Most of the time, high angular resolution observations of dust polarization revealed that the  drops of polarization fraction in observations of cores at coarse angular resolution were partly explained by beam smearing, \ie the fact that the fluctuations of magnetic fields in the plane of the sky could not be resolved in high density zones.  However, while these higher angular resolution observations did tend to detect polarized emission in the holes seen in the low-resolution data, these same high resolution observations saw their own polarization holes at smaller scales, as pointed out in \citet{Galametz2018}.

In observations, this depolarization effect was also quantified thanks to the statistical analysis of both the polarization fraction and the local dispersion of polarization angles \citep{PlanckXIX,PlanckXX}. Assuming perfect alignment efficiency of the dust grains with magnetic field lines, both quantities are correlated with the level of disorder in the magnetic field, and thus both vary as a function of the amount of local fluctuations of the magnetic field. The polarization fraction is sensitive to the cancellation of polarization along the line of sight and hence to the fluctuations of the apparent magnetic field along the line of sight. Conversely, the dispersion of polarization angles provides the fluctuation of the apparent magnetic field orientations in the plane of the sky. Assuming the fluctuations of the magnetic field lines are isotropic, the product of these two quantities gives access to the dust grain alignment efficiency: \citet{Planck2018XII} analysed these statistical estimators as a function of column density from the diffuse interstellar medium (ISM) to molecular clouds, probing column density up to $10^{22}\textrm{cm}^{-2}$. They found no variation of dust grain alignment efficiency with varying conditions typical of these environments. More recently, \citet{Reissl2020} applied this statistical tool to simulations of the diffuse ISM in order to quantify the relative influence that radiative torque intensity and gas pressure have on the dust grain alignment efficiency. They found no significant differences in dust grain alignment efficiency when analyzing polarization from perfectly aligned dust grains versus those aligned by RATs in these environments embedded in the interstellar radiation field.

Given the uncertain validity of RATs in high density environments where irradiation is much less homogeneous and is shifted to long wavelengths, the dense parts of protostellar cores represent regions of interest. While ALMA has recently produced a large number of high-sensitivity observations of young stellar objects, the thermal dust emission emitted by the youngest sources, known as prestellar cores, is heavily spatially filtered by ALMA and thus hardly detected \citep{Dunham2016}, rendering investigations of their polarized dust emission at high spatial resolution challenging.
However, once these cores initiate their gravitational collapse, because their densest regions become warm enough to dissociate $\rm H_2$, a compact structure forms around the nascent embedded protostar, which enables interferometric observations. These youngest protostellar objects, known as Class 0 protostars \citep{Andre2000,Andre2014}, are engaged in a short but vigorous accretion phase during which the central protostar will gather most of its final mass, triggering also ejection of material in the form of bipolar outflows visible in molecular emission lines.
These sources are ideal for our study because during this phase, most of the thermal dust emission is emitted by the envelope surrounding the central embryo; during the (later) Class I phase, the envelope has already been largely accreted/dissipated. Here, we focus on the ALMA dust polarization observations of Class 0 envelopes to optimize the number of detections. Most of these recent observations have shown that specific regions of the cores such as the walls of the bipolar outflow cavities, potential magnetized accretion streamers, and core equatorial plane are preferentially polarized \citep{Hull2017a,Hull2017b,Maury2018,Sadavoy2018a,Sadavoy2018b,Kwon2019,LeGouellec2019a,Takahashi2019,Hull2020a}. The specific locations of the recovered polarized emission raise questions regarding the local conditions required to align a significant fraction of dust grains along magnetic lines, and thus to produce the level of polarized emission observed. Our goal in this paper is to investigate the statistical behavior of polarized dust emission observed with ALMA within protostellar envelopes, by adapting and applying some of the tools previously developed to characterize polarized dust emission at cloud scales.

The paper is structured as follows. In Section \ref{sec:presentation}, we introduce the statistical tools we use to analyse the polarized dust emission detected by ALMA in young protostars, and we present the sample of ALMA observations of Class 0 objects that are the focus of this statistical study. We present the methodology and results of the statistics in Section \ref{sec:analysis}. Finally, we discuss the results we obtain regarding the dust grain alignment in young protostellar objects in Section \ref{sec:disc}, along with comparisons between the ALMA observations and synthetic observations of MHD simulations. We draw our conclusions in Section \ref{sec:conclu}.

\section{ALMA Observations of Class 0 protostars}
\label{sec:presentation}

\subsection{Statistical tools}
\label{sec:persp}

Our objective is to characterize the polarized emission emanating from Class 0 protostellar cores, at envelope scales, targeting the emission from circumstellar material at radii of $\sim 10-2000$ au. The properties of the linear polarization of thermal dust emission are expressed by the Stokes parameters $Q$ and $U$. Stokes $I$ represents the total intensity. We will denote the polarized intensity $P$ (defined as $P = \sqrt{Q^2+U^2}$, which we systematically debias, see Section \ref{sec:method}), the polarization fraction $\Pf$ (defined as $\Pf = P/I$), and the polarization position angle $\phi$ (defined as $\phi=0.5\arctan{\frac{U}{Q}}$).


In the diffuse ISM and molecular clouds, \citet{PlanckXIX,PlanckXX,Planck2018XII} found a correlation between the local dispersion of the polarization position angle and the polarization fraction. Note that a similar correlation was found in \citet{Alves2008}, using optical background-starlight polarization observations of the Pipe Nebula. The polarization angle dispersion function $\S$, which quantifies the local (non)-uniformity of the polarization angle, is defined as follows:
\begin{equation}
\label{equ:S_def}
S(r,\delta)\,=\,\sqrt{\frac{1}{N}\sum^{N}_{i=1}{[\phi(r+\delta_i)-\phi(r)]^2}}\,\,,
\end{equation}
where the angle dispersion is calculated at a given position $r$ and for a given neighborhood $\delta$, which is also known as the lag. The lag describes the area over which the dispersion of polarization angles is derived, and thus corresponds to the characteristic length scale at which we quantify the disorganization of polarization position angles. The computation is performed on $N$ neighboring pixels contained in an annulus centered on $r$, having inner and outer radii of $\delta/2$ and $3\delta/2$, respectively; each of the $N$ pixels is indexed by $i$, and located at $r+\delta_i$ \citep{Planck2018XII}. \citet{Planck2018XII} developed an analytical model (briefly described in Appendix \ref{app:Planck_model}) that relates the two quantities $\S$ and $\Pf$. They found, among other results, that $S\,\propto\,{\Pf}^{-1}$ in the diffuse ISM and molecular clouds: this correlation is shown as a red solid line in our plots. Exploring the evolution of the quantity $\StimesP$---which is a proxy for grain alignment efficiency---as a function of the column density and the dust temperature, \citet{Planck2018XII} did not detect a significant drop of efficiency with increasing column density. We apply a modified version of this technique to ALMA observations assuming the dust grains are aligned with the ambient magnetic field at the typical scales of a protostellar core.

The dispersion of the polarization position angles $\S$ gives us information about the level of disorder in the magnetic field projected in the plane of the sky; the higher the value of $\S$, the more disorganized the apparent magnetic field. $\S$ will saturate at the value of $\pi/\sqrt{N}$, as $(\phi=0^{\circ})\equiv(\phi=180^{\circ})$ in a polarization map. Note that here we distinguish between the disorganization of the apparent magnetic field lines as seen by the observer and the actual (3-dimensional) turbulent component of the magnetic field. Indeed, if a given, moderately turbulent magnetic field is oriented closer to the line of sight, the observed dispersion $\S$ will be larger. In contrast, a uniform magnetic field will have dispersion $\S$ close to zero, regardless of the line of sight.  These facts limit the capability of $\S$ to trace the turbulent component of the magnetic field. However, we should also note that the value of the polarized intensity $P$ (and thus $\Pf$) directly depends on the orientation of the magnetic with respect to the line of sight. Thus, in the extreme line-of-sight cases where $\S$ reaches high values, $P$ may drop below the detection limit.

The polarization fraction $\Pf$ is another tool linked with the disorganization of the magnetic field. A disorganized magnetic field along the line of sight will result in a low value of $\Pf$ as seen by the observer. Consequently, assuming an isotropic turbulent component of the magnetic field, and given the caveats listed above, $\S$ and $\Pf$ are directly linked to level of disorder in the magnetic field lines in a core.

The aim of the study presented here is to search for, compare, and interpret the possible physical causes for a correlation between $\S$ and $\Pf$ in Class 0 protostars, toward which polarized dust emission was observed with ALMA in recent years.

\subsection{ALMA observations of polarized dust emission in Class 0 protostellar cores}
\label{sec:source_sample}

\begin{table*}[!tbph]
\centering
\small
\caption[]{Summary of analyzed ALMA observations}
\setlength{\tabcolsep}{0.4em} 
\begin{tabular}{p{0.2\linewidth}cccccccccc}
\hline \hline \noalign{\smallskip}
Name & $\theta_{\textrm{res}}^{a}$&${\sigma_{I}}^{b}$& ${\sigma_{P}}^{b}$& $N_{\textrm{H}_2}^c$  & $\lambda$ &$P_{\textrm{frac,max}}^d$ &MRS$^e$ &Reference$^f$\\
&au&$\frac{\textrm{mJy}}{\textrm{beam}}$&$\frac{\textrm{mJy}}{\textrm{beam}}$ &$10^{23}\,\textrm{cm}^{-2}$&mm&&$10^3$au&\\
\noalign{\smallskip}  \hline
\noalign{\smallskip}
Serpens Emb 6  &161&0.05&0.060  &0.46--67&0.87&$19\%\pm4$&2.5\phantom{00}& \citet{Hull2017b}\\
\noalign{\smallskip}
\hline
\noalign{\smallskip}
Serpens Emb 8  &161&0.07&0.025 &0.11--9.7&\multirow{2}{*}{0.87}&$42\%\pm9$&\multirow{2}{*}{2.5\phantom{00}}&\citet{Hull2017a}\\
\noalign{\smallskip}
Serpens Emb 8(N) &161&0.08&0.035&0.28--8.7&&\phantom{0}$39\%\pm11$&& \citet{LeGouellec2019a} \\
\noalign{\smallskip}
\hline
\noalign{\smallskip}
BHR71 IRS1  &200&0.30&0.025 &0.13--14&\multirow{2}{*}{1.3\phantom{0}}&$24\%\pm5$& \multirow{2}{*}{4.3\phantom{00}}& \multirow{2}{*}{ \citet{Hull2020a}}\\
\noalign{\smallskip}
BHR71 IRS2  &200&0.15&0.025&0.071--14&&$19\%\pm4$&&  \\
\noalign{\smallskip}
\hline
\noalign{\smallskip}
\multirow{3}{*}{B335} & 101 &0.06&0.002& 0.063--3.7&1.3\phantom{0}&$26\%\pm6$&2.2\phantom{00}&
\multirow{2}{*}{\citet{Maury2018},}\\
&\phantom{0}48 &0.04&0.010&0.15--5.7 &1.5\phantom{0}&$25\%\pm7$&0.7\phantom{00}&  \multirow{2}{*}{   Maury et al. in prep }
\\
& 122 &0.03&0.005&0.17--3.1&3.0\phantom{0}&$30\%\pm7$&3.0\phantom{00}& \\
\noalign{\smallskip}
\hline
\noalign{\smallskip}
IRAS 16293A/B & \phantom{0}36 &0.28&0.025&0.73--61& 1.3\phantom{0} &$35\%\pm7$&0.59\phantom{0}& \citet{Sadavoy2018b}\\
\noalign{\smallskip}
\hline
\noalign{\smallskip}
VLA 1623A/B  & \phantom{0}34 &0.09&0.027 &0.84--59& 1.3\phantom{0}  &$21\%\pm7$&0.62\phantom{0}& \citet{Sadavoy2018a}\\
\noalign{\smallskip}
\hline
\noalign{\smallskip}
L1448 IRS2 & 134 &0.85&0.020&0.21--4.8&1.3\phantom{0}&$32\%\pm6$&1.5\phantom{00}& \citet{Kwon2019}\\
\noalign{\smallskip}
\hline
\noalign{\smallskip}
\multirow{2}{*}{OMC3 MMS6$^g$} & \phantom{0}56 &0.23&0.022&3.7--145&\multirow{2}{*}{1.3\phantom{0}}&$24\%\pm5$&0.43\phantom{0}&\multirow{2}{*}{ \citet{Takahashi2019}}\\
&   \phantom{0}12 &0.08&0.020&43--167&&$24\%\pm4$&0.095&\\
\noalign{\smallskip}
\hline
\noalign{\smallskip}
\multirow{2}{*}{IRAS4A} &112&0.4&0.045 &0.87--62&1.3\phantom{0}&$41\%\pm8$&1.2\phantom{00}&\multirow{2}{*}{\citet{Ko2019}}\\
&\phantom{0}47&0.50&0.046&1.7--65&0.87 &$36\%\pm7$&0.8\phantom{00}&\\
\noalign{\smallskip}
\hline
\smallskip
\end{tabular}
\vspace*{-0.1in}
\caption*{\footnotesize
This table presents the details of the ALMA observations.  For further information about the individual sources, see Table \ref{t.obs}. \\
$^a$ Angular resolution in au. We took the effective synthesized beam size (where the beam is the resolution elements) of the ALMA maps.\\
$^{b}$ rms noise in the maps of total intensity $I$ and polarized intensity $P$.\\
$^c$ Typical $\textrm{H}_2$ column densities probed in the protostellar envelope by the ALMA observations, selecting the pixels with total intensity values greater than 5 times the rms noise in the Stokes $I$ map. The column density of each pixel was derived following the method described in Section \ref{sec:source_sample}. Each pointing in BHR 71 contains both of the components of the wide binary. Thus, the corresponding ranges in column density include all emission from both BHR 71 IRS1 and IRS2.\\
$^d$ The maximum value of polarization fraction in the core, selecting the pixels with total intensity values greater than 5 times the rms noise in the Stokes $I$ map. \\
$^e$ Maximum recoverable scale.\\
$^f$ Reference of the publication(s) presenting the corresponding ALMA polarization dataset(s).\\
$^g$ \citet{Takahashi2019} presented the ALMA observations of OMC3 MMS6 in two separate datasets, as the angular resolutions of the two datasets were very different, and thus probe distinct regions of the envelope.
}
\vspace{-0.4cm}
\label{t:sources}
\end{table*}

In its polarized mode, ALMA produces visibility measurement sets of the three Stokes parameters $I$, $Q$, and $U$, which can be imaged and combined to produce maps of the polarized dust emission.
We gathered publicly available ALMA dust polarization observations toward nearby, low- and intermediate-mass Class 0 protostars.
Since the statistical tools we use require a large number of statistically independent measurements at the typical scale of the object studied, we selected observations with the most extended polarized dust emission. The regions of interest in these protostellar cores correspond to the inner envelope scales ($\sim$10--2000 au). Therefore, we selected the ALMA datasets whose polarized dust emission was observed with combinations of sensitivity and angular resolution that allow us to detect low levels of polarized emission beyond the peak of continuum emission, at these inner envelopes typical scales. We present the resulting sample in Table \ref{t:sources}.

We use the three Stokes maps provided by the authors of the corresponding publications (see Table \ref{t:sources}) to create the polarized emission maps. In the case of NGC1333 IRAS4A, however, because these data were not yet published at the time we started our investigations, we calibrated and imaged these observations ourselves. We produced the polarized dust continuum images using the task \texttt{tclean} in version 5.4 of CASA \citep{McMullin2007}.  We applied four rounds of consecutive phase-only self-calibration, using the total intensity (Stokes $I$) solutions as the model, with a Briggs weighting parameter of \texttt{robust}\,=\,1. The three Stokes parameters $I$, $Q$, and $U$ were cleaned separately after the last round of self-calibration using an appropriate residual threshold and number of iterations.
In order to calculate appropriate thresholds for the data (see the method developed in Section \ref{sec:method}), we require an homogeneous level of noise across the individual fields of view, and thus we do not perform any primary beam correction at this step of the analysis. However, the total intensity maps are primary beam corrected before deriving the column density whose ranges are reported in Table \ref{t:sources}.

\begin{figure*}[!tbh]
\centering
\vspace{-0.1cm}
\subfigure{\includegraphics[scale=0.40,clip,trim= 1.2cm 1.5cm 1.8cm 1.8cm]{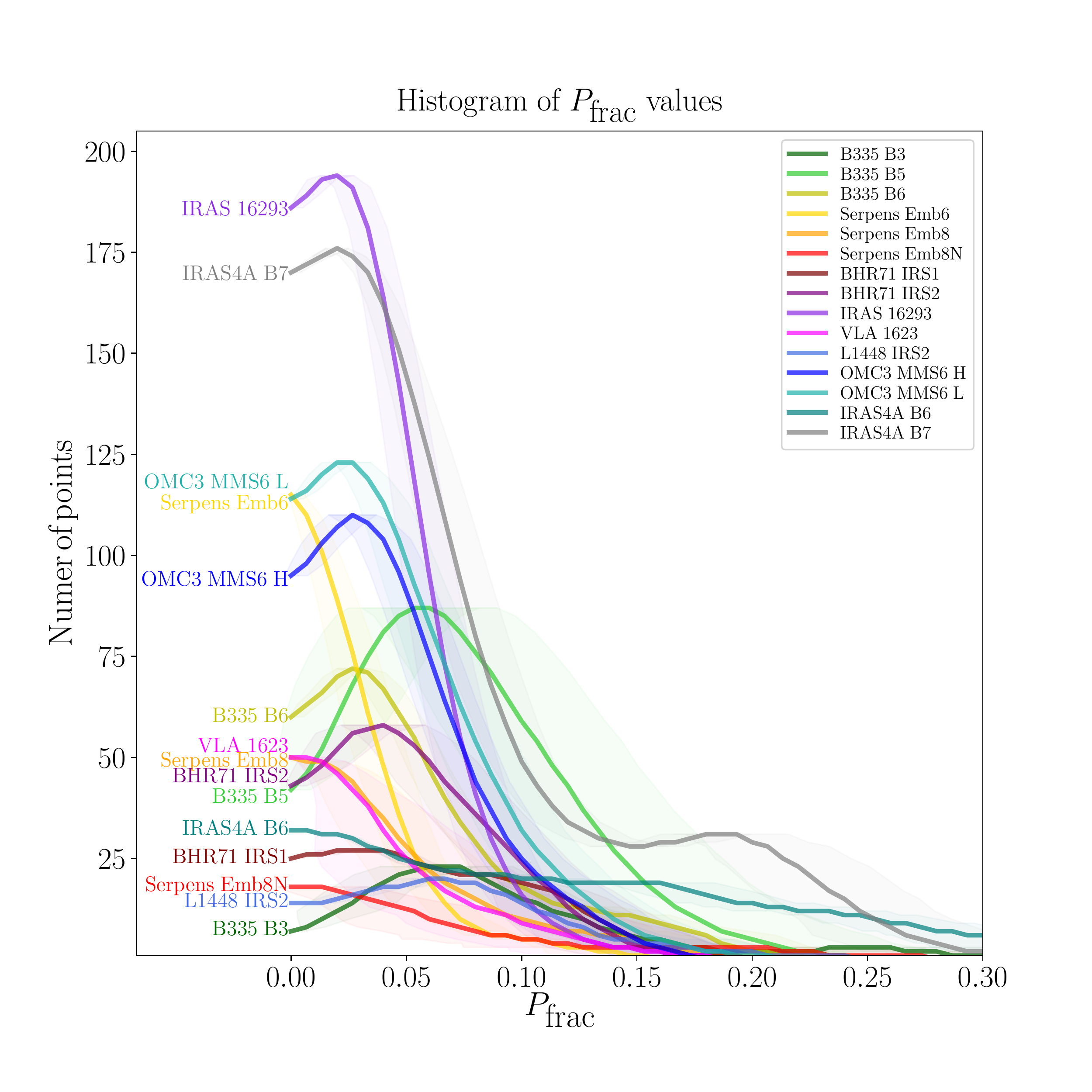}}
\subfigure{\includegraphics[scale=0.40,clip,trim= 1.2cm 1.5cm 1.8cm 1.8cm]{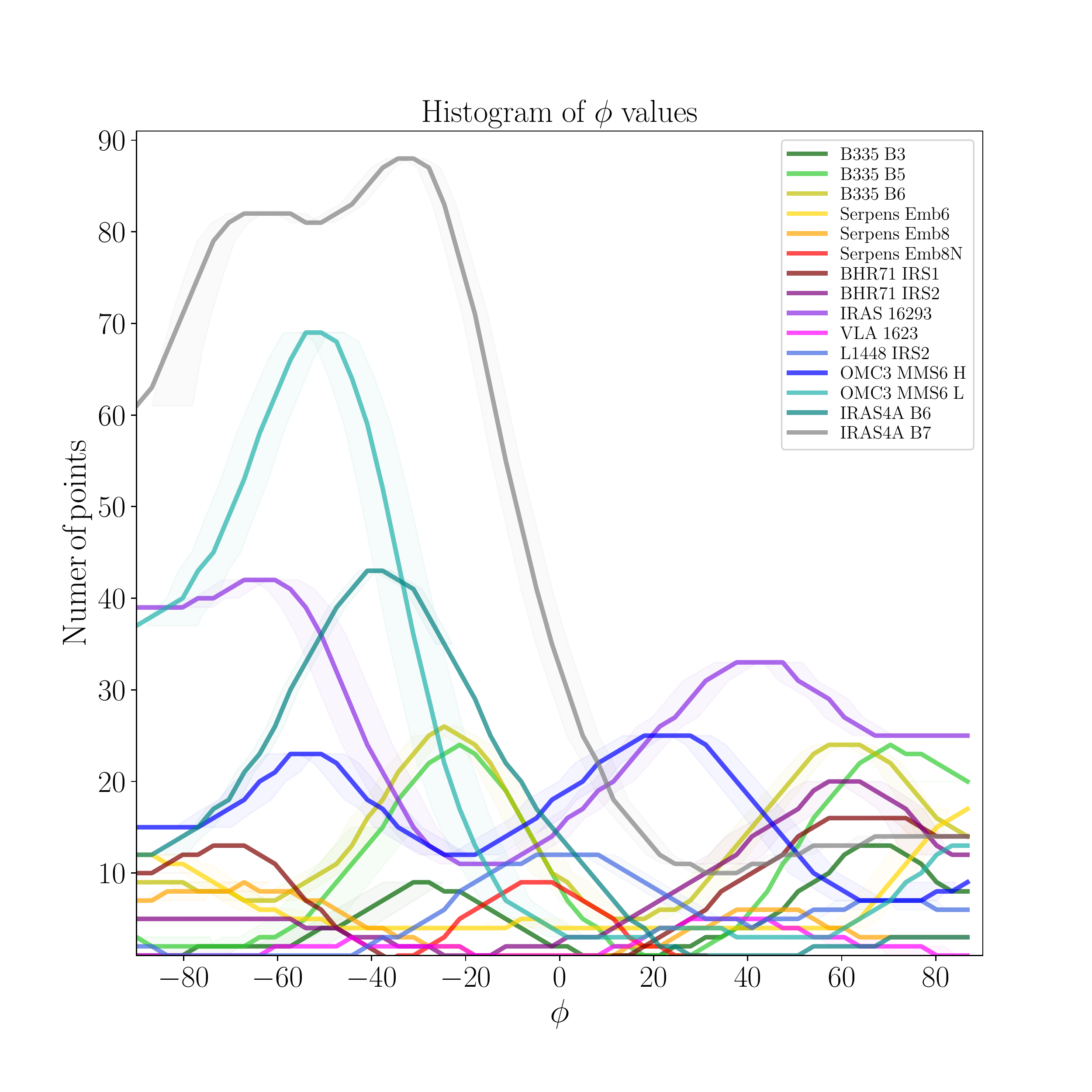}}
\caption[]{\footnotesize Histograms of polarization fraction $\Pf$ (\textit{left}) and polarization position angle $\phi$ (\textit{right}) for all the sources of our sample. The histogram lines have been smoothed with a 1D-Gaussian kernel of a size of 0.2$\%$ in $\Pf$ and $2^{\circ}$ in $\phi$. The shaded areas correspond to the mean of the uncertainty in the values of $\Pf$ and $\phi$ within each bin of the histograms. Note that among the two datasets we have toward OMC3 MMS6, ``OMC3 MMS6 H'' and ``OMC3 MMS6 L'' denote the high and low resolution observations, respectively.}
\label{fig:hist_pa_pfrac}
\end{figure*}

It is crucial, when building polarized dust emission maps from the combination of the Stokes maps, to have a robust assessment of the rms noise levels in each of these maps. This is particularly important to consider in our statistical measurements so that we do not introduce noise bias,
since values of polarization fraction $\Pf$ can be affected significantly when dividing by Stokes $I$ values that are uncertain.
Here, we compute the rms noise values in each of the three Stokes maps $I$, $Q$, and $U$ ($\sigma_I$, $\sigma_Q$, and $\sigma_U$ respectively) by measuring the root mean square in an area without strong emission. We notice that typically $\sigma_Q\,\approx\,\sigma_U$, so we use a single value $\sigma_P\,\equiv\,\sigma_Q\,\approx\,\sigma_U$.
We present in Figure \ref{fig:hist_pa_pfrac} the distribution of $\Pf$ and polarization position angles $\phi$ in the region where the Stokes $I$ is >\,5\,$\sigma_I$, from all individual maps of all sources at each wavelength. In these histograms, the uncertainties in $\Pf$, and $\phi$ (in radians) are showed as shaded areas, and are calculated as follows:

\begin{equation}
\sigma_{\Pf}=\Pf\sqrt{{ \left( \frac{\sigma_P}{P}\right) } ^2+{\left( \frac{\sigma_I}{I}\right)}^2}\,\,,
\end{equation}

\begin{equation}
\sigma_\phi = \frac{1}{2} \frac{\sigma_P}{P}\,\,.
\end{equation}

Finally, assuming the dust emission recovered in the ALMA observations (at scales of 10--2000\,au) is optically thin, we calculate the column density from the total intensity dust emission maps as follows:
\begin{equation}
N_{\textrm{H}_2}=\frac{S_{\nu}d^2}{A\mu_{\textrm{H}_2} m_H  \kappa_\nu B_\nu(T_\textrm{d}(r))}\,\,,
\end{equation}
where $S_{\nu}$ the flux density measured, $d$ is the distance to the source (see Table \ref{t:sources}), $B_\nu(T_\textrm{d}(r))$ is the \plk function at the frequency $\nu$ of our observations for dust of a given temperature $T_\textrm{d}(r)$ (see below), $\kappa_\nu$ is the opacity  at a specific wavelength taken from \citet{Ossenkopf1994}, $m_H$ is the mass of a hydrogen atom, $\mu_{\textrm{H}_2}$ is the mean molecular weight per hydrogen molecule ($\mu_{\textrm{H}_2}=2.8$ for gas composed of 71\% hydrogen, 27\% helium, and 2\% metals by mass; \citealt{Kauffmann2008}), and $A$ is the area over which we calculate the flux density. We assume a gas-to-dust ratio of 100. The value of the dust temperature at a radius $r$ from the position of the protostellar embryo (assumed to coincide with the peak position of the dust continuum emission in the ALMA Stokes $I$ map), can be estimated assuming that only the central protostellar object heats the dust in the inner envelope, following \citet{Terebey1993} and \citet{Motte2001}:

\begin{equation}
T_d(r) = 38\,K\,\left(\frac{r}{100\,\textrm{au}}\right)^{-0.4} \left(\frac{L_\textrm{int}}{1 L_\odot}\right)^{0.2}\,\,,
\end{equation}
where $L_\textrm{int}$ is internal luminosity of the protostar, which is directly linked to the protostellar accretion luminosity. While for some of these sources---Serpens Emb 8, Serpens Emb 8(N), B335, L1448 IRS2, and NGC1333 IRAS4A---the internal luminosities are known from the \textit{Herschel} Gould Belt survey \citep[]{Andre2010} and were used in \citet{Maury2019,Maret2020,Belloche2020}, for others we derived the internal luminosities using archival fluxes at 70 $\mu$m from \textit{Herschel} PACS using the relation from \citet{Dunham2008} (see Table \ref{t.obs}).
We find that the ALMA observations are sensitive to material in the inner envelope with typical column densities $\sim 10^{22} - 10^{25}$ cm$^{-2}$. The individual ranges of column densities probed in each map are reported in Table \ref{t:sources}.

Finally, while most of the polarized dust emission toward the sample of sources we present is caused by thermal emission of dust grains aligned with respect to the magnetic field, this may not be the case where the dust emission becomes optically thick and where the radiation from dust is highly anisotropic (such as protoplanetary disks); in these regions the polarized dust emission can be caused by the self-scattering of thermal dust emission \citep{Kataoka2015}. Within the sample of sources we present, two of them have been clearly identified as having polarized dust emission due to self-scattering in their inner region; these are the two Ophiuchus sources, IRAS 16293A/B and VLA 1623A/B \citep{Sadavoy2018a,Sadavoy2018b,Sadavoy2019}.
We estimate that only 2\% of the pixels could be contaminated in IRAS16293A/B, whereas up to 40\% of the pixels could contain polarized emission mostly due to self-scattering (based on the pattern of polarization position angles) in VLA1623A/B. Thus, we exclude these pixels from our analysis. Moreover, in our sample the dust emission is also optically thick in the inner 100 au region of IRAS4A \citep{Ko2019}: this represents $<1\%$ of the pixels in both our observations at 1.3 and 0.87 mm. We also exclude these pixels from our analysis. Note that although \citet{Kwon2019} and \citet{Takahashi2019} disfavoured self-scattering as the cause of the linear polarization at the very center of the L1448 IRS2 and OMC3 MMS6 cores, respectively, we cannot rule it out. However, if scattering were present in these sources, it would only affect the few pixels at the peak of the dust continuum emission.

\section{Analysis of polarized dust emission in Class 0 protostellar cores}
\label{sec:analysis}

\subsection{Applying Planck statistical tools to interferometric observations}
\label{sec:method}

We aim to apply the statistical tools developed for the analysis of the \plk maps of the polarized ISM to interferometric ALMA observations.  We compare the statistical properties of dust polarization in the dense regions of protostellar cores with the properties found in larger-scale star-forming clouds. However, using interferometric data requires us to adapt the \plk collaboration's methods for investigating large-scale maps.  For example, unlike ALMA observations, \plk observations are not affected by spatial filtering. We treat the ALMA polarization products as follows. We regrid the maps of the three Stokes parameters $I$, $Q$, and $U$ to a Nyquist sampling, with exactly 4 pixels per beam in terms of area. We then calculate the polarization angle dispersion function $\S$ in each pixel $i$ of the Stokes maps, with respect to each of its $n=8$ nearest neighbouring pixels $j$ as follows:
\begin{equation}
S(\delta)_i\,=\,\sqrt{\frac{1}{n}\sum^n_{j=1}{\left[ \frac{1}{2}\textrm{arctan} \frac{Q(j)U(i)-U(j)Q(i)}{Q(i)Q(j)+U(i)U(j)} \right]^2 } }\,\,.
\end{equation}
Considering the sampling described above, the equivalent $\delta$ parameter (see also Equation \ref{equ:S_def}) is approximately $1/2$ of a beam width (comparable to the value chosen in \citealt{Planck2018XII}). Note that the measured value of $\S$ scales with the pixel gridding. Indeed, at a given angular resolution, changing the gridding pattern (\ie how many pixels a beam contains) to a fewer number of pixels per beam leads to a measurement of $\S$ that covers a larger area, and thus the lag $\delta$ is larger. This in turn causes us to quantify the disorganization of the magnetic field across a larger physical area. As the angular resolution of the observations is fixed, increasing the lag will cause $\S$ to increase, as we lose spatial coherence in the apparent magnetic field, which in turn causes an increase in the calculated level of disorganization in the apparent magnetic field. We perform the same analysis with different gridding and choose the value of 4 pixels per beam area in order to strike a balance between statistical accuracy (\ie using a large number of points) and independence of the individual points.
Finally, as explained in Section \ref{sec:persp}, the way the dispersion $\S$ is derived causes the distribution to saturate, \ie a completely random distribution of polarization angles will produce a maximum value of $\S$ of $\pi/\sqrt{n}\,\sim\,63^\circ$ (as we chose $n=8$).

While \citet{Planck2018XII} produced covariance maps and were able to assess finely the noise properties at different spatial scales, interferometric maps are severely affected by imaging systematics such as the limited dynamic range of the images. Furthermore, Stokes $I$ images tend to be much more dynamic range limited than Stokes $Q$ and $U$ images. In addition, the sources lie at different distances, which leads us to probe different angular extents. Finally, the data we analyze are heterogeneous in their $uv$-coverage and sensitivity. Therefore, the noise is neither spatially homogeneous nor correlated in the ALMA maps. Consequently, we compute the rms noise $\sigma$ in each Stokes map, using regions close to the observation pointing center but devoid of emission. This is how we define the signal-to-noise ratio (S/N) of polarized intensity, \ie $P/\sigma_P$. Note when creating these polarized emission maps, one must to correct for the bias that occurs at low S/N levels: to do so and to construct fully sampled $P$ maps, we follow the method from \citet{WardleKronberg1974} (see also \citealt{Hull2015b} for an application of this method to interferometric data).

We follow the method introduced in \citet{Planck2018XII} to compute a pixel-selection criterion in order to test appropriately the correlation between $\S$ and $\Pf$ in our objects. This pixel-selection is a cutoff based on Stokes $I$, which allows us to remove the noise-biased data. We obtain this cutoff by analyzing the average S/N of the polarized intensity $P$, which typically increases with increasing Stokes $I$. When this average S/N of $P$, plotted as a function of the total intensity Stokes $I$ for each dataset, meets the value S/N\,=\,5, we use the corresponding value of Stokes $I$ as the pixel-selection cutoff for the given dataset. We show an example in Figure \ref{fig:SNR_B335} (\textit{top panel}), where we plot the S/N of the polarized intensity map as a function of the total intensity for the 1.3\,mm observations of the B335 core. The vertical dotted line, which denotes this cutoff value of Stokes $I$, thus corresponds to the value of Stokes $I$ where $\langle P/\sigma_P \rangle _I\,\geq\,5$. We then take all points lying above this cutoff in Stokes $I$ and form the sample to which we will apply our statistical method. Note that if this method provides a threshold limit of Stokes $I$ below $5\,\sigma_I$, we chose $5\,\sigma_I$ as the cutoff for the dataset.

As shown in the lower panel of Figure \ref{fig:SNR_B335}, the values of $\StimesP$ diverge at Stokes $I$ values lower than the aforementioned threshold, as a consequence of the noise-bias of $\S$ and $\Pf$ at low values of Stokes $I$. It is important to note that we have not performed any selection based on the $P$ values; we select only on the $I$ values in order to keep the pixels exhibiting low polarized intensity that contain the information of depolarization, which is essential for our statistical analysis. As an example, in Figure \ref{fig:B335_ex_all_maps} we show the maps of Stokes $I$, polarized intensity $P$, polarization fraction $\Pf$, and dispersion of polarization angles $\S$ from the 1.3 mm observations of the B335 core. The contours indicate both the threshold in Stokes $I$ found with the method introduced above, as well as the $5\sigma_I$ level. Similar maps of all sources can be seen in Figure \ref{fig:S_I_pol_maps_sources}.


\def\scaleSP{0.40}
\begin{figure}[!tbph]
\centering
\hspace*{-0.1em}
\subfigure{\includegraphics[scale=\scaleSP,clip,trim= 0.57cm 3.1cm 2.4cm 3cm]{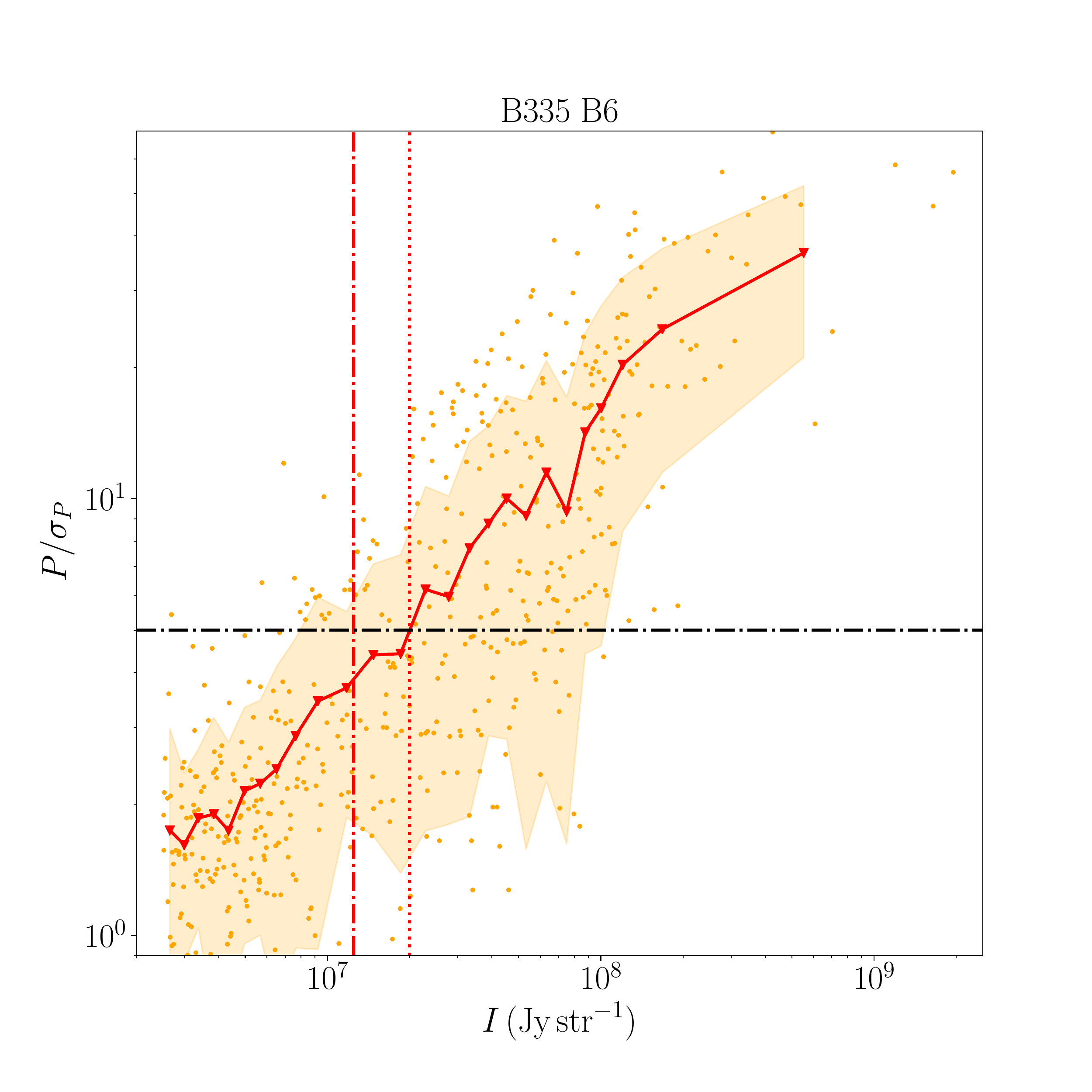}
}\vspace*{-0.9em}
\subfigure{
\includegraphics[scale=\scaleSP,clip,trim= 0.8cm 1.2cm 2.5cm 3.0cm]{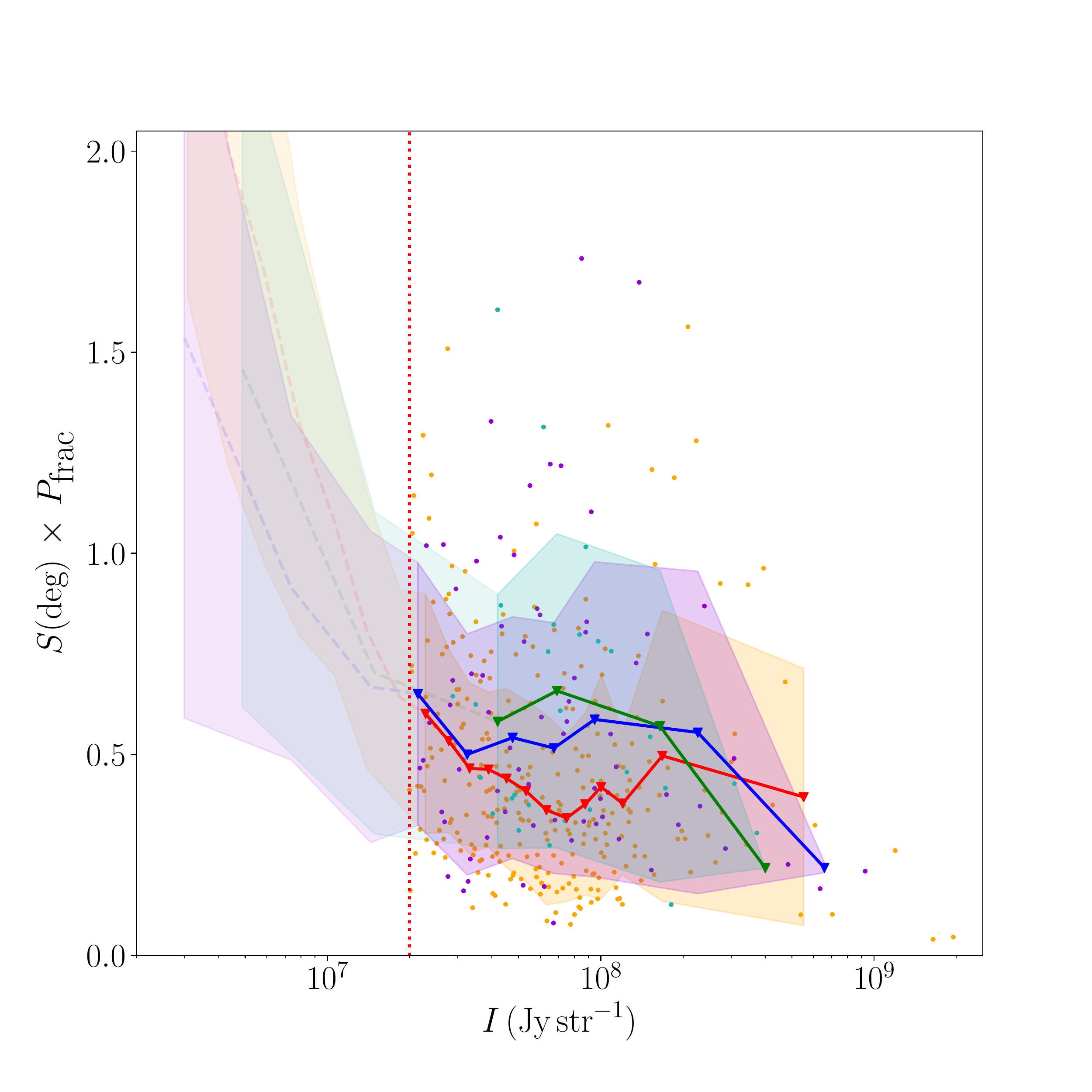}}
\vspace{-0.4cm}
\caption[]{\footnotesize An illustration of how we compute the pixel-selection cutoff in Stokes $I$ in the 1.3\,mm observations of the B335 core. \textit{Top}: Evolution of the S/N of the polarized intensity, \ie $P/\sigma_P$, as a function of the total intensity Stokes $I$ in Jy\,str$^{-1}$. The dot-dashed horizontal black line is at the value of $P/\sigma_P = 5$. The dotted vertical line is the selected cutoff in Stokes $I$ described in Section \ref{sec:method}. The dot-dashed vertical line is the 5$\sigma_I$ value. The solid line is the running mean, which is calculated along the Stokes $I$; the shaded area represents $\pm$ the standard deviation of the Gaussian fit performed on each bin. \textit{Bottom}:  $\StimesP$ for the selected pixels as a function of the total intensity. To the left of the cutoff in Stokes $I$ (the red dotted line, plotted as in the top panel), the points are no longer plotted and the running mean turns in a translucent dashed line. Each color corresponds to an angular resolution: red is the original resolution, whereas blue and green are 4\,$\times$ and 9\,$\times$ lower resolution (in terms of beam area), respectively. Note that, as expected, one see that decreasing the resolution, and thus increasing the spatial length of the lag, causes on the dispersion $\S$ to increase as well, on average (see Section \ref{sec:persp}).}
\label{fig:SNR_B335}
\end{figure}

\def\scaleSP{0.54}
\begin{figure*}[!tbph]
\centering
\vspace{-0.1cm}
\subfigure{\includegraphics[scale=\scaleSP,clip,trim= 0cm 2.8cm 3.8cm 2.6cm]{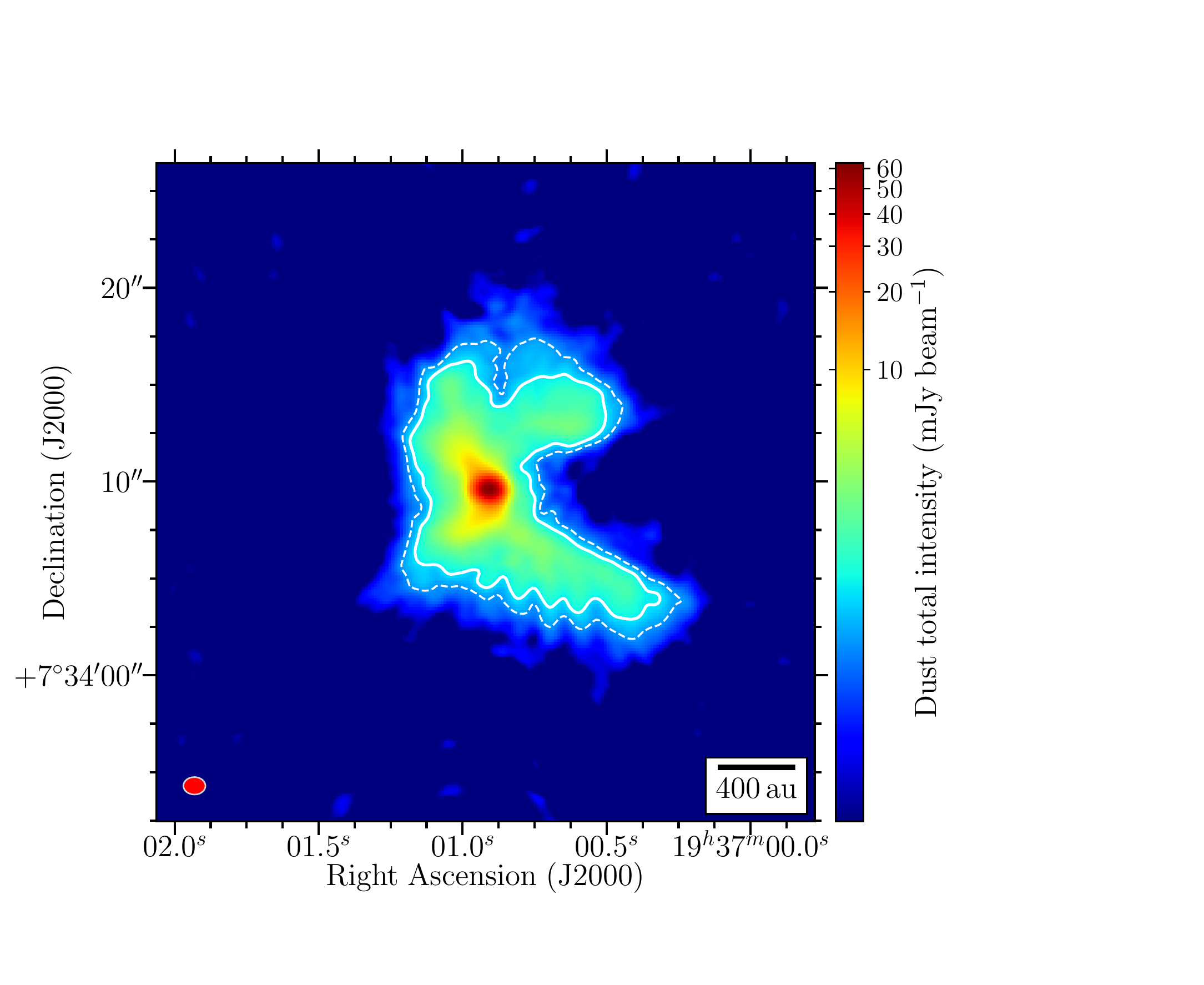}}
\subfigure{\includegraphics[scale=\scaleSP,clip,trim= 2.7cm 2.8cm 3.8cm 2.6cm]{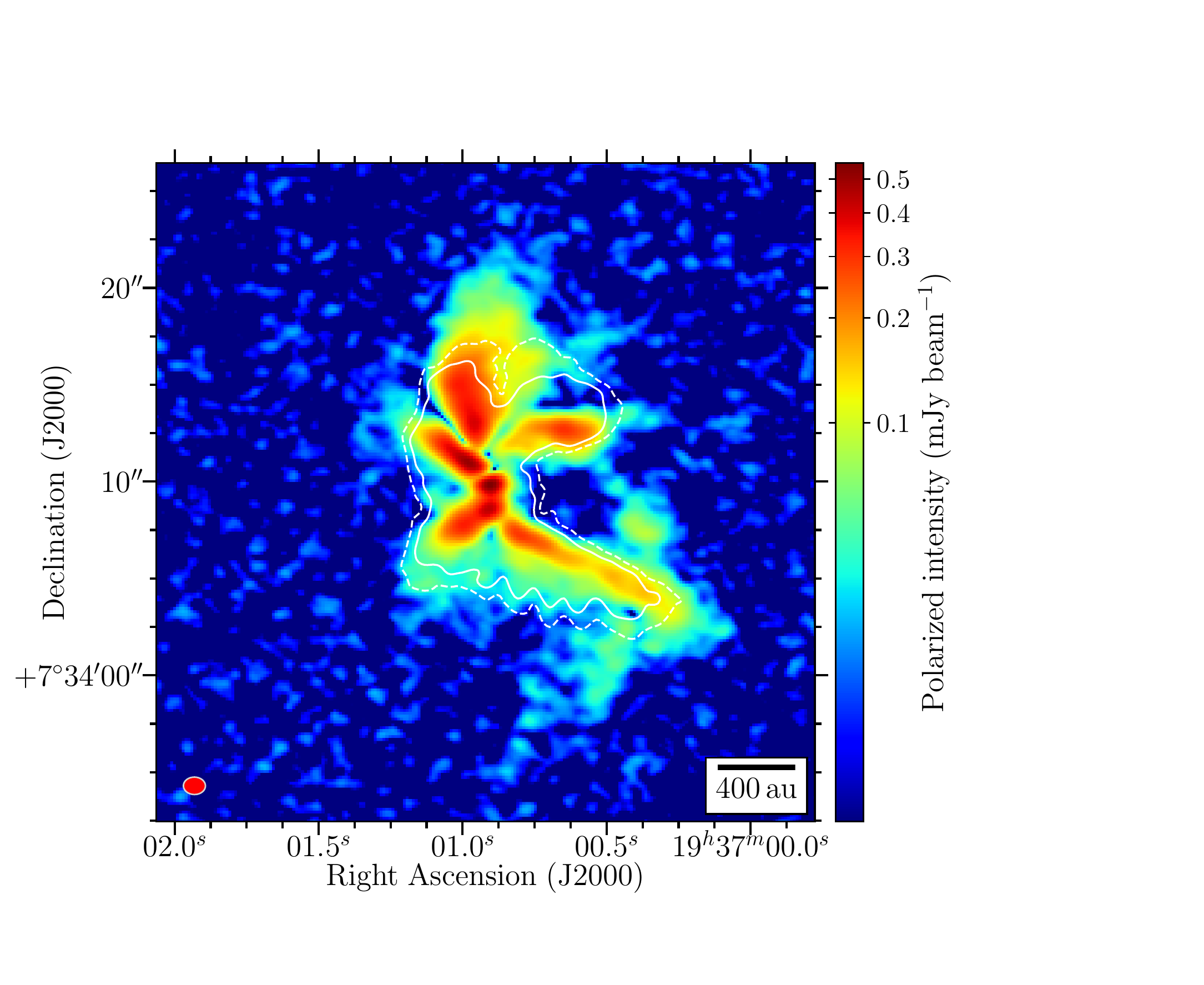}}
\subfigure{\includegraphics[scale=\scaleSP,clip,trim= 0cm 1cm 3.8cm 2.6cm]{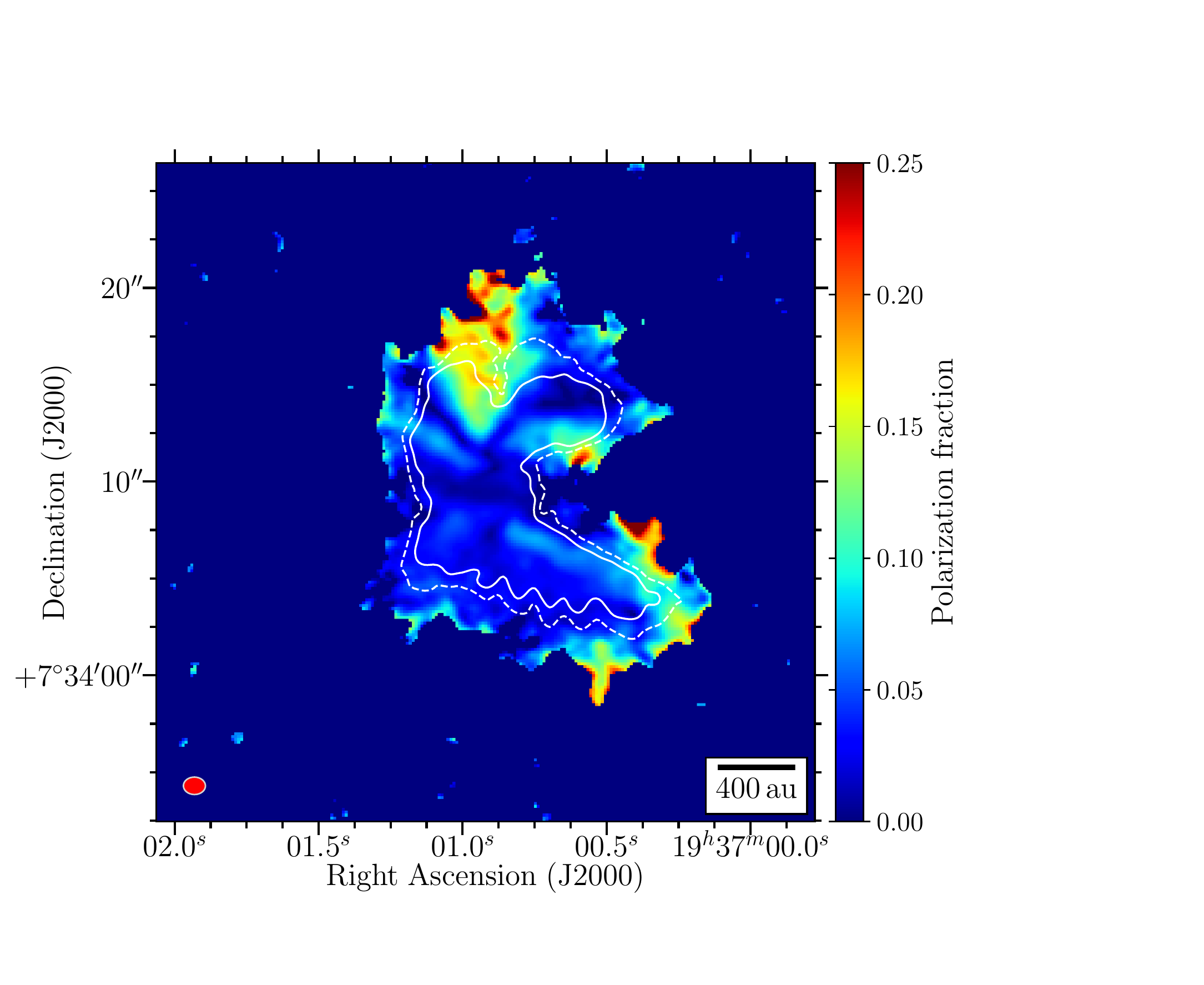}}
\subfigure{\includegraphics[scale=\scaleSP,clip,trim= 2.76cm 1cm 3.8cm 2.6cm]{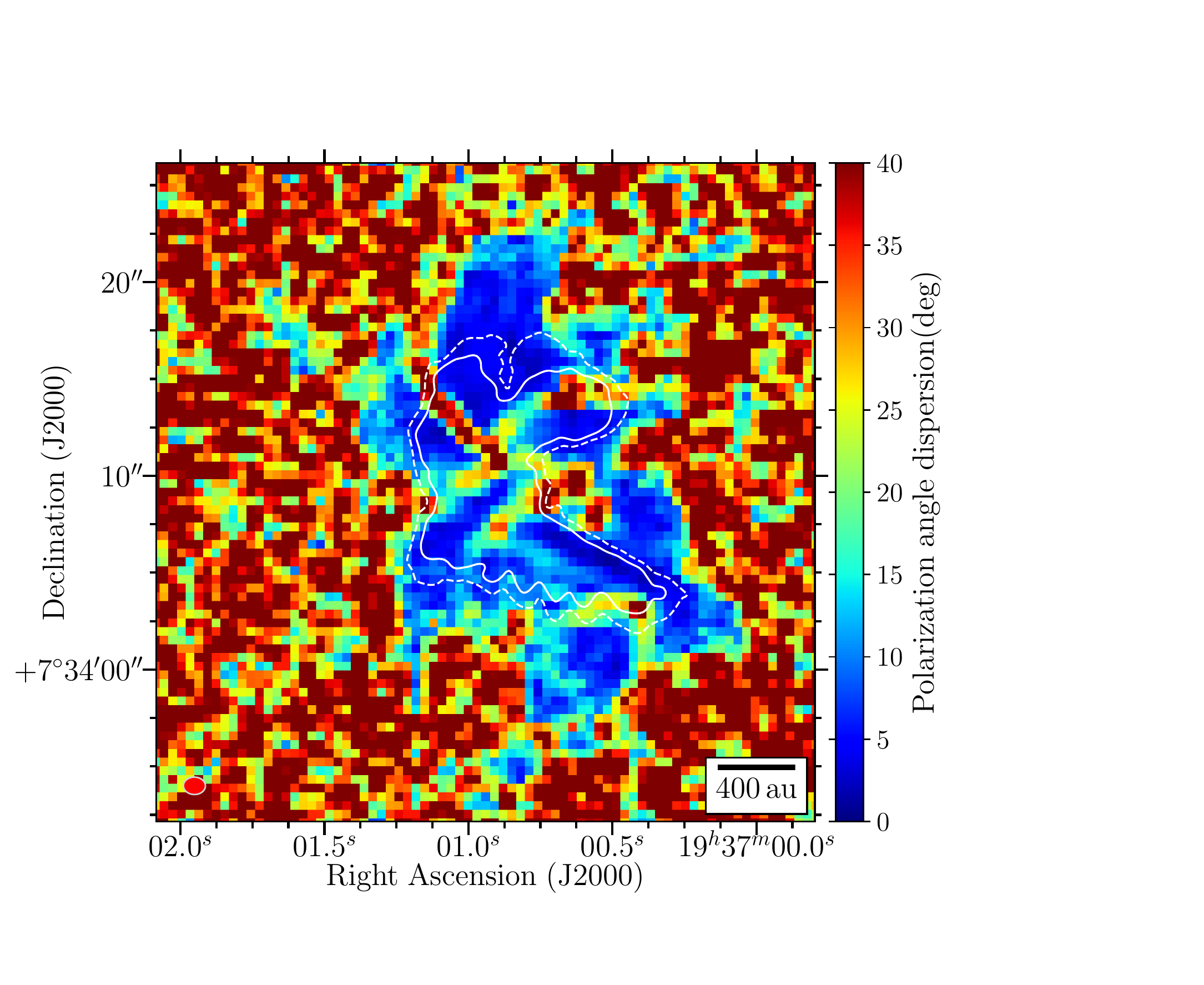}}
\vspace{-0.1cm}
\caption[]{\footnotesize Maps from the 1.3\,mm observations of the B335 core. \textit{Top-left.} Total intensity (Stokes $I$) thermal dust emission in color scale. The emission is shown from 3$\sigma_I$ where $\sigma_I$ is the rms noise in the Stokes $I$ map. \textit{Top-Right.} Polarized intensity $P$ in colorscale. \textit{Bottom-Left.} Polarization fraction $\Pf$ in colorscale, shown where $I\,>\,3\sigma_I$. \textit{Bottom-Right.} Dispersion $\S$ of polarization position angles in color scale; the pixel size corresponds to the pixel size considered in the statistics. The dashed white contour represents the 5$\sigma_I$ level.  The solid white contour represents the threshold level of Stokes $I$ calculated as described in Section \ref{sec:method}, above which the mean S/N of $P$\,>\,5. The beam size is 1$\farcs$14 $\times$ 0$\farcs$90, with a position angle of 89.1$^\circ$.}
\label{fig:B335_ex_all_maps}
\end{figure*}

\subsection{Results from the statistical analysis of the polarized dust emission in protostellar envelopes}
\label{sec:stats_results}

We present here the outcome of the statistical analysis of the polarized dust emission from the sample of ALMA observations presented in Table \ref{t:sources}. In Figure \ref{fig:scatter_S_Pfrac_Alma} we show he distribution of the dispersion of the polarization angles $\S$ as a function of the polarization fraction $\Pf$ in the 15 maps probing the dust emission in protostellar cores.
In these plots, the running mean of $\Pf$ (shown as black points and line) shows the average trend and evolution of $\S$ with $\Pf$. In particular, one can clearly see the area of the distribution affected by the saturation effect of $\S$ described above, and that the distribution is linear in the logarithmic two-dimensional (2D) space outside of this saturated area. In each distribution, the points are coloured based on their Stokes $I$ value. The relationship between $\S$ and $\Pf$ observed by \plk at cloud scales is reported in each diagram as a red line, for reference.
We find a global trend similar to the \plk findings, with high values of polarization fraction $\Pf$ associated with low dispersion in polarization angles $\S$ in regions of faint Stokes $I$ values. Conversely, we see high $\S$ and low $\Pf$ in regions with bright Stokes $I$. We list in Table \ref{t:alphas_values} the values of the power law indexes $\alpha$ derived from the fit to the $\mathcal{S} \propto \Pf^{-\alpha}$ relation in each individual core. These values range from $\alpha =0.523 \pm 0.094$ to $\alpha = 0.866 \pm 0.040$.

\begin{table}[!tbph]
\centering
\small
\caption[]{Power law index $\alpha$ of the correlation between $\S$ and $\Pf$ as $\mathcal{S} \propto \Pf^{-\alpha}$}
\setlength{\tabcolsep}{0.13em} 
\begin{tabular}{p{0.34\linewidth}ccccc}
\hline \hline \noalign{\smallskip}
Source name & $\lambda$  & No. of pts.& cutoff &$\alpha$ \\
& (mm) & &$\frac{\textrm{mJy}}{\textrm{beam}}$&\\
\noalign{\smallskip}  \hline
\noalign{\smallskip}
Serpens Emb 6  & 0.87& 314& 13.3 & 0.523 $\pm$ 0.094\\
\noalign{\smallskip}
\hline
\noalign{\smallskip}
Serpens Emb 8  &0.87& 206&2.85&0.616 $\pm$ 0.114 \\
\noalign{\smallskip}
Serpens Emb 8(N) &0.87& 42&5.63& 0.648 $\pm$ 0.277\\
\noalign{\smallskip}
\hline
\noalign{\smallskip}
BHR71 IRS1  &1.3\phantom{0}& 500&1.50& 0.594 $\pm$ 0.114 \\
\noalign{\smallskip}
BHR71 IRS2  &1.3\phantom{0}& 197&2.58& 0.599 $\pm$ 0.129 \\
\noalign{\smallskip}
\hline
\noalign{\smallskip}
\multirow{3}{*}{B335} &1.3\phantom{0}& 835&4.80&0.678 $\pm$ 0.070 \\
&1.5\phantom{0}& 114&1.11& 0.601 $\pm$ 0.223 \\
& 3.0\phantom{0}& 288&0.20& 0.792 $\pm$  0.197 \\
\noalign{\smallskip}
\hline
\noalign{\smallskip}
IRAS 16293A/B &1.3\phantom{0}& 1241&3.00& 0.517 $\pm$  0.073 \\
\noalign{\smallskip}
\hline
\noalign{\smallskip}
VLA 1623A/B  &1.3\phantom{0}& 61 &7.15& 0.639 $\pm$  0.447 \\
\noalign{\smallskip}
\hline
\noalign{\smallskip}
L1448 IRS2 & 1.3\phantom{0}& 358 &0.48& 0.675 $\pm$ 0.101 \\
\noalign{\smallskip}
\hline
\noalign{\smallskip}
\multirow{2}{*}{OMC3 MMS6} &1.3\phantom{0}&1323&1.39& 0.756 $\pm$ 0.062 \\
&1.3\phantom{0}&873&1.54& 0.685 $\pm$  0.073 \\
\noalign{\smallskip}
\hline
\noalign{\smallskip}
\multirow{2}{*}{IRAS4A} & 1.3\phantom{0}&890&2.00& 0.828 $\pm$  0.065 \\
&0.87& 2756&2.50& 0.864 $\pm$  0.041 \\
\noalign{\smallskip}
\hline
\smallskip
\end{tabular}
\caption*{\footnotesize Results from the correlations presented in Figure \ref{fig:scatter_S_Pfrac_Alma}. We list the values of the power law indexes $\alpha$ and associated uncertainties obtained from the linear fits, as well as the wavelength of observations $\lambda$, the number of points selected in each case, and the cutoff in Stokes $I$ applied}.
\vspace{-0.4 cm}
\label{t:alphas_values}
\end{table}

\def\scaleSP{0.34}
\def\extansion{}

\begin{figure}[!tbph]
\centering
\vspace{-0.1cm}
\subfigure{\includegraphics[scale=\scaleSP,clip,trim= 1.6cm 1.4cm 3cm 1.8cm]{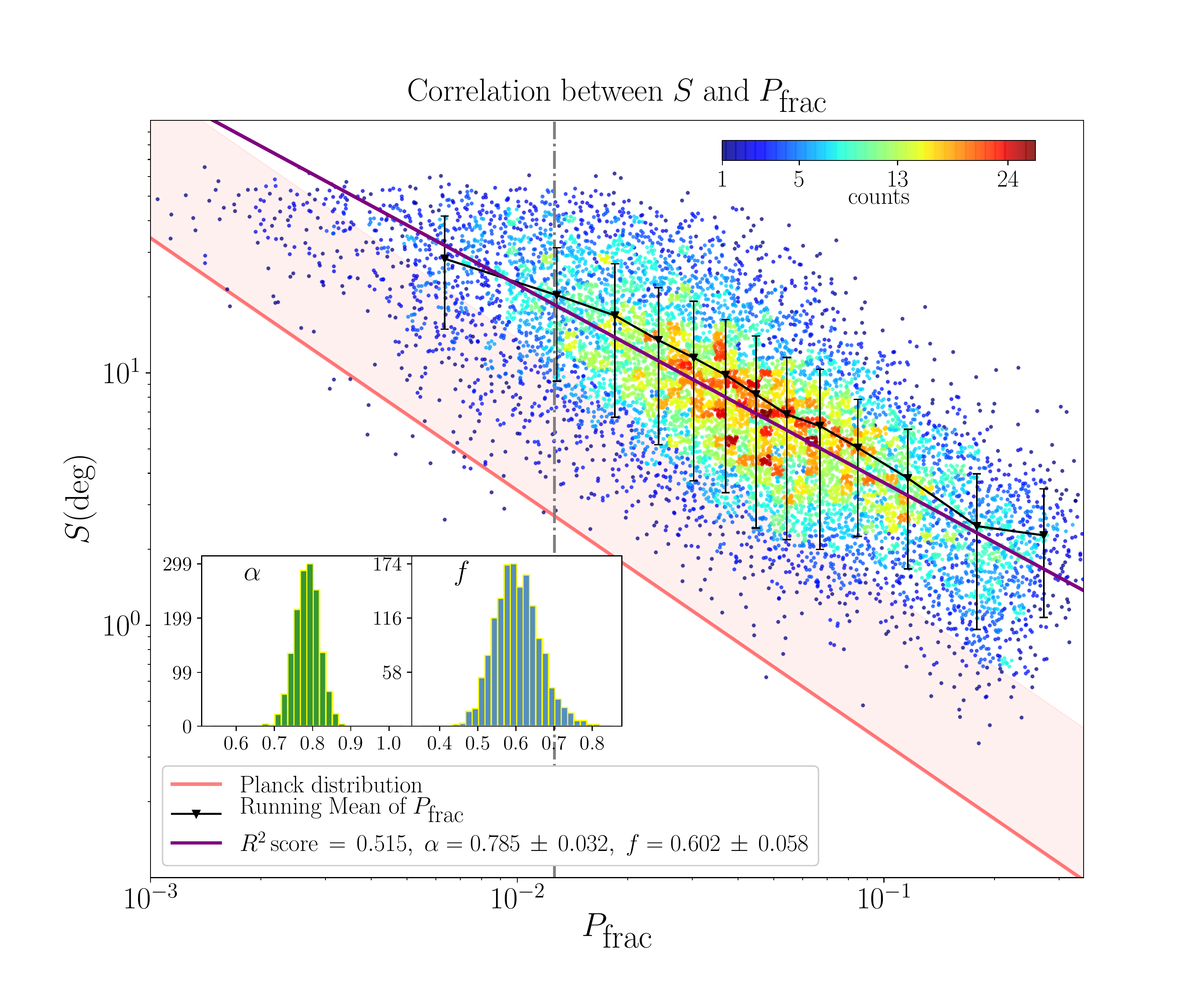}
}
\caption[]{\footnotesize Distributions of the dispersion of polarization position angles $\S$ as a function of the polarization fraction $\Pf$ from all of the datasets merged together. The points were selected according the method developed in Section \ref{sec:method}. The color scale represents the number density of points in the plot. The solid black line and black points represent the running mean of $\Pf$; the associated black error bars are $\pm$ the standard deviation of each bin. We plot the linear fit in purple, which is a linear regression. We take into account the saturation of $\S$ in the derivation of the linear fit by applying a threshold in polarization fraction, indicated by the vertical dot-dashed grey line. This threshold denotes the $\Pf$ level beyond which the distribution is linear. The solid red line corresponds to the \plk correlation from \citet{Planck2018XII}, which we scaled down to the highest angular resolution of our ALMA observations. The red shaded area extends up to the same \plk correlation, scaled down this time to the largest field of view of our ALMA observations. As we gather all of the ALMA observations at their various angular resolutions, this red shaded area encompasses all of the corresponding scalings of the \plk correlation. The two parameters $f$ and $\alpha$ are derived from the linear fit, where the analytical correlation is as follows: $S\,=\,f/{P^{\,\alpha}_\textrm{frac}}$. The histograms in the two little subplots show the distributions of the values of $\alpha$ and $f$ values derived from a large number of randomly chosen sub-samples of points. We calculate the uncertainties in $f$ and $\alpha$ as standard deviations of Gaussian fits to those histograms.}
\vspace{-0.2cm}
\label{fig:scatter_S_Pfrac_Alma_merged}
\end{figure}

\begin{figure*}[!tbph]
\centering
\hspace{-0.5cm}
\begin{tabular}{cc}
\includegraphics[scale=0.23,clip,trim= 1.3cm 1.2cm 2.5cm 1.9cm]{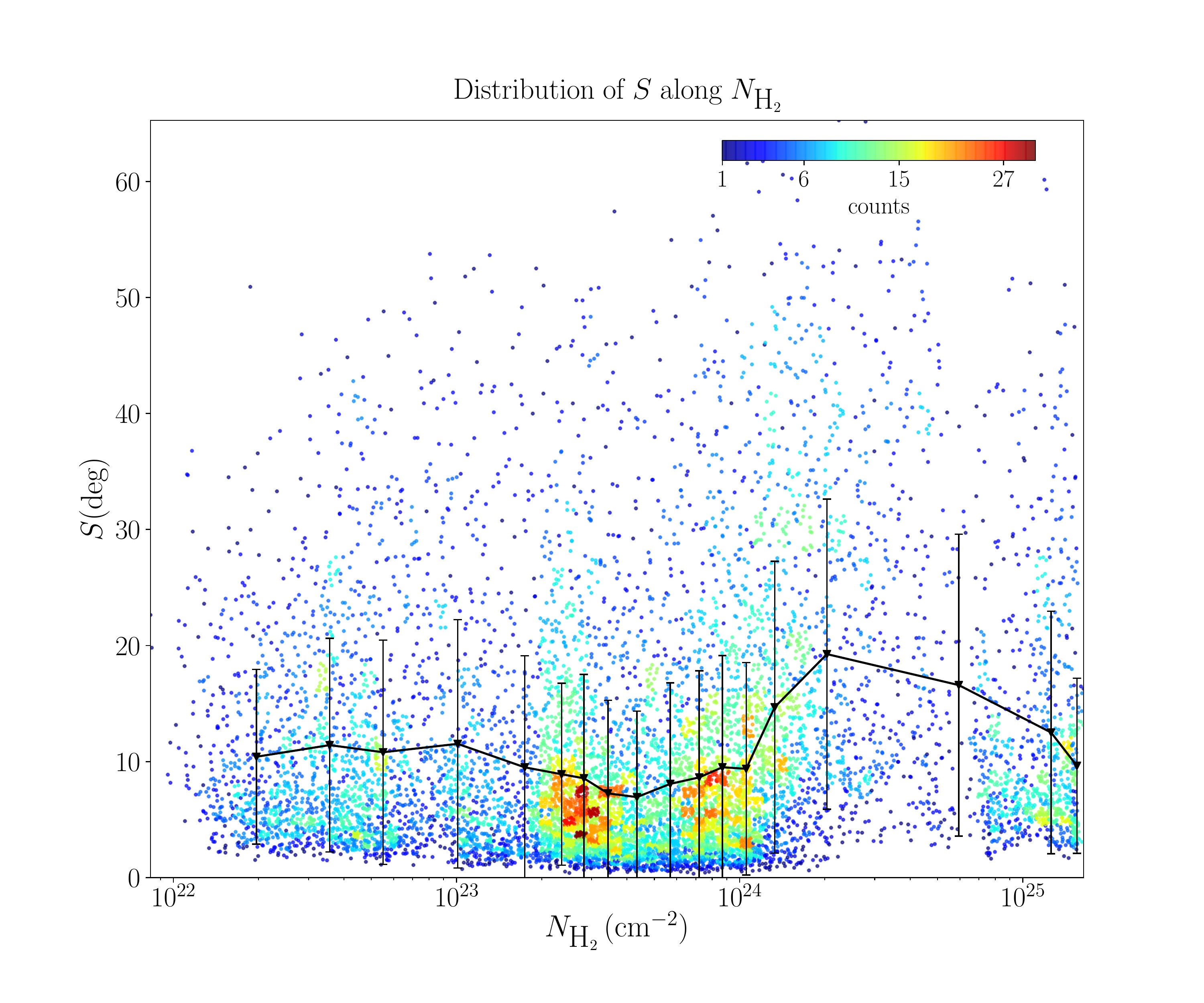}
&\hspace{-0.2cm}\multirow{2}{*}[4.6cm]{\hspace{-0.2cm}\includegraphics[scale=0.464,clip,trim= 1.5cm 1.2cm 2.5cm 1.9cm]{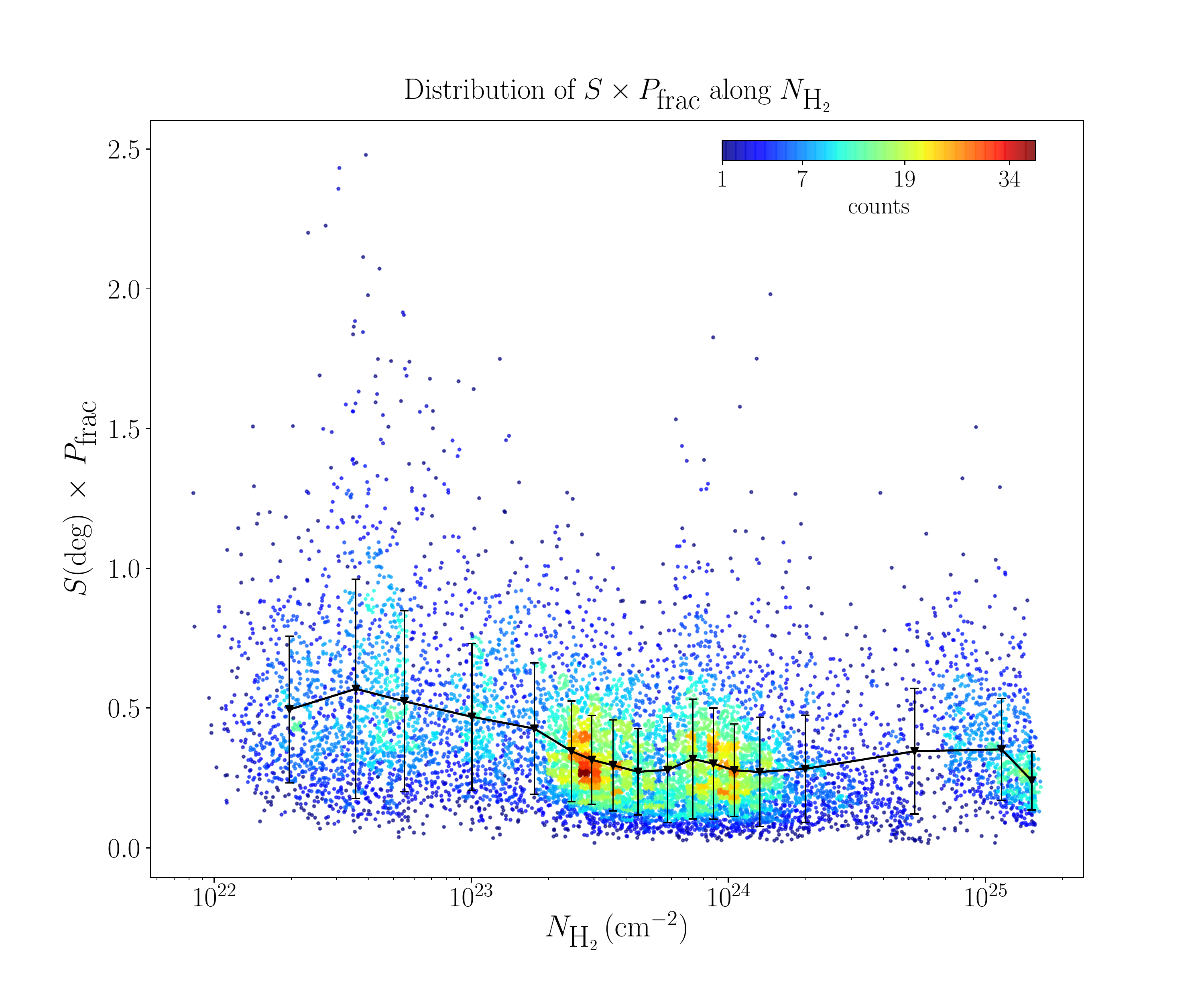}}\\ 
\includegraphics[scale=0.23,clip,trim= 1.3cm 1.2cm 2.5cm 1.9cm]{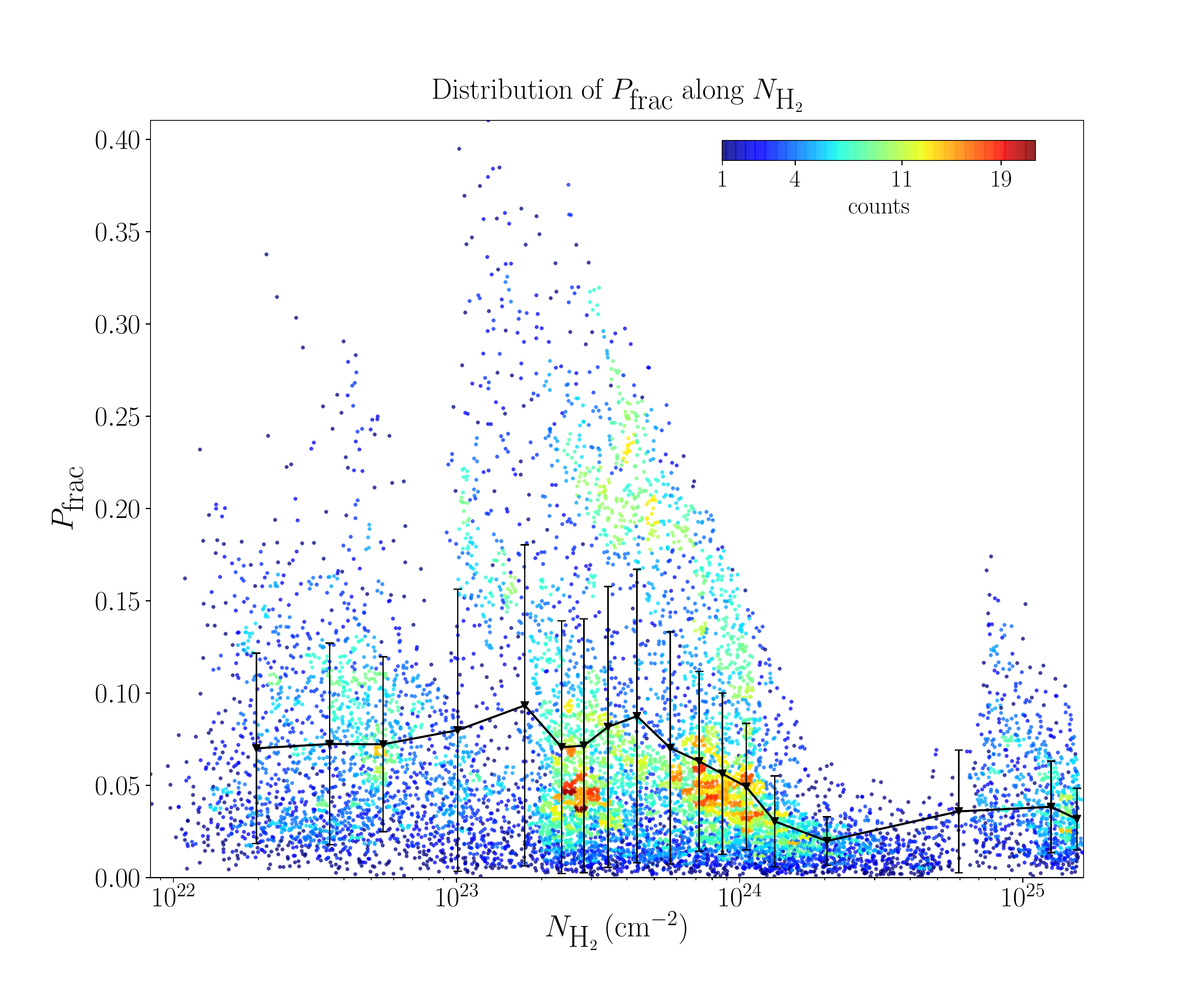}&
\end{tabular}
\caption[]{\footnotesize Distributions of the dispersion of polarization position angles $\S$ (\textit{top-left}), the polarization fraction $\Pf$ (\textit{bottom-left}), and $\StimesP$ (\textit{right}), as a function of the column density $N_{\textrm{H}_2}$, where the data from all the cores are merged. The color scale represents number density of points in the plots. The solid black line and black points represent the running mean of $\S$, $\Pf$, and $\StimesP$; the associated black error bars are $\pm$ the standard deviation of each bin.}
\vspace{0.2cm}
\label{fig:scatter_coldens}
\end{figure*}

In order to take full advantage of the statistical power of our methodology and to discuss global properties of the polarized dust emission in Class 0 protostars, we have merged all data from each of the 15 ALMA observations. Figure \ref{fig:scatter_S_Pfrac_Alma_merged} shows the merged distribution of $\S$ as a function of $\Pf$, along with the linear fit previously described, which is defined by the two parameters $\alpha$ and $f$ such that $S\,=\,f/{\Pf^{\alpha}}$. Note that at low values of $\Pf$, the distribution of $\S$ flattens because of the saturation of $\S$ for high dispersion values. This is an artifact arising from the definition of $\S$, and thus these points should be excluded from the linear fit. To do so, we establish a threshold in $\Pf$ of $1.3\%$, indicated by the vertical dot-dashed grey line in Figure \ref{fig:scatter_S_Pfrac_Alma_merged}, which denotes the $\Pf$ level beyond which the distribution is linear. We calculate this threshold by determining where the $\alpha$ value from the linear fit would no longer have changed if we had moved the threshold up in polarization fraction. We obtain a power law index $\alpha\,=\,\slope\,\pm\,\dslope$, which is flatter than the results and the analytical correlation found with \plk observations at larger scales, where $S\,\propto\,\Pf^{-1}$.

Merging all the ALMA data does not enable us to investigate the relation between $\S$ and $\Pf$ with respect to Stokes $I$ because of the heterogeneous properties of the sources and observations (\eg the wavelength of the observations). Thus, we use the column density (calculated as described in Section \ref{sec:source_sample}) of the individual lines of sight in order to collect all the data points in a single plot.
Figure \ref{fig:scatter_coldens} presents the variation of $\S$, $\Pf$, and $\StimesP$ as a function of the local column density $N_{\textrm{H}_2}$ in the envelopes. Figure \ref{fig:scatter_coldens_norma} shows the same distribution of points, but where all column density values are normalized to the maximum column density in the map including all optically thin lines of sight (\eg excluding highly extinct lines of sight where polarized dust emission could be severely contaminated by self-scattering)\footnote{Note that in VLA1623 for example, the highest column density is probed in one of these line-of-sights and reaches values of $10^{25}$cm$^{-2}$.}. We notice a global trend in the merged data shown in Figure \ref{fig:scatter_coldens_norma} of increasing $\S$ and decreasing $\Pf$ with increasing column density. Figure \ref{fig:trends_S_Pfrac_Alma} presents the results of linear regressions on the trends of $\S$ and $\Pf$ as a function of $N_{\textrm{H}_2}$ in individual cores: the resulting polar-law indices and R-squared values suggest a significant decrease of $\Pf$ with increasing $N_{\textrm{H}_2}$ across the sample, and hints of increasing $\S$ with column density, although these latter trends are noisier. This suggests that the behavior of the disorganization of the apparent magnetic field evolves in the same way from the outer to the inner core, despite the widely varying ranges of column density in each core (see Table \ref{t:sources}).

\section{Discussion}
\label{sec:disc}

\subsection{Comparing the statistical properties of the polarized dust emission in protostellar cores with those in star-forming clouds}

The statistical analysis of the polarized dust emission from the sample of 15 datasets analyzed (see Table \ref{t:sources}) reveals a significant correlation between the dispersion of polarization angles $\S$ observed in the plane of the sky and the polarization fraction $\Pf$ measured in each line of sight in these 11 protostellar envelopes. However, the $\mathcal{S} \propto \Pf^{-\slope}$ relationship we find at core scales is shallower than the $\mathcal{S} \propto \Pf^{-1}$ relationship found at larger scales in the \plk observations of star-forming clouds \citep{Planck2018XII}. Moreover, we obtain on average higher values of $\S$ and $\Pf$ than those found in the lower density molecular clouds probed by \plk. Here we discuss possible origins of the different polarization properties at protostellar scales versus cloud scales. We start by investigating the nature of the disorganized component of the magnetic field (Section \ref{sec:turbu_B_compo}); we then address the different intrinsic spatial scales of total intensity versus polarized emission (Section \ref{sec:scales_of_emission}), and how interferometric filtering may affect observed polarization properties (Section \ref{sec:spatial_filtering}).

\subsubsection{What physics governs the disorganized component of the magnetic field?}
\label{sec:turbu_B_compo}

The correlation between $\S$ and $\Pf$ is governed by the level of disorganization of the apparent magnetic field lines projected on the plane of the sky. The magnetic field is also linked with the kinematics of the gas, assuming the gas is well coupled to the field. The polarization is detected as long as the main orientation of the magnetic field is not along the line of sight (see Section \ref{sec:persp}), which is unlikely to be common considering the relatively high polarization fractions observed in the protostellar envelopes considered here.

\begin{figure*}[!tbph]
\centering
\hspace{-0.5cm}
\begin{tabular}{cc}
\includegraphics[scale=0.23,clip,trim= 1.3cm 1.2cm 2.5cm 1.9cm]{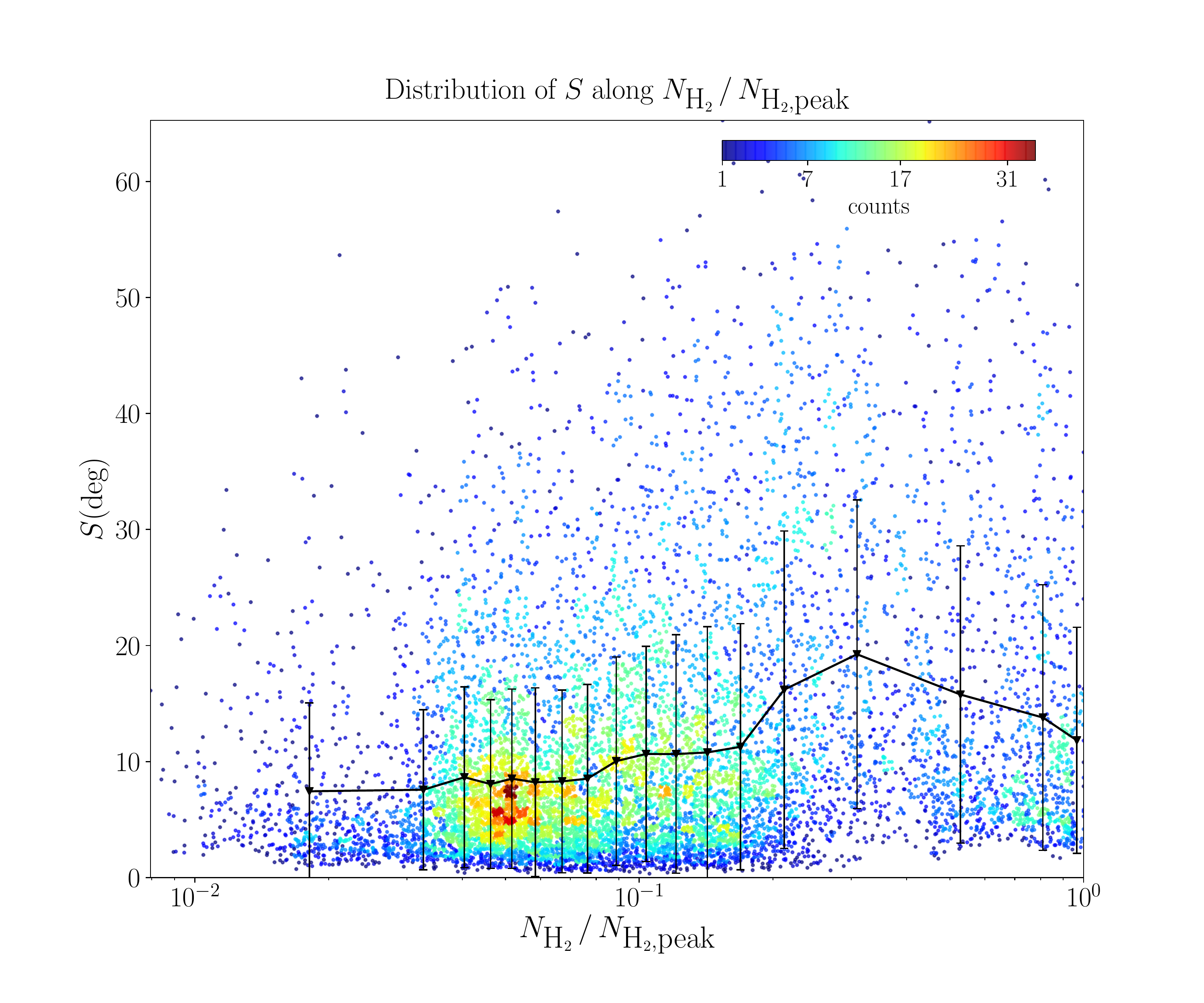}
&\hspace{-0.2cm}\multirow{2}{*}[4.6cm]{\hspace{-0.2cm}\includegraphics[scale=0.464,clip,trim= 1.3cm 1.2cm 2.5cm 1.9cm]{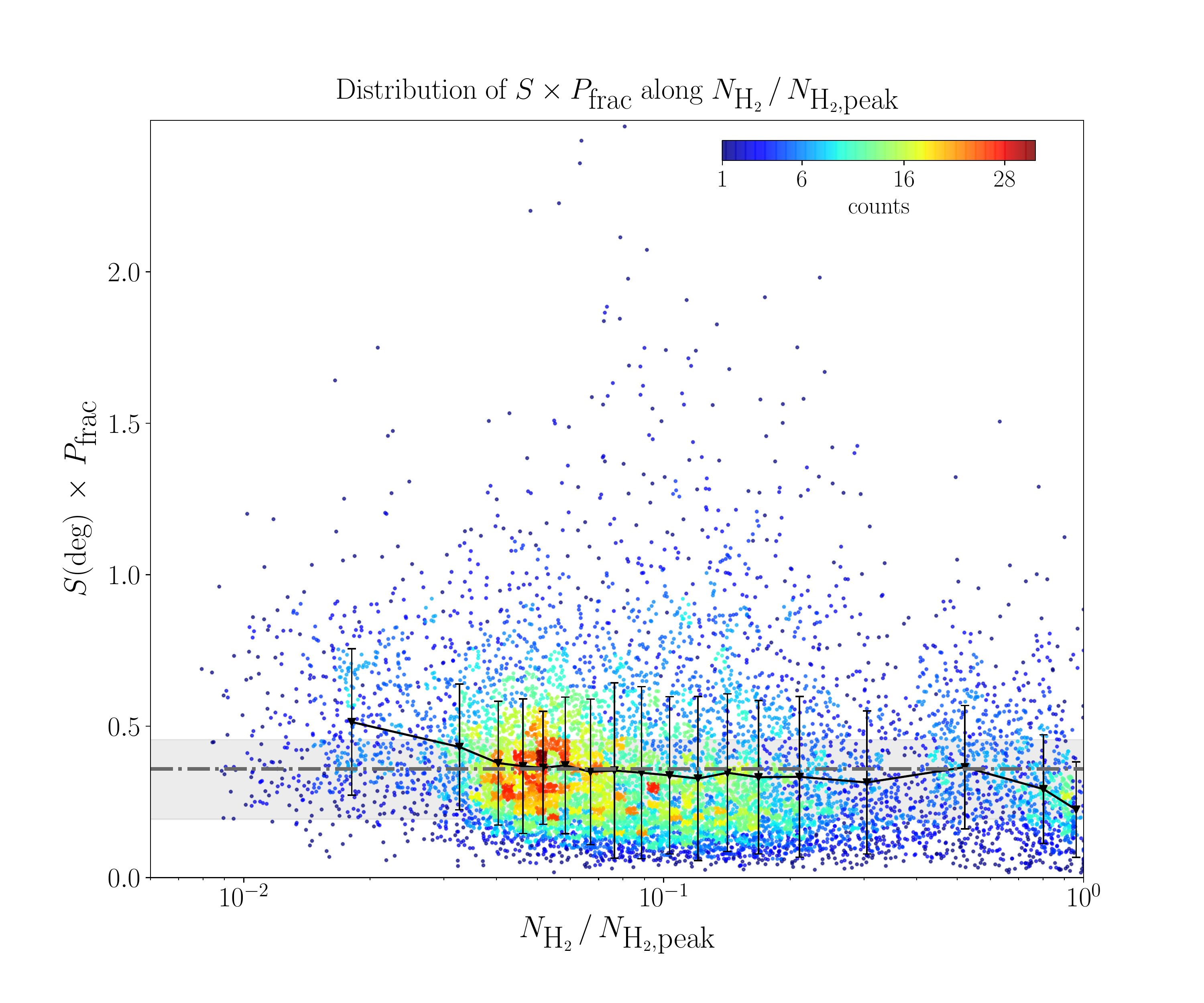}}\\ 
\includegraphics[scale=0.23,clip,trim= 1.3cm 1.2cm 2.5cm 1.9cm]{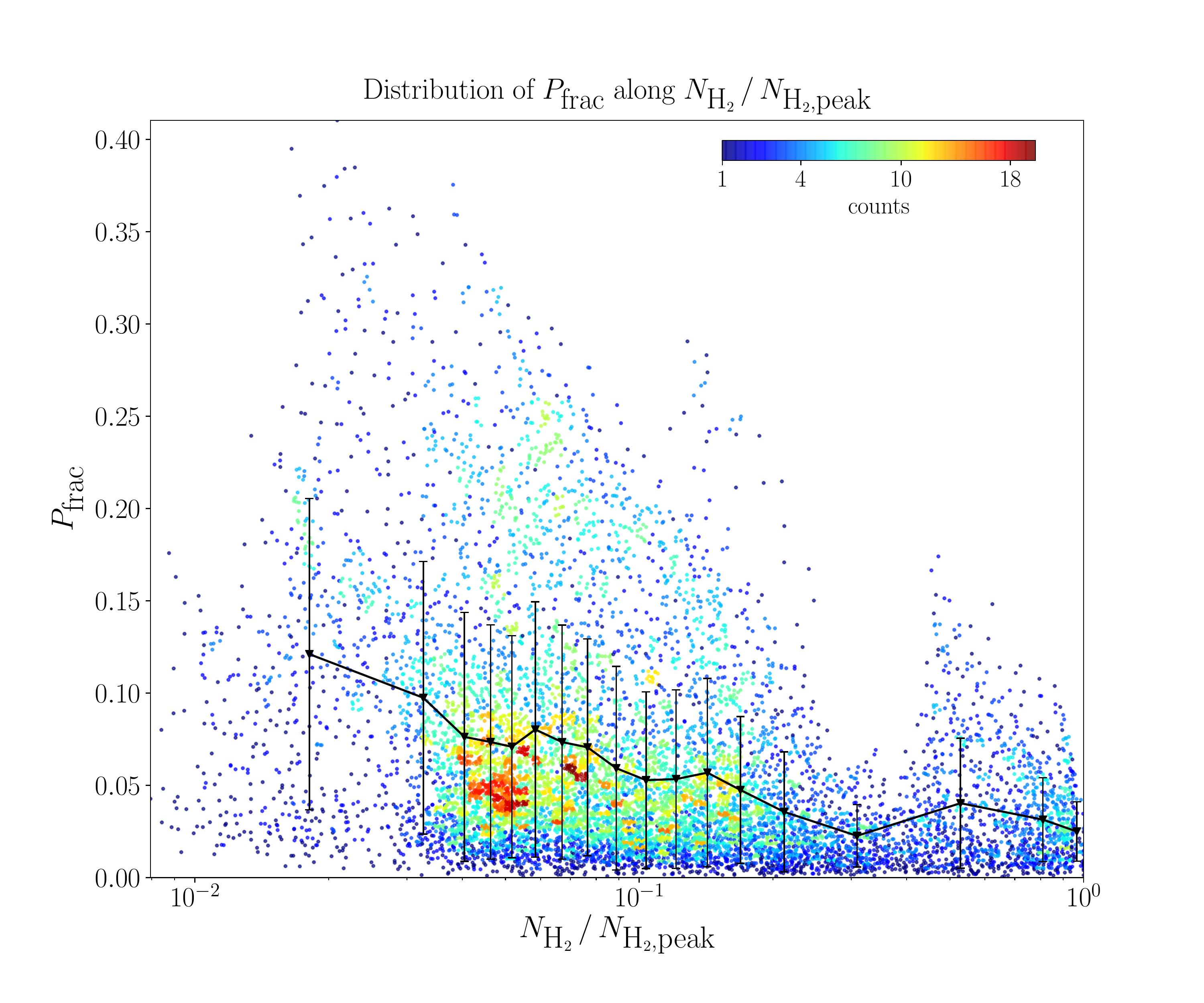}&
\end{tabular}
\caption[]{\footnotesize Same as Figure \ref{fig:scatter_coldens}, but where the column density in each core has been normalized to the column density peak  $N_{\textrm{H}_2\textrm{,peak}}$. The horizontal dot-dashed grey line indicates the mean of all the $\StimesP$ values, \SPracValue. These uncertainty values are represented by the shaded area, which spans the range between the first and third quartiles of the $\StimesP$ distribution.}
\label{fig:scatter_coldens_norma}
\end{figure*}

The differences in the power law index $\alpha$ relating $\S$ and $\Pf$ between the \plk results at cloud scales and the correlation found at core scales with ALMA data (see our correlation and the red line in Figure \ref{fig:scatter_S_Pfrac_Alma_merged}) may be caused by different natures of the disorganized components of the magnetic field at these two spatial scales, where local physical conditions are very different.
In the analytical model of \citealt{Planck2018XII}, the function $f_{\rm m}(\delta)$, which depends on the lag, quantifies the disorganized component of the magnetic field relative to its uniform component. Using the dependence of $f_{\rm m}(\delta)$ as a function of $\delta$, one cannot adequately perform the extrapolation of the correlation's intercept value we found (which we denote as $f$ in Figure \ref{fig:scatter_S_Pfrac_Alma_merged}) between the \plk and ALMA scales, because the underlying analytical model used to express the dependence of this function on the scale relies on the hypothesis that the disorganized component of the magnetic field is isotropic, which in turn reflects the properties of the turbulent cascade at work in the diffuse ISM.
\footnote{Note that our sources are at different distances and have been observed at different resolutions. Therefore, the variation of the lag $\delta$ among all of our 15 datasets causes the parameter $f_{\rm m}(\delta)$ to vary. However, this variation does not affect our results as the dependence of the variations of $f_{\rm m}(\delta)$ caused by the different values of $\delta$ fall within the uncertainty of our ALMA correlation (the standard deviation of all the extrapolated values of $f_{\rm m}(\delta)$ from the fifteen datasets is 0.004). Nevertheless, all the processes responsible for the disorder in the magnetic field are not necessarily associated with a peculiar physical scale, but rather are phenomena that act across a range of scales, from the envelope to the disk. In consequence, the lag does not represent any intrinsic turbulent length scale within a core.} This model would predict values of $f_{\rm m}(\delta)$ (and thus levels of turbulence) that are too small at core scales; the vertical shift between the red line and our correlation confirms this point (Figure \ref{fig:scatter_S_Pfrac_Alma_merged}). For typical low-mass cores, the contribution of the turbulence from the ISM is expected to be negligible. However, cores are observed to be turbulent at some level: typical linewidths are subsonic in the $\sim$\,1000\,au-scale inner envelopes of Class 0 protostellar cores \citep{Gaudel2020}, to trans-sonic at low-mass star-forming cores scales \citep{Friesen2017,Keown2017}. In addition, within these cores, it is expected that the turbulent component of an initially homogeneous magnetic field at core scales would originate from gravo-turbulence induced by collapse motions \citep{Vazquez-Semadeni2012,Mocz2017,Ballesteros-Paredes2018,Vazquez-Semadeni2019} and outflow phenomena \citep{Zhang2005,Arce2007,Plunkett2013,Frank2014,Plunkett2015}.

Moreover, note that an adaption of the \citet{Planck2018XII} analytical model (the original version of which included multiple layers of turbulence along the line of sight; see Appendix \ref{app:Planck_model}) using the specific case of a single-layer model of randomly oriented magnetic field predicts $\mathcal{S} \propto \Pf^{-0.5}$. It is therefore possible that the flatter correlation between $\S$ and $\Pf$ observed in cores is due to a smaller number of contributing layers along the line of sight, resulting in an overall less turbulent component of the apparent magnetic field compared with that produced by the multi-scale turbulence at work in the ISM.

The two left panels of Figure \ref{fig:scatter_coldens_norma} show the evolution of $\S$ and $\Pf$ as a function of the normalized envelope column density $N_{\textrm{H}_2}$ in the envelopes. It seems that $\S$ and $\Pf$ show opposite trends, which are the result of an increase in the fluctuations in the apparent magnetic field with increasing column density. Note that a similar trend in $\Pf$ was found with increasing column density in the diffuse ISM of the Vela C molecular cloud \citep{Fissel2016}. In spite of this intrinsic increase of complexity of the apparent magnetic field with increasing local column density, we still detect substantially organized magnetic fields, as shown by the relatively high values of polarization fraction observed in cores even at high column densities ($>3\%$ at $N_{\textrm{H}_2}>10^{24}$ cm$^{-2}$). In addition, despite of the fact that at the core scale, the main sources of the magnetic field disorder are the dynamical phenomena occurring in the core (e.g., gravitational collapse, outflows, rotation), we tend to detect strongly polarized emission linked to organized magnetic fields in regions associated with infalling material.

The angular resolution remains an important factor in the statistical analysis of dust polarization observations, because depolarization effects can occur if the resolution of the observations is not high enough to resolve the characteristic length scales of the phenomena driving the small-scale magnetic field morphologies both along the line of sight as well as in the plane of the sky. Beyond the heterogeneity in the characteristics of the ALMA observations that we analyze (such as the angular resolutions and the dynamic range), at the scales we probe here, the magnetic field strength, ionization fraction, gravitational potential, and gas kinematics will affect how an initially uniform magnetic field at envelope scales will develop a complex topology. Given the simple assumption that the gravitational potential is isotropic, considering that the typical spatial resolution we have is on the order of or smaller than the typical Jeans length at the envelope densities we probe, the typical spatial scales at which gravity is expected to significantly distort the magnetic field lines are mostly resolved at the scales (a few beams) where we compute the dispersion $\S$. Nevertheless, if the magnetic field is highly complex at smaller scales than the ones we probe, then indeed $\Pf$ drops and conversely $\S$ rises toward its highest values.

\subsubsection{On the differences in the intrinsic scales of the total intensity (Stokes $I$) and polarized (Stokes $Q$ and $U$) emission}
\label{sec:scales_of_emission}

The spatial distributions of both the polarized and unpolarized emission in the plane of the sky show characteristics that are likely to affect the polarization fraction toward protostellar cores, and thus the statistical results we present in this paper.

A qualitative view of typical ALMA maps of the dust emission from cores often reveals that the emission in Stokes $Q$ and $U$ looks sharper and more extended that in Stokes $I$. We have therefore examined the spatial power spectra, which quantify the power present at each spatial scale, of the Stokes $I$, $Q$, and $U$ emission in each of our 15 datasets (see Appendix \ref{app:power_spectrum}). Each spectrum is normalized to its maximum value, which allows us to compare the relative power of the emission as a function of spatial scale.
We find that, generally, once normalized to its maximum value, the power in the Stokes $Q$ and $U$ maps tends to be larger than the power in Stokes $I$ sometimes by more than one order of magnitude. A larger fraction of the total polarized power resides at larger spatial scales, which explains why the polarized intensity maps appears less peaked than the total intensity maps.
This effect could be due to severe dynamic-range limitations and image recovery artifacts affecting the Stokes $I$ maps.
However, we stress that the discrepancies in power between $Q$ and $U$ versus $I$, in the majority of the sources, are more and more significant as we probe larger spatial scales. Since all Stokes are from the same electromagnetic waves received by the same interferometer, if the polarized and unpolarized emission originally had similar spatial distributions, there would be no reason that interferometric filtering would create such differences in the power recovered at different angular scales for the different Stokes parameters. Thus, it is likely that these power spectra reflect the different intrinsic (``true'') spatial distribution of polarized and total dust emission at typical core scales probed with ALMA observations.

\begin{figure*}[!tbh]
\centering
\vspace{-0.1cm}
\subfigure{\includegraphics[scale=0.38,clip,trim= 0.8cm 1.5cm 1.8cm 2.0cm]{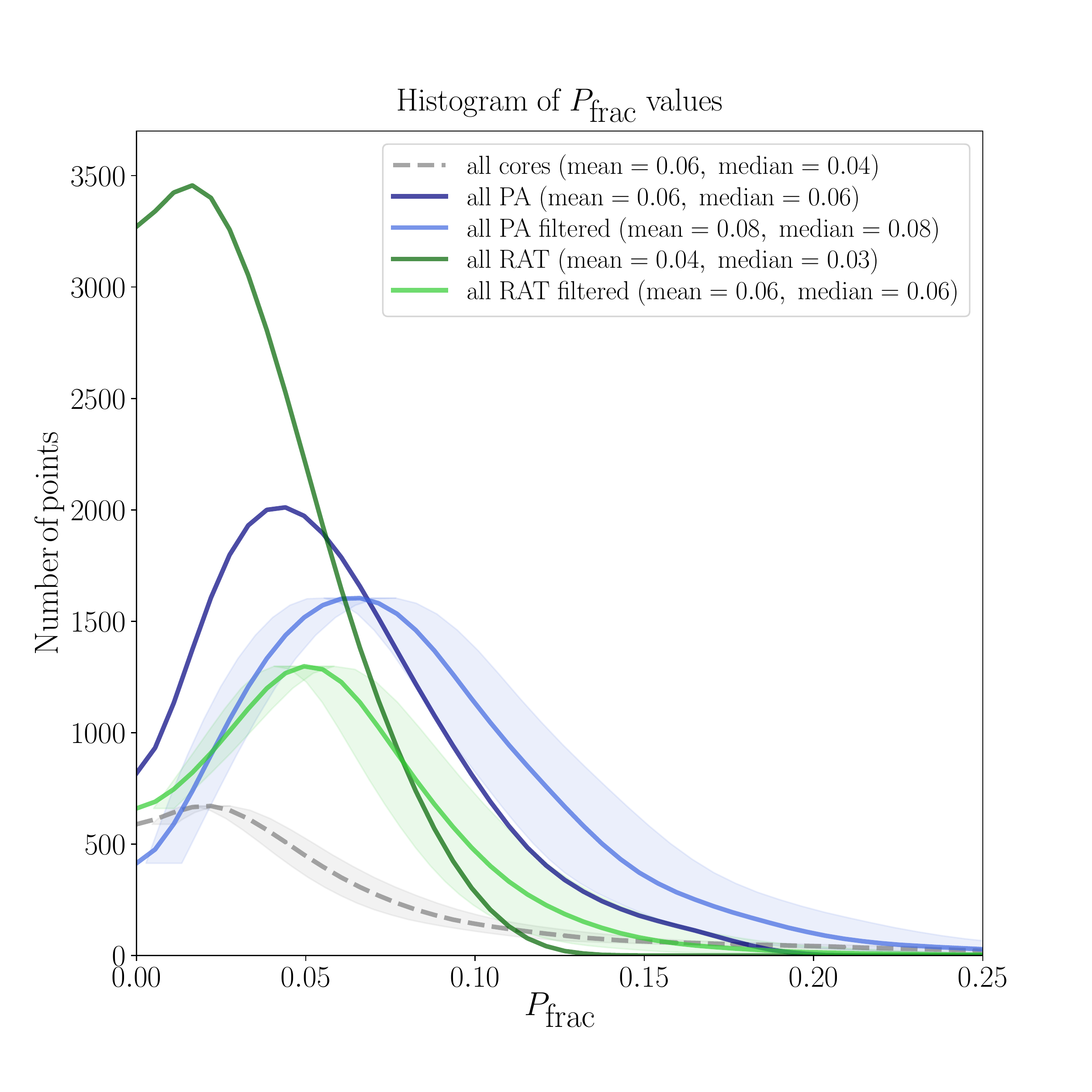}}
\subfigure{\includegraphics[scale=0.38,clip,trim= 0.8cm 1.5cm 1.8cm 2.0cm]{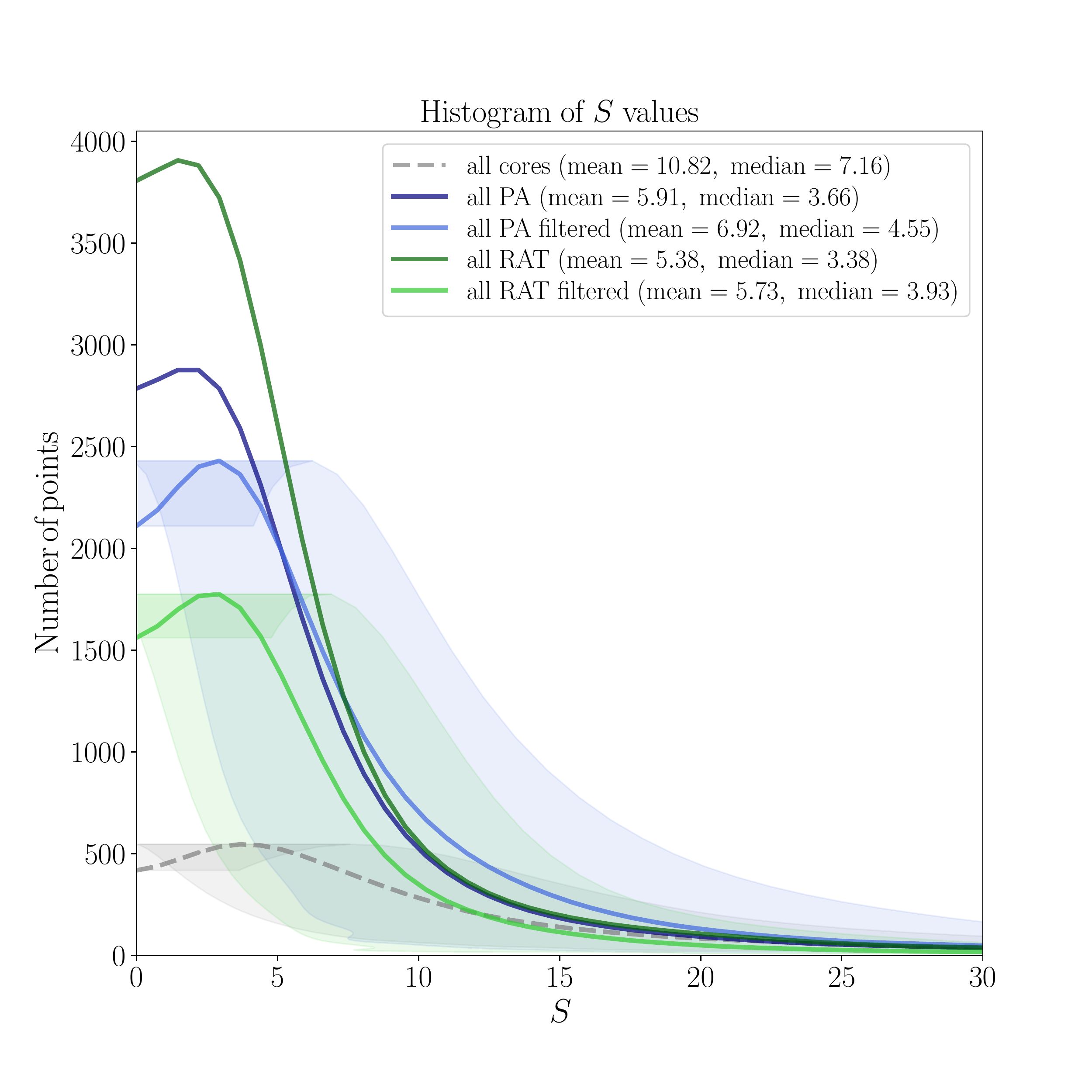}}
\caption[]{\footnotesize Histograms of polarization fraction $\Pf$ (\textit{left}) and dispersion of polarization angles $\S$ (\textit{right}) for all ALMA cores (dashed lines), and for all simulations (solid lines; with grain alignment via perfect alignment [PA] or radiative torques [RATs], and with or without filtering). The histogram lines have been smoothed with a 1D-Gaussian kernel of a size of 0.2$\%$ in $\Pf$ and $2^{\circ}$ in $\S$. In both panels, the shaded areas correspond to the mean of the uncertainty in $\Pf$ and $\S$ within each bin of the histogram}. In the right panel, we do plot the errors in $\S$, derived following \citet{Alina2016}. We do not show uncertainties for the synthetically observed simulations, as they have not been filtered by the CASA simulator.
\label{fig:hist_simu}
\end{figure*}

The differences in the spatial distribution of power between Stokes maps towards protostellar cores could be the underlying cause of the high values of polarization fractions observed at lowest observed column densities, which correspond to the largest radii in the envelopes (indeed, see the trend of Figure \ref{fig:scatter_coldens_norma} and the core-by-core analysis of Figure \ref{fig:trends_S_Pfrac_Alma} where we see that the high polarization fraction values correspond to low column density values). While models do not predict such high levels of polarization (see Section \ref{sec:spatial_filtering} and the simulation and radiative transfer presented in Appendix \ref{app:maps_synth}), this may contribute to the shallower correlation we find in the $\S$ versus $\Pf$ relation from the ALMA observations with respect to the trend obtained with large-scale cloud observations: indeed, a decrease in the the highest values of $\Pf$ would result in a relation closer to $\S\,\propto\,\Pf^{-1}$, found in \plk and predicted by their analytical model.

Confirmation of this result will require further investigation, as it is crucial to understand how much the scales of emission differ between polarized and total intensity emission. Quantifying this would allow us to remove the biases in the values of polarization fraction derived from interferometric observations.

\subsubsection{On the effect of spatial filtering on statistical polarization properties}
\label{sec:spatial_filtering}

One major issue we face with the ALMA dust polarization observations is spatial filtering by the interferometer, which removes the scales of emission that are not included in the $uv$-coverage of the dataset. In contrast to the statistical analysis of dust polarization performed with single dish instruments such as \plk  \citep{PlanckXIX,PlanckXX,Planck2018XII}, BLASTPOL in \citep{Fissel2016}, and SCUPOL  \citep{Poidevin2013}, our analysis of interferometric data requires us to characterize how the filtering alters the polarization quantities we use in our statistics. With this aim, we use a set of synthetically observed non-ideal MHD simulations computed with RAMSES \citep{Teyssier2002,Fromang2006} that follow the gravitational collapse of cores whose range of initial mass and turbulence reproduce the main characteristics of the sources from our sample. The set consists of six simulations of collapsing cores (with total masses of 30, 60, and 100 $M_\odot$). We perform radiative transfer on these models using the POLARIS code \citep{Reissl2016}, which produces the Stokes $I$, $Q$, and $U$ maps and assumes either that a constant fraction of the dust grains are perfectly aligned everywhere (perfect alignment, known as ``PA'' hereafter) or that paramagnetic grains are aligned via radiative torques, known as ``RATs'' hereafter \citep[e.g.,][]{Lazarian2007}. Note that the hypothesis of perfect alignment is not physical, and we do not aim to reproduce or interpret the polarized dust emission from Class 0 envelopes as resulting from perfect alignment. However, while we recognize that an hypothesis of perfect dust alignment is not a physical model but a phenomenological one, it has been suggested that the properties of dust polarization at the larger scales of the diffuse ISM (especially the results of $\StimesP$) can be explained and reproduced with perfect alignment \citep{Planck2018XII,Seifried2020}. In the first part of our discussion we aim to compare our results with those obtained at larger scales, and thus perfect alignment remains an interesting point of comparison with RATs, and is a useful benchmark to compare how different physical models of grain alignment affect the statistical properties of the polarized emission. In addition, a case where the grains are perfectly aligned is only taking into account the source-specific geometrical effects governing the resulting polarization maps, and thus is useful to understand where alignment drops or is suppressed. We present all the details of the simulations and the radiative transfer calculations in Appendix \ref{app:maps_synth}. In order to produce realistic synthetic observations to compare with the ALMA datasets, we use the CASA simulator (with the typical ALMA $uv$-coverage of these observations) to implement the effects of interferometric filtering and atmospheric noise on the POLARIS synthetic emission maps.

\begin{figure*}[!tbph]
\centering
\subfigure{
\includegraphics[scale=0.395 ,clip,trim= 0.8cm 1.2cm 2.1cm 1.8cm]{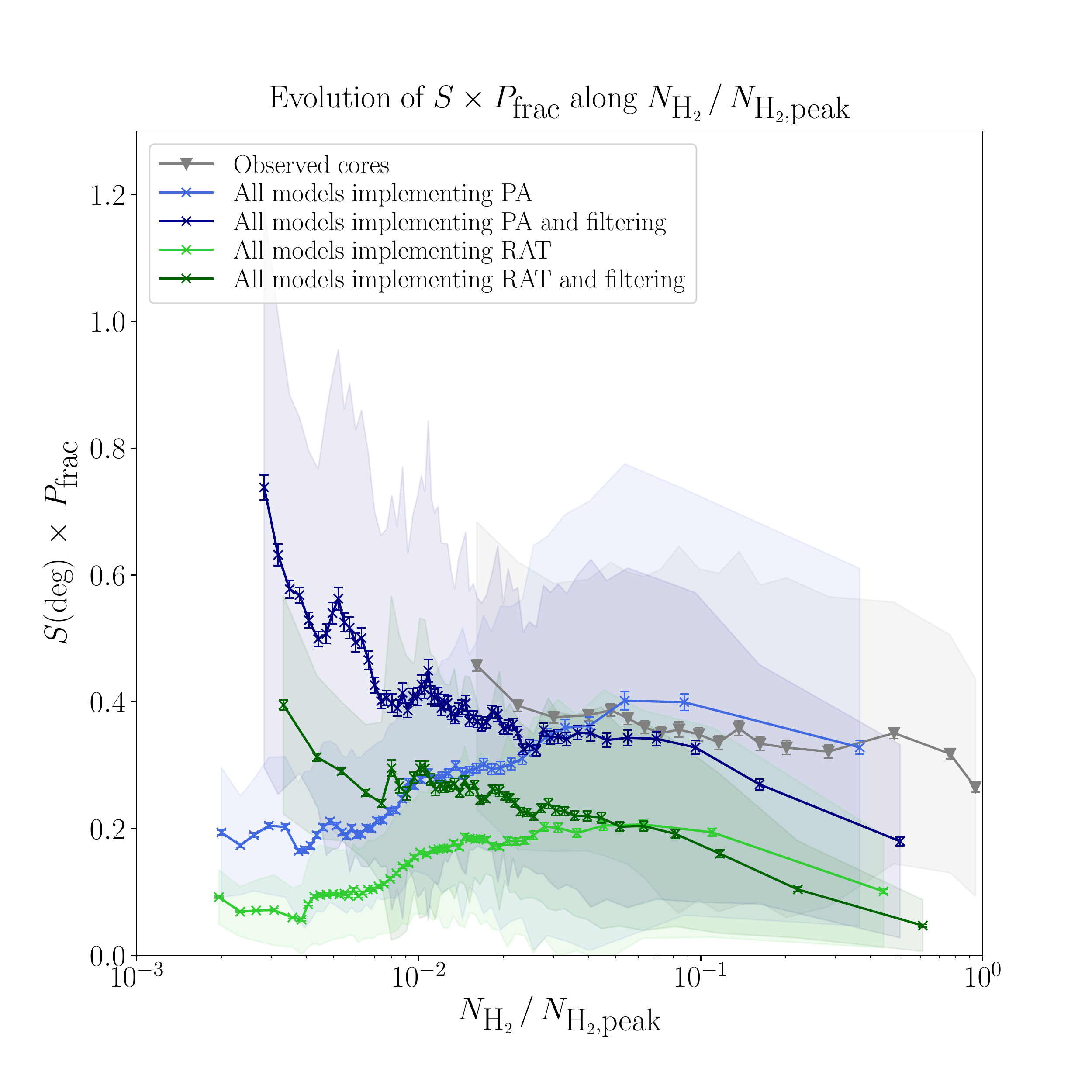}}
\subfigure{
\includegraphics[scale=0.395 ,clip,trim= 0.8cm 1.2cm 2.1cm 1.8cm]{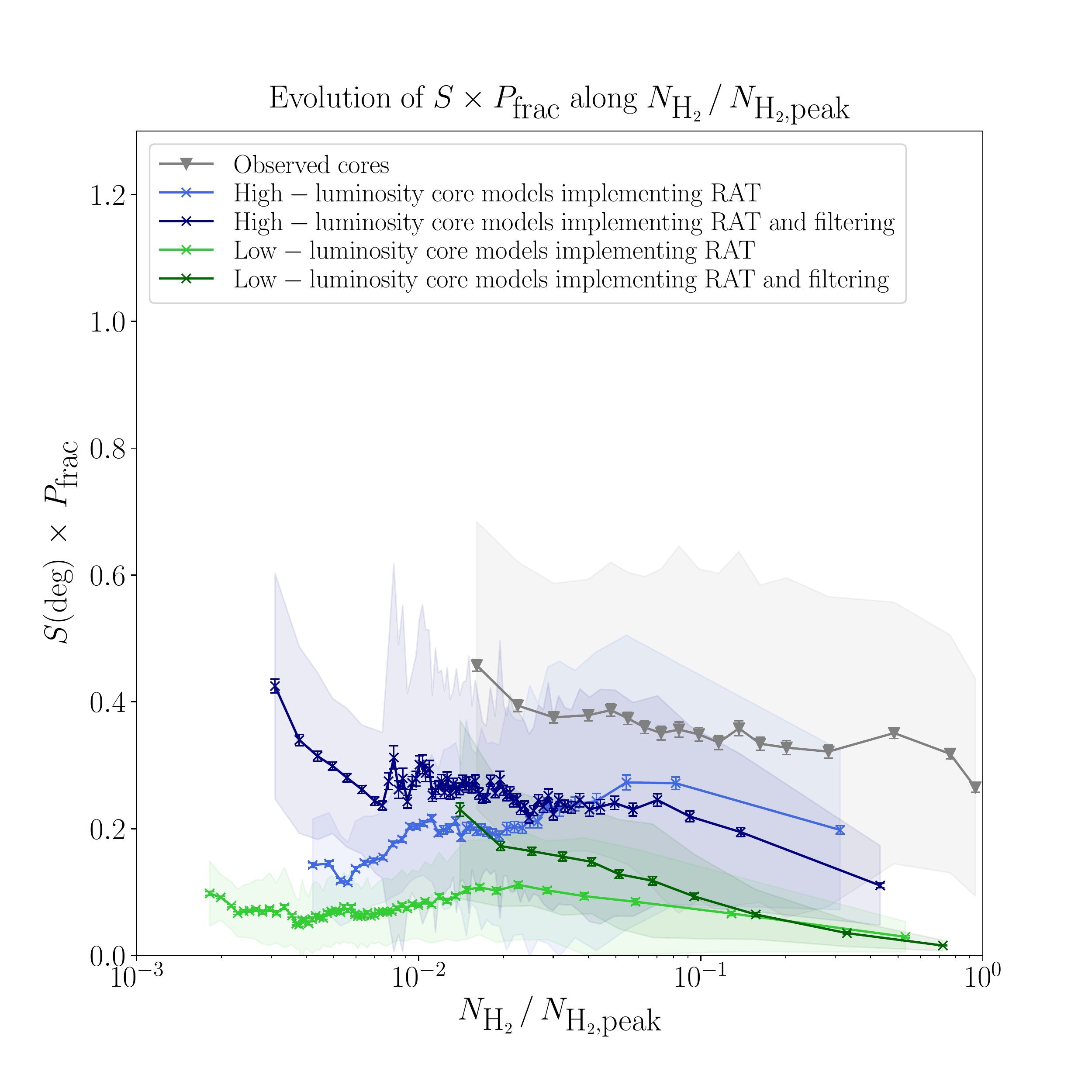}
}
\caption[]{\footnotesize Observed distributions of the mean values of $\StimesP$ as a function of the column density $N_{\textrm{H}_2}$ (normalized by its maximum value, $N_{\textrm{H}_2\textrm{,peak}}$) of all the ALMA cores (triangles) and of all the models (crosses). \textit{Left:} the four lines representing the simulations correspond to results from all the simulations merged together, using RATs or PA, both filtered and not filtered. \textit{Right:} we focus on the simulations implementing RATs only, both filtered and not filtered, separating the three simulations with low protostellar accretion luminosity from the three with high accretion luminosity. The shaded areas represent $\pm$ the standard deviation of the Gaussian fit performed on each bin of points. The error bars correspond to these standard deviation values divided by the square root of the number of points in each bin.}
\label{fig:scatter_coldens_norma_simu}
\end{figure*}

In Figure \ref{fig:hist_simu} we present the histograms of $\S$ and $\Pf$ (where all simulations have been merged) before and after filtering, assuming RATs or PA \footnote{Note that merging all the simulations does not change the result, as each simulation of the six we present in Appendix \ref{app:maps_synth} sees their values of $\S$ and $\Pf$ increase. This increase is also seen in Figures \ref{fig:scatter_S_Pfrac_simu_all_wturbu} and \ref{fig:scatter_S_Pfrac_simu_all_woturbu}. The effects of filtering in the case of the three massive simulations (Figure \ref{fig:scatter_S_Pfrac_simu_all_woturbu}) are  marginal, most likely because these cores are very bright and exhibit magnetic fields that are on average more organised than the three less massive simulations.}.
Note that if no spatial filtering is applied to the synthetic maps, it is difficult to reproduce the rather high values of polarization fraction typically observed in ALMA observations of protostellar cores using models that only include grain alignment via RATs (despite the fact that we include relatively large grains in our calculations; see details in Appendix \ref{app:maps_synth}); this was pointed out previously in \citet{Valdivia2019}.
We find that spatial filtering systematically causes the entire distributions of the dispersion of polarization angles $\S$ and the polarization fraction $\Pf$ to increase. These increases also translate into an increase in the mean values of $\StimesP$ (see Figure \ref{fig:scatter_coldens_norma_simu} for the evolution of $\StimesP$ as a function of the column density, in different simulations and implementing grain alignment via both PA and RATs).
In addition, one can see that the effect of filtering in Figure \ref{fig:scatter_coldens_norma_simu} seems to be stronger at low column densities.  This makes sense because the low column density regions lie at large scales (within the envelope probed by our ALMA observations), where the power spectra in the Stokes $I$ versus Stokes $Q$ and $U$ maps show large discrepancies (see Section \ref{sec:scales_of_emission}).  It is therefore possible that some of the high values of polarization fraction $\Pf$ found in the ALMA dust polarization observations of protostars may be related to the spatial filtering of the intrinsically different spatial distributions of the polarized versus total intensity emission.

While there is significant uncertainty in the reliability of the calculated values of $\Pf$ given our analysis of both filtering and power spectra, we note that the statistical behavior of $\Pf$ corresponds to what is predicted by theory and models. In addition, the distributions of $\Pf$ we present in Figure \ref{fig:hist_pa_pfrac} peak at reasonable values ($\sim$\,5$\%$); the high values of $\Pf$ that we find in the ALMA observations are located in the tail of the $\Pf$ distribution.

\begin{figure*}[!tbph]
\centering
\vspace{-0.1cm}
\subfigure{\includegraphics[scale=\scaleSP,clip,trim= 1.6cm 1.3cm 3cm 2.1cm]{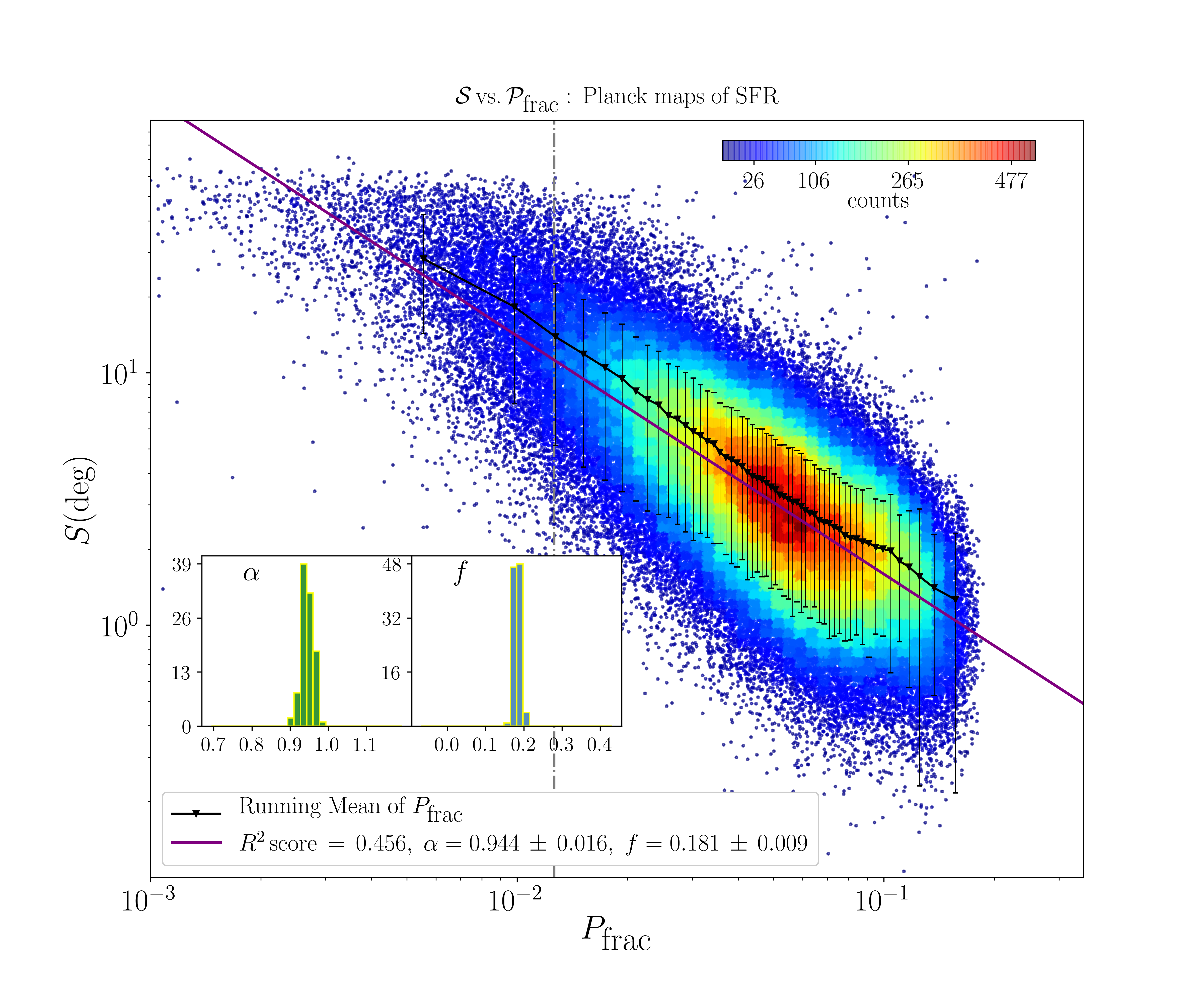}}
\subfigure{\includegraphics[scale=\scaleSP,clip,trim= 1.6cm 1.4cm 3cm 2.1cm]{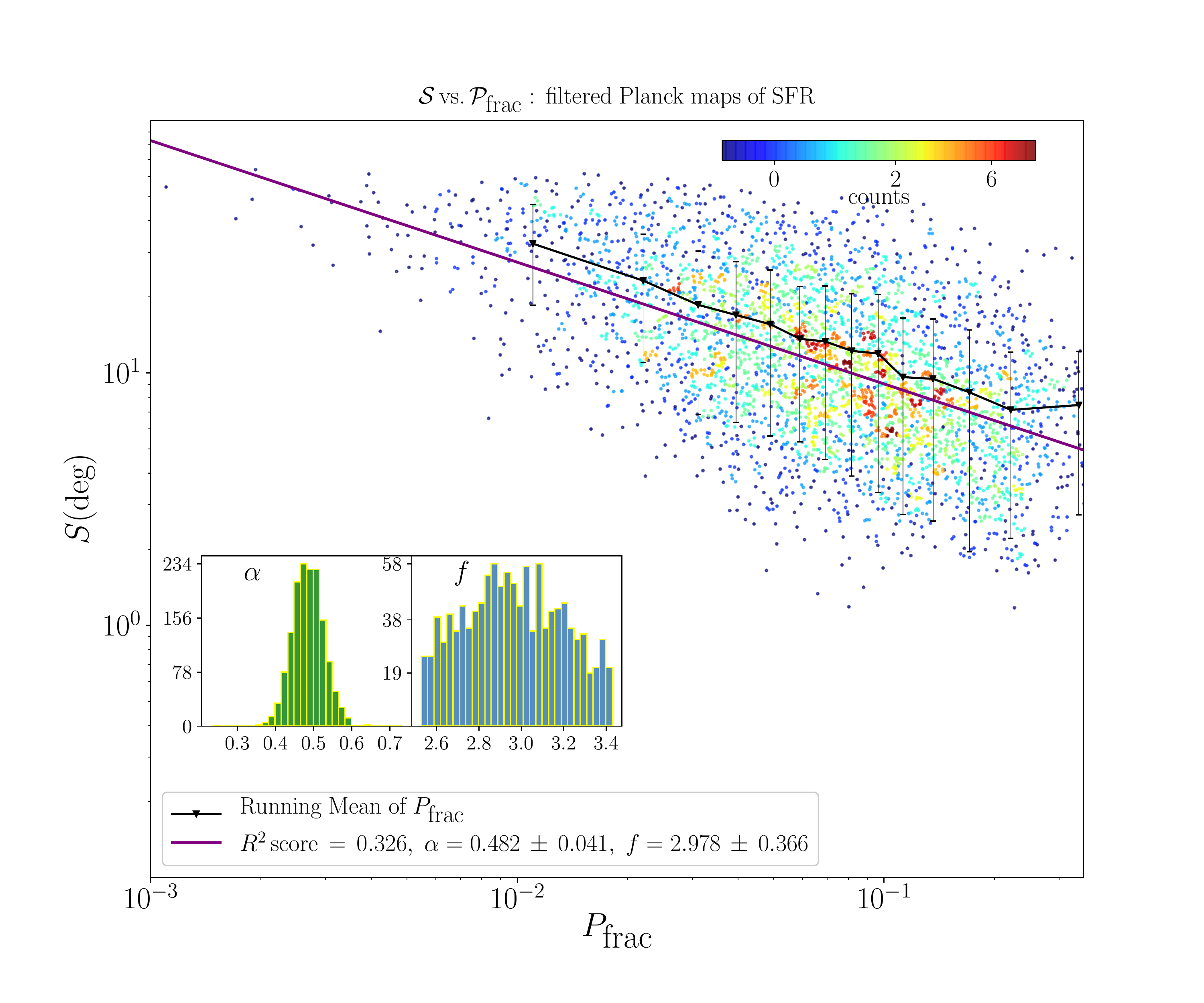}}
\caption[]{\footnotesize Distributions of the dispersion of polarization position angles $\S$ as a function of the polarization fraction $\Pf$ from star formation regions observed by \plk without filtering (\textit{left}), and with filtering (\textit{right}). Same as Figure \ref{fig:scatter_S_Pfrac_Alma_merged}. Interferometric filtering degrades the quality of the $\S$ versus $\Pf$ correlation (smaller $R^2$) and affects its power law index $\alpha$, which flattens from 0.94 before filtering to 0.48 after filtering.}
\label{fig:scatter_S_Pfrac_Planck}
\end{figure*}

As a thought experiment, we apply the same kind of filtering that ALMA produces to the \plk maps by artificially placing them further away so each map has the angular size of the ALMA field of view at 870\,$\mu$m. We then synthetically observe them with the CASA simulator using a combination of ALMA antenna configurations similar to those used to observe the sample of cores analyzed here. We find that this simple exercise indeed confirms the change of the power law index of the correlation between $\S$ and $\Pf$: from an initial 0.94 before filtering to 0.48 after filtering (see Figure \ref{fig:scatter_S_Pfrac_Planck}). This change is drastic, and the power law index we obtain is much smaller than the ones we obtain from the ALMA observations. This can be explained by the fact that the emission from the \plk observations corresponds to very diffuse regions, and thus the filtering is removes a significant part of the initial flux in the three Stokes maps, which yields a more dramatic effect of the spatial filtering on the recovered correlation between $\S$ and $\Pf$.

\subsection{On the dust grain alignment efficiency inside a Class 0 protostellar core}

In this section we discuss how our statistical analysis of polarized dust emission properties has improved our understanding of the dust grain alignment process at work in Class 0 protostellar envelopes. We mainly focus on the interpretation of the evolution of $\StimesP$ as a function of column density, both in ALMA observations (Sections \ref{sec:SxP_alma} and \ref{sec:low_high_cav_planeq}) and synthetic observations of MHD models of protostellar evolution (Section \ref{sec:RATs_simu}). Finally, we investigate how the statistical properties of the polarized emission from Class 0 protostellar cores may be explained by different dust grain characteristics (Section \ref{sec:grain_proper}) or additional grain alignment mechanisms (Section \ref{sec:other_mecha}).

\subsubsection{$\StimesP$ suggests no strong radial dependence of the average dust grain alignment efficiency in protostellar envelopes}
\label{sec:SxP_alma}

According to the analytical model developed for the ISM by the \plk collaboration (see Appendix \ref{app:Planck_model}), the product $\StimesP$ is a proxy for the maximum dust grain alignment efficiency $\Pf_{\rm, max}$\footnote{For a brief discussion of a recently developed alternative method for evaluating the dust grain alignment efficiency, see Appendix \ref{app:PI_attempt}.}, and is statistically independent of the magnetic field configuration. This value of $\Pf_{\rm, max}$ is influenced by a variety of parameters, such as the collisional de-alignment of grains by gas particles (which scales with density); the dust grain size, shape, and composition; and the local irradiation conditions. We stress that the absolute average values of $\StimesP$ that we present here cannot be compared directly with the values derived from the \plk data because of possibly different physical origins of the turbulent component of the magnetic field at ISM versus core scales. However, if the turbulent component of the magnetic field is still on average isotropic at core scales, then, as $\S$ and $\Pf$ are inversely dependent on the disorganization level of the apparent magnetic field, it is reasonable to assume that $\StimesP$ traces the intrinsic capability of dust grains to align themselves with the local magnetic field.

We show in Figure \ref{fig:scatter_coldens_norma} that the product $\StimesP$ obtained with all the ALMA dust polarization observations is remarkably constant as a function of column density in protostars, with an average value of \SPracValue. Despite the increasing complexity of the magnetic field topology from core to disk scales, the drastically different local physical conditions (\eg density, pressure, temperature, and irradiation conditions), the flat profile of the average $\StimesP$ over two orders of magnitude in column density suggests that, within the statistical uncertainties reported in Figure \ref{fig:scatter_coldens_norma}, the grain alignment efficiency remains approximately constant throughout a protostellar envelope. This is reminiscent of the \plk results in star-forming clouds: it suggest that both in the ISM and in cores, the dust grain alignment mechanism(s) at work do not appear to be very sensitive to local physical conditions.

We stress that in the range of column densities accessible to ALMA, the models implementing RATs (when all averaged together) show a decrease of a factor of two in $\StimesP$ relative to what we see from the ALMA data (see Figure \ref{fig:scatter_coldens_norma_simu} left). In the sections that follow, we explore possible reasons (e.g., different local environmental conditions) for the discrepancy between our ALMA results and the models implementing grain alignment via RATs.


\subsubsection{The effects of environmental conditions on dust grain alignment efficiency}
\label{sec:low_high_cav_planeq}

Our findings that the average grain alignment efficiency does not strongly depend on the local column density in protostellar cores is, however, at odds with the expected behavior of grain alignment with respect to the quite inhomogeneous local conditions in cores. For example, the rise of gas pressure and density near the center of the protostellar core, which causes the gaseous damping timescale to decrease, is a crucial factor that theoretically leads to a loss of dust grain alignment efficiency \citep{Reissl2020}. Furthermore, observations have revealed that radiation, presumably caused by accretion processes near the central protostar, causes enhanced polarized emission along the cavity walls of bipolar outflows \citep{Hull2017b,Maury2018,LeGouellec2019a,Hull2020a}. Finally, indications of larger dust grain size with respect to the ISM dust grain population have been found in embedded objects \citep{Miotello2014,Valdivia2019,Galametz2019,LeGouellec2019a,Hull2020a}. This suggests that, in the context of the RAT alignment mechanism, these phenomena may counter-balance one another, thus precluding a significant variation of alignment efficiency as a function of column density. Then the constant trend of $\StimesP$ could be due to averaging all the observations from our sample, whereas the individual protostars may have very different local conditions at a given normalized column densities because, \eg their luminosity and absolute densities are different. In addition, note that the statistical weights (in terms of number of independent points, see Table \ref{t:alphas_values}) of each observation are very different. Therefore, averaging all the observations may cause observations with larger weights to mask the results of those with lower weights.

Hence, to examine in more detail some of these physical processes that are thought to cause the efficiency of RATs to vary, we perform two separations in our datasets. First, we separate the ``low'' and ``high'' luminosity cores (\ie high [bolometric luminosity >\,10$\,L_\odot$]: NGC1333 IRAS4A, OMC3 MMS6, Serpens Emb 6, IRAS 16293, BHR71; low  [bolometric luminosity <\,10$\,L_\odot$]: Serpens Emb 8, Serpens Emb 8(N), B335, VLA 1623, L1448 IRS2) in order to investigate whether the strength of the central source of irradiation affects the dust grain alignment efficiency in the entire core. Note that the value of 10$\,L_\odot$ is arbitrary, and simply allow us to separate the sample into two bins in luminosity. Second, we separate the emission from the outflow cavities versus that from the envelope on each side of the outflow by taking the emission from inside and outside of the cone of the bipolar outflow (see Figure \ref{fig:S_I_pol_maps_sources} and Appendix \ref{app:ALMA_cores}).
While the magnetic field lines in outflow cavity walls are very organized (having small values of $\S$), which contributes to the observed high values of polarization fraction, the enhancement of polarized emission in these regions seems to be linked with irradiation conditions favourable to dust grain alignment via RATs. Indeed, detections of CCH spectral line emission---a molecular tracer known to be sensitive to irradiation---toward outflow cavity walls in B335 \citep{Imai2016} and in Serpens Emb 8(N) \citep{LeGouellec2019a} support this hypothesis. On the contrary, the envelope emission not associated with the outflow is not expected to have such favorable irradiation conditions because of the larger amount of dense envelope material located at all depths along the photon propagation path through the envelope.

\begin{figure}[!tbph]
\centering
\includegraphics[scale=0.4 ,clip,trim= 0.8cm 1.1cm 2cm 1.8cm]{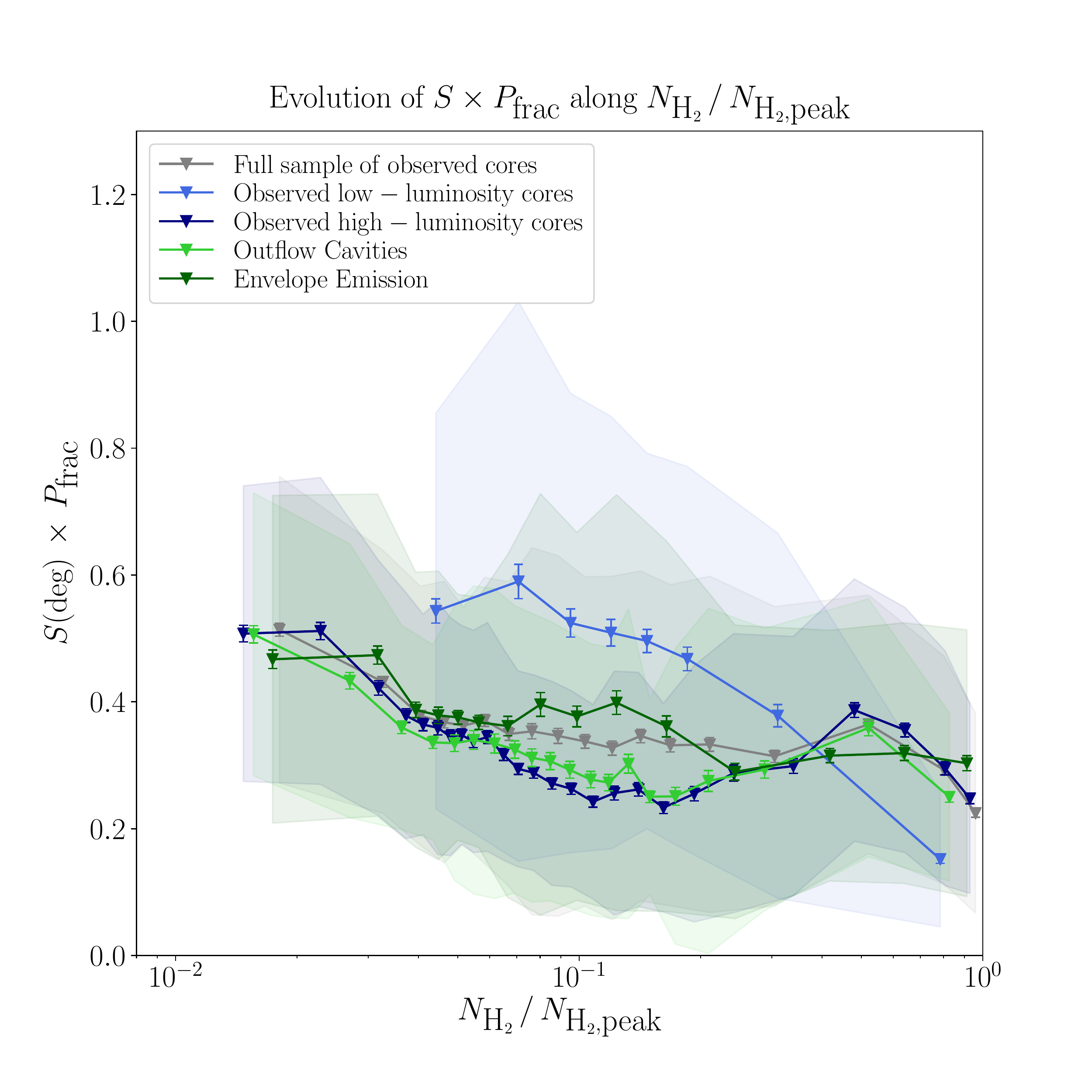}
\caption[]{\footnotesize Observed distributions of  $\StimesP$  as a function of the column density $N_{\textrm{H}_2}$, which is normalized to its maximum value $N_{\textrm{H}_2\textrm{,peak}}$, for all the cores, and for other cases we tried, including separating high- and low-luminosity cores, as well as outflow cavity walls versus envelope emission. The solid lines are the running means, and the shaded areas are $\pm$ the standard deviation of each bin of points. The error bars correspond to these standard deviation values divided by the square root of the number of points in each bin.}
\label{fig:scatter_coldens_norma_attempts}
\end{figure}

In Figure \ref{fig:scatter_coldens_norma_attempts} we present, for each of these four cases, the evolution of $\StimesP$ as a function of the normalized column density. The trends of $\StimesP$ in outflow cavity versus envelope emission are very similar. Thus, despite very different irradiation conditions in the outflow cavities and the surrounding envelope, these differences seem insufficient to cause observable changes in the grain alignment efficiency between these two regions. Assuming grains are aligned via RATs, the grains embedded in the envelope thus must still receive amounts of anisotropic irradiation that are sufficient to align grains. However, we do see significant differences in the trends of $\StimesP$ between the low- and high-luminosity cores. The distribution from the high-luminosity cores follows the constant trend from the two previous cases (\ie outflow cavity and envelope emission), as well as the trend seen when all the datasets are merged together. On the other hand, the distribution from the low-luminosity cores shows a clear decrease in $\StimesP$ as a function of column density, which indicates less efficient grain alignment in the highest density regions of these cores. As the product $\StimesP$ is also dependent on the function $f_\textrm{m}(\delta)$ (see Section \ref{sec:turbu_B_compo}), this may indicate that these low-luminosity cores are subject to different amounts of the gravo-turbulent motions responsible for the correlation between $\S$ and $\Pf$. However, while this would modify the average values of $\S$ and $\Pf$, it is unclear that this would result in such a strong decrease of $\StimesP$ with the column density, and there are no obvious reasons why only the low-luminosity cores would be affected.

A possible explanation for this decrease of $\StimesP$ toward low-luminosity cores is the amount of irradiation received by dust grains with respect to their position in the protostellar envelope. Indeed, the sub-sample of low-luminosity cores tend to have smaller envelope masses than the high-luminosity cores. Therefore, the optically thin regions of dust emission in the low-luminosity cores tend be closer to the protostellar embryo. However, as the irradiation emanating from the low-luminosity protostellar embryo at the center of the envelope is smaller relatively to the high-luminosity protostars, dust grains located close to the peak in column density may be less efficiently aligned in the low-luminosity cores, which causes the drop we see of $\StimesP$ toward these sources. On the contrary, the thermal dust emission emanating from the outer envelope of the low-luminosity cores, \ie where $10^{-2}$\,<\,$N_{\textrm{H}_2}/N_{\textrm{H}_2\textrm{,peak}}$\,<\,$10^{-1}$, may correspond to column densities that are low enough to still be permeated by the interstellar radiation field. This would increase dust grain alignment efficiency and could potentially justify the on-average higher $\StimesP$ values of the low- versus the high-luminosity cores, in this range of normalized column density. In addition, the higher irradiation emanating from the central protostar of the high-luminosity cores may propagate far enough to maintain a relatively high grain alignment efficiency, even at larger envelope radii.


Finally, note that the overall larger column densities in more massive protostars (which is the case for the high-luminosity cores of our sample) also leads to a larger amount of material being optically thick: it is possible that a decrease of $\StimesP$ also happens in these cores, but within the inner envelope radii where dust emission becomes optically thick (and is thus hidden) at (sub)-mm wavelengths. For example, the central region $\sim$\,200 au of Serpens Emb 6 \citep{LeGouellec2019a} was found to be likely optically thick at 870\,$\mu$m; \citet{Takahashi2019} also reported an optically thick $\sim$\,200 au central region in the high-luminosity source OMC3 MMS6. However, \citet{LeGouellec2019a} reported optically thin central regions in the low-luminosity cores Serpens Emb 8 and Serpens Emb 8(N). We examine the maps of integrated optical depth in the synthetic observations of our numerical protostellar models and find that, indeed, the central regions (i.e., the inner $\sim\,$100 au in the three low-luminosity simulations, and the inner $\sim\,$200 au of the high-luminosity simulations) are optically thick. Nevertheless, the trends of $\StimesP$ as a function of column density for the simulations implementing RATs show relatively flat profiles for both the low- and high-luminosity cases (Figure \ref{fig:scatter_coldens_norma_simu}).

\subsubsection{The role of protostellar luminosity in aligning grains via RATs in MHD models}
\label{sec:RATs_simu}

Analyzing the statistics from the simulations using the mean of $\StimesP$ estimator as a function of $N_{\textrm{H}_2}$ yields the result shown in Figure \ref{fig:scatter_coldens_norma_simu}, where we compare in the left panel the observational distribution presented in Figure \ref{fig:scatter_coldens_norma} with the distributions from all the simulations, implementing RATs or PA, both filtered and not filtered. Similarly, in the right panel of Figure \ref{fig:scatter_coldens_norma_simu} we compare the same observational distribution with the distributions from the simulations implementing RATs, separating the three simulations with more massive cores and hence higher luminosities from the three others with lower masses and lower luminosities.As we index the irradiation emanating from the central protostar directly to its mass, the three more massive simulations have a stronger radiation field in the core. We find that, in the case of RATs, $\StimesP$ is on average higher in high-luminosity sources than in  low-luminosity sources (Figure \ref{fig:scatter_coldens_norma_simu} right panel). This shows that the statistical estimator $\StimesP$ is sensitive to dust grain alignment efficiency, as expected from RAT theory, which predicts that grain alignment efficiency scales with the strength of the local radiation field.  Finally, note that in the case of perfect alignment, as all of the susceptible grains are aligned with the magnetic field, we see even higher average values of $\StimesP$ (Figure \ref{fig:scatter_coldens_norma_simu} left panel).

Note that the three massive simulations correspond to much higher-mass cores that those in our sample of Class 0 protostellar cores. Moreover, these simulations do not have initial turbulence, and are quite axisymmetric at the time steps we choose, which may be inconsistent with the requirement of the analytical model that the disorganized component of the magnetic field be isotropic in order to justify tracing dust grain alignment efficiency with $\StimesP$. We stress that, however, at all column densities probed in our sample of protostars, the observed $\StimesP$ values are overall larger than those predicted by models with dust grains aligned via RATs, but are consistent with $\StimesP$ values predicted by models where grains are perfectly aligned (see the left panel of Figure \ref{fig:scatter_coldens_norma_simu}). The perfect alignment hypothesis allows us to estimate the typical $\StimesP$ values produced by the combination of perfect local alignment and geometrical effects along the line of sight and in the plane of the sky, and in consequence suggests that the grain alignment efficiency in protostellar envelopes is higher than the efficiency produced by standard RATs alone. Only models implementing RATs in high-luminosity cores (see Figure \ref{fig:scatter_coldens_norma_simu}, right panel) produce values that are marginally consistent with the observed values of $\StimesP$ in our sample of protostars. Our results thus suggest that the efficiency of grain alignment via RATs does not match most observations, and highlight the importance of investigating the potentially key role of protostellar irradiation, in our future efforts to reproduce the observed $\StimesP$. We stress, however, that implementing different dust-grain properties (see Section \ref{sec:grain_proper}) as well as different grain alignment mechanisms (see Section \ref{sec:other_mecha}) could potentially allow models to approach the values of $\StimesP$ seen in ALMA observations.

\subsubsection{Different dust grain properties}
\label{sec:grain_proper}

Other potential origins for the differences in dust polarization properties in protostellar environments with respect to the ISM, are the different dust properties, such as, e.g., dust grain size, structure, or even composition.
The plethora of dust polarization detections toward the densest regions of young protostellar cores indicate that dust grains are still aligned down to very small scales ($\sim$\,100\,au) close to the embedded protostar, where dust grain characteristics may not be well constrained.


Estimations of photon-penetration length scales in sub-millimeter wavelength ALMA observations of protostellar cores have revealed that, given the wavelength of the radiation impinging on the dust grains, the detected dust polarization should emanate from dust grains larger than the (sub-)micron-sized dust grain size expected in a typical ISM population \citep{LeGouellec2019a,Hull2020a}. Radiative transfer studies assuming dust grains aligned by RATs of simulations of low-mass collapsing cores have shown that the typical amount of polarization detected in observations can be reproduced by simulations if the implemented maximum dust grain size exceeds $10\,\mu$m \citep{Valdivia2019}. In addition, indications of such large grains have been found in multi-wavelength observations of protostellar envelopes in studies analyzing dust grain emissivities \citep{Miotello2014,Galametz2019}. Finally, while the typical elongation of dust grains in star-forming environments is unconstrained observationally, grain alignment may strongly depend on this parameter; further efforts that use dust models to produce predictive observational tests would help further constrain the effect of grain elongation.

Finally, grain alignment efficiency closer to the observed values traced by $\StimesP$ in protostellar envelopes (see Section \ref{sec:RATs_simu}, where we find that the observed level of dust grain alignment efficiency is not reproduced by standard RATs in the radiative transfer calculations of our models) may be reached if we change the paramagnetic properties of the dust grains used in our radiative transfer calculations. Assuming RATs are the main mechanism aligning grains in protostellar environments, considering dust grains with super-paramagnetic inclusions allows RATs to align more grains \citep{Hoang2016}. Note that this modified RAT theory was tested in models of the diffuse interstellar medium by \citet{Reissl2020}, who found that RATs acting on super-paramagnetic grains produce values of grain alignment efficiency very similar to those obtained when grains are perfectly aligned.



\subsubsection{Other grain alignment mechanisms}
\label{sec:other_mecha}

Despite a global agreement in the trends, the statistical properties of polarized dust emission seen in ALMA observations of protostellar cores cannot be fully reproduced by the synthetic observations of MHD simulations of young, collapsing protostellar cores. The distributions of $\S$ versus $\Pf$ from the simulations (see Figures \ref{fig:scatter_S_Pfrac_simu_all_wturbu} and \ref{fig:scatter_S_Pfrac_simu_all_woturbu}) show clear trends, but they do not match those found from the ALMA observations (Figure \ref{fig:scatter_S_Pfrac_Alma_merged}). As mentioned in Section \ref{sec:grain_proper}, we lack detailed understanding of dust grain properties in the protostellar envelopes probed by ALMA observations. In addition, although we demonstrate the influence of irradiation on the efficiency of grain alignment via RATs in the $\StimesP$ analysis of our models (see Section \ref{sec:RATs_simu}), we see in Section \ref{sec:low_high_cav_planeq} that no major differences in dust grain alignment efficiency were found between two regions of the cores that should experience different irradiation conditions: the outflow cavity walls, and the regions in the envelope not associated with the outflow. Finally, the average values of $\StimesP$ from ALMA observations do not seem to match the values obtained from models implementing RATs, but rather show a better match with perfect alignment.

One possibility to explain the aforementioned issues, assuming our protostellar MHD models accurately represent the environments of young protostars, is that we may not fully understand all of the mechanisms causing the linear polarization we detect, and that an additional mechanism(s) may be dominant over RATs when the latter are no longer effective. The dynamical context of some dust polarization observations, especially in the outflow cavity walls or accretion streamers, may favor other theories of grain alignment such as Mechanical Alignment Torques (MATs). Introduced in \citet{Hoang2018}, the MAT theory describes the alignment of dust grains with respect to the magnetic field orientation via mechanical torques induced by supersonic gas-dust drift (in an outflow, for example). It is, however, not yet clear how one could identify the occurrence of this new dust grain alignment mechanism in the objects we study here. In addition, the dust grain size distribution may be affected by RAdiative Torque Disruption (RATD; \citealt{Hoang2019NatAs,Hoang2019}), which predicts that the dust-grain size distribution will shift to smaller values due to the disruption of large aggregates that are spun-up to suprathermal rotation speeds and thus broken apart by radiative torques.


The statistical analysis of the dust polarization we perform in this article, thanks to the tools of $\S$ and $\Pf$, is able to characterize the processes at work in the alignment of dust grains in the envelopes of Class 0 protostellar cores, such as the role played by the radiation field. However, additional methods must be developed in order to identify the potential occurrence of recently proposed grain alignment mechanisms.

\section{Conclusions}
\label{sec:conclu}

We perform a statistical analysis of the polarized dust emission emanating from a sample of fifteen ALMA observations toward eleven Class 0 protostellar cores (namely: Serpens Emb 6, Serpens Emb 8, Serpens Emb 8(N), BHR71 IRS1, BHR71 IRS2, B335, IRAS 16293, VLA 1623, L1448 IRS2, OMC3 MMS6, and NGC1333 IRAS4A), at wavelengths ranging from 870\,$\mu$m to 3\,mm. The conclusions we draw are as follows.

\begin{enumerate}

\item We find a significant correlation between the dispersion of polarization angles $\S$ and polarization fraction $\Pf$ in the polarized dust emission from protostellar envelopes, with a resulting correlation of $\mathcal{S} \propto \Pf^{-\slope}$. This correlation is sensitive to the morphology of the turbulent component of the magnetic field and to other intrinsic characteristics of the polarized emission. This correlation found in the ALMA cores has a smaller power law index than the correlation found at larger scales in the $\plk$ observations of star-forming clouds, where they found $S\propto\Pf^{-1}$. This could be a consequence of the different nature of turbulence in Class 0 sources versus the ISM (i.e., due to gravitational infall, rotation, and outflowing motions); or due to interferometric filtering, which produces artificially high $\Pf$ in ALMA observations; as well as of the possibility that grain alignment varies with local conditions, which are significantly different between the star-forming molecular clouds and protostellar cores. Finally, our observations and their comparison to synthetic observations of protostellar models suggest that additional alignment mechanisms may be at work in protostars (see point 6).

\item We find that the flattening of the correlation between $\S$ and $\Pf$ in our ALMA results versus the larger-scale \plk results can be reconciled with the \plk analytical model if it is modified to include only one layer of randomly oriented magnetic field to represent the turbulent component of the field.
This results in an overall less turbulent component of the apparent magnetic field compared with that produced by the multi-scale turbulence at work in the ISM.

\item The product $\StimesP$, which is sensitive to dust grain alignment efficiency, shows a constant profile as a function of column density in the sample of cores analyzed, with a constant value of \SPracValue. This suggests that the grain alignment mechanism producing the polarisation observed at millimeter wavelengths, over 3 orders of magnitude in column density (from $\NH=10^{22}$\,\cmsq\ to $\NH=10^{25}$\,\cmsq), may not depend strongly on the local conditions such as gas density and temperature.

\item We examine the statistical properties of polarized dust emission emanating from the outflow cavity walls versus the regions of the envelope not associated with the outflow. These regions are expected to experience drastically different irradiation conditions. We do not find any obvious difference in dust grain alignment efficiency between the two.

\item However, we find hints that, contrary to the highest luminosity cores in our sample, the lowest luminosity sources experience a decrease of their dust grain alignment efficiency at higher column densities. The environmental conditions in the central regions of the envelopes are indeed expected to disfavor the alignment of dust grains via the Radiative Alignment Torque (RAT) mechanism, as a result of the lower level of irradiation emanating from the central protostar of the low-luminosity cores relative to the high-luminosity cores. The density of the outer envelope of these low-luminosity cores may be tenuous enough to be permeated by the interstellar radiation field, thus increasing dust grain alignment efficiency with increasing radii. Finally, the higher irradiation emanating from the central protostar of the high-luminosity cores may propagate far enough to maintain a relatively high grain alignment efficiency, even at larger envelope radii.

\item We use synthetic observations of the polarized dust emission in a small sample of outputs from non-ideal MHD simulations of protostellar collapse. We apply the $\StimesP$ analysis to these synthetic maps of polarized dust emission, assuming either grain alignment via Radiative Alignment Torques (RATs) or perfect alignment (PA; i.e., alignment of all susceptible grains), and we show that the statistical estimators used in our work seem to be sensitive to the overall efficiency of grain alignment. Furthermore, our $\StimesP$ analysis of the simulations implementing RATs suggests that the average value of this estimator is sensitive to the radiation field strength in the core. Finally, the simulations with perfect alignment yield on average higher $\StimesP$ values than those implementing RATs.

\item When implementing RAT alignment in our radiative transfer calculations, we do not reproduce with our simulations the $\S$ versus $\Pf$ statistics obtained from the ALMA observations. This may suggest that the simulations are not fully adequate representations of the Class 0 protostellar envelopes in our observations, or that the $\S$ versus $\Pf$ correlation is not sensitive to the details of the physical mechanism(s) aligning the dust grains. The values of $\StimesP$ obtained from the ALMA observations seems to lie among the values predicted by PA, and are significantly higher than those found in models including RATs alone, especially at high column density. This suggests that, to be able to reproduce the dust alignment efficiency found in cores, one needs either more efficient RATs than the classical RATs with paramagnetic grains, an extra alignment mechanism(s), or different irradiation conditions than those assumed in models.


\item Our results suggest that the continuum and polarized dust emission in the ALMA observations have different intrinsic spatial scales, which affects the statistics. We show that the differences in emitting power of the different Stokes parameters as a function of spatial scale can produce artificially high $\Pf$, especially at large scales where Stokes $I$ has on average less power with respect to Stokes $Q$ and $U$. Finally, this work on synthetic observations suggests that interferometric filtering biases the values of $\S$ and $\Pf$, causing artificially high values of both.

\end{enumerate}

While the work we present here has shed light on the physics of dust grain alignment in Class 0 protostellar cores, many open questions remain about the details of the physical environment at envelope scales. Future investigations involving detailed comparisons of the observations of cores with those reproduced by simulations, alongside observations of chemical tracers associated with the polarized dust emission, will illuminate the role played by the local conditions in producing the polarization observed at small scales in Class 0 protostellar envelopes.

~\\
{
\small
\textit{Acknowledgments}
V.J.M.L.G. acknowledges the support of the ESO Studentship Program.
A.J.M. acknowledges support from the Joint ALMA Observatory Visitor Program, and ESO Visitor program.
J.M.G. is supported by the Spanish grant AYA2017-84390-C2-R (AEI/FEDER, UE).
C.L.H.H. acknowledges the support of both the NAOJ Fellowship as well as JSPS KAKENHI grants 18K13586 and 20K14527.
This work has benefited from the support of the European Research Council under the Horizon 2020 Framework Programme (Starting Grant MagneticYSOs with grant agreement no. 679937).
V.J.M.L.G. thanks Maud Galametz and Javier Abril for the precious help they offered for the statistical analysis.
This paper makes use of the following ALMA data: ADS/JAO.ALMA\#2013.1.00726.S, ADS/JAO.ALMA\#2017.1.00655.S, ADS/JAO.ALMA\#2013.1.01380.S, ADS/JAO.ALMA\#2017.1.01392.S, ADS/JAO.ALMA\#2018.1.01873.S, ADS/JAO.ALMA\#2015.1.01112.S, ADS/JAO.ALMA\#2016.1.00604.S, ADS/JAO.ALMA\#2015.1.00341.S, ADS/JAO.ALMA\#2016.1.01089.S, ADS/JAO.ALMA\#2015.1.00546.S. ALMA is a partnership of ESO (representing its member states), NSF (USA) and NINS (Japan), together with NRC (Canada), MOST and ASIAA (Taiwan), and KASI (Republic of Korea), in cooperation with the Republic of Chile. The Joint ALMA Observatory is operated by ESO, AUI/NRAO and NAOJ.
The National Radio Astronomy Observatory is a facility of the National Science Foundation operated under cooperative agreement by Associated Universities, Inc.\\

\textit{Facilities:} ALMA.

\textit{Software:} APLpy, an open-source plotting package for Python hosted at \url{http://aplpy.github.com} \citep{Robitaille2012}. CASA \citep{McMullin2007}. Astropy \citep{Astropy2018}.

}


\begin{appendix}

\addcontentsline{toc}{section}{Appendix}
\renewcommand{\thesection}{\Alph{section}}

\clearpage
\section{Planck Analytical Model}
\label{app:Planck_model}

\citet{Planck2018XII} developed an analytical model able to reproduce the phenomenological properties of polarized dust emission. They assumed the total emission arises from a small number $N$ of independent layers, each of them emitting a fraction $1/N$ of the total intensity. The magnetic field was described as the sum of a uniform and an isotropic turbulent component. This model is based on a few essential parameters, including the maximum polarization fraction $\mathcal{P}_{\textrm{frac,max}}$ (which will tell us about the intrinsic capability of the grains to align themselves with respect to the magnetic field), the ratio of the standard deviation of the turbulent magnetic field to the magnitude of the ordered field $f_\textrm{m}$, and the spectral index $\alpha_\textrm{M}$ of this turbulent component. At a given location, the analytical relationship between the dispersion of polarization angles $\S$ and the polarization fraction $\Pf$ was found to be the following:
\begin{equation}
\S^2(\delta)\,=\,\frac{f^{2}_{\textrm{m}}(\delta)}{3N}\frac{\mathcal{P}^2_\textrm{frac,max}}{\mathcal{P}^2_\textrm{frac}}\,\mathcal{A}\,\,,
\end{equation}
with
\begin{equation}
\mathcal{A}\,=\,\sum^{N}_{i=1}{\left(\sin^2{2\Delta\phi_i}\sin^2{\Gamma_i}+\cos^2{2\Delta\phi_i}\cos^2{\Gamma_i}\right)}\,\,,
\end{equation}
where $\phi$ is the polarization position angle, $\Delta\phi_i = \phi-\phi_i$, $\delta$ is the lag (as introduced in Section \ref{sec:method}, the lag describes the surface over which the dispersion $\S$ is derived, and thus corresponds to the characteristic length scale at which we quantify the disorganization of polarization position angles), and $\Gamma_i$ is the inclination angle of the magnetic field $\overrightarrow{B_i}$ in a given layer $i$ with respect to the plane of the sky. The value of $\mathcal{A}$ is approximated as follows:
\begin{equation}
\langle\mathcal{A}\rangle_{\Pf} \sim 1/\sqrt{2}\,\,,
\end{equation}
such that we obtain:
\begin{equation}
\langle\S\left(\delta\right)\rangle_{\Pf} \approx \frac{f_{\textrm{m}}(\delta)}{\sqrt{6 N}}\frac{\mathcal{P}_\textrm{frac,max}}{\Pf}\,\,,
\end{equation}
where factor $f_{\textrm{m}}(\delta)$ represents the typical relative fluctuation of the magnetic field at the scales corresponding to the annulus between $\delta/2$ and $3\delta/2$. This factor was
defined as follows:
\begin{equation}
{f_{\textrm{m}}(\delta)}\,=\,\frac{{\sigma_{B_i}}(\delta)}{B_i}\,\,,
\end{equation}
where ${{\sigma_{B_i}}(\delta)}$ is the fluctuation of the magnetic field  $\overrightarrow{B_i}$. This function was modeled as follows:
\begin{equation}
{f_{\textrm{m}}(\delta)}\,=\,0.164\,f_\textrm{M}\,\left(\frac{\omega}{160^{'}}\right)^{-1-\alpha_\textrm{M}/2}\,\,,
\end{equation}
where $\omega$ is the full width half maximum of the spatial resolution of the observations. The parameter values used in \citet{Planck2018XII} are: $\alpha_{\textrm{M}}=-2.36$, $f_\textrm{M}=0.9$, $N\,=\,7$, and $P_\textrm{frac,max}\,=\,0.26$. Using these values yielded the following analytical relation:
\begin{equation}
\langle\S\left(\delta\right)\rangle_{\Pf} \,=\, \frac{0.339}{{\Pf}}\,\left(\frac{\omega}{160^{'}}\right)^{0.18}
\end{equation}
In the plots where we relate $\S$ and $\Pf$, we plot this relation in red, using the analytical coefficient of 0.339. The \plk team found a coefficient of 0.31 from their observations.

When considering the results of the \plk team, this analytical model yields a dispersion of polarization angles $\S$ that is proportional to $\Pf^{-1}$. However, for our study, the specific case of $N=1$ is relevant. This gives:

\begin{equation}
\S^2(\delta)\,=\,\frac{f^{2}_{\textrm{m}}(\delta)}{3}\frac{\mathcal{P}^2_\textrm{frac,max}}{\mathcal{P}^2_\textrm{frac}}\,\cos^2{\Gamma}\,\,.
\end{equation}
In this specific case:
\begin{equation}
\Pf\,=\,\mathcal{P}_\textrm{frac,max}\,\cos^2{\Gamma}\,\,,
\end{equation}
and thus we obtain:
\begin{equation}
\S(\delta)\,=\,f_{\textrm{m}}(\delta)\sqrt{\frac{\mathcal{P}_\textrm{frac,max}}{3\Pf}}\,\,.
\end{equation}
Consequently, in the specific case of $N=1$, $\S$ is proportional to $\Pf^{-1/2}$ in the analytical model.

\clearpage
\section{ALMA cores}
\label{app:ALMA_cores}

\begin{figure}[!tbph]
\centering
\vspace{-0.5cm}
\includegraphics[scale=0.35,clip,angle=-35,trim= 5cm 1.5cm 5cm 0.5cm]{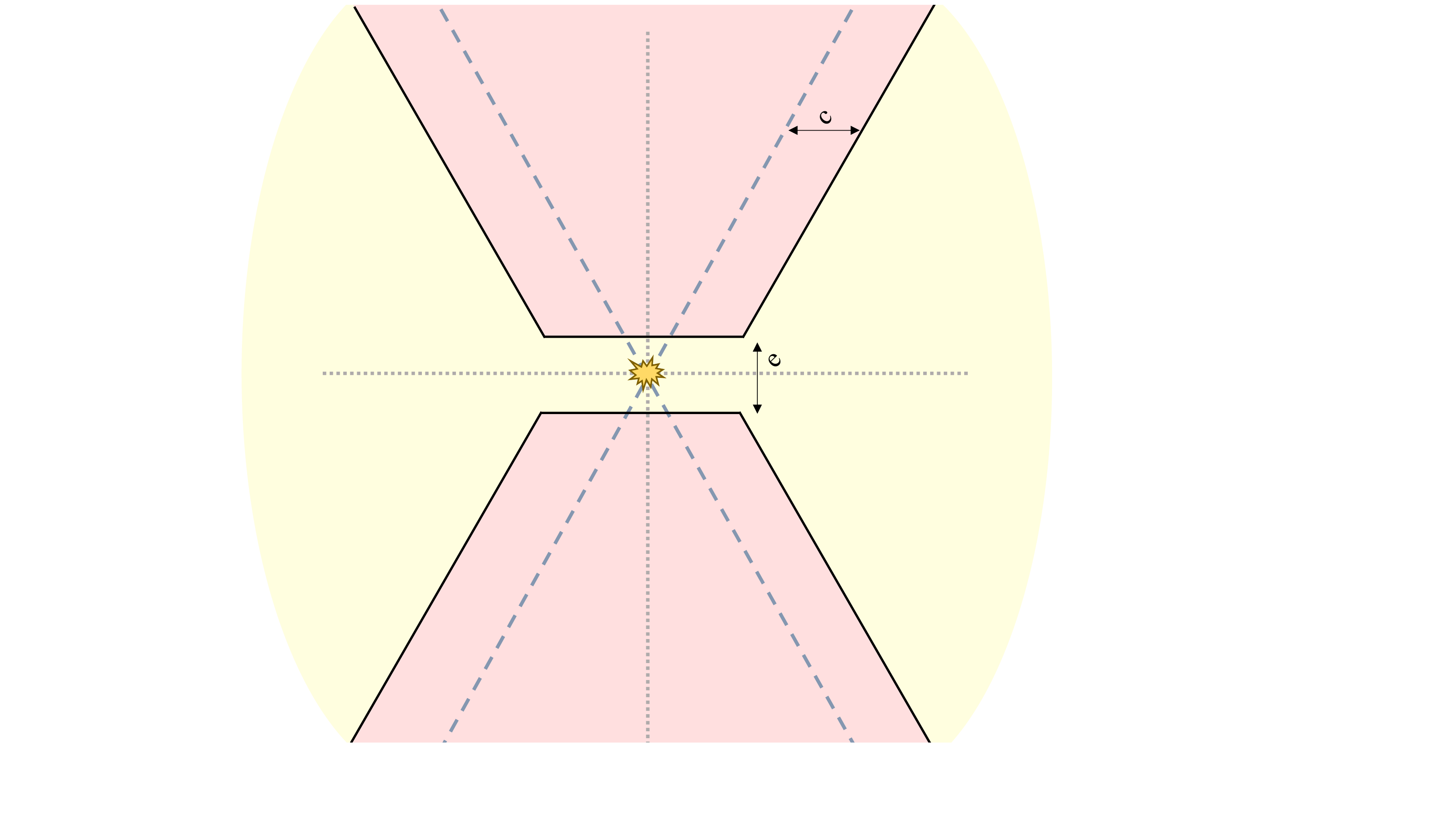}
\vspace{-1cm}
\caption[]{\small Scheme of the separation we have performed in our polarized intensity maps. The red and yellow areas are defined as the outflow cavity and the envelope emission not associated with the outflow cavities, respectively.}
\label{fig:scheme_outflow_cavity}
\end{figure}

In Table \ref{t.obs} we list the details and the coordinates of each source in our sample. In Table \ref{t.cav_eq} we present the outflow properties of each source.
In Figure \ref{fig:S_I_pol_maps_sources} we present the maps of Stokes $I$, polarized intensity $P$, and polarization angle dispersion $\S$, for all of the ALMA observations. The white dotted lines show the separation between the outflow cavities and the envelope emission not associated with the outflow cavities. We characterize this separation in Figure \ref{fig:scheme_outflow_cavity}, where $e$ and $c$ denote the thickness of the equatorial planes (separating the two lobes of the outflow) and the outflow cavity walls, respectively, as fit by eye using the polarized intensity maps. The outflow position angles and opening angles are taken from the literature.

Figure \ref{fig:scatter_S_Pfrac_Alma} shows the $\S$ vs. $\Pf$ correlations for all the datasets.

\begin{table*}[!tbph]
\centering
\small
\caption[]{Details of individual sources}
\label{t.obs}
\setlength{\tabcolsep}{0.4em} 
\begin{tabular}{p{0.25\linewidth}cccccccc}
\hline \hline \noalign{\smallskip}
Name & $\alpha_{\textrm{J2000}}$ & $\delta_{\textrm{J2000}}$ &$M_\textrm{env}$$^a$ & $L_{\textrm{bol}}$$^a$& $L_{\textrm{int}}$$^b$ & Dist.$^c$&Features$^d$\\
&&&$M_\odot$&$L_\odot$&$L_\odot$&pc&\\
\noalign{\smallskip}  \hline
\noalign{\smallskip}
Serpens Emb 6  & 18:29:49.81 & \phantom{-}01:15:20.41& 27.2\phantom{0} &134.6&111.3$^*$&484&C,E\\
\noalign{\smallskip}
\hline
\noalign{\smallskip}
Serpens Emb 8 & 18:29:48.09 & \phantom{-}01:16:43.30 & \multirow{2}{*}{12.8\phantom{0}}&\multirow{2}{*}{\phantom{00}7.3}&\phantom{0}13.6$^\dagger$\phantom{$^*$} &\multirow{2}{*}{484}&E\\
\noalign{\smallskip}
Serpens Emb 8(N) & 18:29:48.73 & \phantom{-}01:16:55.61& &&\phantom{00}1.3$^*$&&C  \\
\noalign{\smallskip}
\hline
\noalign{\smallskip}
BHR71 IRS1 & 12:01:36.51 & –65:08:49.31& \phantom{0}4.6\phantom{0} &\phantom{0}14.7&\multirow{2}{*}{\phantom{0}11.8$^*$\phantom{$^*$}}&\multirow{2}{*}{200} & E\\
\noalign{\smallskip}
BHR71 IRS2 &  12:01:34.04 & –65:08:47.87 &-- &\phantom{00}1.7& &&C \\
\noalign{\smallskip}
\hline
\noalign{\smallskip}
B335 &   19:37:00.91&  \phantom{-}07:34:09.60&\phantom{0}1.44&\phantom{00}2.7&\phantom{00}1.4\phantom{$^*$}&165& C,E \\
\noalign{\smallskip}
\hline
\noalign{\smallskip}
IRAS 16293A & 16:32:22.87 & –24:28:36.63 &\multirow{2}{*}{\phantom{0}4.3\phantom{0}} &\multirow{2}{*}{\phantom{0}30.2}&\multirow{2}{*}{\phantom{0}19$^*$\phantom{.0}} & \multirow{2}{*}{144}& \multirow{2}{*}{E} \\
\noalign{\smallskip}
IRAS 16293B & 16:32:22.61 & –24:28:32.61 & & & &\\
\noalign{\smallskip}
\hline
\noalign{\smallskip}
VLA1623A/B & 16:26:26.35 & –24:24:30.55 & \phantom{0}0.8\phantom{0} &\phantom{00}4.4 &\phantom{00}1.2$^*$& 144 &E\\
\noalign{\smallskip}
\hline
\noalign{\smallskip}
L1448 IRS2 & 03:25:22.41 & \phantom{-}30:45:13.21 & \phantom{0}1.3\phantom{0} & \phantom{00}7.0 &\phantom{00}4.7\phantom{$^*$}&294& C,E\\
\noalign{\smallskip}
\hline
\noalign{\smallskip}
OMC3 MMS6 &05:35:23.42 & –05:01:30.53  &36\phantom{.00} &\phantom{0}38.4 & \phantom{0}27.3$^*$&432&C,E\\
\noalign{\smallskip}
\hline
\noalign{\smallskip}
NGC1333 IRAS4A &03:29:10.55&\phantom{-}31:13:31.00&12.3\phantom{0}&\phantom{0}14.2&\phantom{00}4.7\phantom{$^*$}&294&C,E \\
\noalign{\smallskip}
\hline
\smallskip
\end{tabular}
\vspace*{-0.1in}
\caption*{\small
We scale the masses and luminosities found in the literature to the distances we adopt.\\
$^a$ Envelope mass $M_\textrm{env}$ and bolometric luminosity $L_{\textrm{bol}}$ values for Serpens Emb 6 are from observations that encompass the whole Emb 6 core \citep{Enoch2011,Kristensen2012}.  The values calculated for Serpens Emb 8 and Emb 8(N) include both sources together \citep{Enoch2009b,Enoch2011}. \citet{Tobin2019} for BHR71. \citet{Kurono2013,Maury2018} for B335. \citet{Pineda2012, Jorgensen2016} for IRAS 16293A/B. \citet{Froebrich2005,Karska2018} for VLA1623A/B. \citet{Sadavoy2014,Karska2018} for L1448 IRS2. \citet{Karska2018,Galametz2019} for IRAS4A. \citet{Chini1997,Manoj2016} for OMC3 MMS6. \\
$^b$ Internal Luminosity. Values with $^*$ are calculated via the relation from \citet{Dunham2008} using fluxes from archival data of \textit{Herschel} PACS at 70\,$\mu$m. \citet{Galametz2019} for Serpens Emb 8. \citet{Yang2017BHR} for BHR 71. \citet{Galametz2018} for B335. \citet{Galametz2019} for L1448 IRS2 and NGC1333 IRAS4A. In the cases where the internal luminosities is quoted for a core that actually consists of two binaries, we index the respective internal luminosity of each protostar with their respective peak in total intensity. \\
$^c$ Distance reported in the literature. \citet{Zucker2019} for the Perseus, Ophiuchus, Serpens, and Orion sources. \citet{Watson2020} for B335. \citet{Seidensticker1989,Bourke2001} for BHR71.\\
$^d$ Main features present in the maps of total and/or polarized intensity. C: Cavities, \ie outflow cavity wall of the bipolar outflow. E: emission from the envelope that is not associated with the bipolar outflow. \\
$^\dagger$ For Serpens Emb 8 $L_{\textrm{bol}}$ < $L_{\textrm{int}}$, which is unphysical.  The two values are derived from studies using two different telescopes: $L_{\textrm{bol}}$ from BOLOCAM \citep{Enoch2009b,Enoch2011}, and $L_{\textrm{int}}$ from \textit{Herschel} \citep{Galametz2019}. However, \citet{Enoch2011} mention the source was saturated at 70\,$\mu$m, so $L_{\textrm{bol}}$ is likely an underestimate.
}
\end{table*}


\begin{table*}[!tbph]
\centering
\small
\caption[]{Outflow details of individual sources}
\setlength{\tabcolsep}{0.4em} 
\begin{tabular}{p{0.25\linewidth}ccccc}
\hline \hline \noalign{\smallskip}
Name & Position angles$^a$ & Opening angles$^a$ & $e$$^b$ & $c$$^c$ & \\
& deg & deg & au & au & \\
\noalign{\smallskip}  \hline
\noalign{\smallskip}
Serpens Emb 6 &135&80&290&350\\
\noalign{\smallskip}
\hline
\noalign{\smallskip}
Serpens Emb 8 &129&34&290&290\\
\noalign{\smallskip}
Serpens Emb 8(N) &--73&14&290&350\\
\noalign{\smallskip}
\hline
\noalign{\smallskip}
BHR71 IRS1  &174&55&600&600\\
\noalign{\smallskip}
BHR71 IRS2 &--31&47&600&600\\
\noalign{\smallskip}
\hline
\noalign{\smallskip}
B335$^d$  &99&63&180&200\\
\noalign{\smallskip}
\hline
\noalign{\smallskip}
IRAS 16293A$^e$  &90,\,135&30,\,30&90&90\\
\noalign{\smallskip}
IRAS 16293B$^e$  &---&---&---&---\\
\noalign{\smallskip}
\hline
\noalign{\smallskip}
VLA1623A/B$^f$  &125&30&60\phantom{0}&60\phantom{0}\\
\noalign{\smallskip}
\hline
\noalign{\smallskip}
L1448 IRS2  &134&70&310&430\\
\noalign{\smallskip}
\hline
\noalign{\smallskip}
OMC3 MMS6$^g$  &171&20&200,\,40&180,\,40\\
\noalign{\smallskip}
\hline
\noalign{\smallskip}
NGC1333 IRAS4A$^h$ &--9,\,19&40,\,40&160,\,160&160,\,160\\
\noalign{\smallskip}
\hline
\smallskip
\end{tabular}
\vspace*{-0.1in}
\caption*{\small
$^a$ Position angles of bipolar outflows and outflow opening angles, measured counterclockwise from North, taken from the literature. If the red- and blueshifted lobes have different position angles, we average them together. \citet{Hull2017b,Aso2019,LeGouellec2019a,Tychoniec2019} for Serpens Emb 6, Emb 8 and Emb 8(N). \citet{Tobin2019} for BHR71 IRS1 and IRS2. \citet{Takahashi2012a,Hull2014} for OMC3 MMS6. \citet{Hull2014,Velusamy2014,Tobin2018} for L1448 IRS2 and B335. \citet{Santangelo2015a,Ching2016,Tobin2018} for IRAS4A. \citet{Santangelo2015b,Murillo2018} for VLA 1623. \citet{Yeh2008,Kristensen2013a,Girart2014,vanderWiel2019} for IRAS 16293.\\
$^b$ Thickness of the equatorial planes, fit by eye.\\
$^c$ Thickness of outflow cavity walls, fit by eye.\\
$^d$ We show the average values of $e$ and $c$ obtained from the observations at the three different wavelengths. \\
$^e$ Both outflows are from IRAS 16293A; IRAS 16293B does not have any detected outflow.
The parameters $e$ and $c$ have been determined by eye thanks to the previously published CO emission. Furthermore, IRAS 16293B lies within one of the outflow cones of IRAS 16293A; however, we consider it to be envelope emission. \\
$^f$ VLA1623A/B do not have clearly identifiable outflow cavities in the polarized dust emission. Therefore we consider all the emission of this source to be coming from the envelope. However, we keep the parameters $e$ and $c$ (determined by eye thanks to the CO emission) in this Table and the white-dotted lines in Figure \ref{fig:S_I_pol_maps_sources} to simply indicate the location and shape of the outflow.\\
$^g$ We report $e$ and $c$ for the two datasets separately, as they have very different angular resolutions. \\
$^h$ We report values separately for the two protostars in this core, IRAS4A1 and IRAS4A2. \\
}
\label{t.cav_eq}
\end{table*}

\def\scaleSP{0.66}

\begin{figure*}[!tbph]
\centering
\vspace{-0.1cm}
\subfigure{\includegraphics[scale=\scaleSP,clip,trim= 1.9cm 3.1cm 1.85cm 2.6cm]{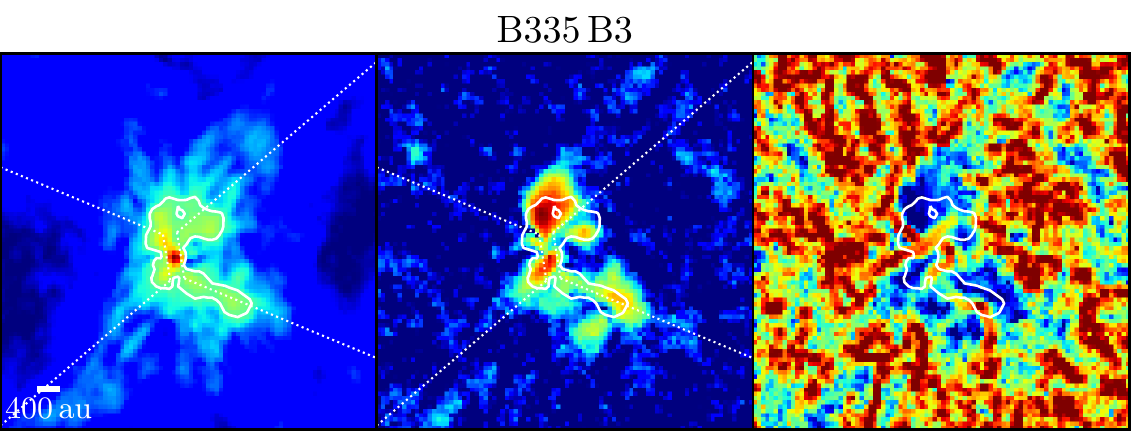}}
\vspace{-0.4cm}
\subfigure{\includegraphics[scale=\scaleSP,clip,trim= 1.9cm 3.1cm 0.2cm 2.6cm]{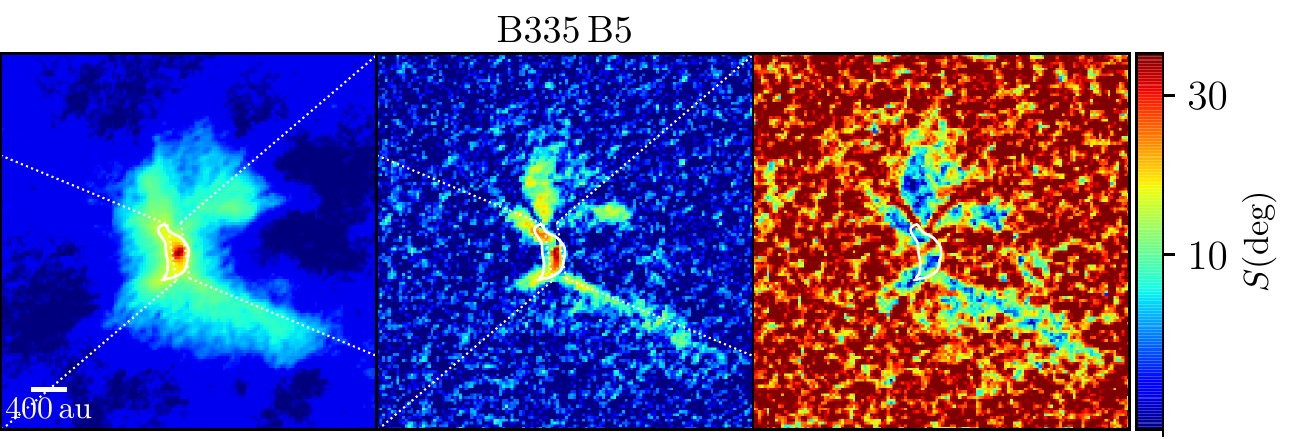}}
\vspace{-0.2cm}
\subfigure{\includegraphics[scale=\scaleSP,clip,trim= 1.9cm 3.1cm 1.85cm 2.6cm]{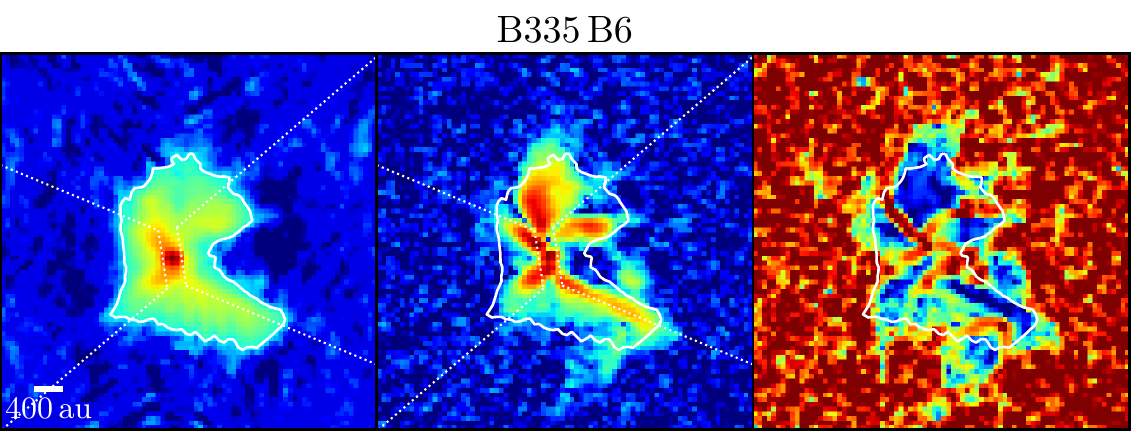}}
\vspace{-0.2cm}
\subfigure{\includegraphics[scale=\scaleSP,clip,trim= 1.9cm 3.1cm 0.2cm 2.6cm]{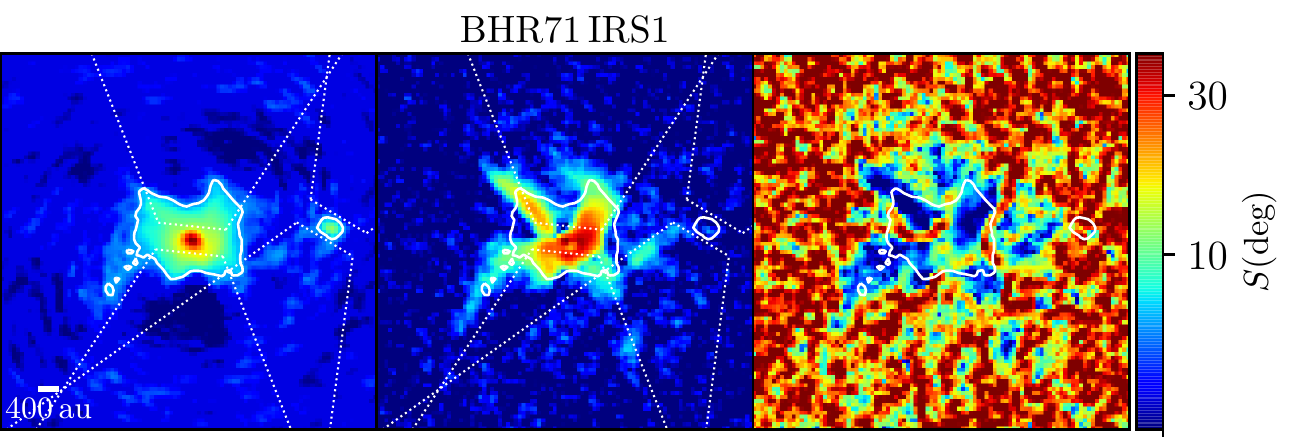}}
\vspace{-0.2cm}
\subfigure{\includegraphics[scale=\scaleSP,clip,trim= 1.9cm 3.1cm 1.85cm 2.6cm]{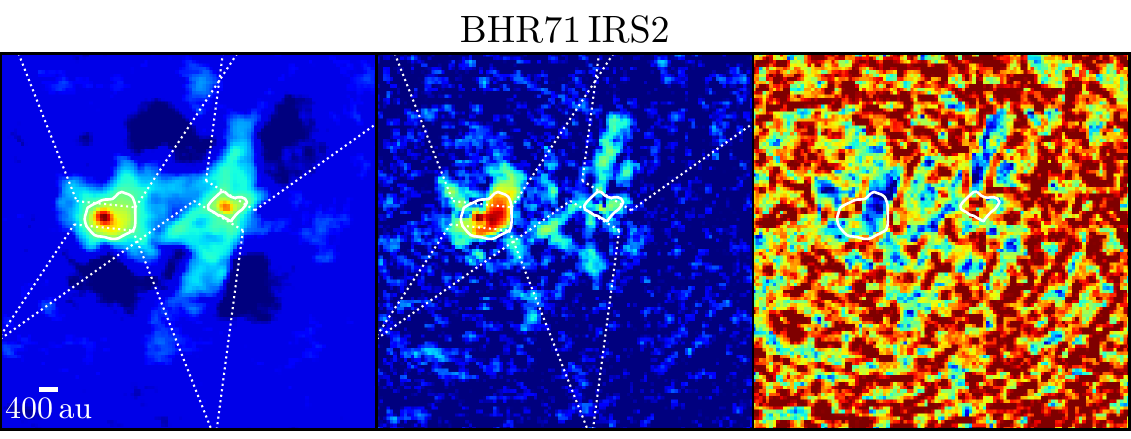}}
\vspace{-0.2cm}
\subfigure{\includegraphics[scale=\scaleSP,clip,trim= 1.9cm 3.1cm 0.2cm 2.6cm]{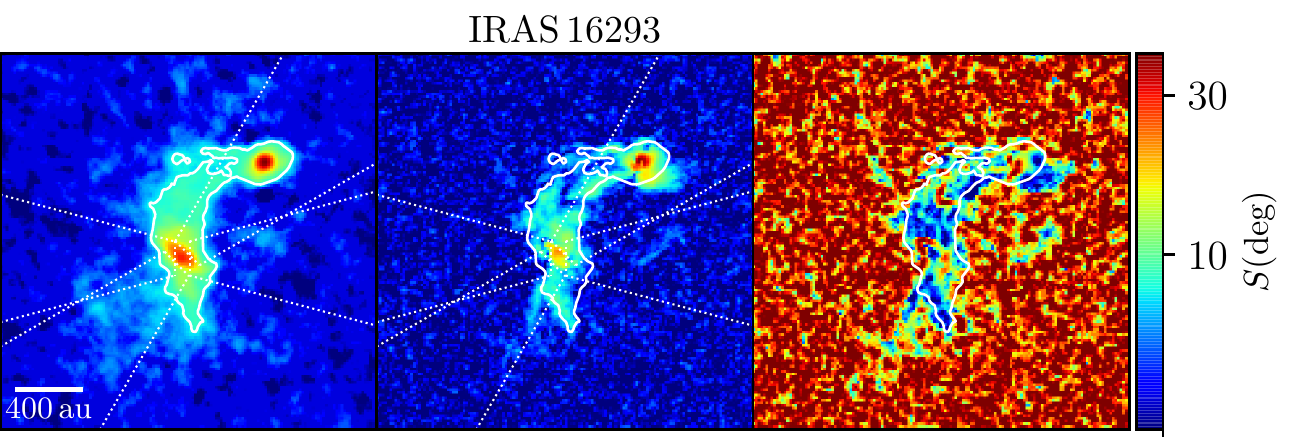}}
\vspace{-0.2cm}
\subfigure{\includegraphics[scale=\scaleSP,clip,trim= 1.9cm 3.1cm 1.85cm 2.6cm]{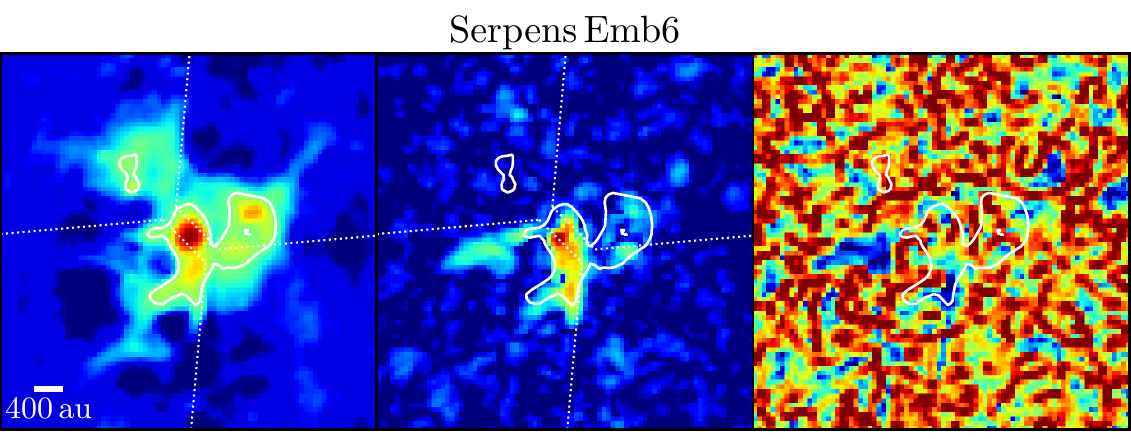}}
\vspace{-0.2cm}
\subfigure{\includegraphics[scale=\scaleSP,clip,trim= 1.9cm 3.1cm 0.2cm 2.6cm]{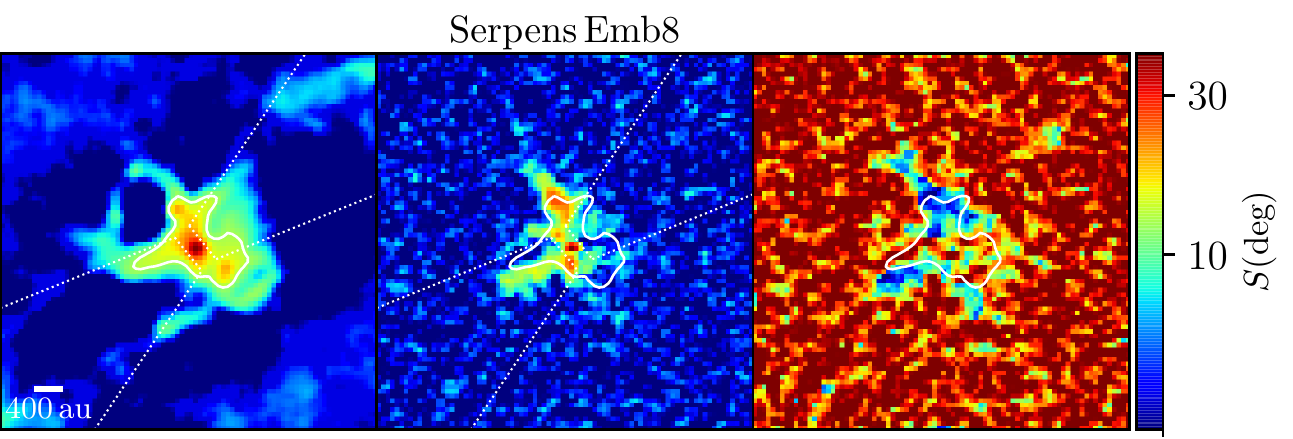}}
\vspace{-0.2cm}
\subfigure{\includegraphics[scale=\scaleSP,clip,trim= 1.9cm 3.1cm 1.85cm 2.6cm]{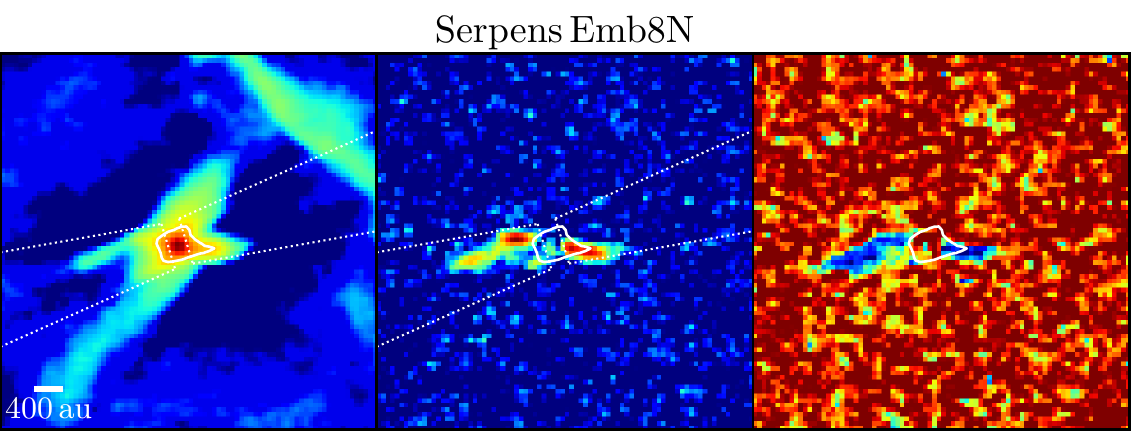}}
\vspace{-0.2cm}
\subfigure{\includegraphics[scale=\scaleSP,clip,trim= 1.9cm 3.1cm 0.2cm 2.6cm]{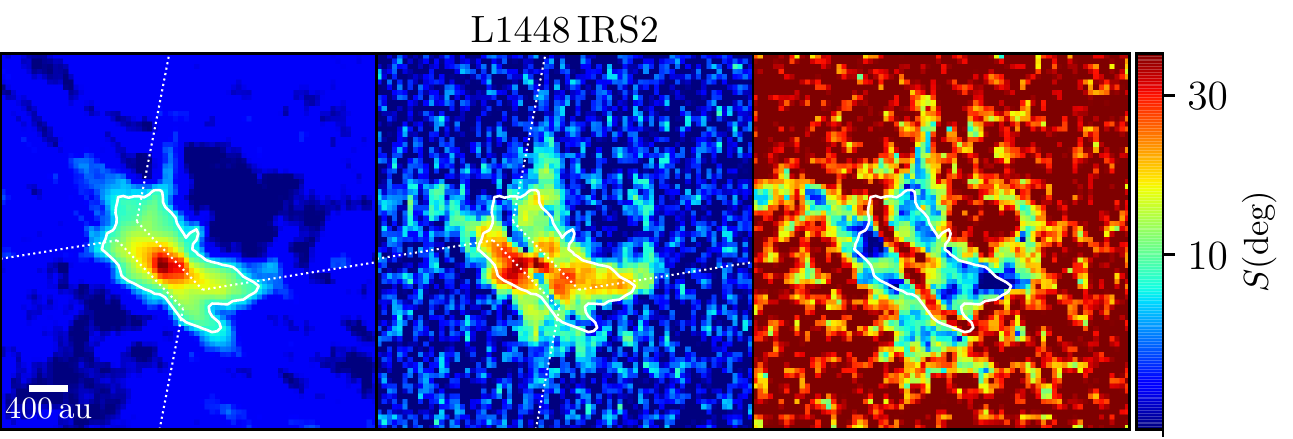}}
\vspace{-0.2cm}
\subfigure{\includegraphics[scale=\scaleSP,clip,trim= 1.9cm 3.1cm 1.85cm 2.6cm]{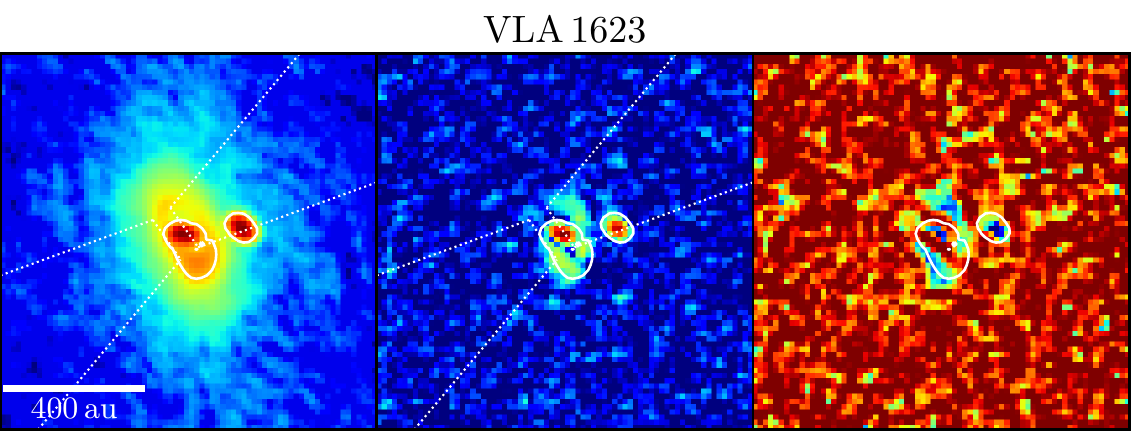}}
\vspace{-0.2cm}
\subfigure{\includegraphics[scale=\scaleSP,clip,trim= 1.9cm 3.1cm 0.2cm 2.6cm]{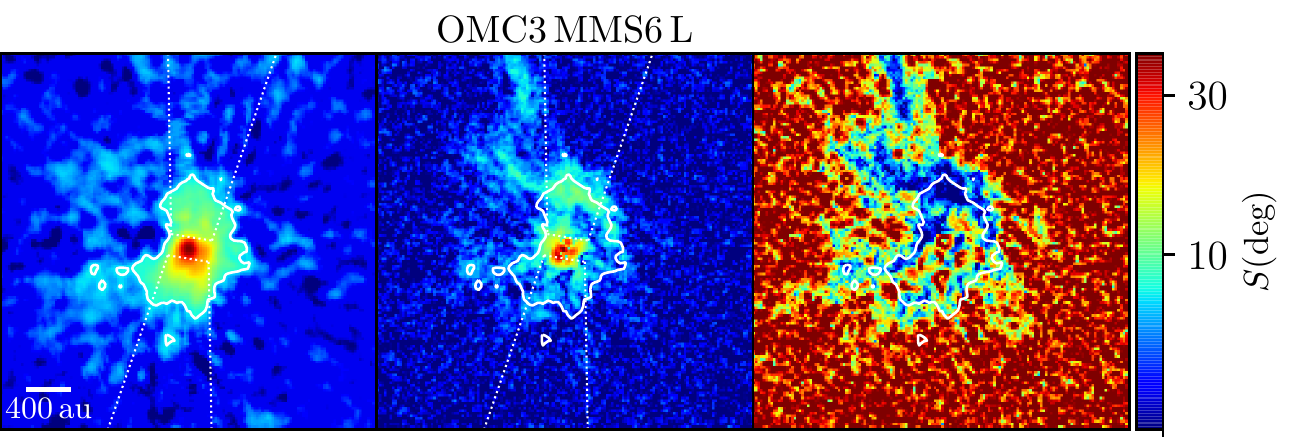}}
\vspace{-0.2cm}
\subfigure{\includegraphics[scale=\scaleSP,clip,trim= 1.9cm 3.1cm 1.85cm 2.6cm]{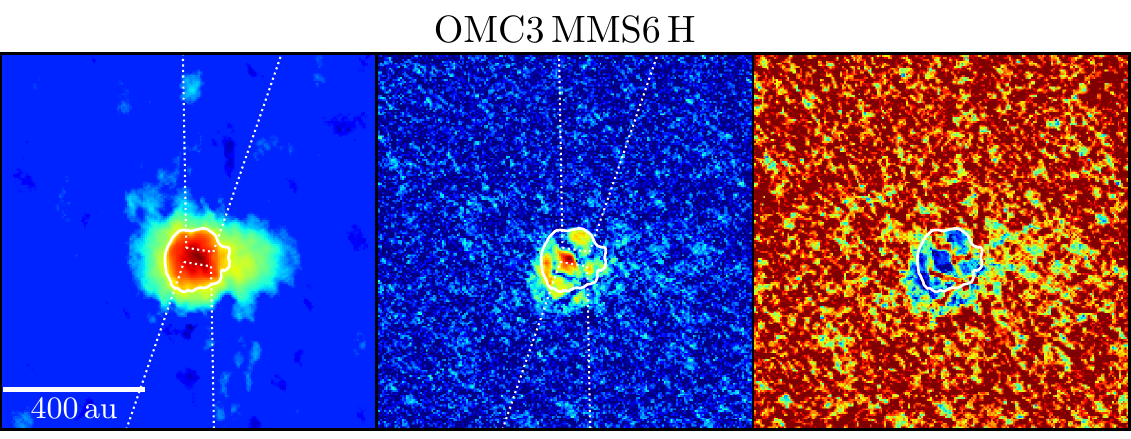}}
\vspace{-0.2cm}
\subfigure{\includegraphics[scale=\scaleSP,clip,trim= 1.9cm 3.1cm 0.2cm 2.6cm]{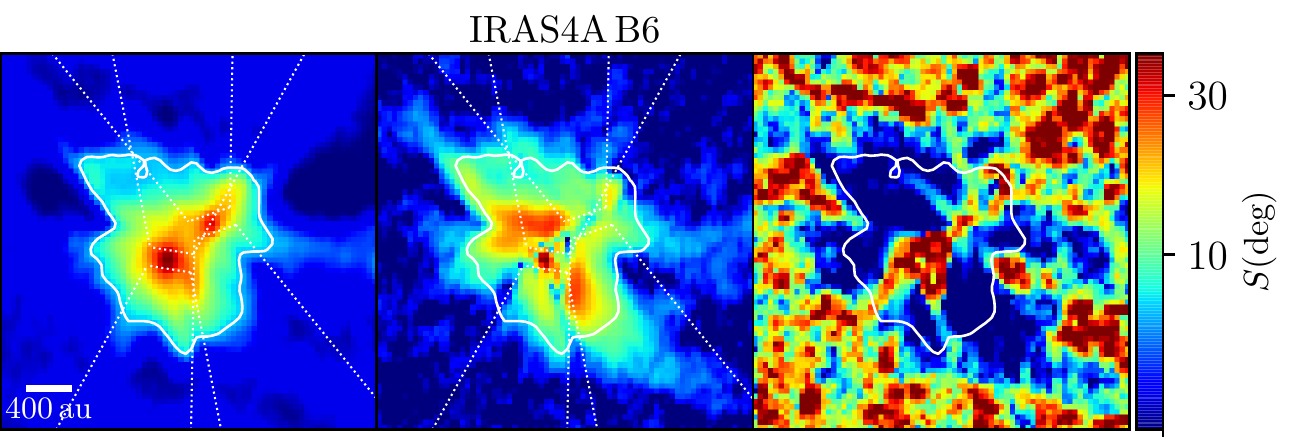}}
\vspace{-0.2cm}
\subfigure{\includegraphics[scale=\scaleSP,clip,trim= 1.9cm 3.1cm 0.2cm 2.6cm]{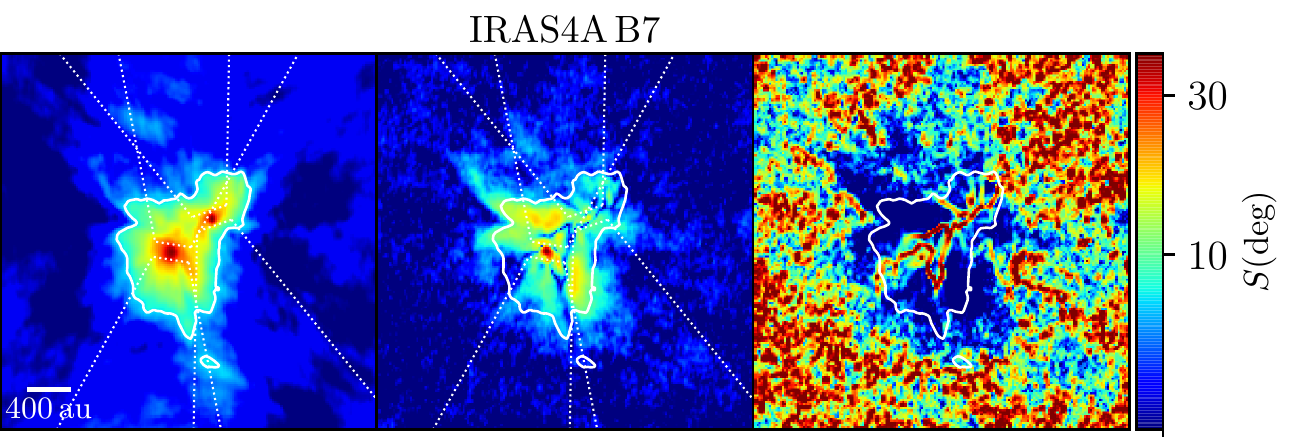}}
\caption[]{\small \textit{Left:} total intensity (Stokes $I$); \textit{center:} polarized intensity $P$; \textit{right:} dispersion of polarization angles $\S$. The solid white contour represents the threshold in Stokes $I$ used to select the data, as described in Section \ref{sec:method}.The white dotted lines denote the cone of the bipolar outflow that we use to separate the statistics in the outflow cavities versus those in the regions of the envelope not associated with the outflow cavities.} (see Figure \ref{fig:scheme_outflow_cavity}).
\label{fig:S_I_pol_maps_sources}
\vspace{0.2cm}
\end{figure*}

\def\scaleSP{0.226}
\def\extansion{}

\begin{figure*}[!tbph]
\centering
\subfigure{\includegraphics[scale=\scaleSP,clip,trim= 1.6cm 2.27cm 3cm 2.9cm]{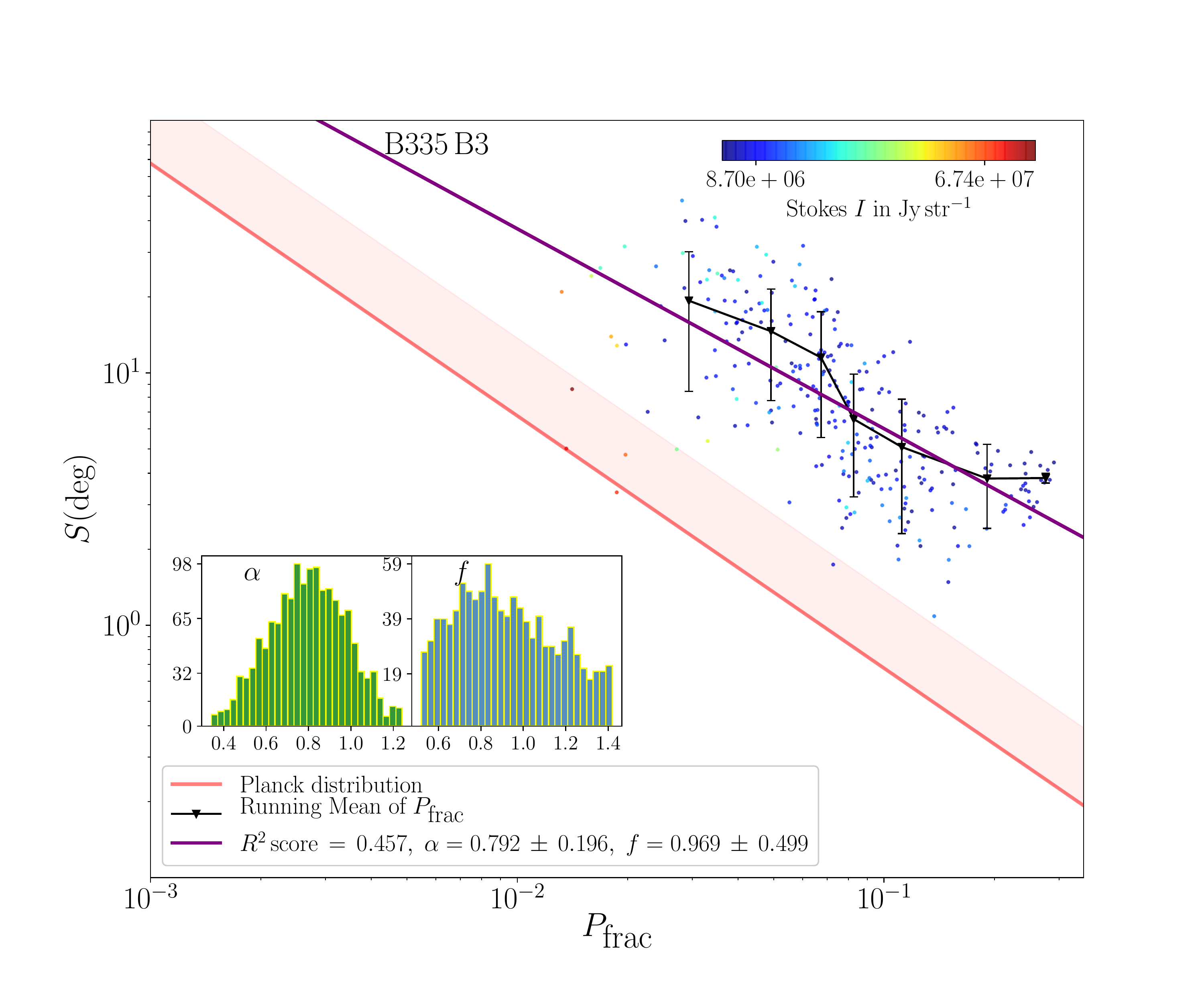}}
\vspace{-0.2cm}
\subfigure{\includegraphics[scale=\scaleSP,clip,trim= 1.6cm 2.27cm 3cm 2.9cm]{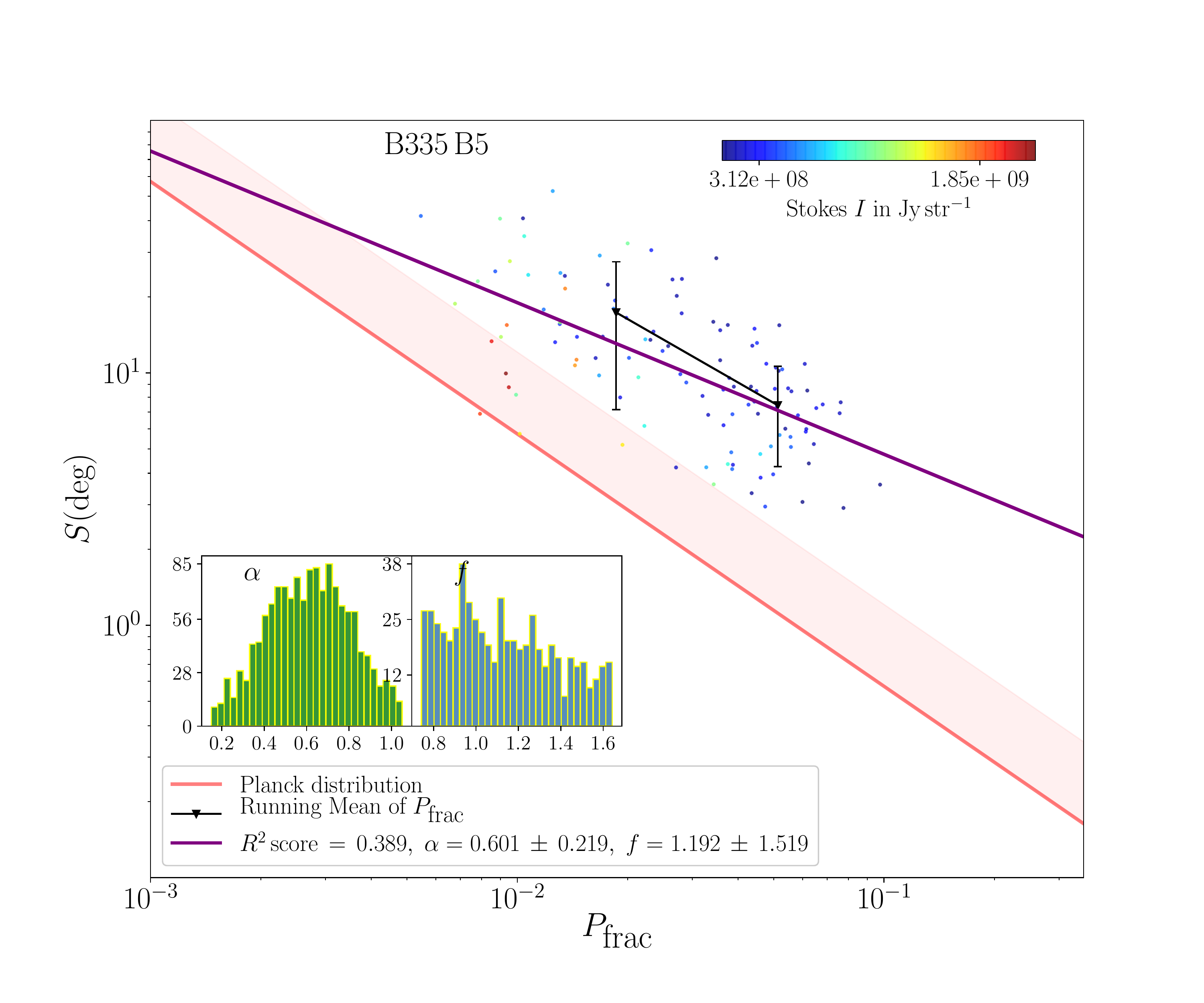}}
\vspace{-0.2cm}
\subfigure{\includegraphics[scale=\scaleSP,clip,trim= 1.6cm 2.27cm 3cm 2.9cm]{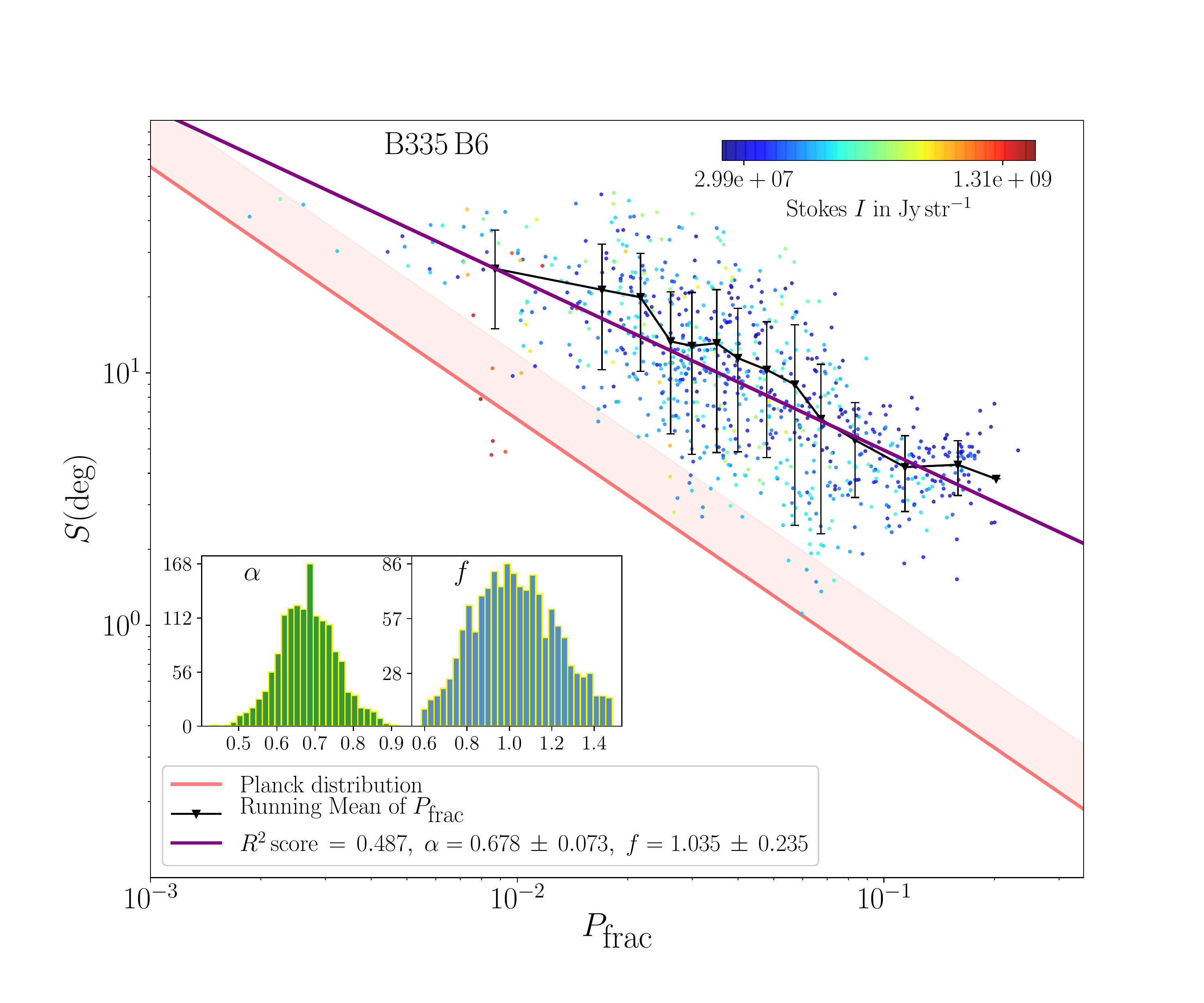}}
\subfigure{\includegraphics[scale=\scaleSP,clip,trim= 1.6cm 2.27cm 3cm 2.9cm]{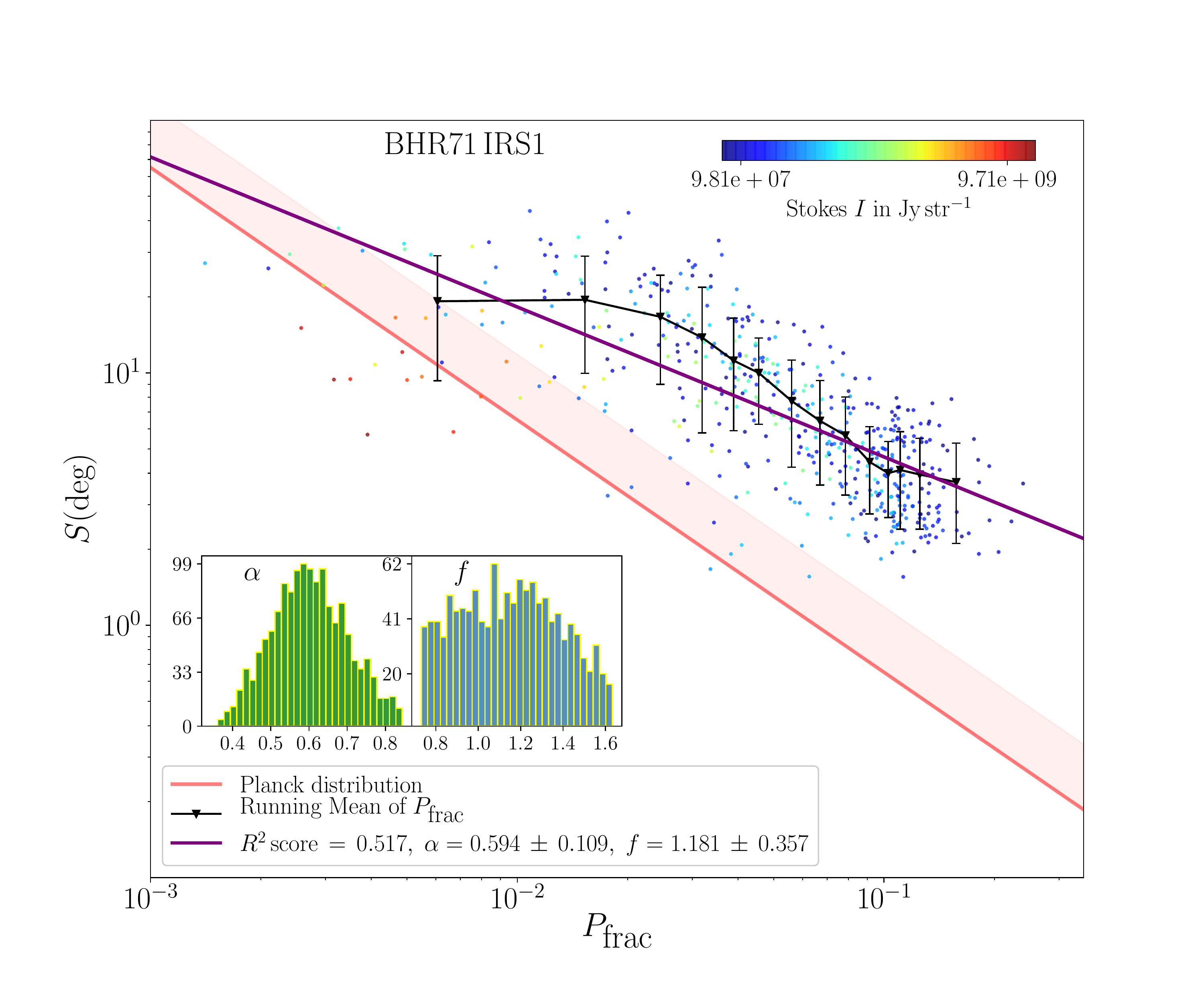}}
\vspace{-0.2cm}
\subfigure{\includegraphics[scale=\scaleSP,clip,trim= 1.6cm 2.27cm 3cm 2.9cm]{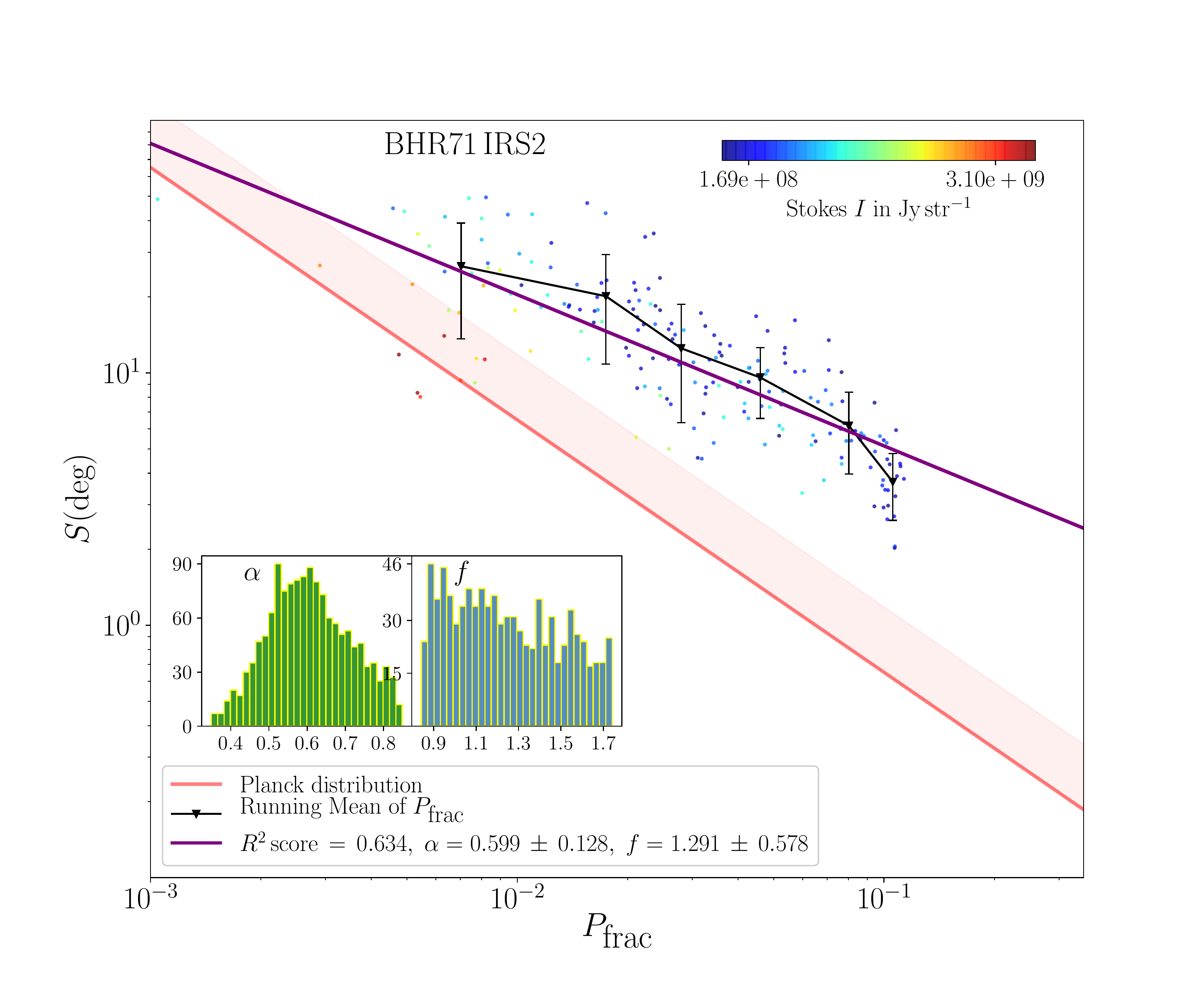}}
\vspace{-0.2cm}
\subfigure{\includegraphics[scale=\scaleSP,clip,trim= 1.6cm 2.27cm 3cm 2.9cm]{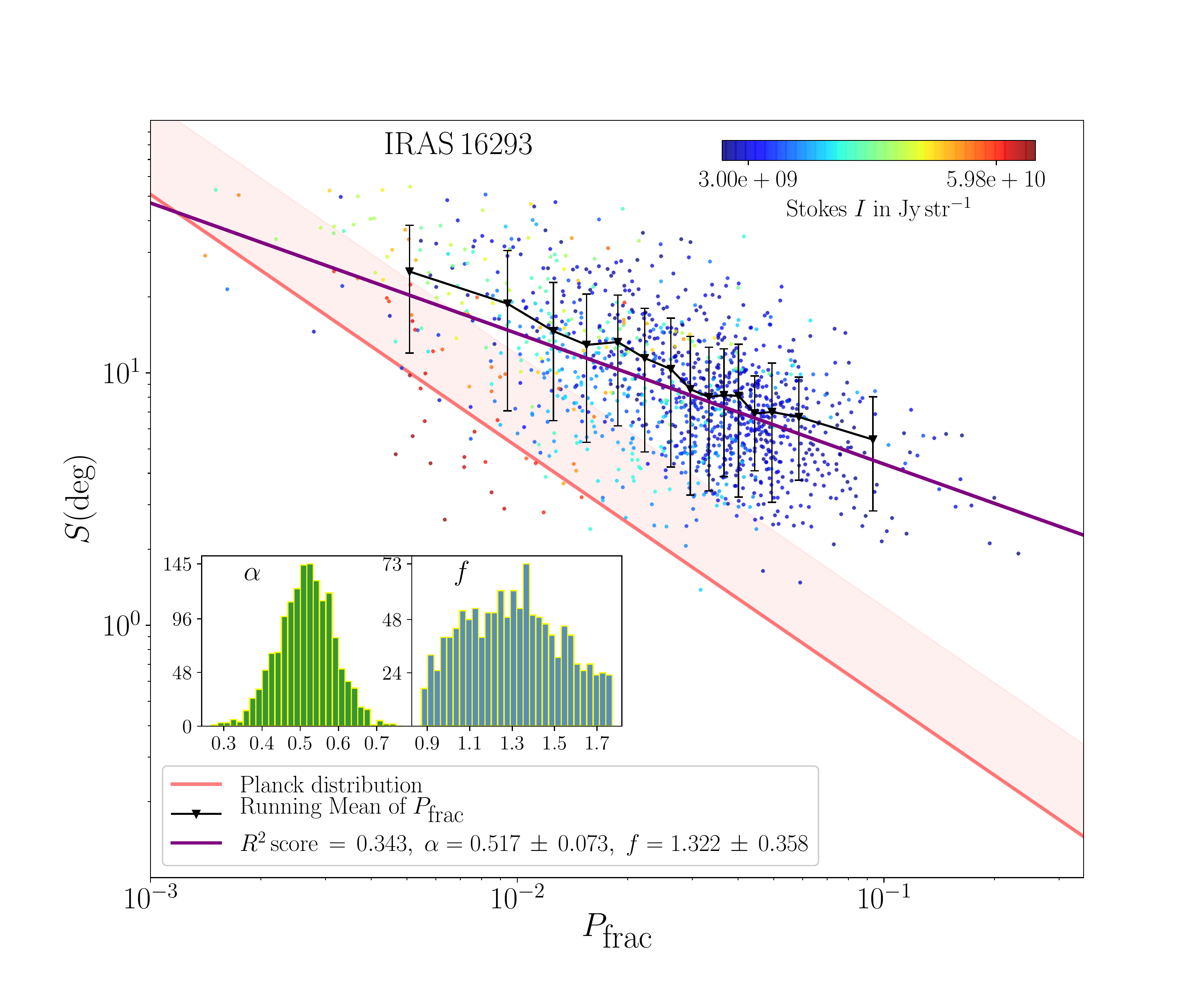}}
\subfigure{\includegraphics[scale=\scaleSP,clip,trim= 1.6cm 2.27cm 3cm 2.9cm]{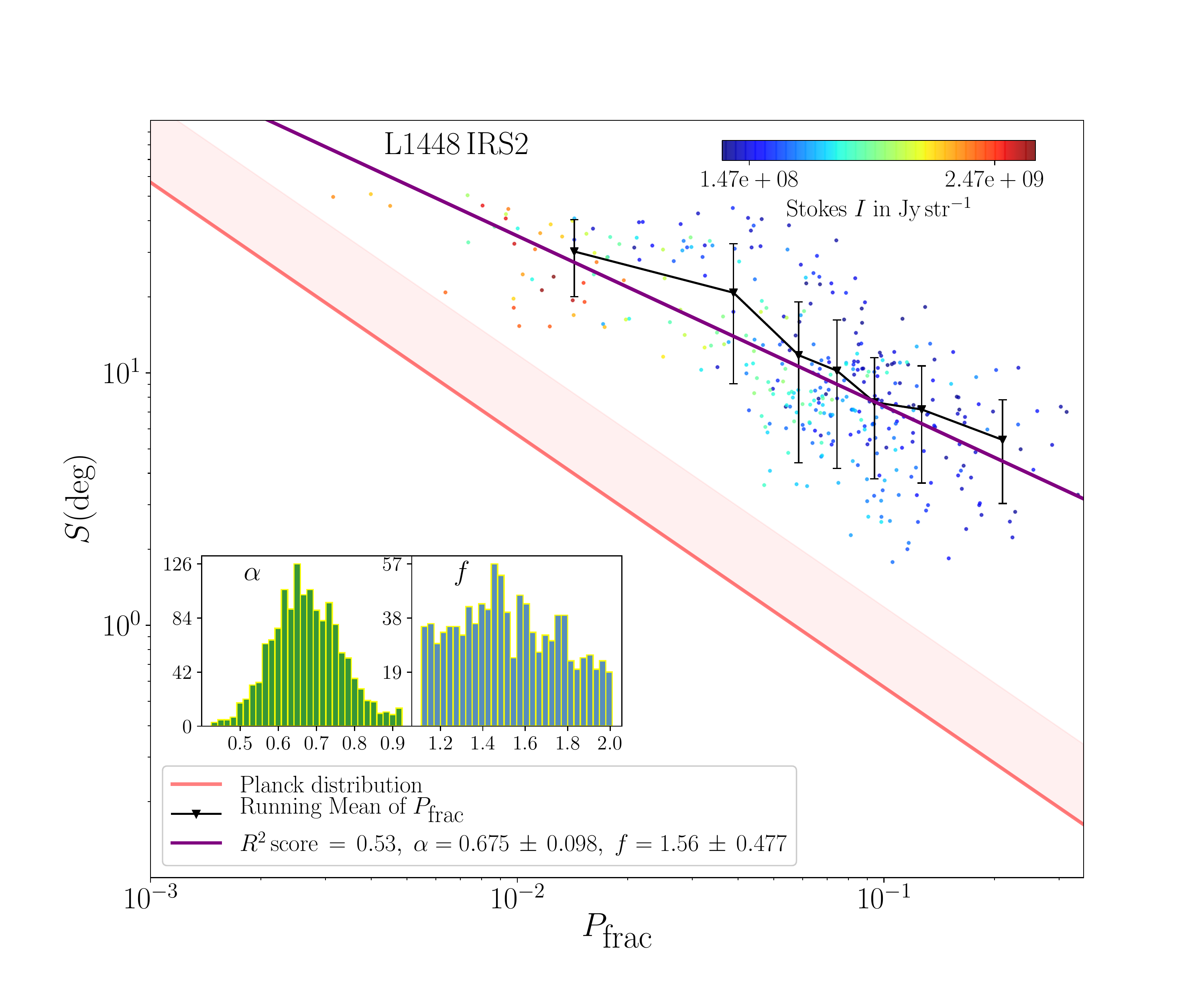}}
\vspace{-0.2cm}
\subfigure{\includegraphics[scale=\scaleSP,clip,trim= 1.6cm 2.27cm 3cm 2.9cm]{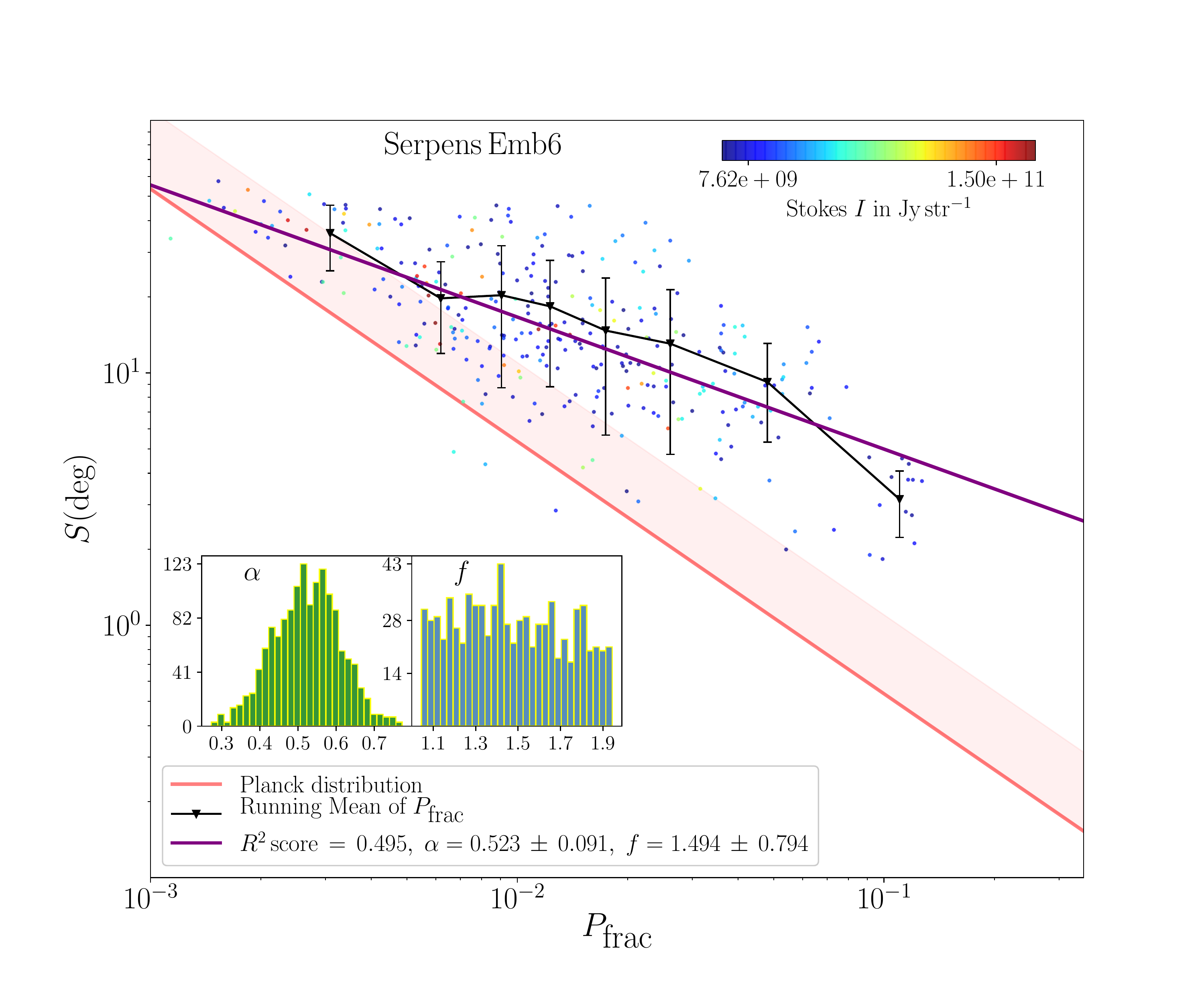}}
\vspace{-0.2cm}
\subfigure{\includegraphics[scale=\scaleSP,clip,trim= 1.6cm 2.27cm 3cm 2.9cm]{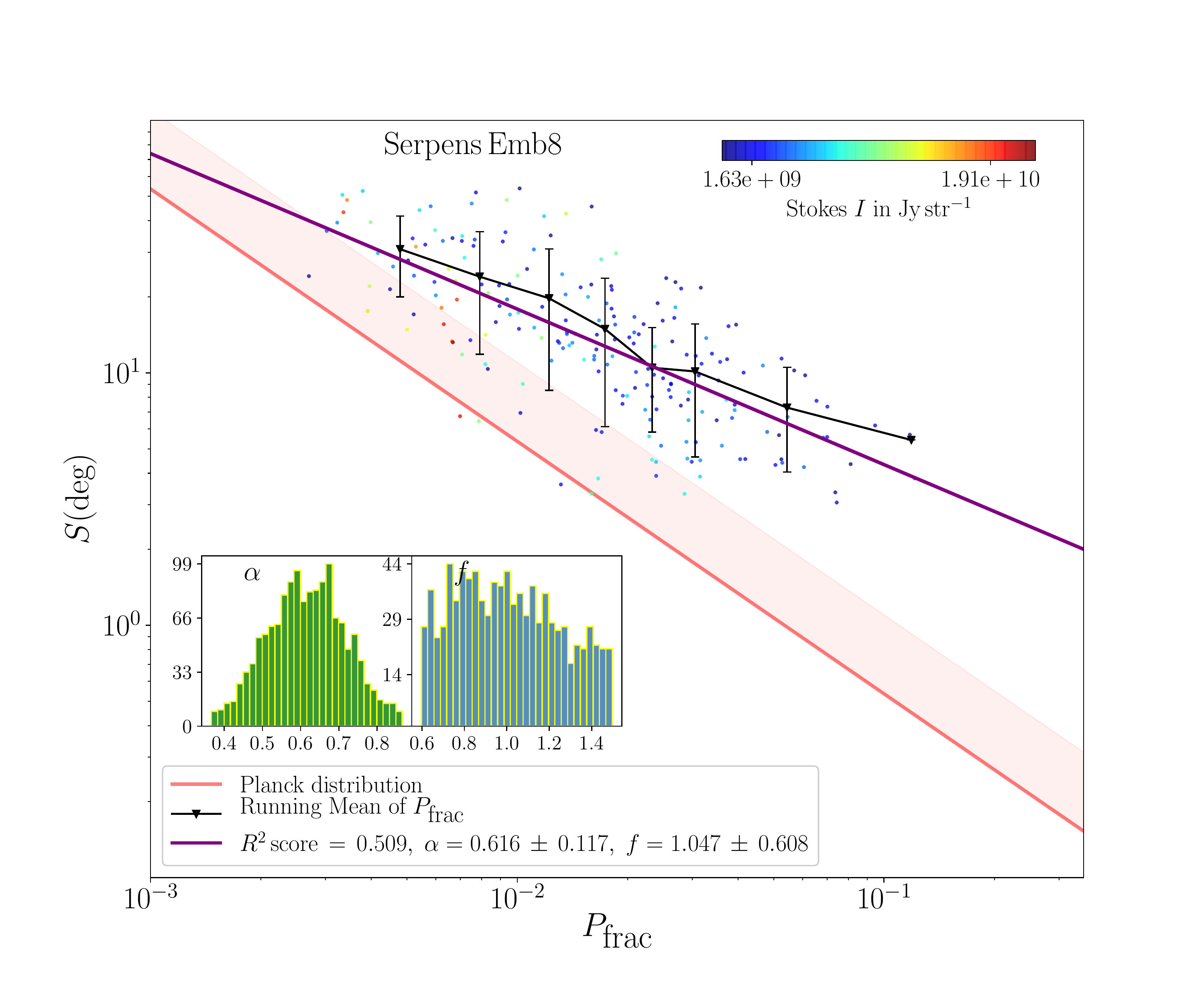}}
\subfigure{\includegraphics[scale=\scaleSP,clip,trim= 1.6cm 2.27cm 3cm 2.9cm]{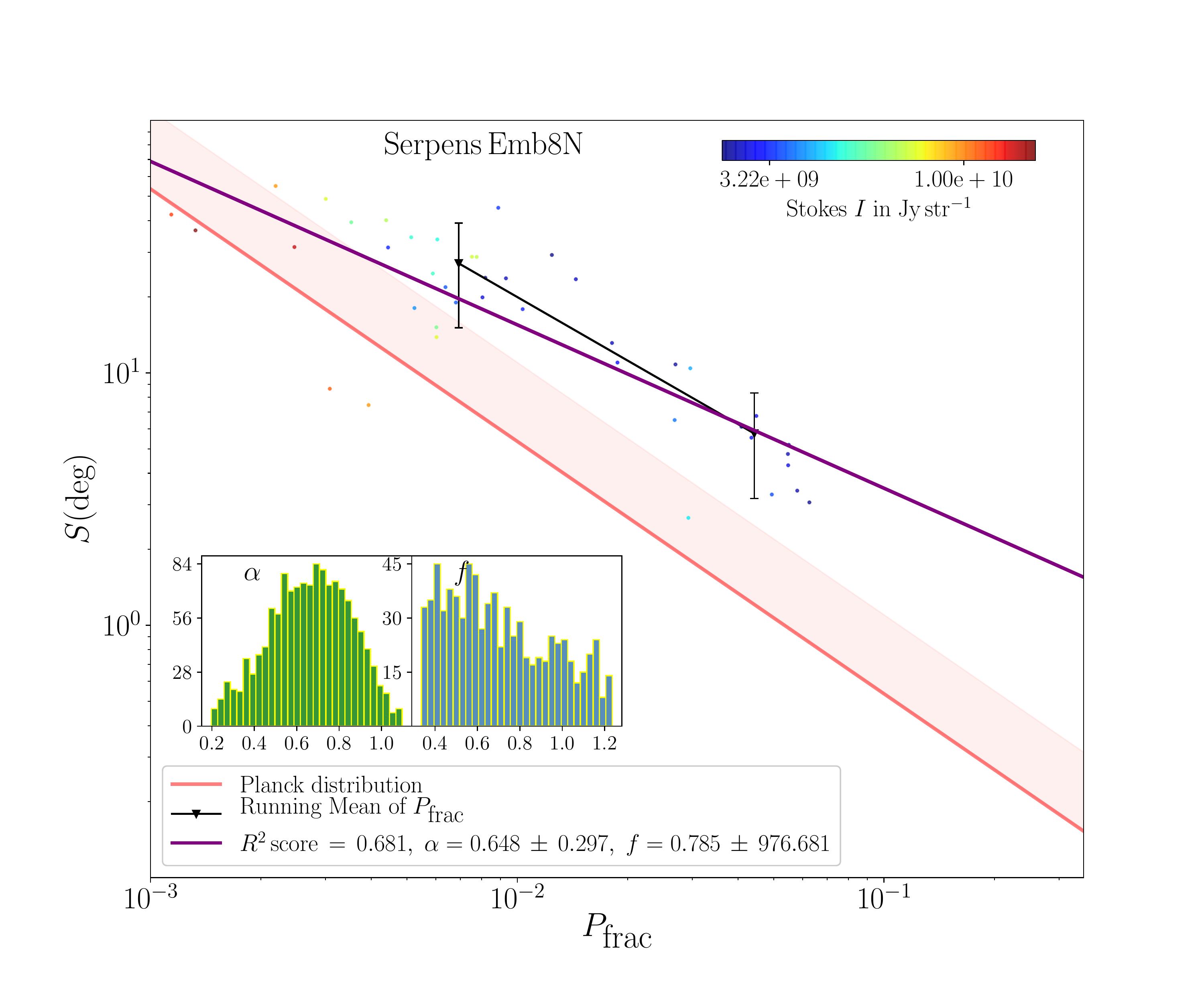}}
\vspace{-0.2cm}
\subfigure{\includegraphics[scale=\scaleSP,clip,trim= 1.6cm 2.27cm 3cm 2.9cm]{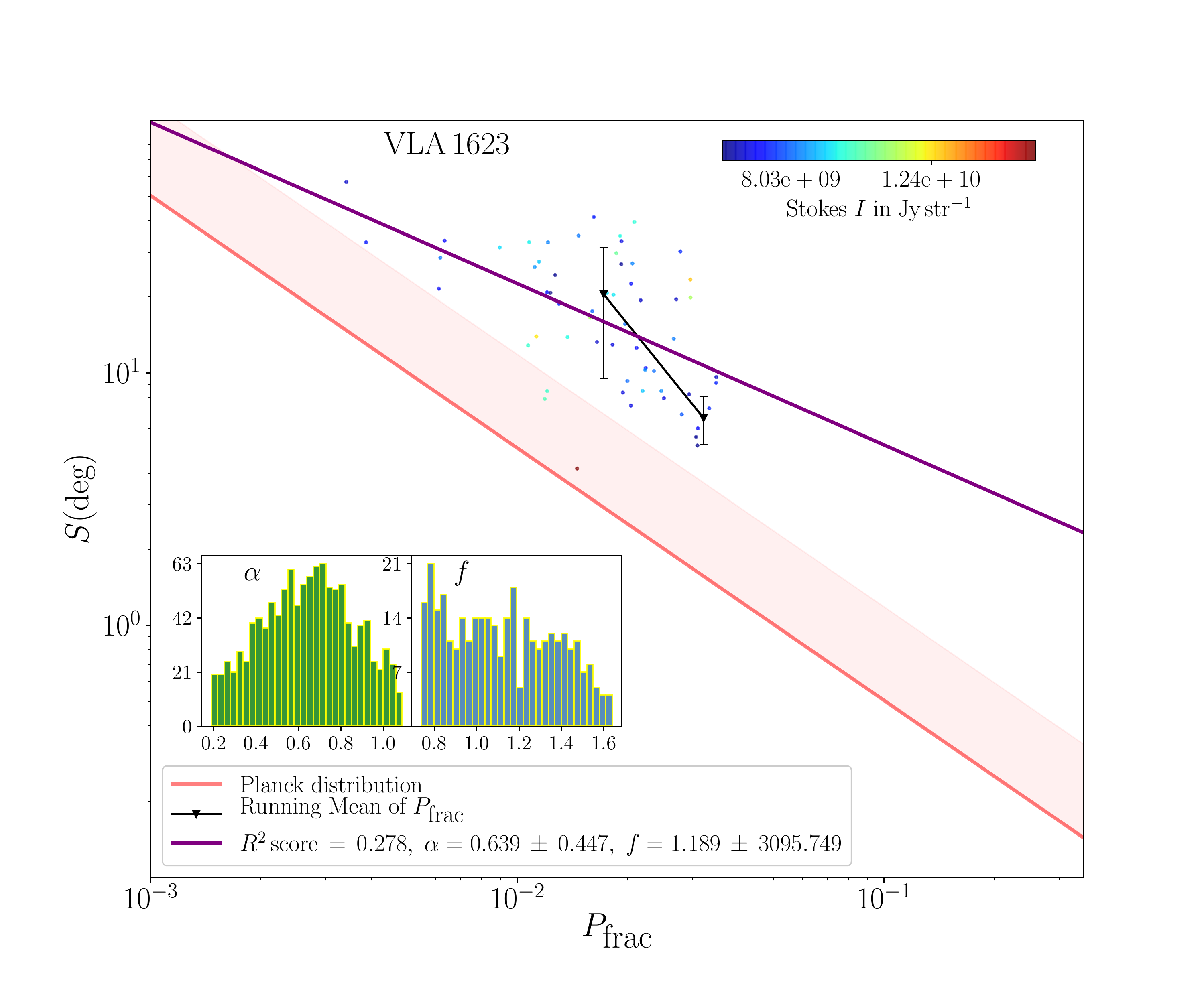}}
\vspace{-0.2cm}
\subfigure{\includegraphics[scale=\scaleSP,clip,trim= 1.6cm 2.27cm 3cm 2.9cm]{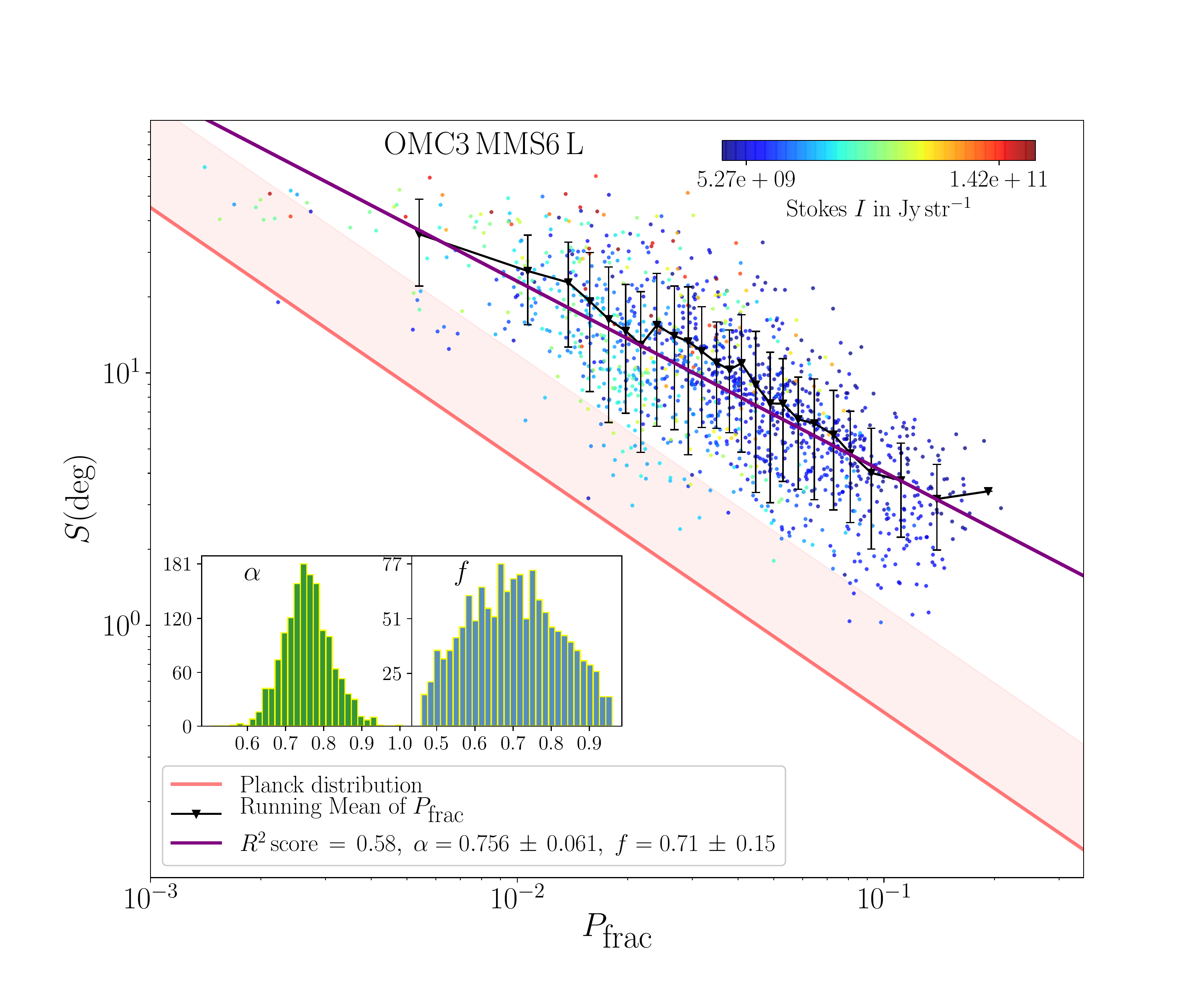}}
\subfigure{\includegraphics[scale=\scaleSP,clip,trim= 1.6cm 1.4cm 3cm 2.9cm]{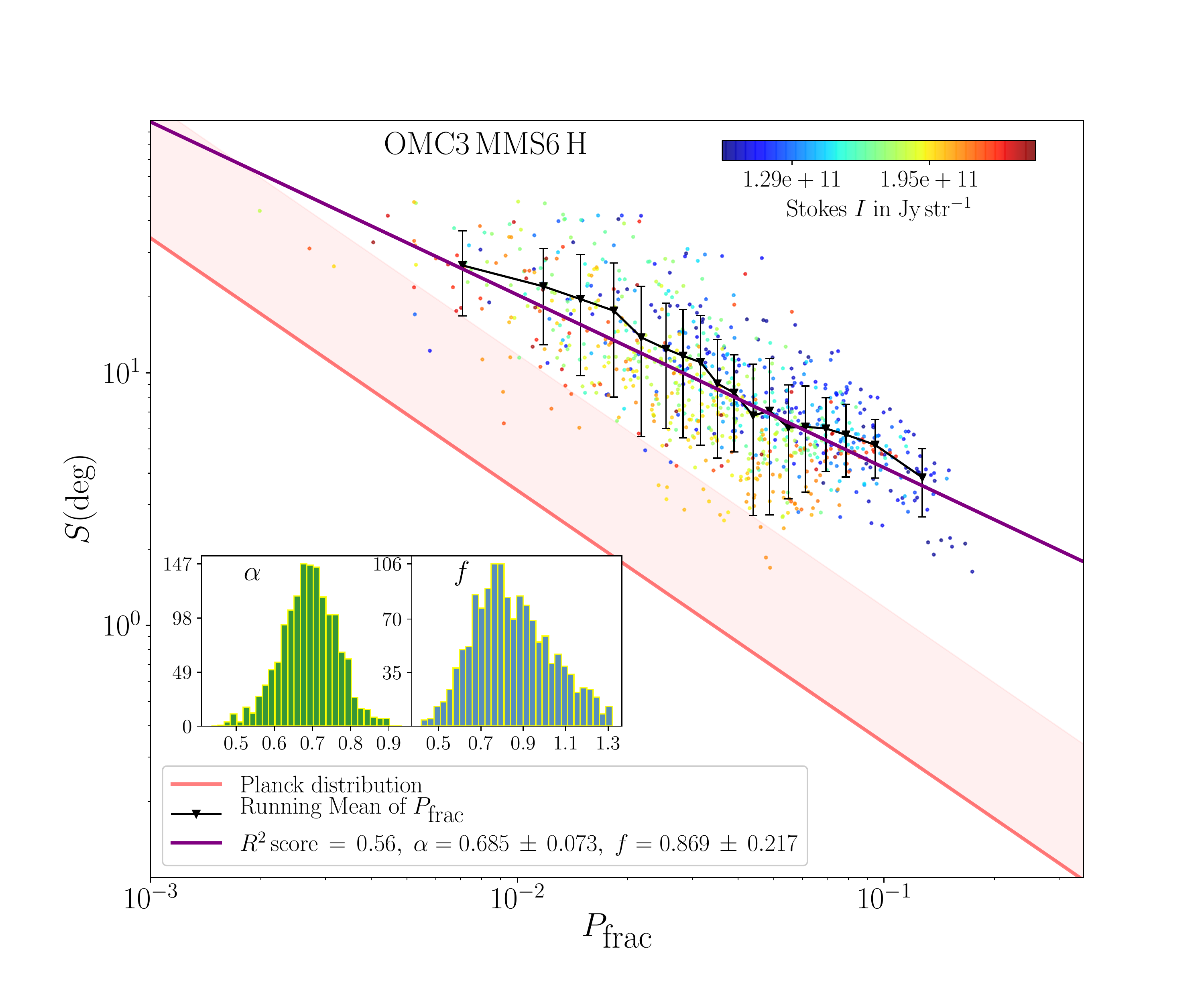}}
\subfigure{\includegraphics[scale=\scaleSP,clip,trim= 1.6cm 1.4cm 3cm 2.9cm]{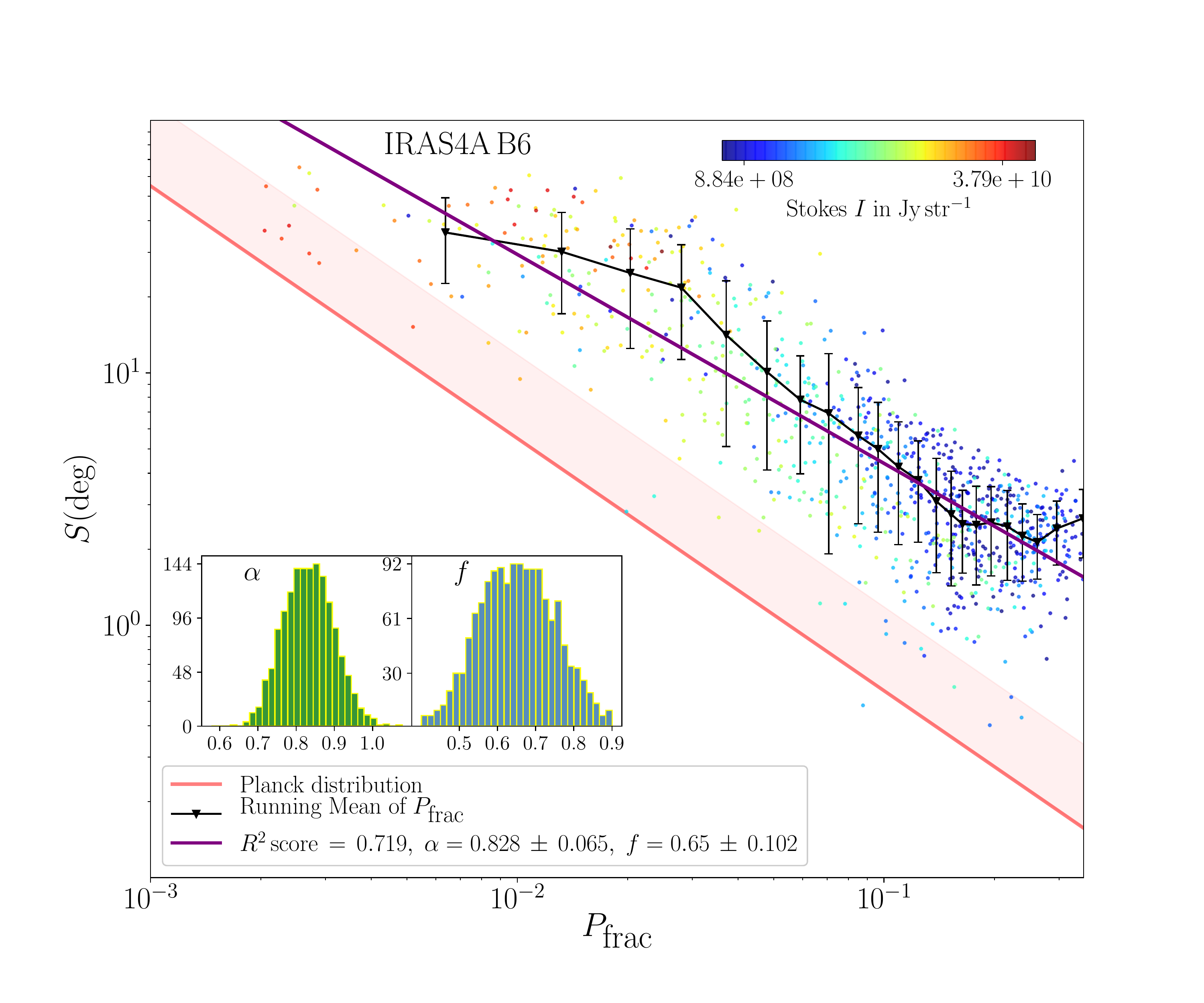}}
\subfigure{\includegraphics[scale=\scaleSP,clip,trim= 1.6cm 1.4cm 3cm 2.9cm]{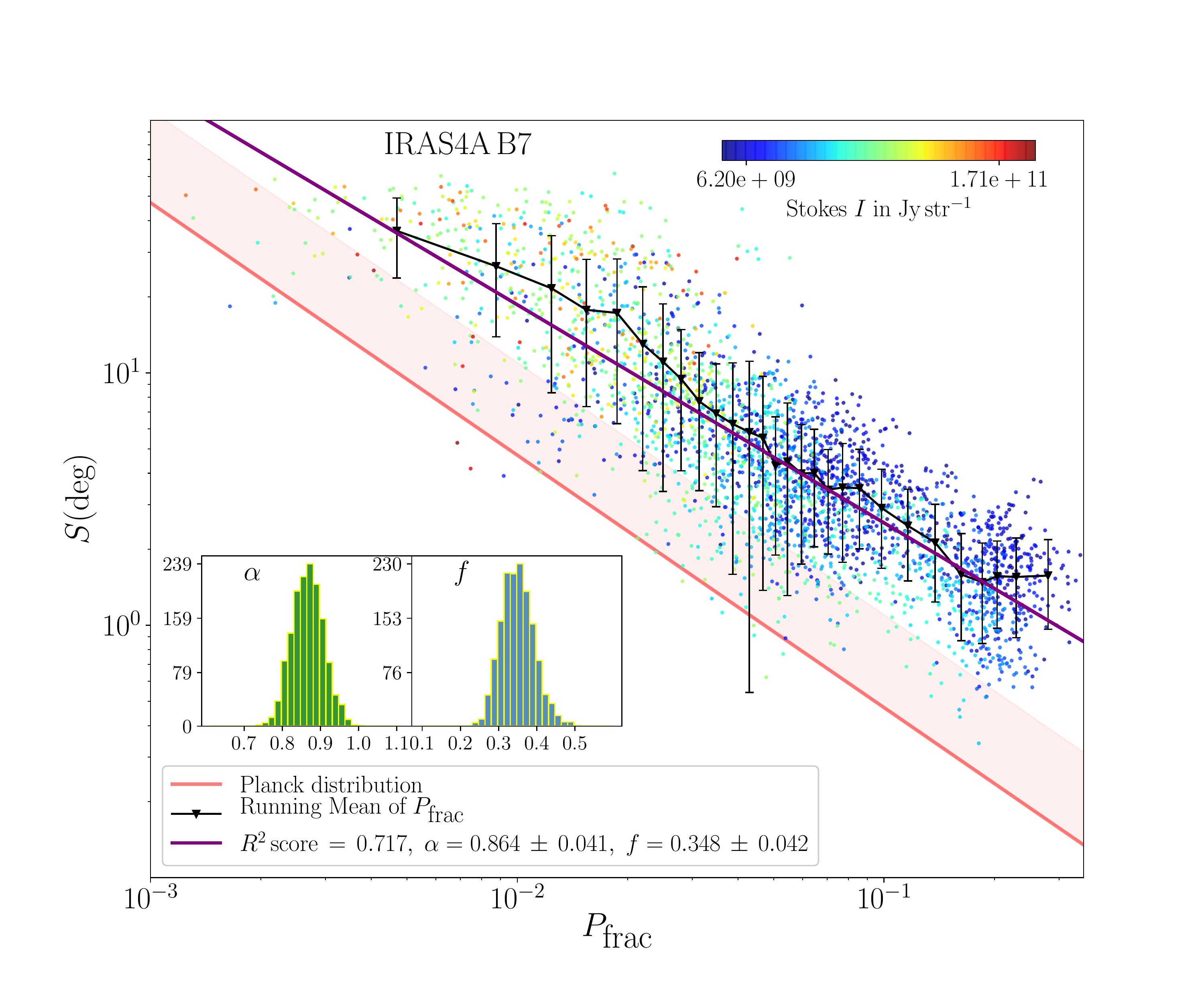}}
\caption[]{\small  Distributions of the dispersion of polarization position angles $\S$ with respect to the polarization fraction $\Pf$ for each of the sources of our sample. Same as Figure \ref{fig:scatter_S_Pfrac_Alma_merged}, except that here the color scale represents the value of Stokes $I$ of the corresponding point, in Jy str$^{-1}$.}
\vspace{0.2cm}
\label{fig:scatter_S_Pfrac_Alma}
\end{figure*}

\def\scaleSP{0.221}
\def\extansion{}

\begin{figure*}[!tbph]
\centering
\subfigure{\includegraphics[scale=\scaleSP,clip,trim= 0.4cm 2.3cm 0.4cm 2cm]{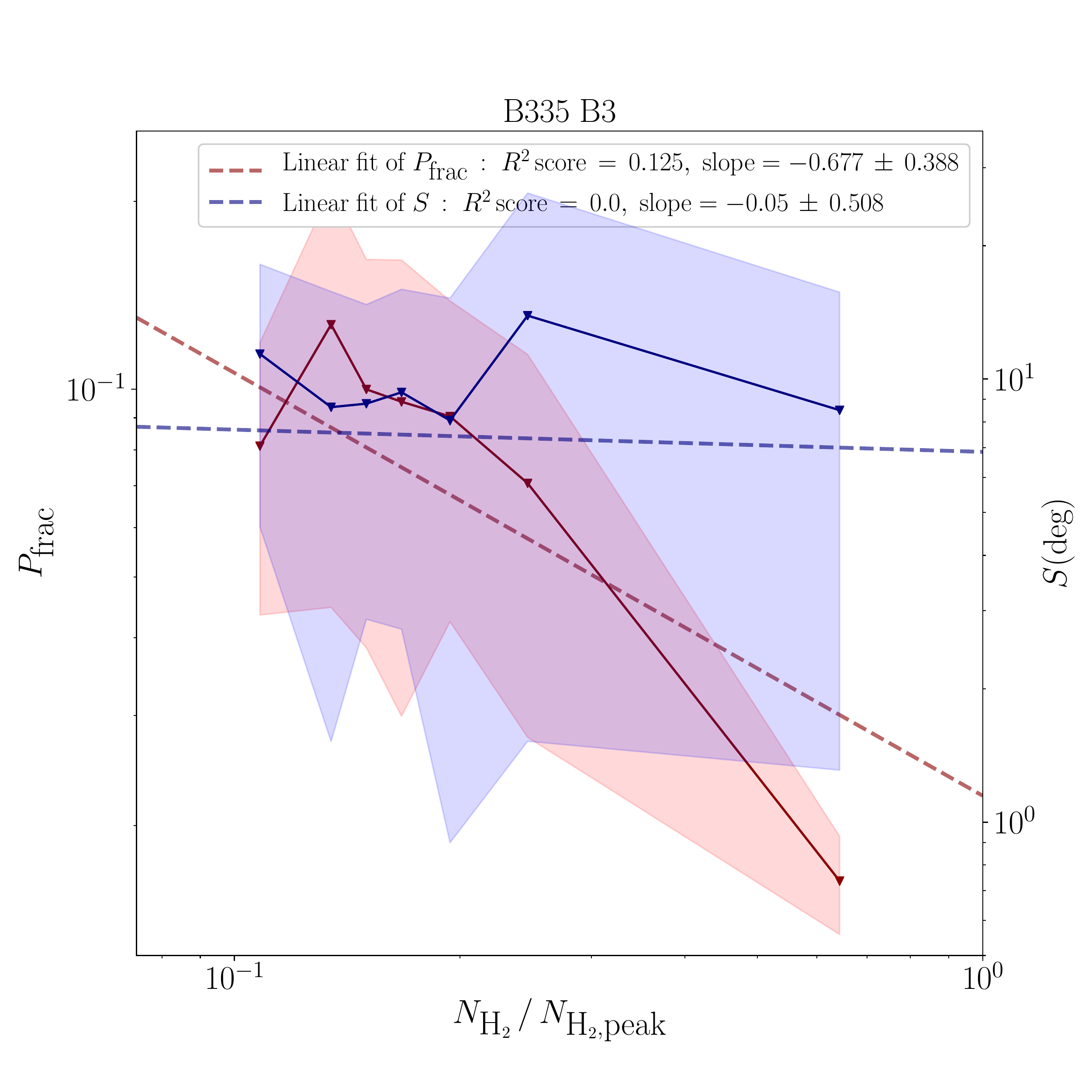}}
\vspace{-0.2cm}
\subfigure{\includegraphics[scale=\scaleSP,clip,trim= 0.4cm 2.3cm 0.4cm 2cm]{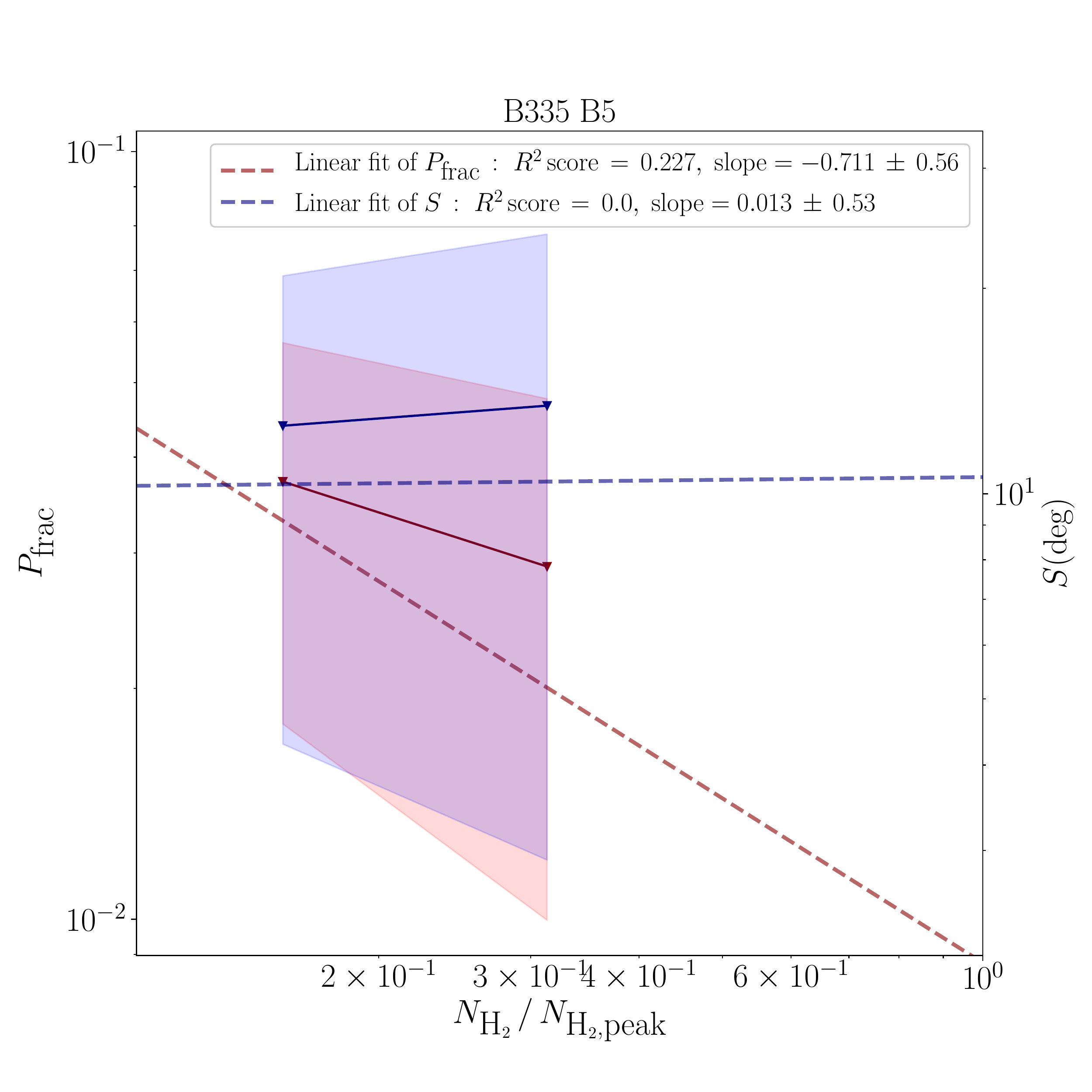}}
\vspace{-0.2cm}
\subfigure{\includegraphics[scale=\scaleSP,clip,trim= 0.4cm 2.3cm 0.4cm 2cm]{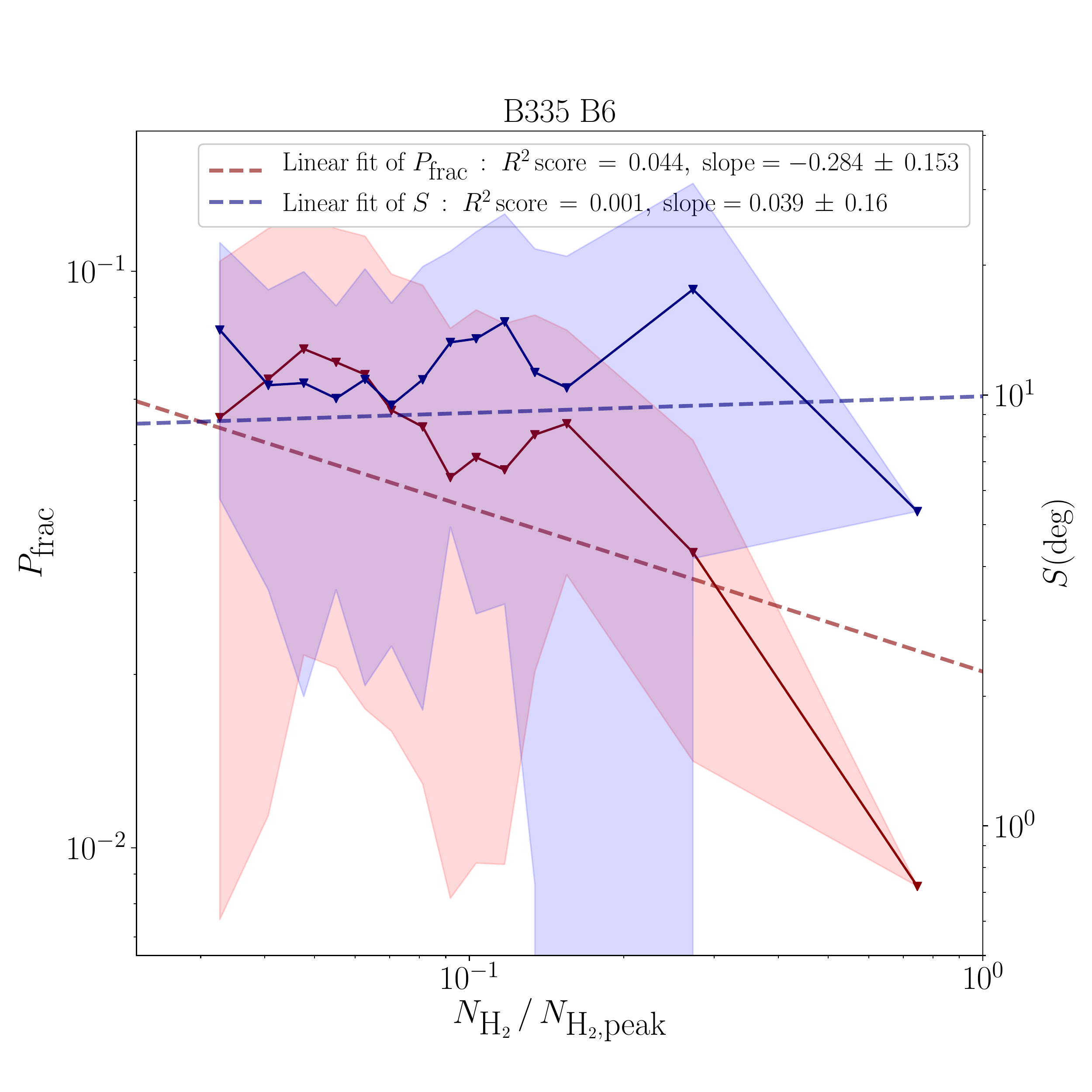}}
\subfigure{\includegraphics[scale=\scaleSP,clip,trim= 0.4cm 2.3cm 0.4cm 2cm]{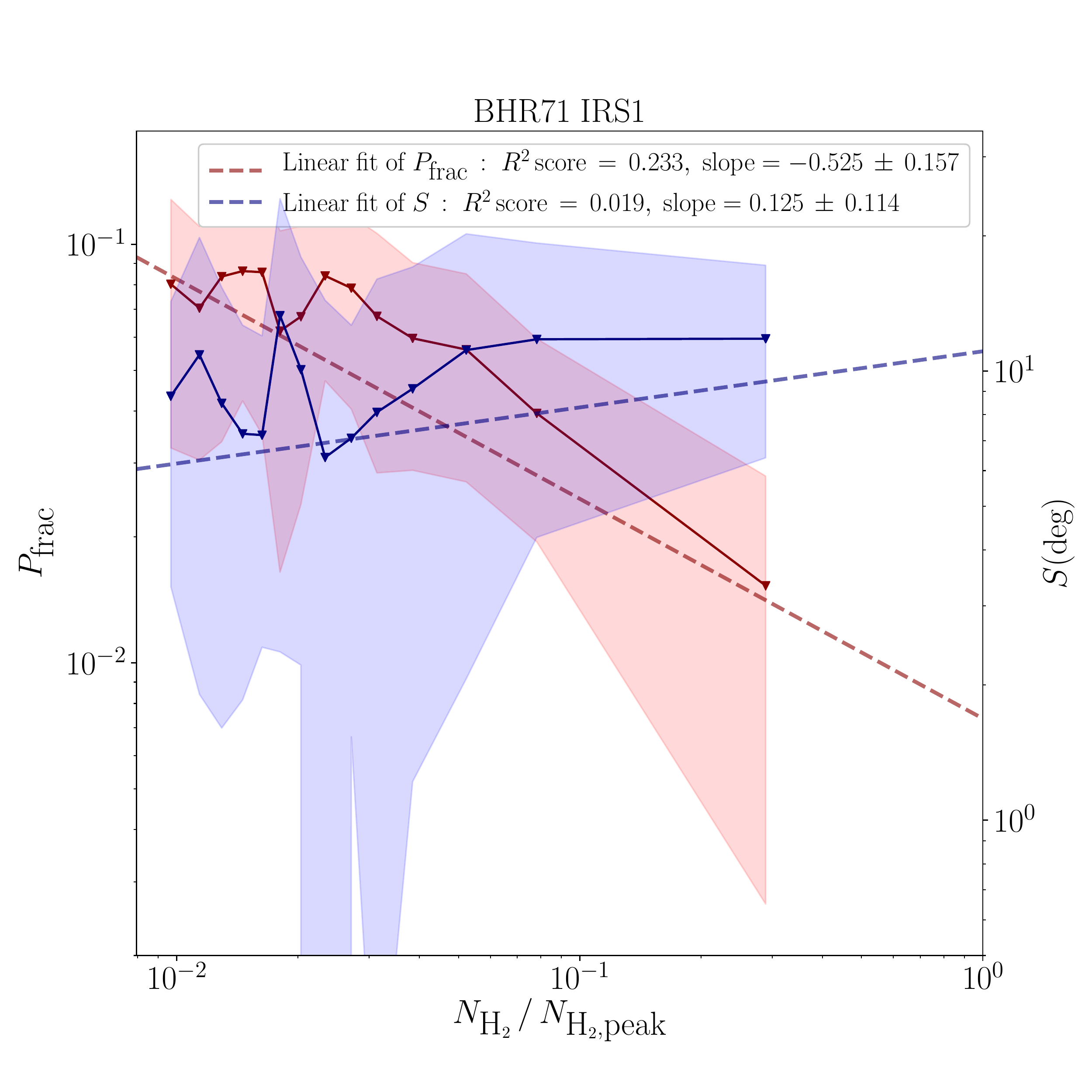}}
\vspace{-0.2cm}
\subfigure{\includegraphics[scale=\scaleSP,clip,trim= 0.4cm 2.3cm 0.4cm 2cm]{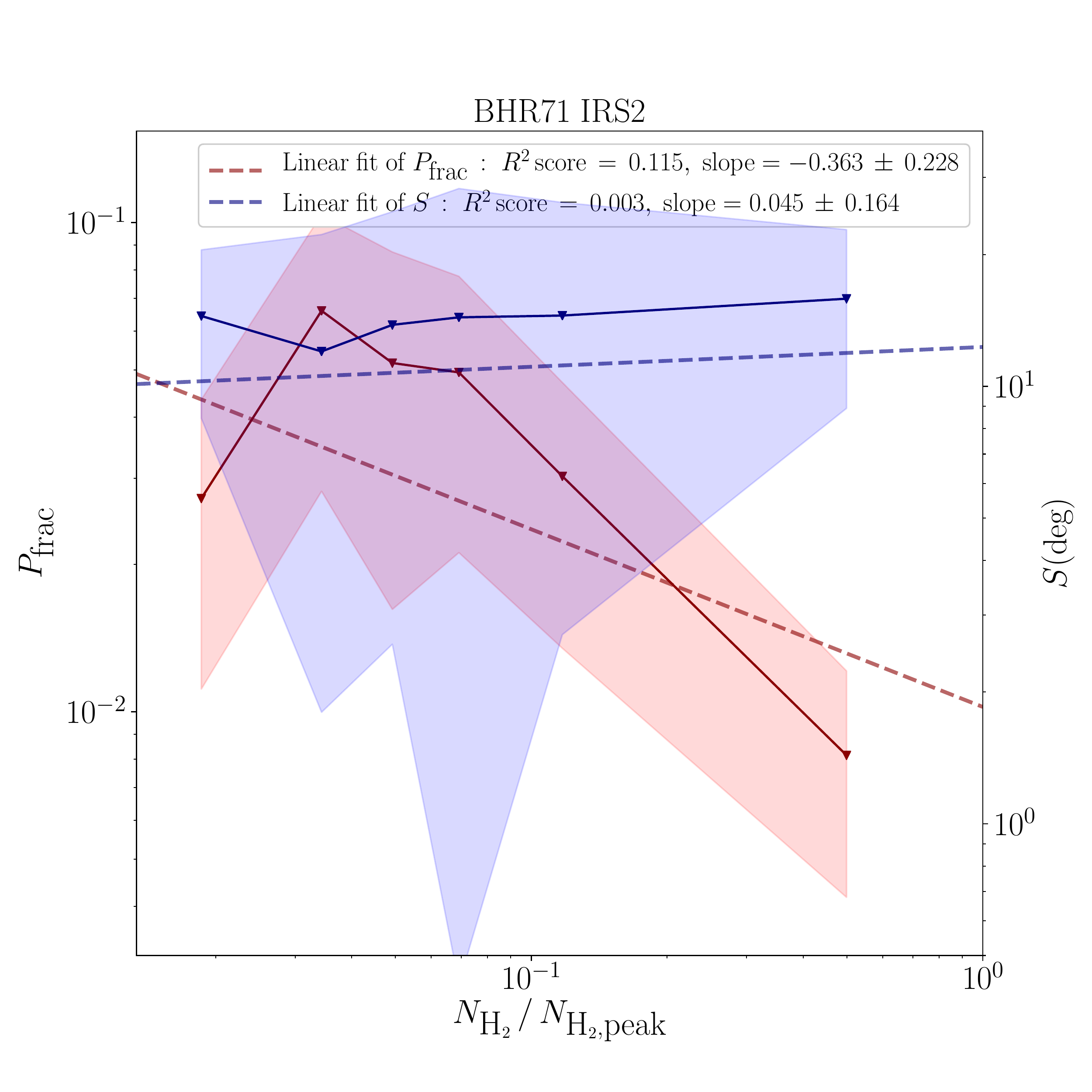}}
\vspace{-0.2cm}
\subfigure{\includegraphics[scale=\scaleSP,clip,trim= 0.4cm 2.3cm 0.4cm 2cm]{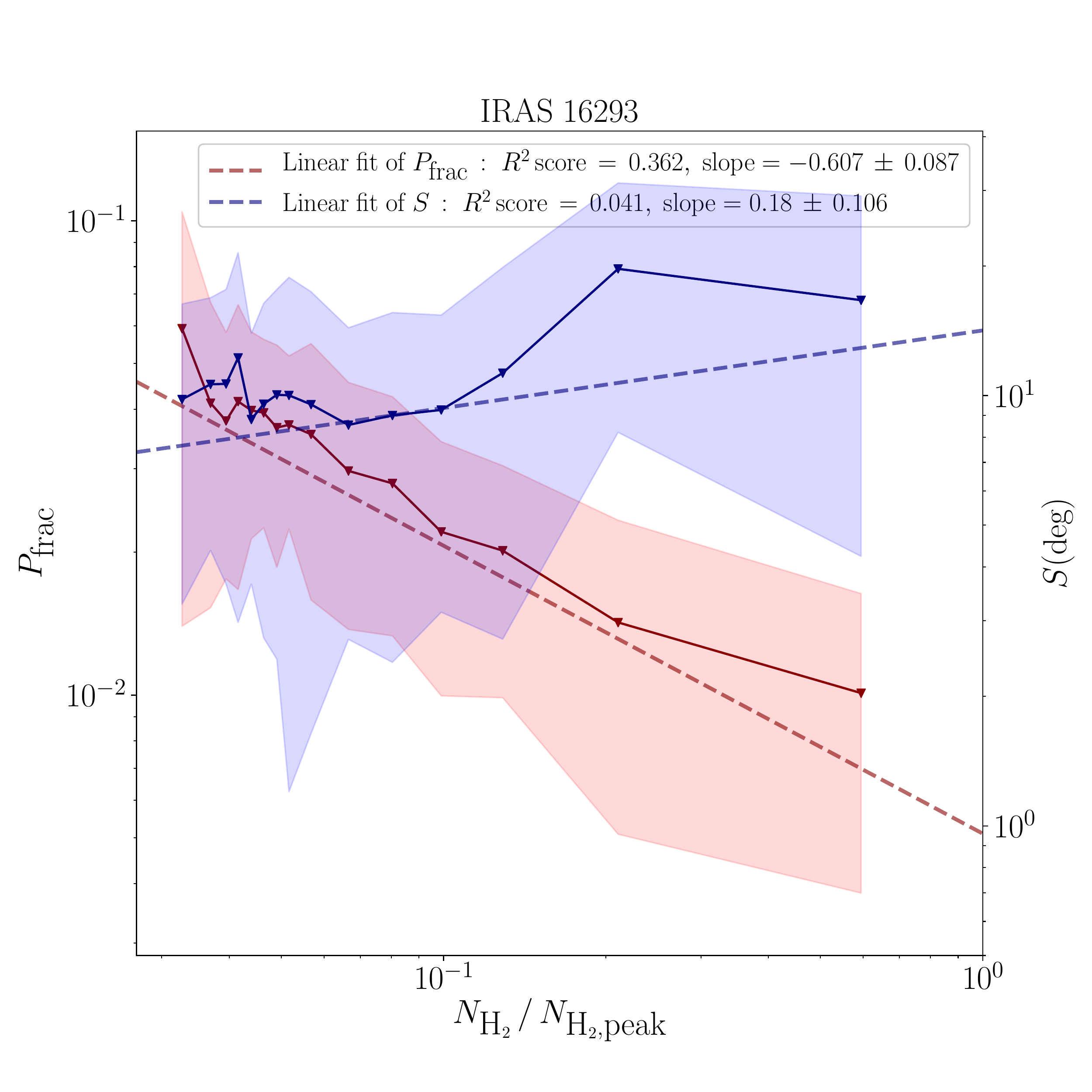}}
\subfigure{\includegraphics[scale=\scaleSP,clip,trim= 0.4cm 2.3cm 0.4cm 2cm]{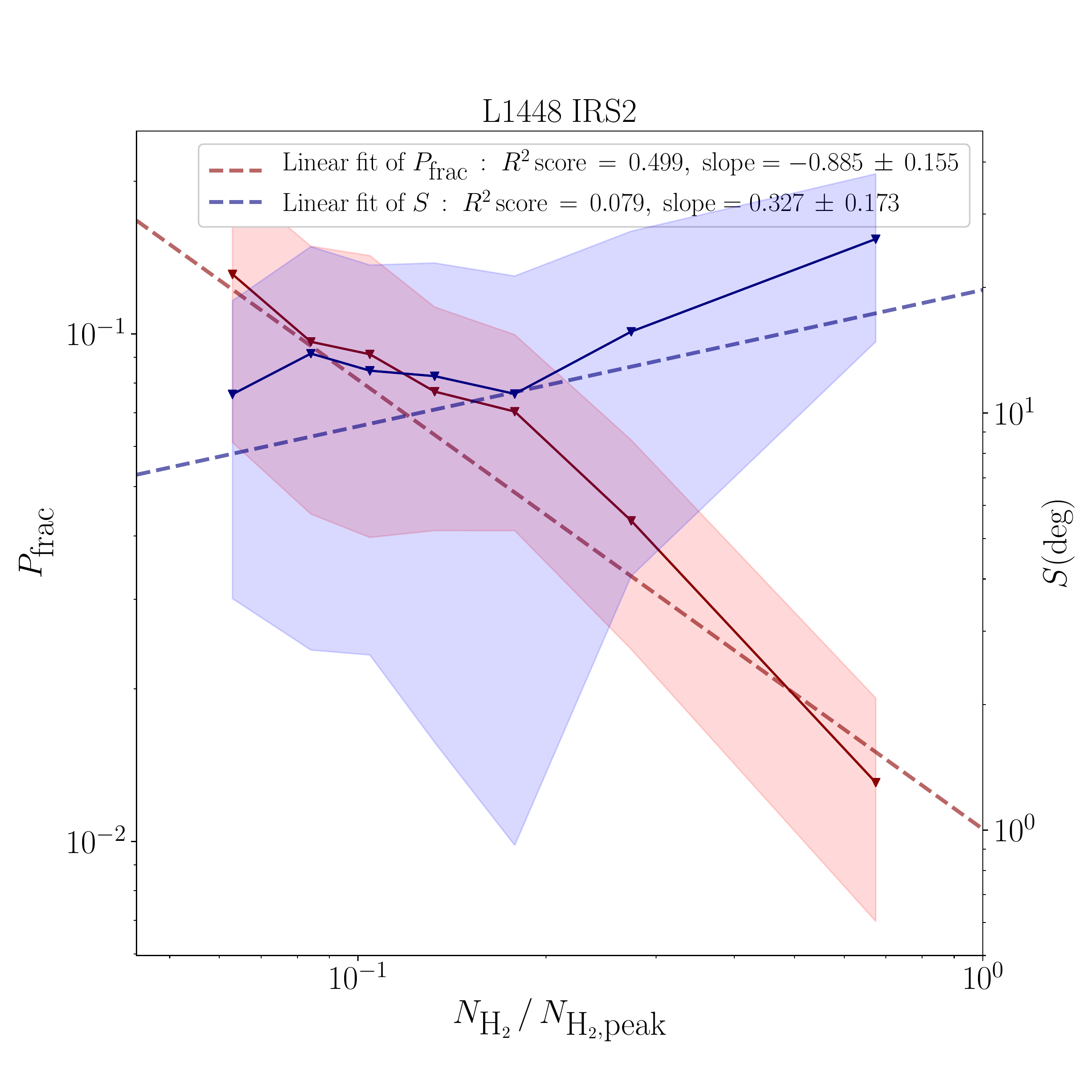}}
\vspace{-0.2cm}
\subfigure{\includegraphics[scale=\scaleSP,clip,trim= 0.4cm 2.3cm 0.4cm 2cm]{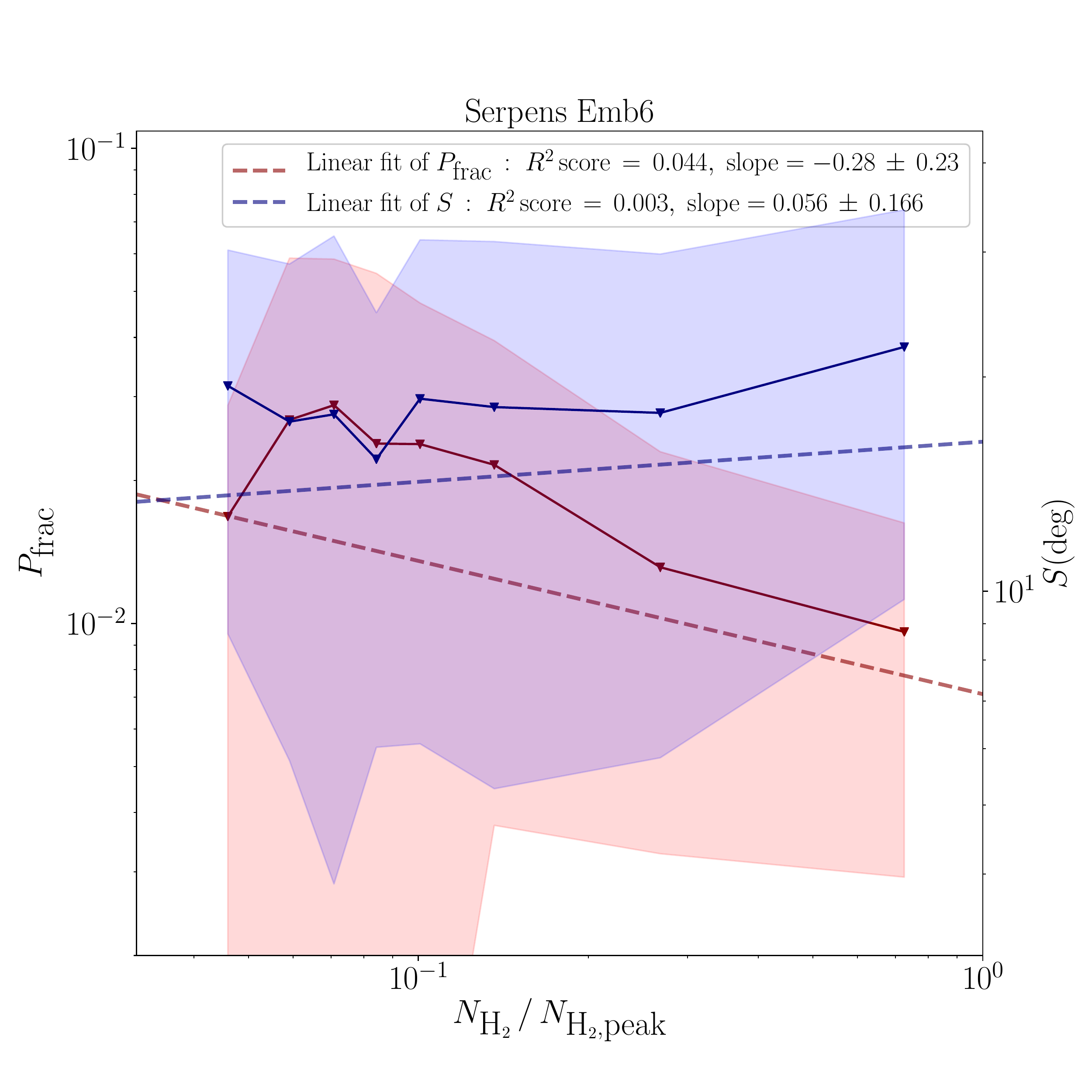}}
\vspace{-0.2cm}
\subfigure{\includegraphics[scale=\scaleSP,clip,trim= 0.4cm 2.3cm 0.4cm 2cm]{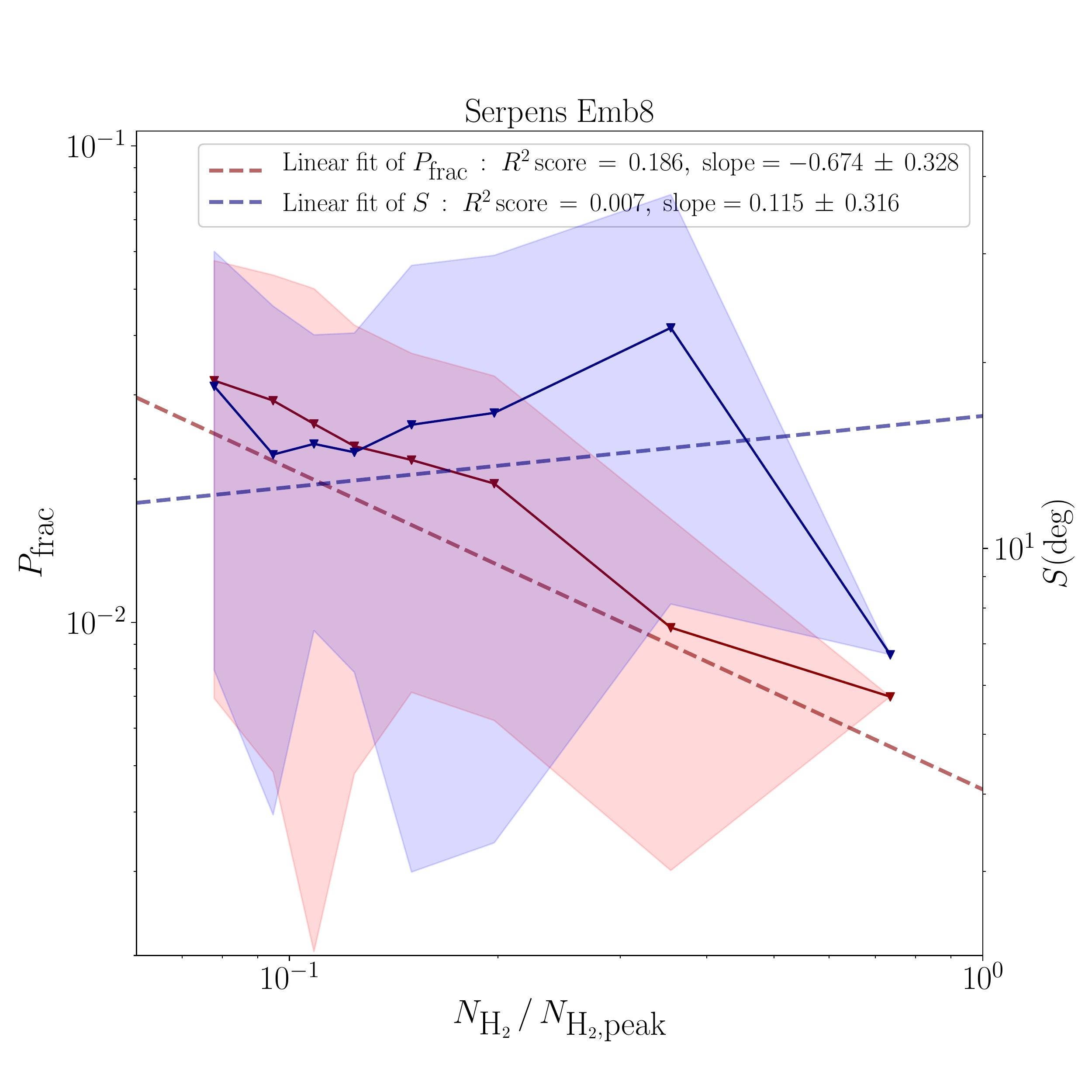}}
\subfigure{\includegraphics[scale=\scaleSP,clip,trim= 0.4cm 2.3cm 0.4cm 2cm]{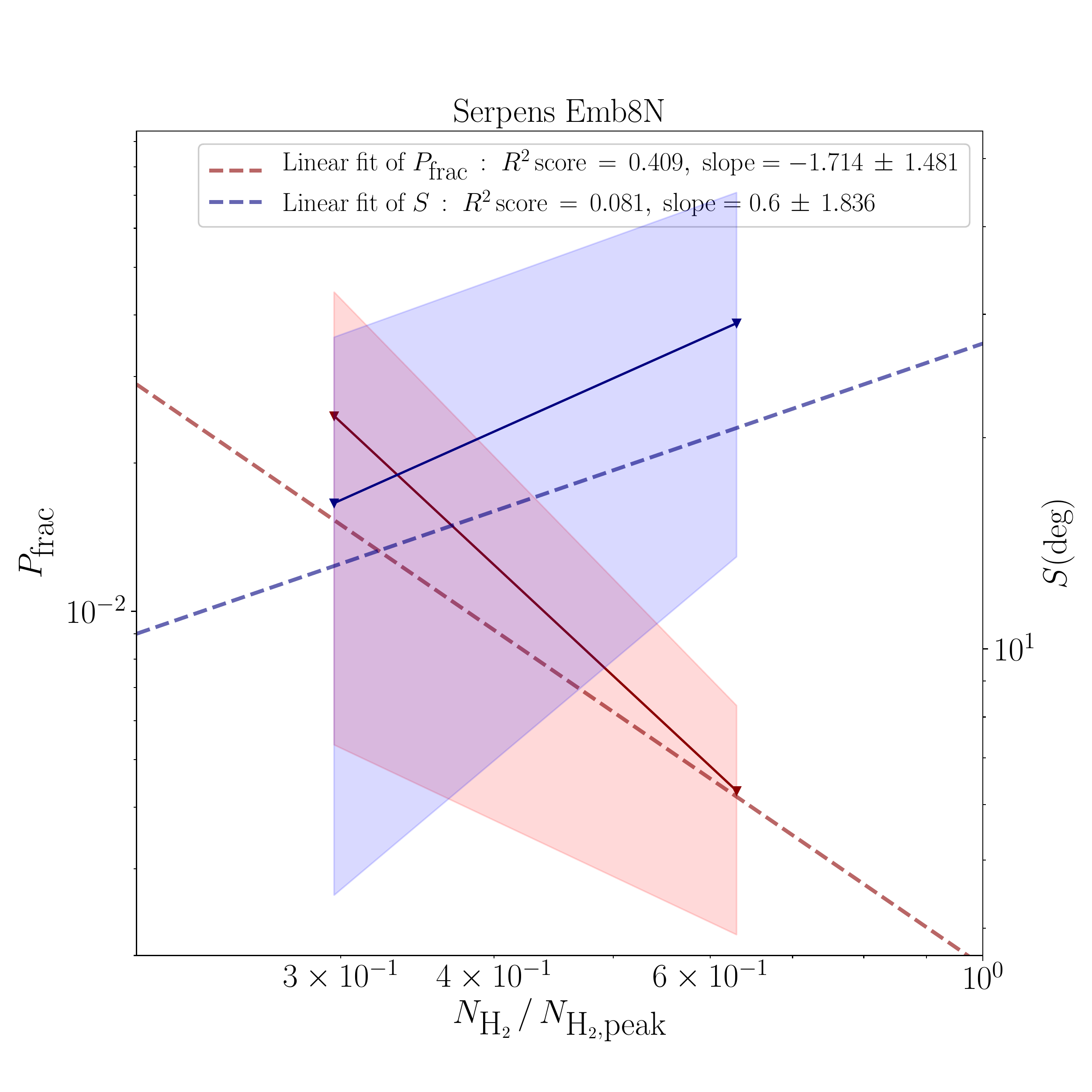}}
\vspace{-0.2cm}
\subfigure{\includegraphics[scale=\scaleSP,clip,trim= 0.4cm 2.3cm 0.4cm 2cm]{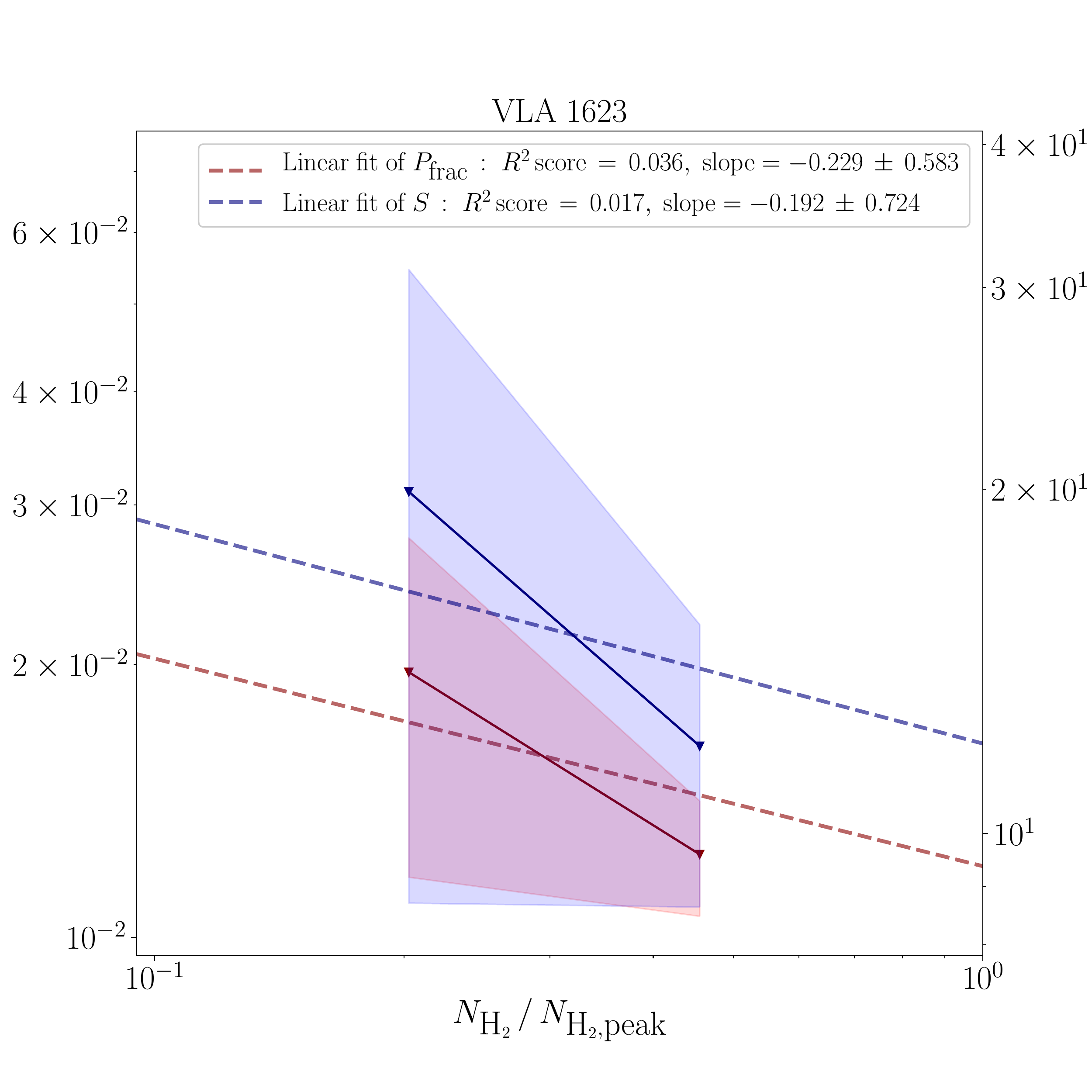}}
\vspace{-0.2cm}
\subfigure{\includegraphics[scale=\scaleSP,clip,trim= 0.4cm 2.3cm 0.4cm 2cm]{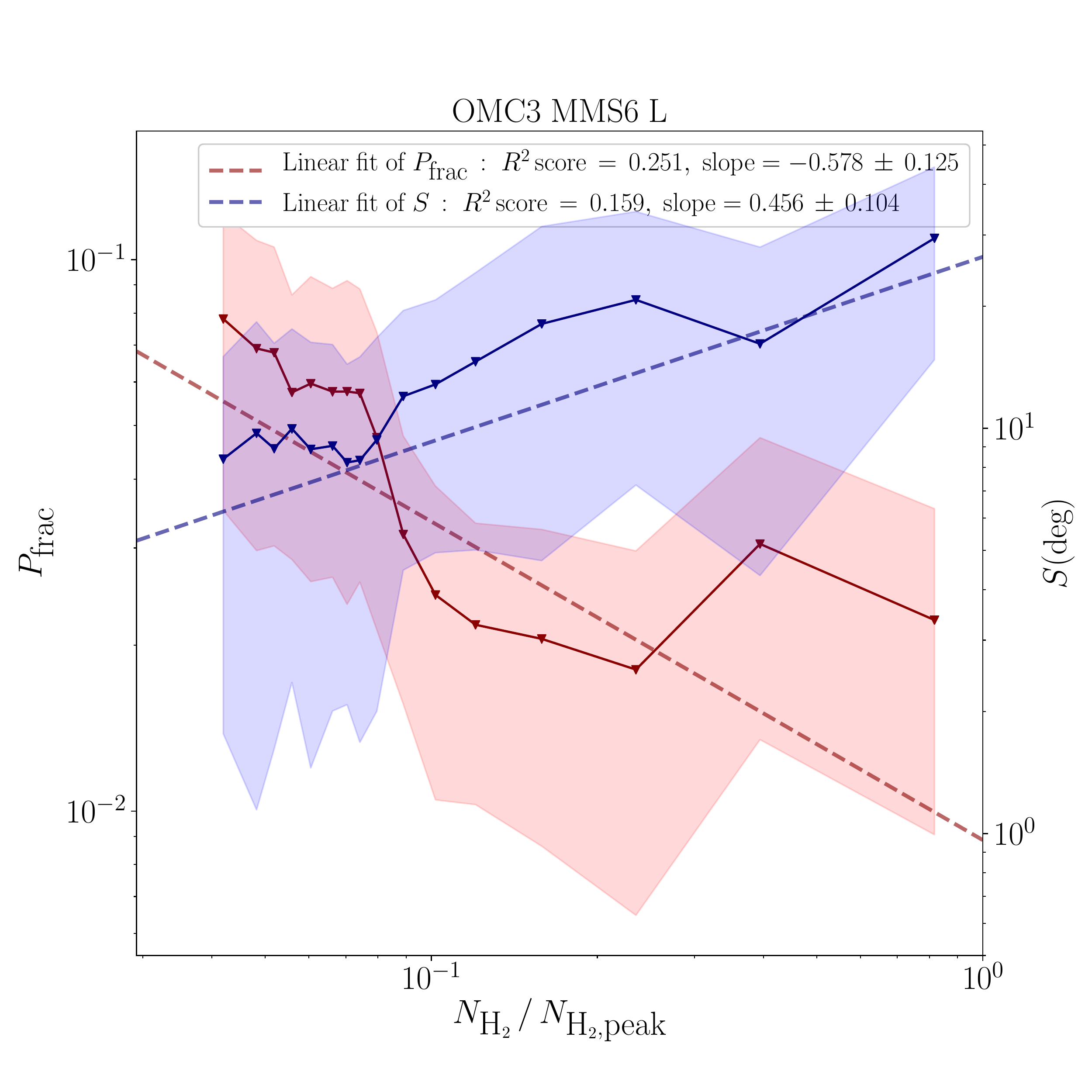}}
\subfigure{\includegraphics[scale=\scaleSP,clip,trim= 0.4cm 1cm 0.4cm 2cm]{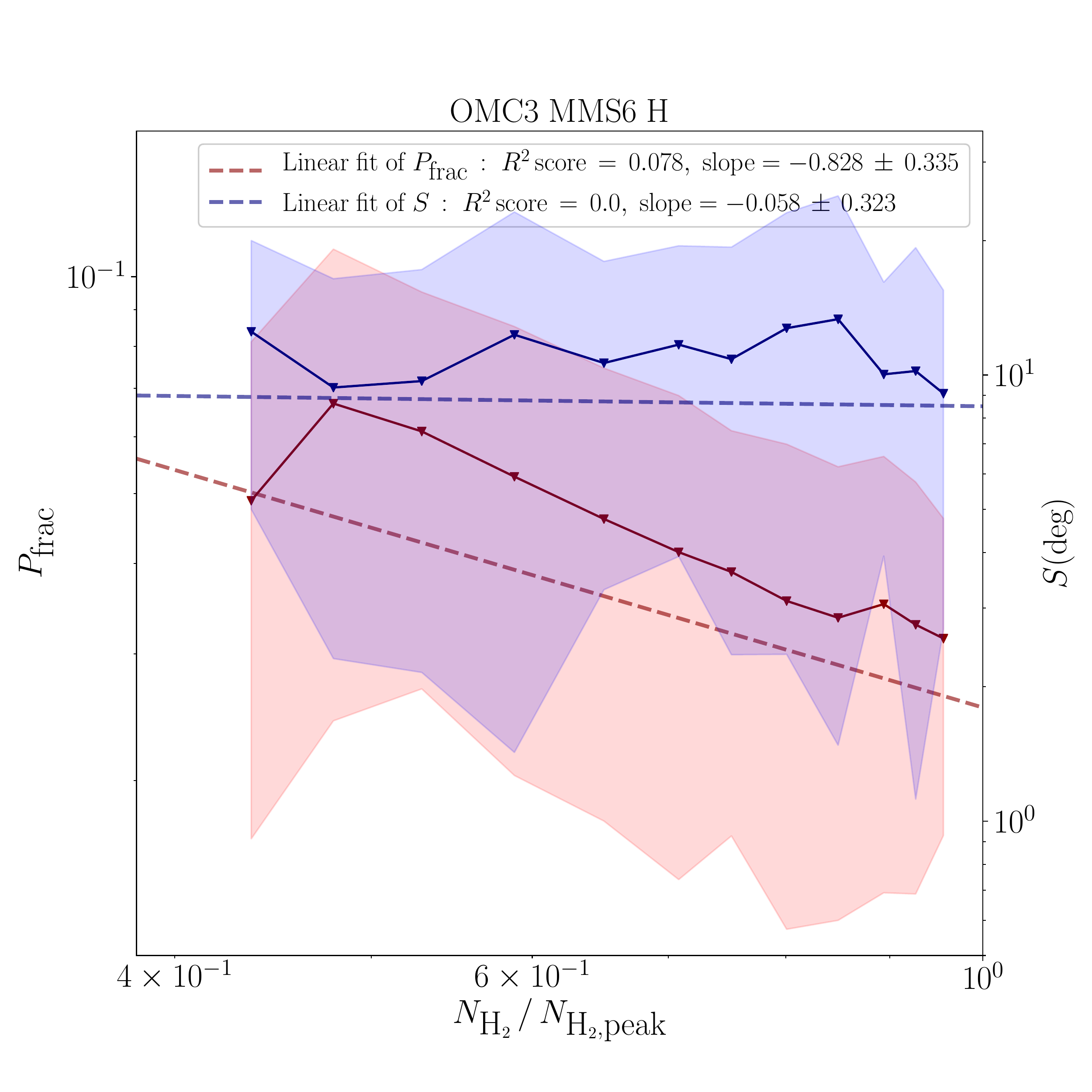}}
\vspace{-0.2cm}
\subfigure{\includegraphics[scale=\scaleSP,clip,trim= 0.4cm 1cm 0.4cm 2cm]{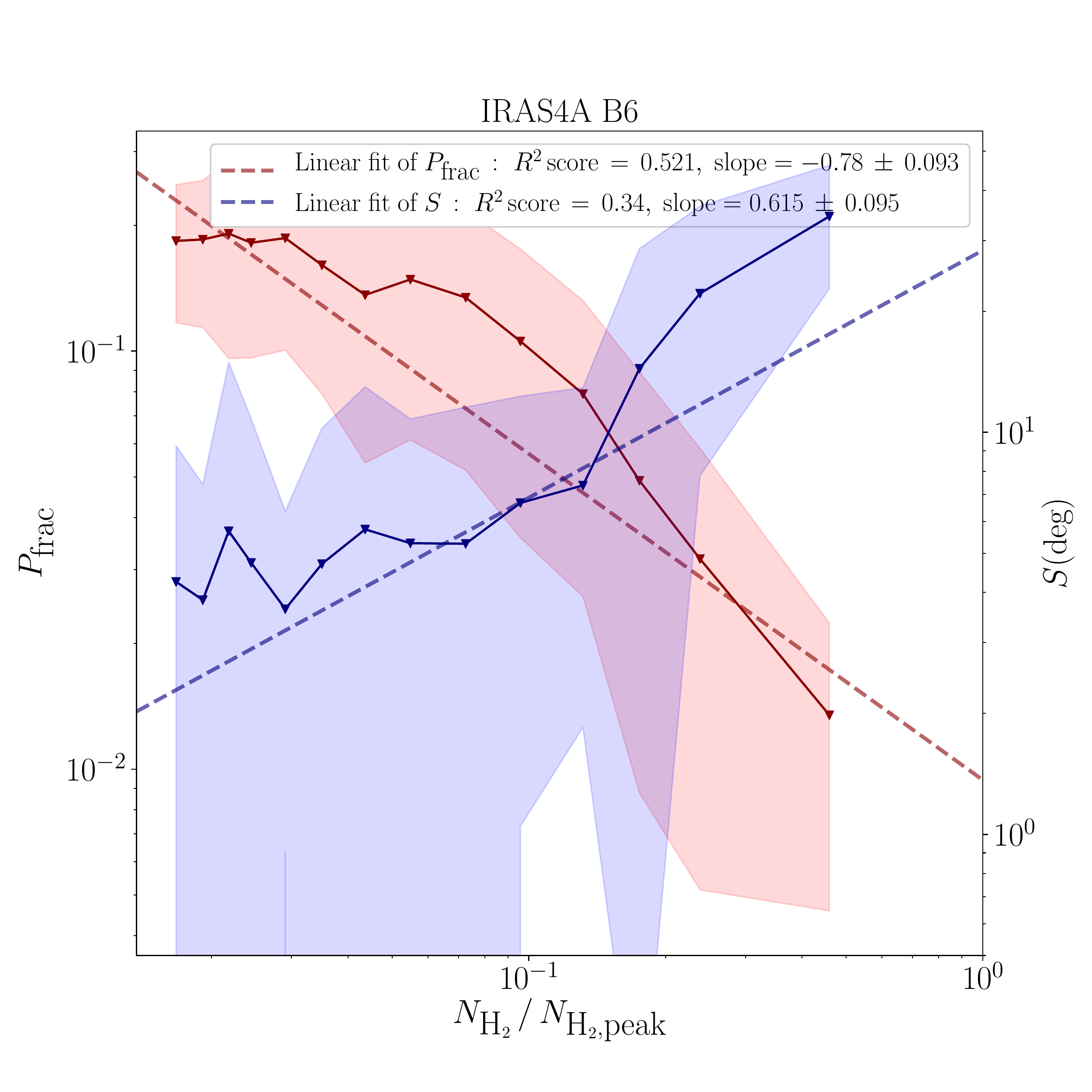}}
\vspace{-0.2cm}
\subfigure{\includegraphics[scale=\scaleSP,clip,trim= 0.4cm 1cm 0.4cm 2cm]{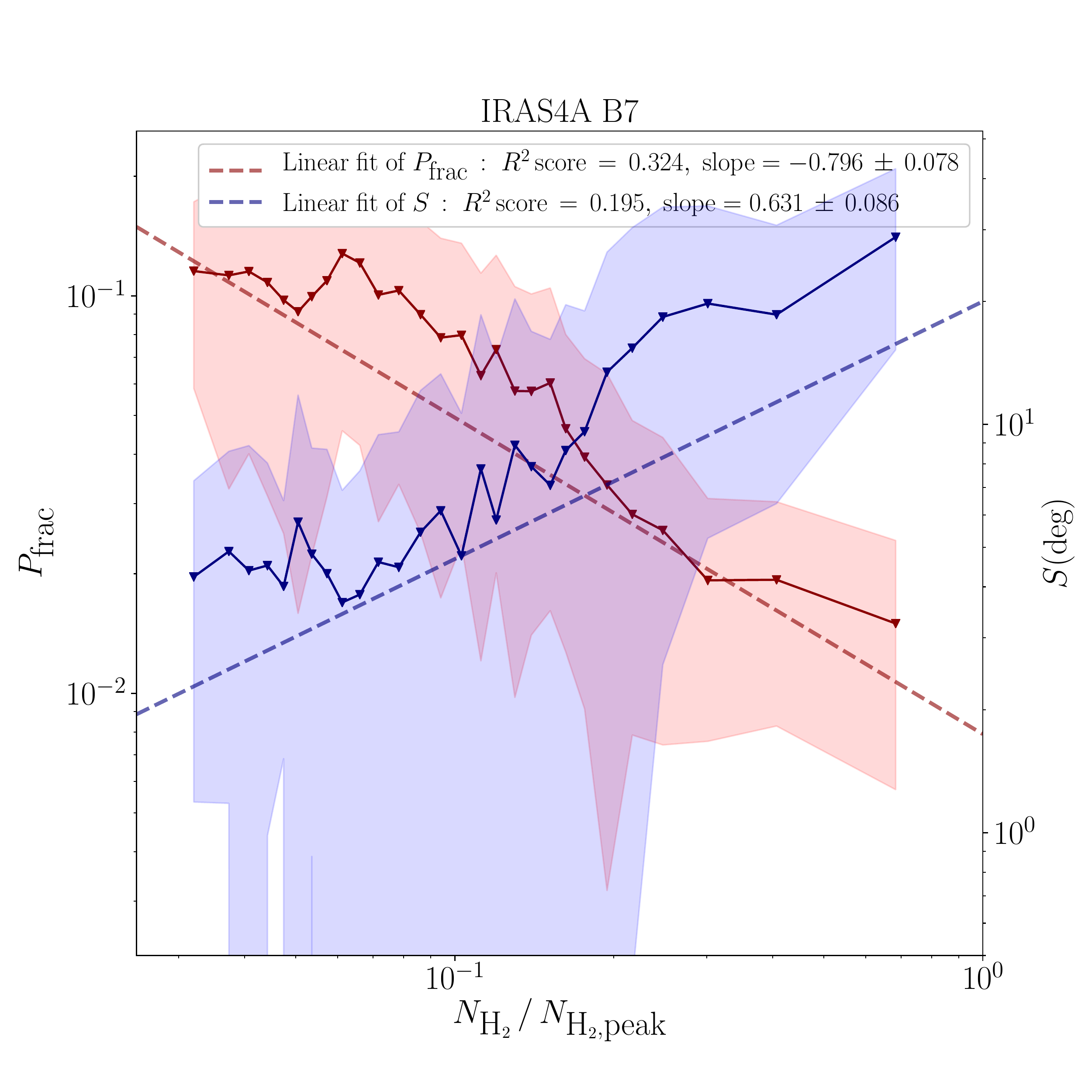}}
\caption[]{\small  Distributions of the dispersion of polarization position angles $\S$ (blue) and the polarization fraction $\Pf$ (red) as a function of the normalized column density $N_{\textrm{H}_2}/N_{\textrm{H}_2\textrm{,peak}}$, for all the cores. The solid lines and points represent the running mean of $\S$ and $\Pf$; the associated shaded areas are $\pm$ the standard deviation of each bin. The dashed lines are the linear fits, which are linear
regressions done in the logarithmic space, whose slopes and uncertainties are calculated the same way as in Figure \ref{fig:scatter_S_Pfrac_Alma_merged}.}
\label{fig:trends_S_Pfrac_Alma}
\end{figure*}

\clearpage
\section{Power spectra as a function of spatial scale in the ALMA observations}
\label{app:power_spectrum}

The power spectra of the maps of the Class 0 sources observed with ALMA are shown in  \ref{fig:scatter_power_spectrum}. To produce these power spectra we perform 2D Fourier transforms of the Stokes $I$, $Q$, and $U$ maps and take their normalized absolute magnitude, after which we calculate the azimuthal average to derive the power with respect to the spatial scale.

We make an attempt to recover the missing flux at large spatial scales of Stokes $I$ with respect to Stokes $Q$ and $U$. To correct the Stokes $I$ power spectrum across a given range of spatial scales, we scale up the flux in Stokes $I$ using the differences between the Stokes $I$ and the Stokes $Q$ and $U$ power spectra at those same spatial scales. The idea behind this correction is to solve the problem of high $\Pf$ values discussed in Sections \ref{sec:scales_of_emission} and \ref{sec:spatial_filtering}. Unfortunately, this simplistic method is not robust; as the Stokes $I$ signal at large scales is buried in noise, we do not properly recover the initial missing flux. Furthermore, this simple method creates artifacts in the Stokes $I$ maps.

Note that we produced power spectra of the Stokes maps from the simulations in the same way that we do for the ALMA observations. We find the same discrepancies between the power in Stokes $I$ and the power in Stokes $Q$ and $U$ at large scales. However, as our simulations do not reproduce rigorously the variety of morphologies we detect in the Class 0 ALMA observations, we do not show these additional plots.

\def\scaleSP{0.215}

\begin{figure*}[!tbph]
\centering
\subfigure{\includegraphics[scale=\scaleSP,clip,trim= 0.5cm 2.2cm 2.5cm 2.2cm]{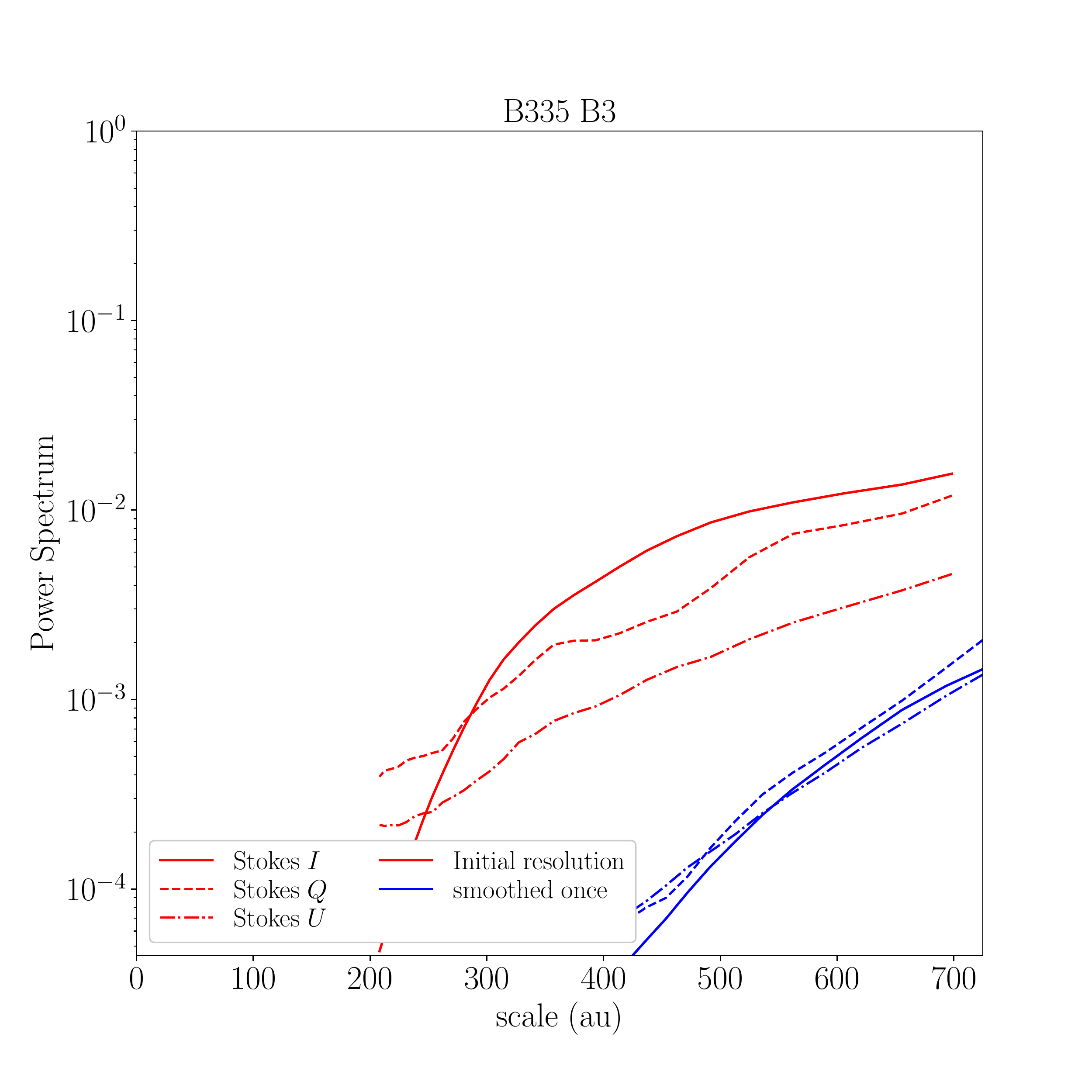}}
\vspace{-0.25cm}
\subfigure{\includegraphics[scale=\scaleSP,clip,trim= 0.5cm 2.2cm 2.5cm 2.2cm]{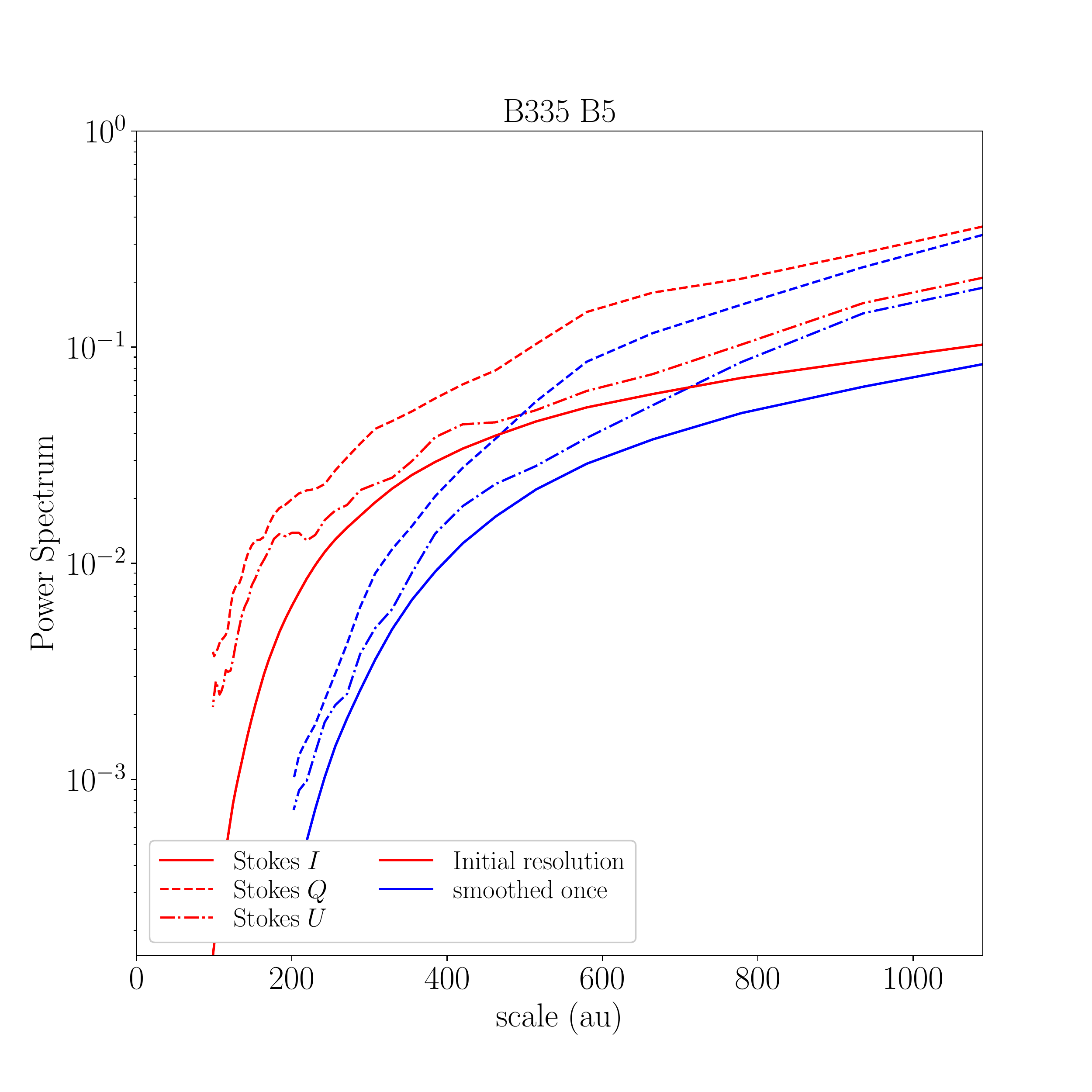}}
\subfigure{\includegraphics[scale=\scaleSP,clip,trim= 0.5cm 2.2cm 2.5cm 2.2cm]{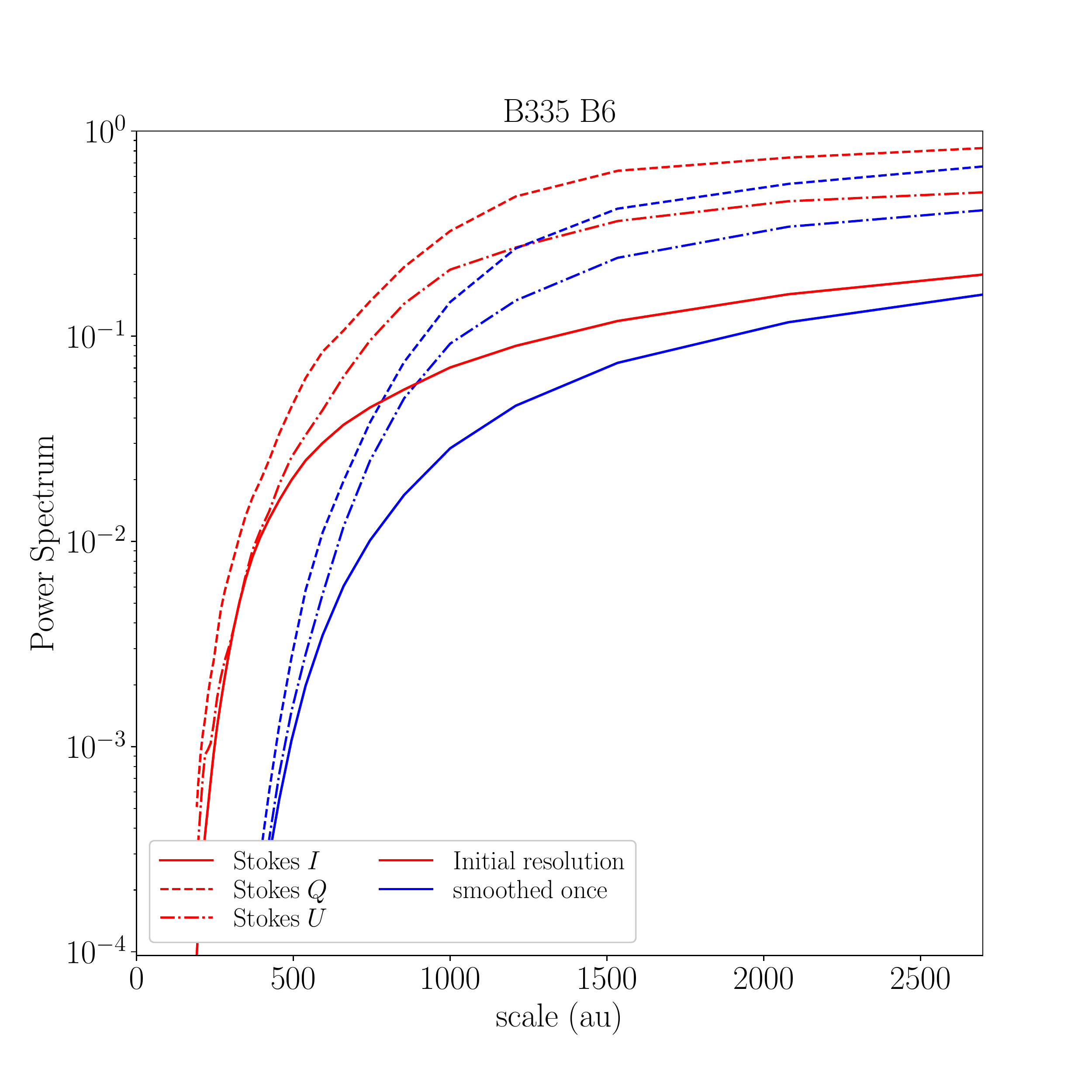}}
\subfigure{\includegraphics[scale=\scaleSP,clip,trim= 0.5cm 2.2cm 2.5cm 2.2cm]{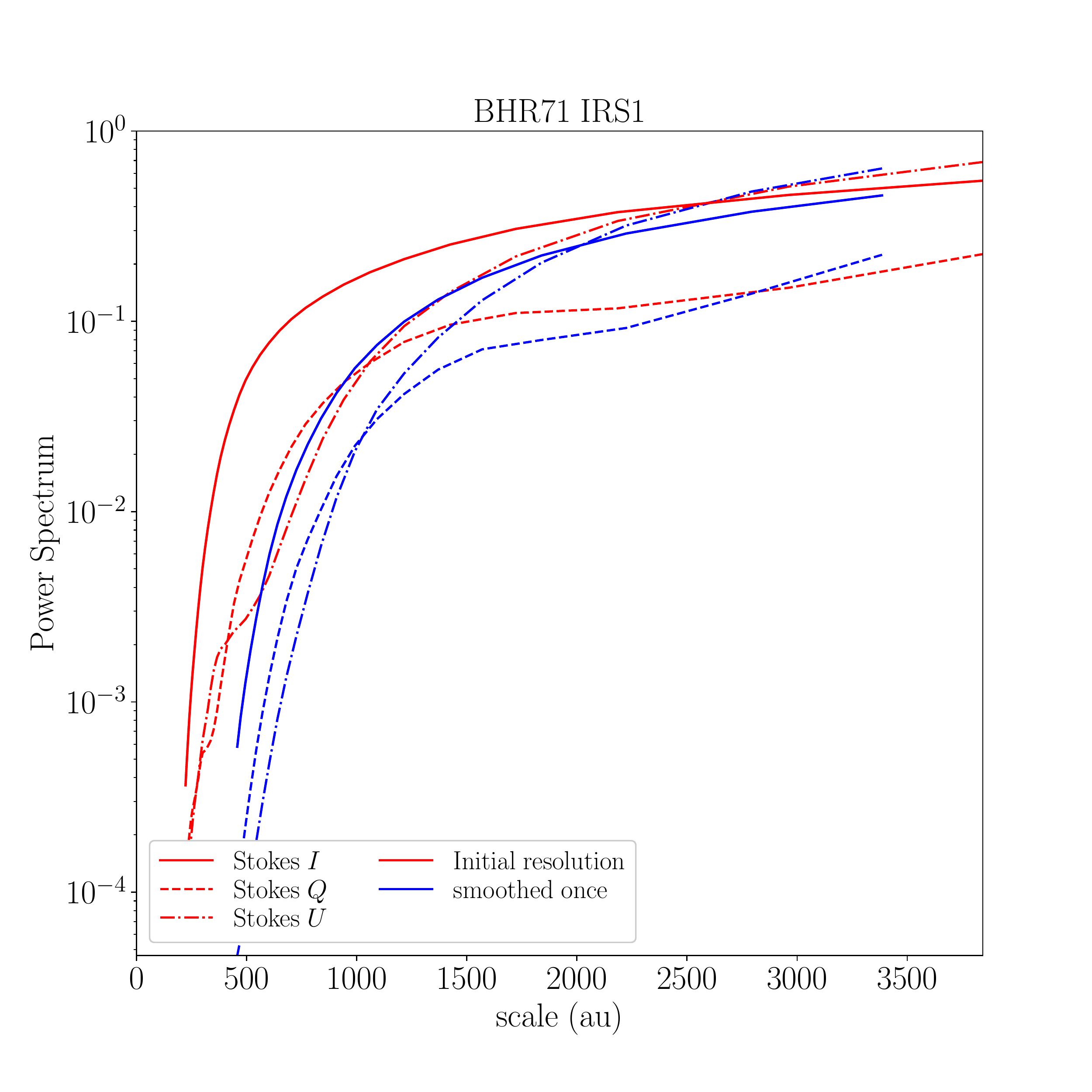}}
\vspace{-0.25cm}
\subfigure{\includegraphics[scale=\scaleSP,clip,trim= 0.5cm 2.2cm 2.5cm 2.2cm]{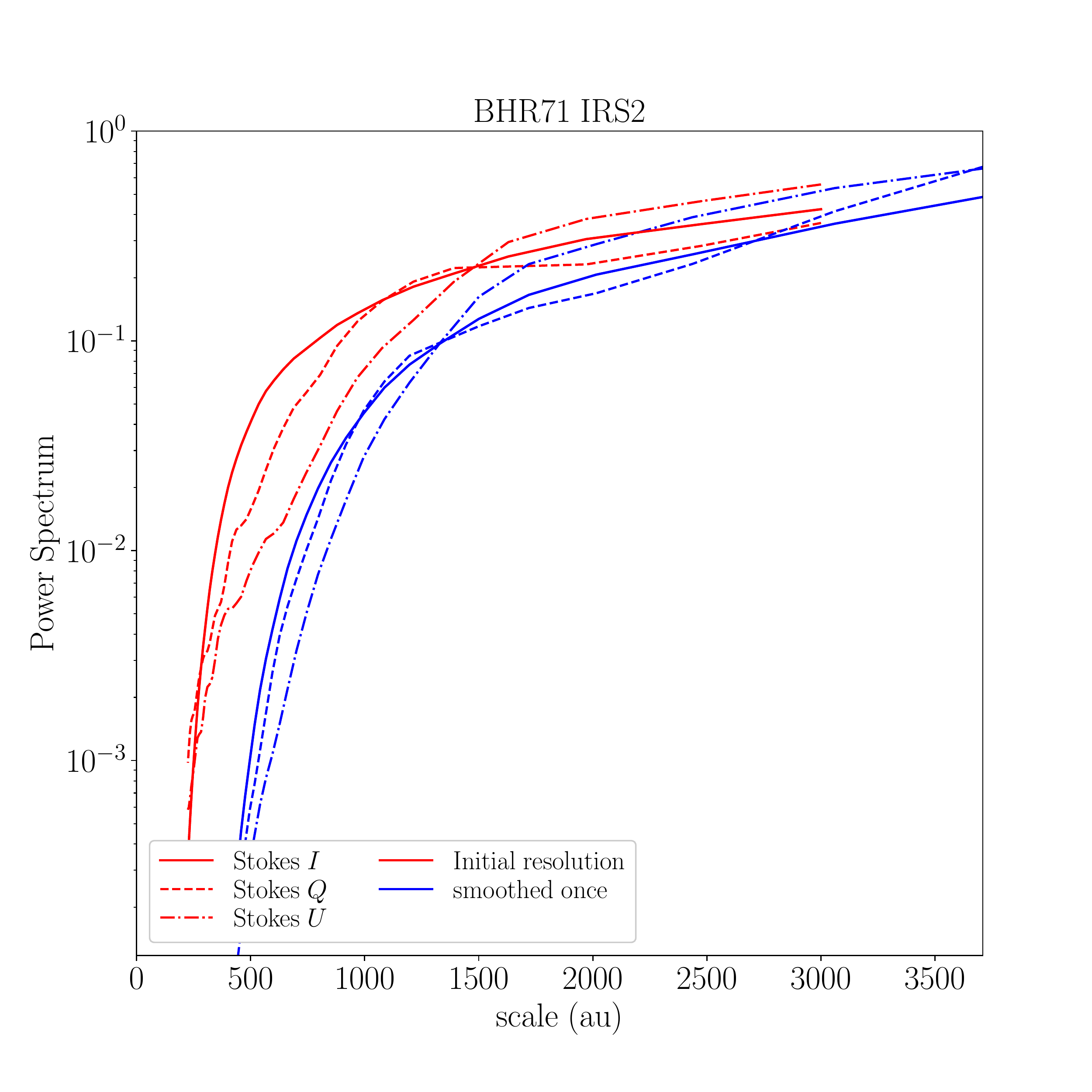}}
\subfigure{\includegraphics[scale=\scaleSP,clip,trim= 0.5cm 2.2cm 2.5cm 2.2cm]{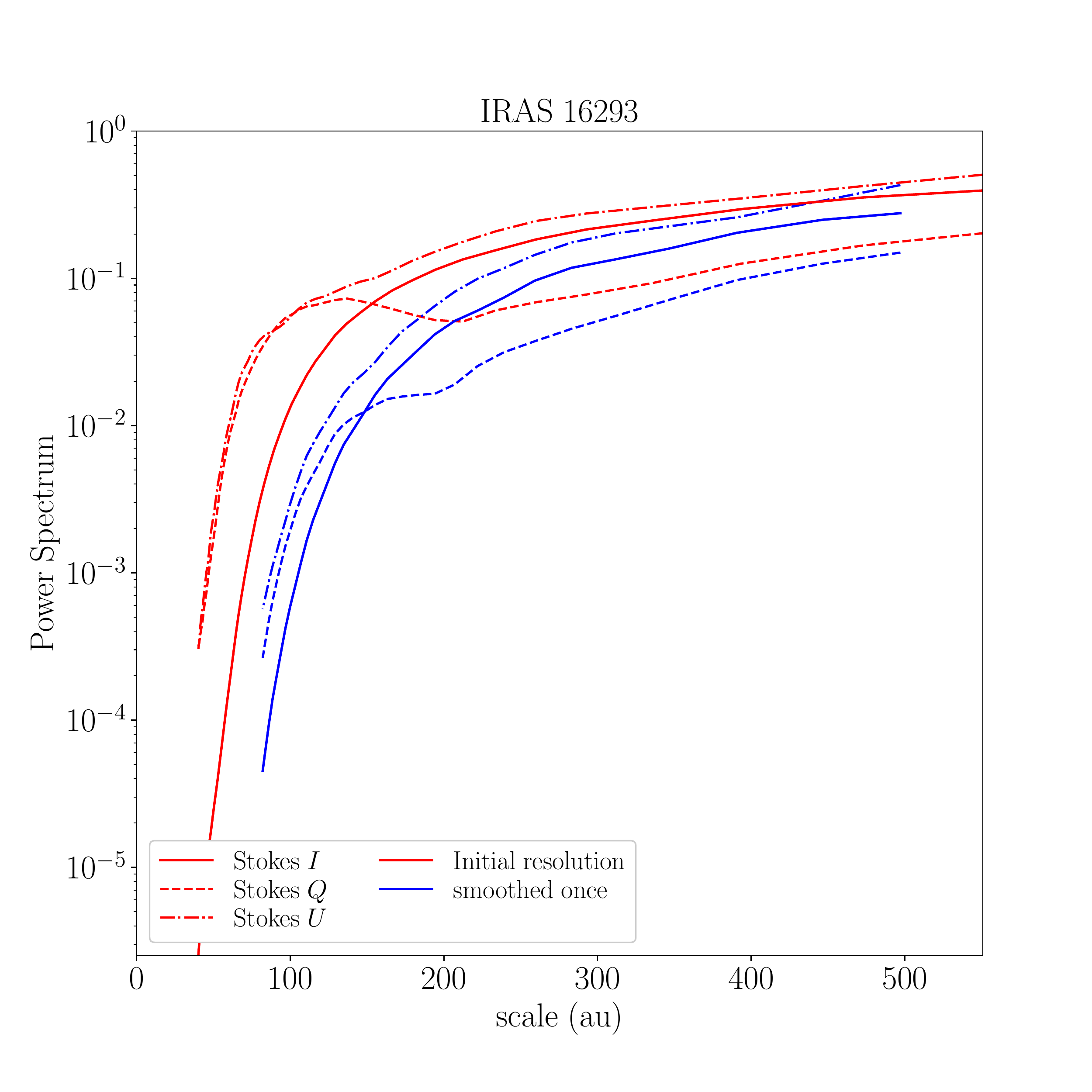}}
\subfigure{\includegraphics[scale=\scaleSP,clip,trim= 0.5cm 2.2cm 2.5cm 2.2cm]{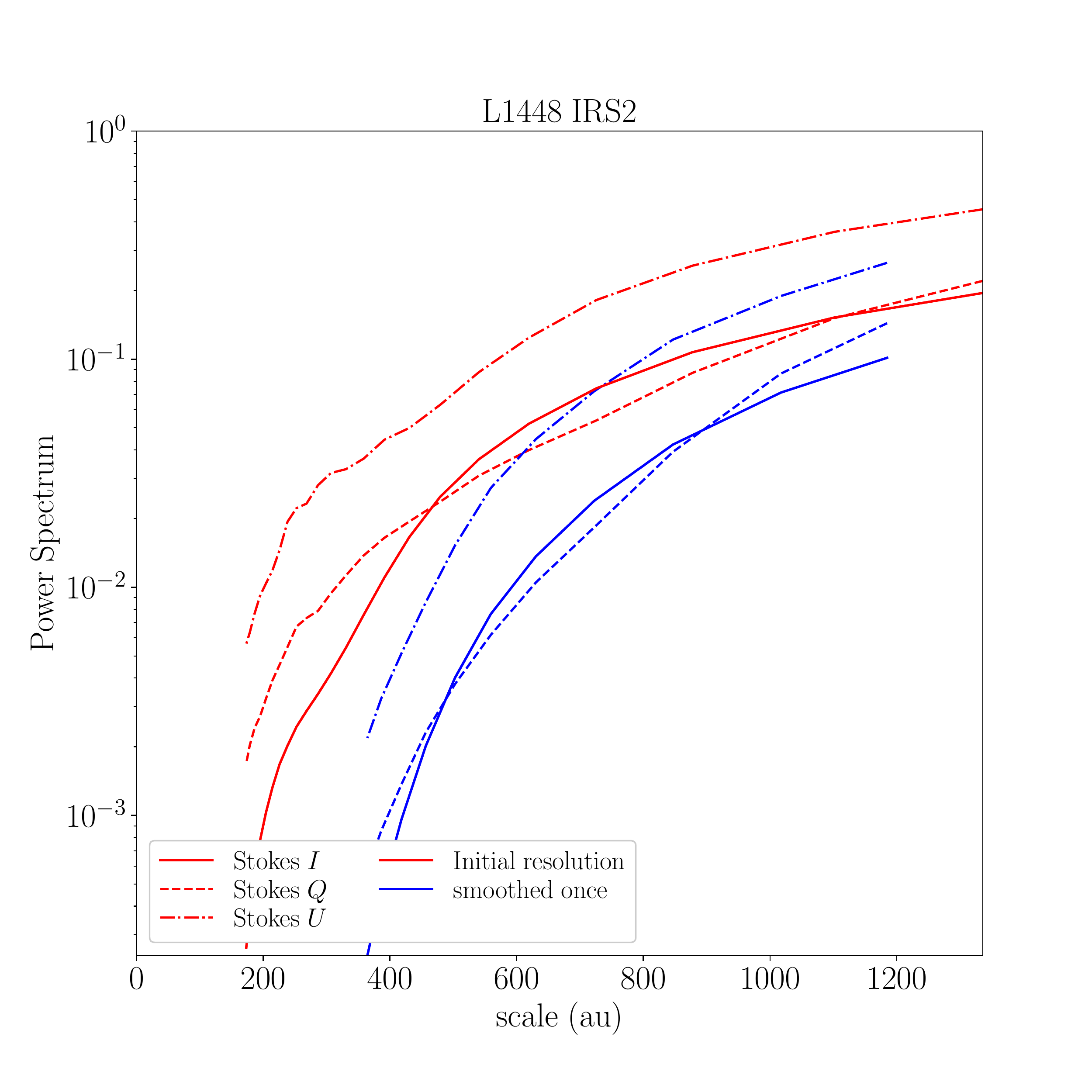}}
\vspace{-0.25cm}
\subfigure{\includegraphics[scale=\scaleSP,clip,trim= 0.5cm 2.2cm 2.5cm 2.2cm]{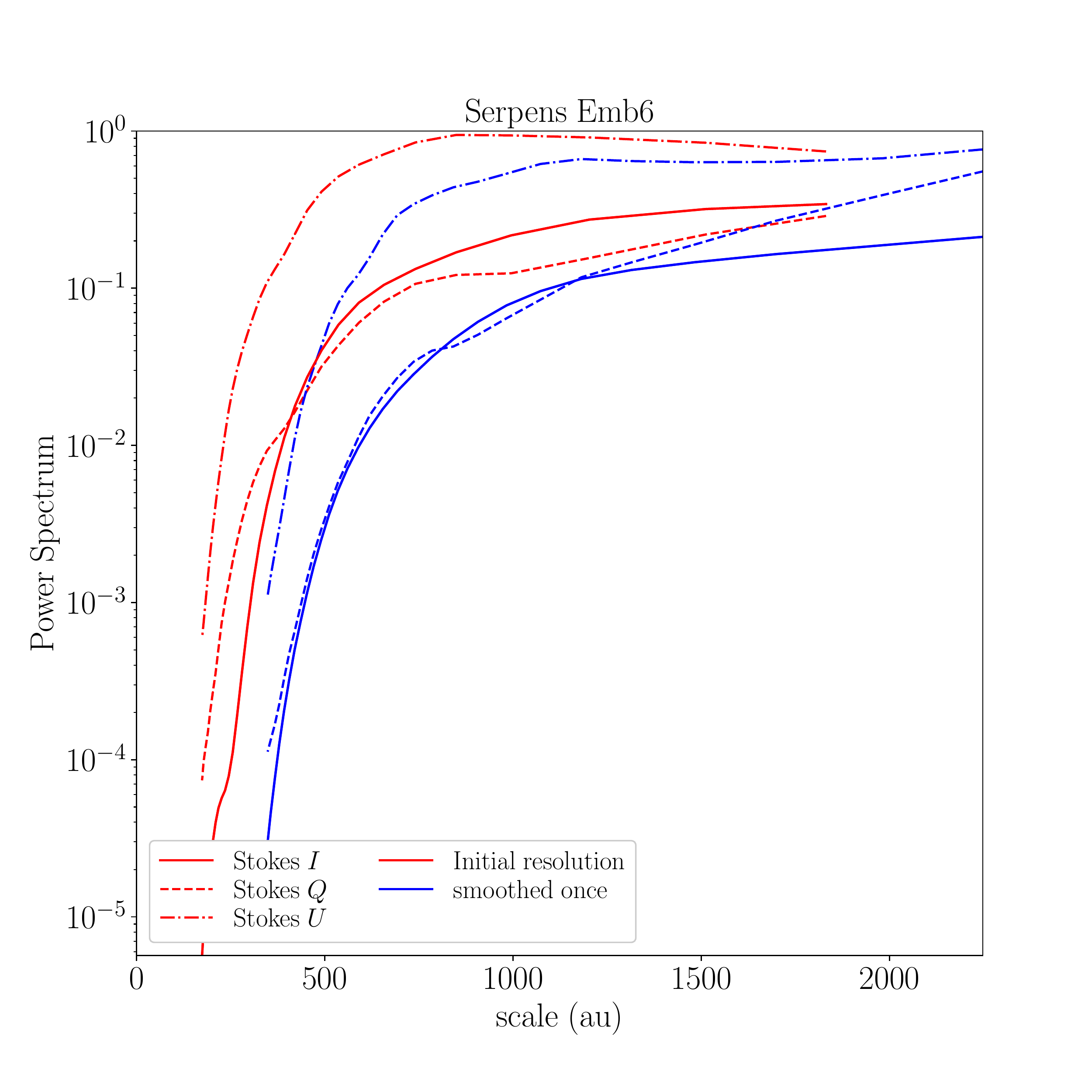}}
\subfigure{\includegraphics[scale=\scaleSP,clip,trim= 0.5cm 2.2cm 2.5cm 2.2cm]{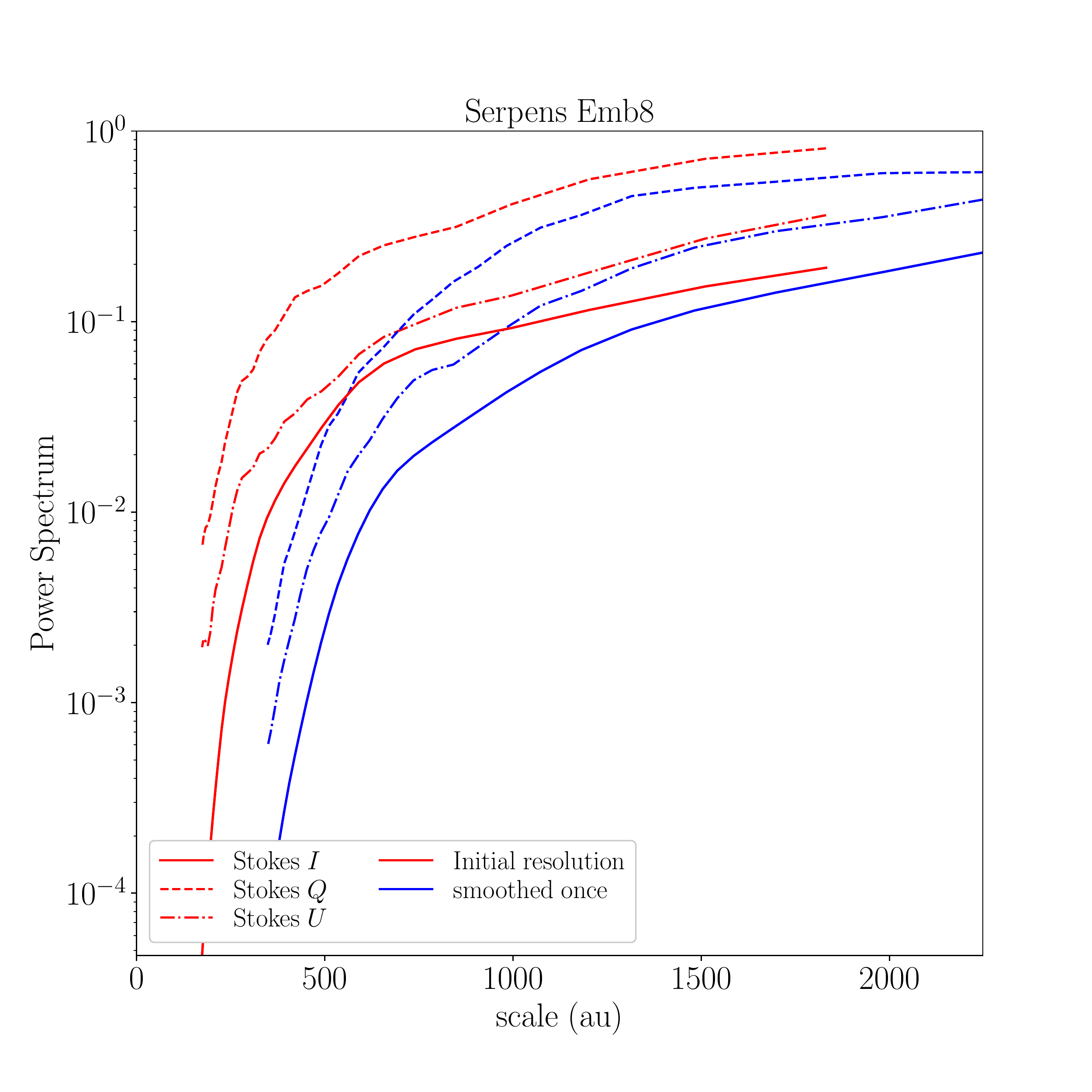}}
\subfigure{\includegraphics[scale=\scaleSP,clip,trim= 0.5cm 2.2cm 2.5cm 2.2cm]{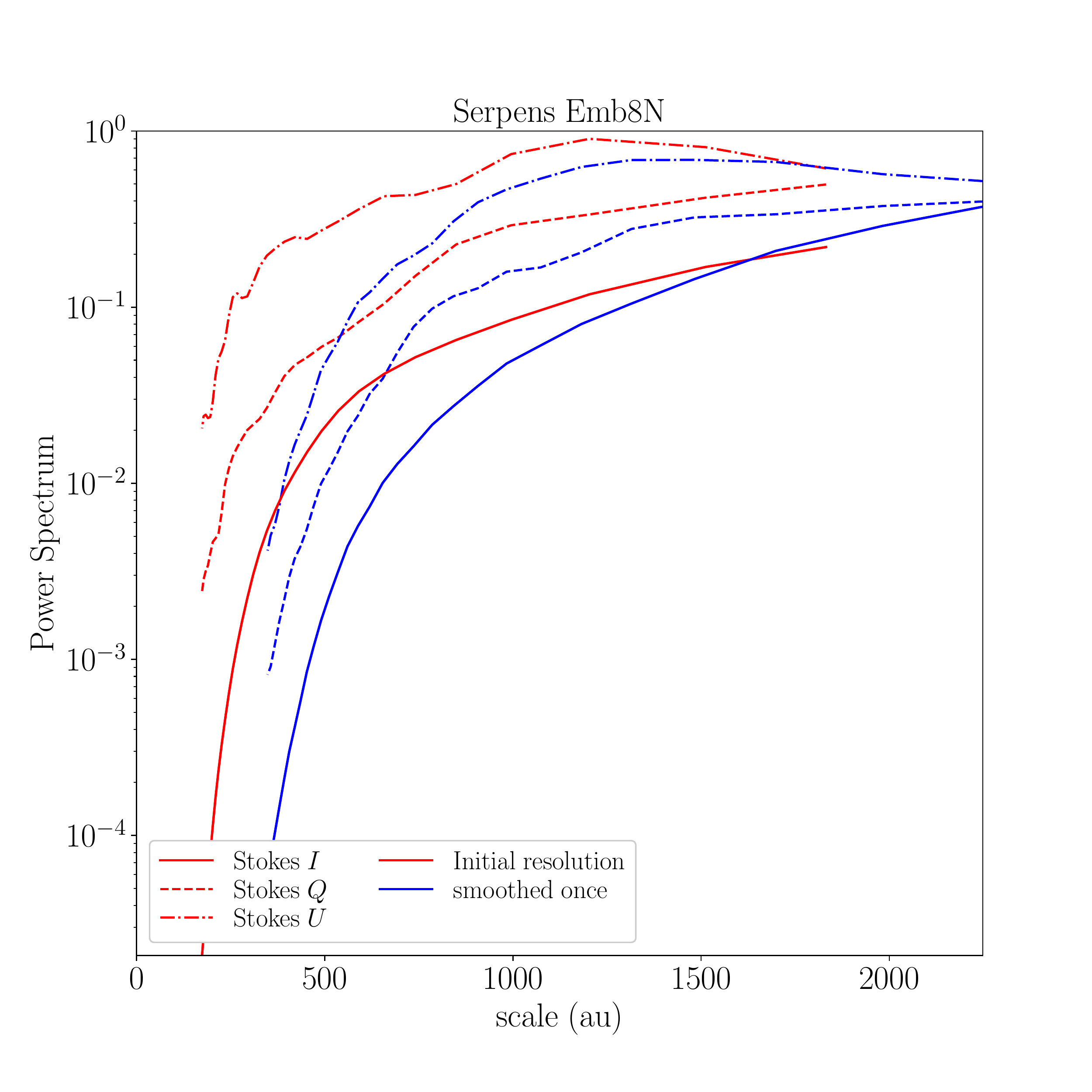}}
\vspace{-0.25cm}
\subfigure{\includegraphics[scale=\scaleSP,clip,trim= 0.5cm 2.2cm 2.5cm 2.2cm]{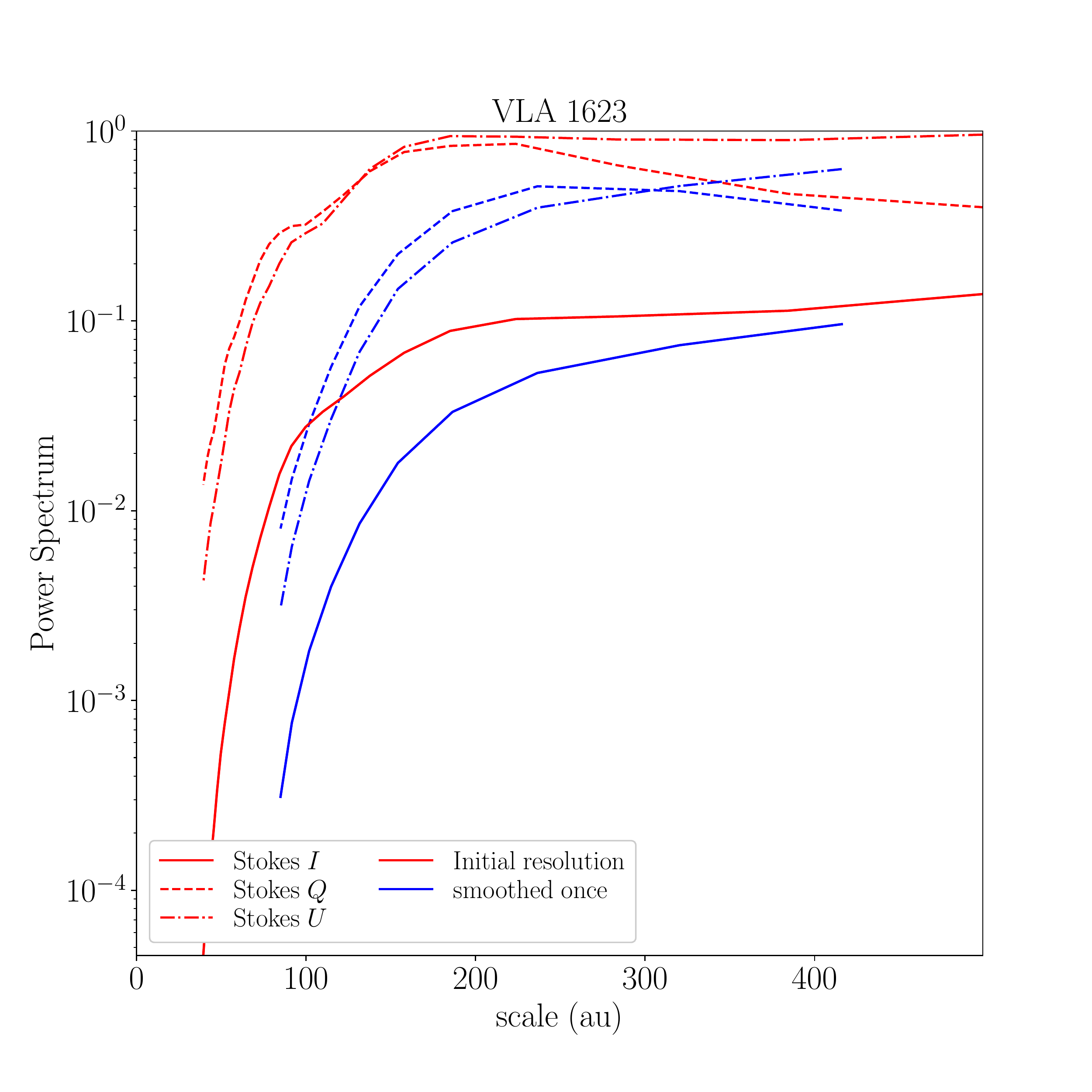}}
\subfigure{\includegraphics[scale=\scaleSP,clip,trim= 0.5cm 2.2cm 2.5cm 2.2cm]{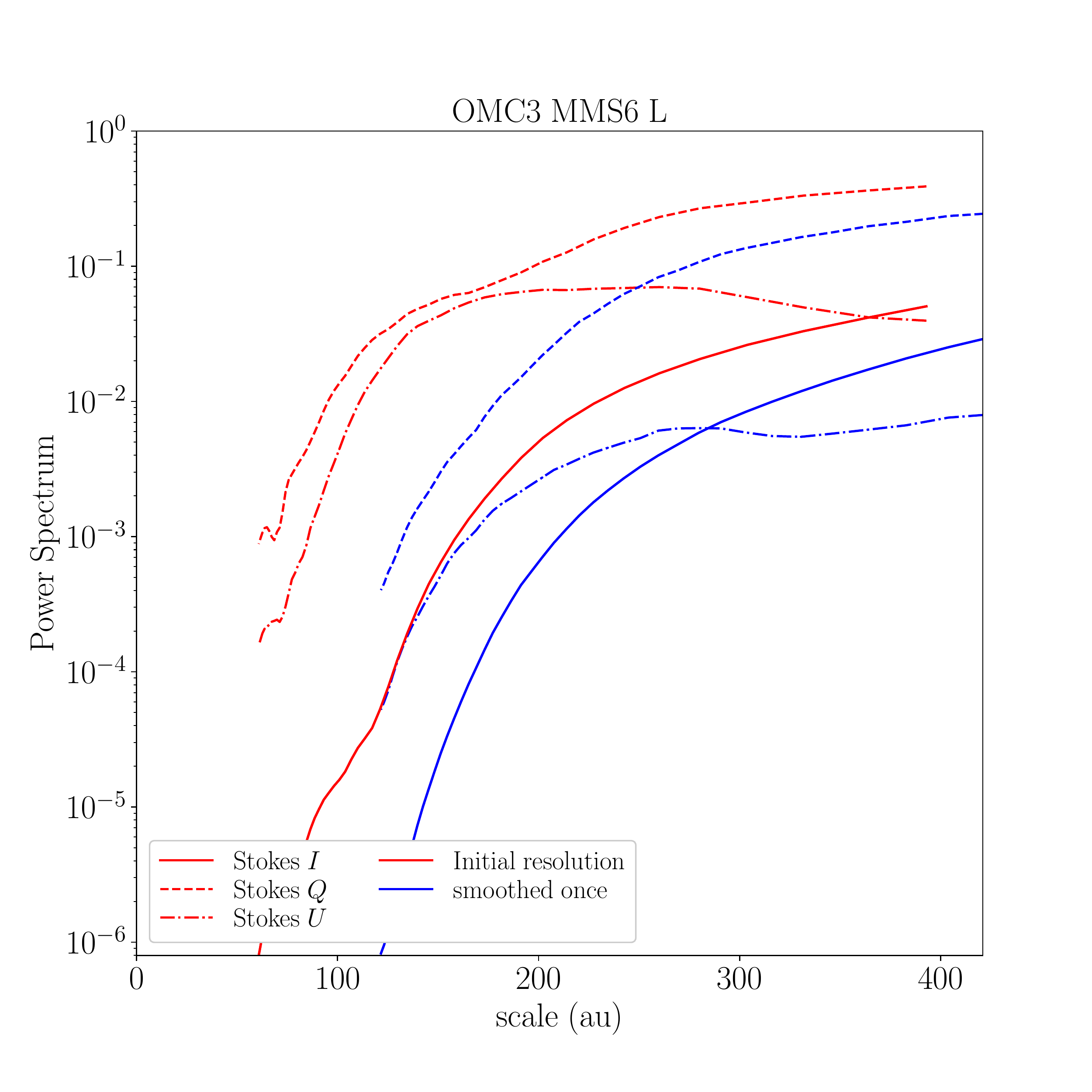}}
\subfigure{\includegraphics[scale=\scaleSP,clip,trim= 0.5cm 1.4cm 2.5cm 2.2cm]{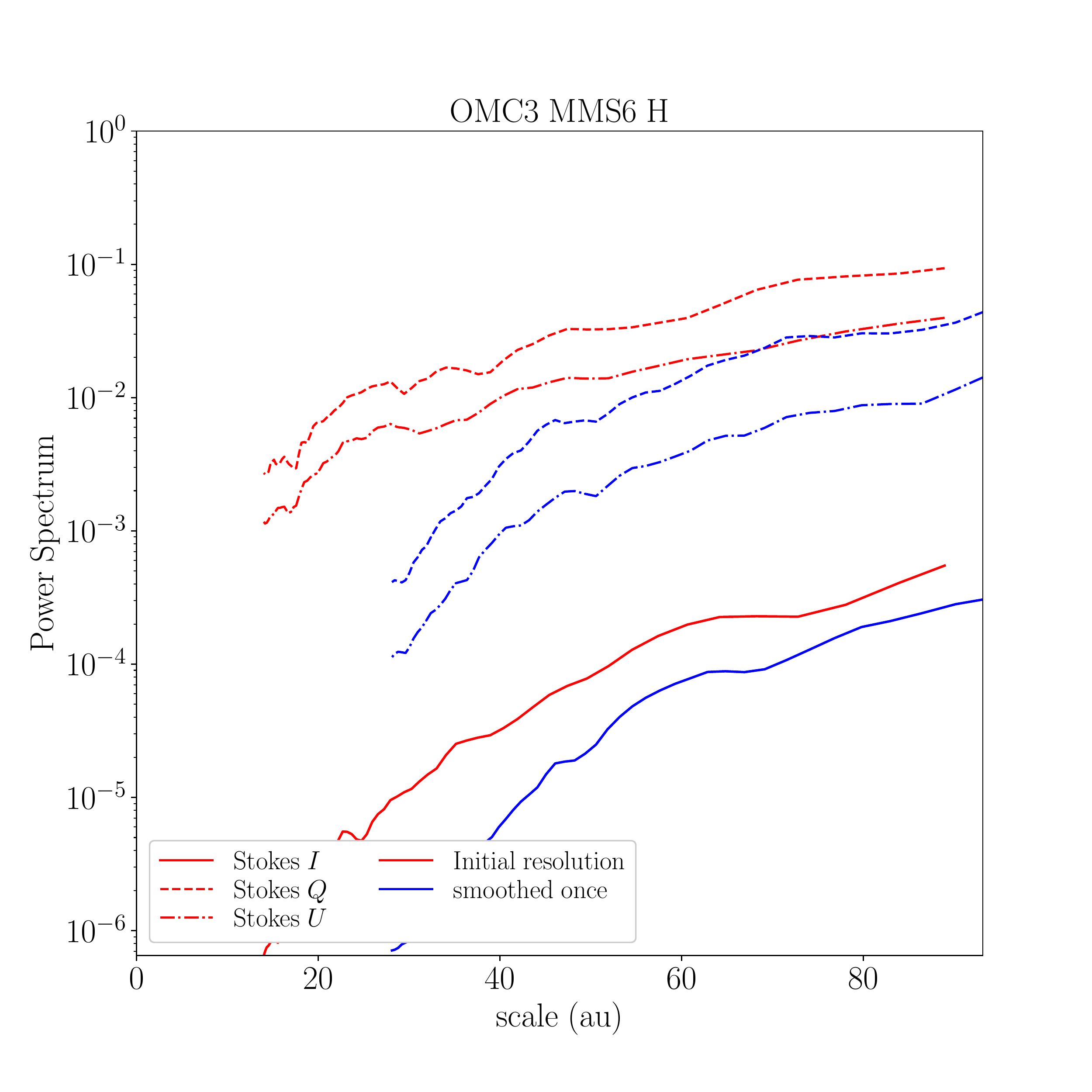}}
\vspace{-0.25cm}
\subfigure{\includegraphics[scale=\scaleSP,clip,trim= 0.5cm 1.4cm 2.5cm 2.2cm]{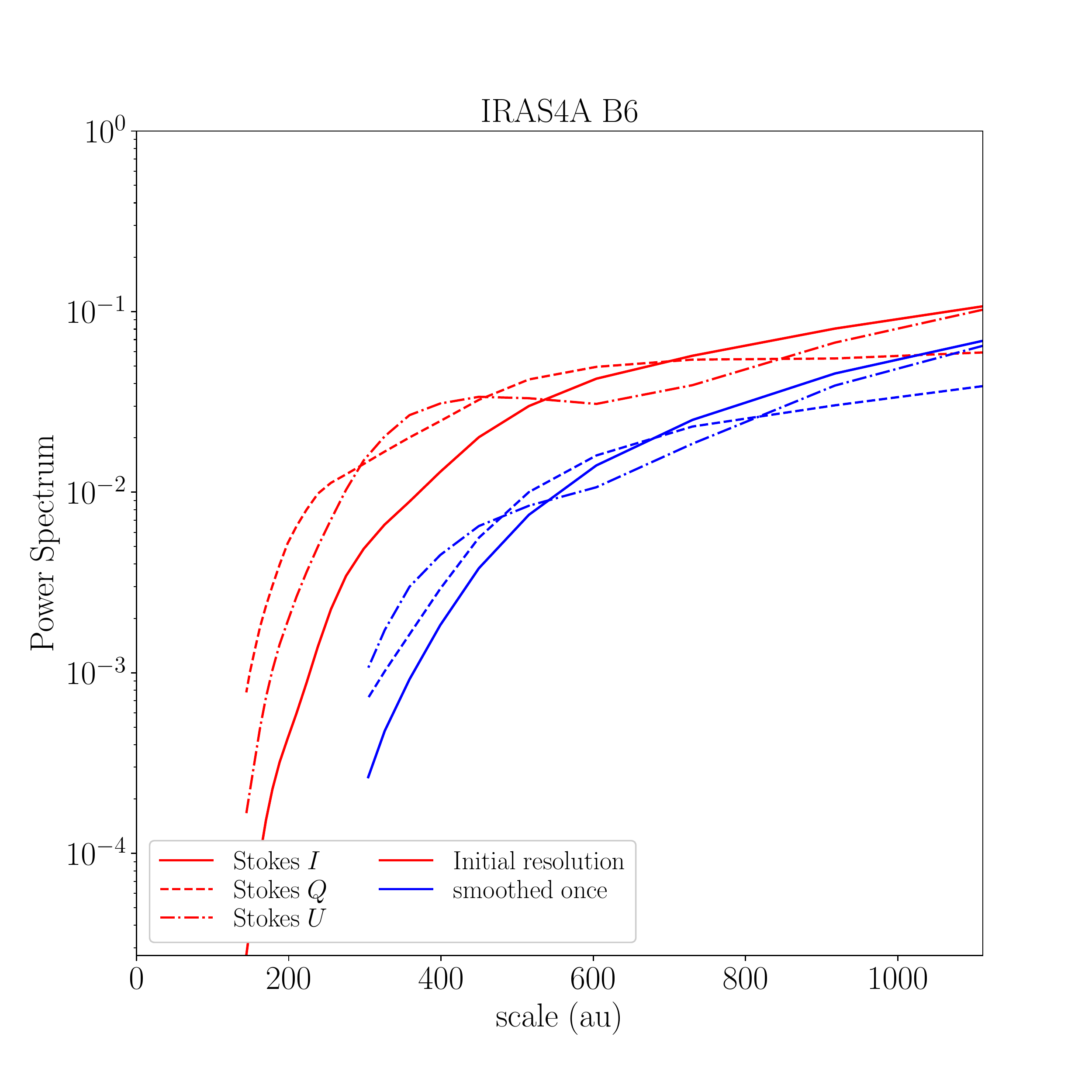}}
\subfigure{\includegraphics[scale=\scaleSP,clip,trim= 0.5cm 1.4cm 2.5cm 2.2cm]{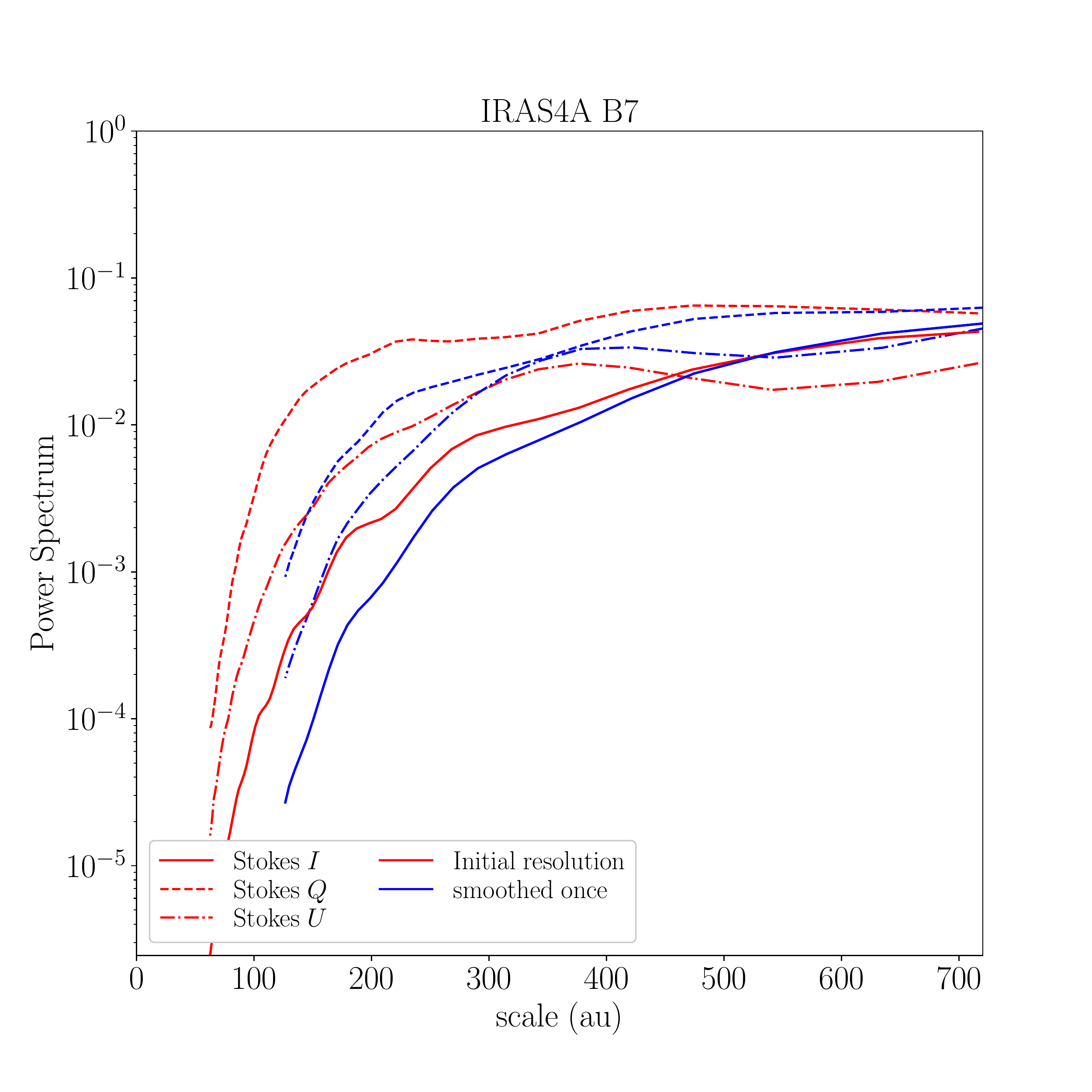}}
\caption[]{\small Normalized power spectra of the three Stokes maps $I$, $Q$, and $U$, toward all the cores of the sample as a function of spatial scale, in au, ranging from the beam size up the maximum recoverable scale. Solid Lines: Stokes $I$. Dashed and dot-dashed lines: Stokes $Q$ and $U$, respectively. Each color corresponds to an angular resolution. Red is the original angular resolution. Blue is four times lower resolution (in terms of beam area).}
\label{fig:scatter_power_spectrum}
\end{figure*}

%

\clearpage
\section{MHD simulations and their synthetic observations}
\label{app:maps_synth}

In order to characterize better the statistics we obtain from our analysis of ALMA dust polarization observations of Class 0 protostars, we perform synthetic observations of non-ideal radiation-magneto-hydrodynamic (MHD) simulations of protostellar collapse, exploring the impact that a range of parameters---such as the dust grain alignment hypothesis, the initial mass and turbulence of the simulation, and the effect of interferometric filtering---have on the statistics from these simulations.

We use six different setups for the simulations performed with the RAMSES code \citep{Teyssier2002,Fromang2006,Commercon2011,Masson2012}, where sink particles are implemented \citep{Krumholz2004,BleulerTeyssier2014}. \citet{Risse2020} and Verliat et al., in preparation present in detail similar simulations; however, their simulations employ a novel radiative transfer method that the simulations we use here do not. Three of the simulations follow the collapse of magnetized, intermediate-mass dense cores without initial turbulence, while the three others follow the collapse of weakly magnetized, low-mass cores with initial turbulence. The idea behind our analysis of these simulations is to choose different physical conditions to represent the variety of environments present in the observed ALMA cores. To this aim, we use as our models six simulation outputs with central stellar masses between 0.5 and 7\,M$_\odot$ that sample randomly a domain in initial mass, magnetic energy, turbulent energy, as is also the case with our observations. The details of models we use can be found in Table \ref{t.simu_details}. In this paper, we do not aim to reproduce or interpret the polarized dust emission from Class 0 envelopes as resulting from perfect alignment. With the goal to assess the statistical properties of polarization, one model already provides enough data points to make a statistical analysis: the inclusion of several models only allow to illustrate that there may be some local conditions (turbulence) affecting slightly the trends, and that irradiation due to the central protostar is key.

We perform radiative transfer calculations on these simulations using the POLARIS code \citep{Reissl2016}, which calculate the local dust temperature and dust grain alignment efficiency of oblong dust grains with respect the magnetic field orientation following the RAT theory developed in \citet{LazarianHoang2007,Hoang2014b}. In each run of POLARIS, we either choose to calculate the grain alignment of each dust grain via RATs, or we employ the Perfect alignment (PA) hypothesis, which assumes that all susceptible grains are aligned with their long axes perpendicular to the local magnetic field orientation. We derive the temperature of the central object from the luminosity of the blackbody, which is indexed to the mass of the sink following the empirical correlation from \citet{Cox_and_Giuli2004}. We also include the interstellar radiation field using a value of $G_0=1$ \citep{Mathis1983}. Note that in order to compute the radiative transfer in a reasonable amount of time, we must delete the mass in the highest density cells surrounding the sink (in a $\sim$\,15\,au diameter region). This may result in an overestimated radiation field in the core, as the photons will not be processed by the material we remove. We assume a gas-to-dust ratio of 100. The dust grain population is composed of 62.5\% astronomical silicates and 37.5\% graphite grains \citep{Mathis1977}; note that this composition governs the ultimate number of aligned grains in the PA regime, as silicates can be aligned with the magnetic field much more easily than graphite/carbonaceous grains \citep[][and references therein]{Andersson2015}. The dust grains are oblate with an aspect ratio of 0.5 \citep{Hildebrand_Dragovan1995} and they follow a standard MRN-like distribution \citep{Mathis1977} with cutoff sizes of $a_\textrm{min}\,=\,2\,\textrm{nm}$ and $a_\textrm{max}\,=\,10\,\mu\textrm{m}$. We choose this latter value as the maximum grain size in POLARIS in light of recent work that has hinted at the presence of grains larger than the typical $\sim$\,0.1\,$\mu$m ISM dust grains in Class 0 envelopes \citep[e.g.,][]{Valdivia2019,LeGouellec2019a,Galametz2019,Hull2020a}. The radiation field resulting from the radiative transfer, impinging on the dust grains in the protostellar envelope, comprises low-energy submillimeter photons whose wavelength need to be comparable to the size of dust grains in order to efficiently align the grains via RATs.

We synthetically observe the MHD simulations with POLARIS at 870 $\mu$m, at a distance of 400\,pc, in maps 8000\,au in size with pixel sizes of 8\,au.
We observed each of the six simulations along two independent, orthogonal lines of sight.
As a result, we analyse twelve different POLARIS synthetic observations, each of which was produced assuming grain alignment via either RATs or PA. Thus, this sample of twelve models and our fifteen ALMA datasets yield a similar numbers of cases to which we can apply our statistical analysis. From POLARIS we obtain the three Stokes parameters $I$, $Q$, and $U$, which we convolve with a 2D Gaussian kernel to smooth out the different resolutions of the cells based on local density, which is due to the use of Adaptive Mesh Refinement (AMR). We choose a pixel size of 8\,au with POLARIS; however, within the 8000\,au core, there are many different cell sizes, which degrades the spatial resolution in some regions of our radiative transfer maps. In order to have independent points while running our statistics, we smooth the resulting Stokes maps to 80\,au resolution, which is the largest cell size in the central region of the synthetic observation. Beyond this central region, the AMR cell sizes are even larger than 80\,au, but we compute the statistics within the central $\sim$\,1500\,au zone, where the AMR cell size is smaller than 80 au. At this point, we obtain in this central zone a first set of ``perfect'' maps on which we calculate the same statistical estimators used to study the polarization properties of our ALMA observations; we denote these perfect synthetic observations, ``without filtering.'' In addition, we use the CASA simulator to interferometrically filter the synthetically observed maps, mimicking ALMA observations. For each simulation, we combine synthetic observations from ALMA configurations C-3, C-5, and C-6, with an exposure time of 6000\,s per antenna configuration. The resulting synthesized beams (resolution elements) of these filtered maps have an effective size of 80\,au. After filtering the maps with the CASA simulator, we compute our statistics in the same way that we do with the ALMA data, using the threshold criterion of Stokes $I$ explained in Section \ref{sec:method}. We denote this latter set of results, ``with filtering.''  Similar to Figure \ref{fig:S_I_pol_maps_sources}, we present in Figures \ref{fig:S_I_pol_maps_simu1} and \ref{fig:S_I_pol_maps_simu2} the maps of Stokes $I$, polarized intensity $P$, and dispersion of polarization angles $\S$ for all the projections of the MHD simulations we used, before and after filtering. In these Figures, we only show the perfect alignment case, as the emission in these maps is clearer than in the case of alignment via RATs.

\begin{table*}[!tbph]
\centering
\small
\caption[]{Simulation information}
\setlength{\tabcolsep}{0.26em} 
\begin{tabular}{p{0.23\linewidth}ccccccc}
\hline \hline \noalign{\smallskip}
Name & Total Mass & Sink Mass & Magnetization ($\mu$) & Turbulent Mach number & Age
\\
&&$M_\odot$&$M_\odot$&&kyr \\
\noalign{\smallskip}  \hline
\noalign{\smallskip}
High-luminosity model I&100&4.1&5&0&39.2\\
\noalign{\smallskip}
\hline
\noalign{\smallskip}
High-luminosity model II&100&3.7&5&0&37.1\\
\noalign{\smallskip}
\hline
\noalign{\smallskip}
High-luminosity model III&\phantom{0}60&7.4&5&0&35.6\\
\noalign{\smallskip}
\hline
\noalign{\smallskip}
Low-luminosity model I&\phantom{0}30&1.3&9&2&148\phantom{.0}\\
\noalign{\smallskip}
\hline
\noalign{\smallskip}
Low-luminosity model II&\phantom{0}30&2.0&9&2&187\phantom{.0}\\
\noalign{\smallskip}
\hline
\noalign{\smallskip}
Low-luminosity model III&\phantom{0}30&0.8&9&2&153\phantom{.0}\\
\noalign{\smallskip}
\hline
\noalign{\smallskip}
\smallskip
\end{tabular}
\caption*{\small All the simulations have a spatial resolution of a few au, and implement ambipolar diffusion. The initial density profile is $\rho \varpropto 1/(1+r^2)$ in runs I, II, III, and is uniform in runs IV, V, and VI. For each of them we select a time step at which the simulation exhibits compact features in density similar to those that we see in the ALMA observations, \ie bright emission from the infalling envelope, disc-like structures, and bipolar outflow cavities.\\
$^a$ The sink mass is that of the largest central sink. This core is fragmenting, and thus there are several other, smaller sinks.\\
$^b$ The jet is implemented by hand, with a speed of 66$\%$ of the escape speed, and an opening angle of 30$^{\circ}$. The corresponding outflowing mass ejected by the sink is 1/3 of the mass accreted by the sink.}
\label{t.simu_details}
\end{table*}

\def\scaleSP{0.65}
\def\extansion{}

\begin{figure*}[!tbph]
\centering
\subfigure{\includegraphics[scale=\scaleSP,clip,trim= 1.9cm 3.1cm 1.85cm 2.65cm]{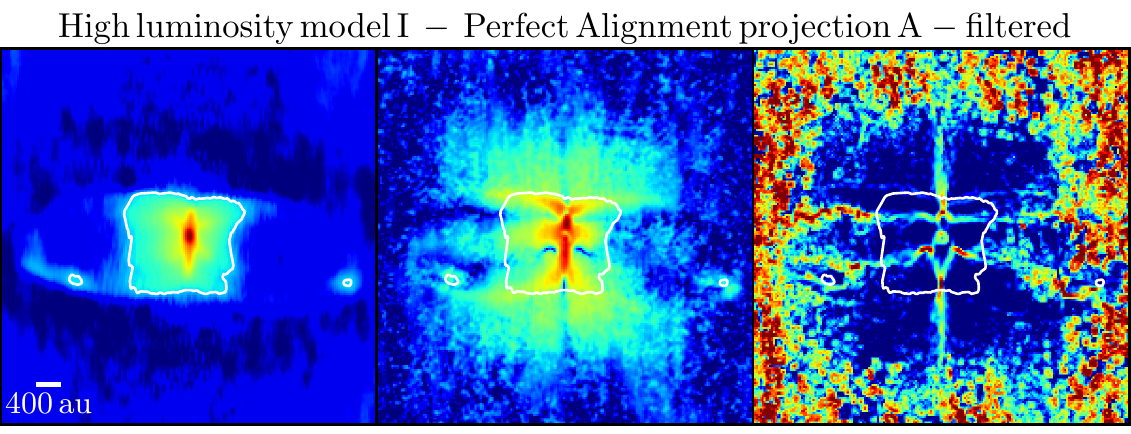}}
\vspace{-0.2cm}
\subfigure{\includegraphics[scale=\scaleSP,clip,trim= 1.9cm 3.1cm 0.2cm 2.65cm]{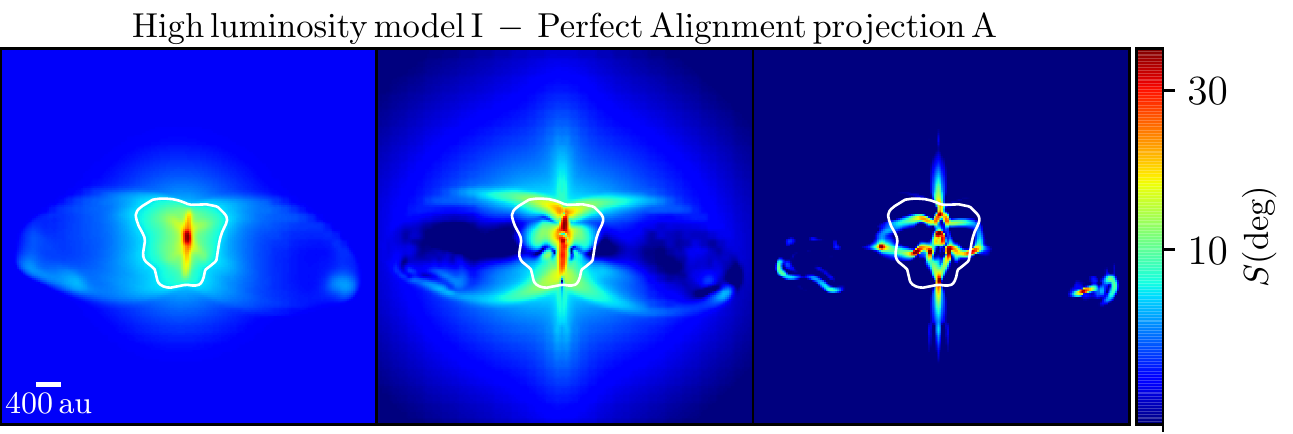}}
\subfigure{\includegraphics[scale=\scaleSP,clip,trim= 1.9cm 3.1cm 1.85cm 2.65cm]{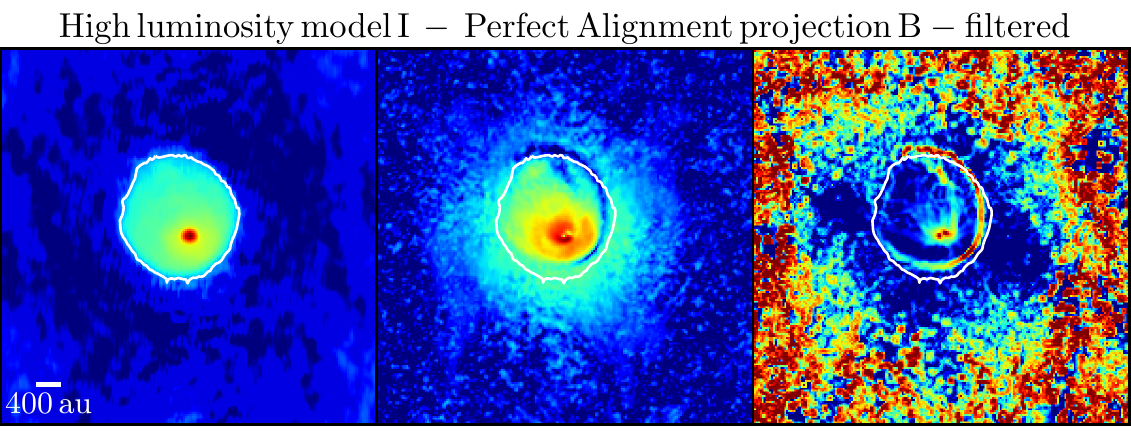}}
\vspace{-0.2cm}
\subfigure{\includegraphics[scale=\scaleSP,clip,trim= 1.9cm 3.1cm 0.2cm 2.65cm]{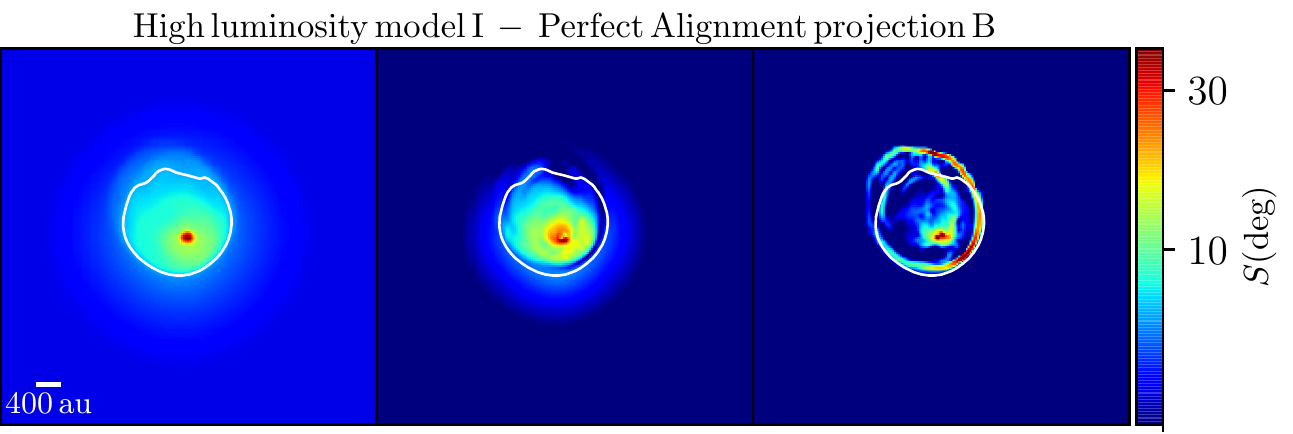}}
\subfigure{\includegraphics[scale=\scaleSP,clip,trim= 1.9cm 3.1cm 1.85cm 2.7cm]{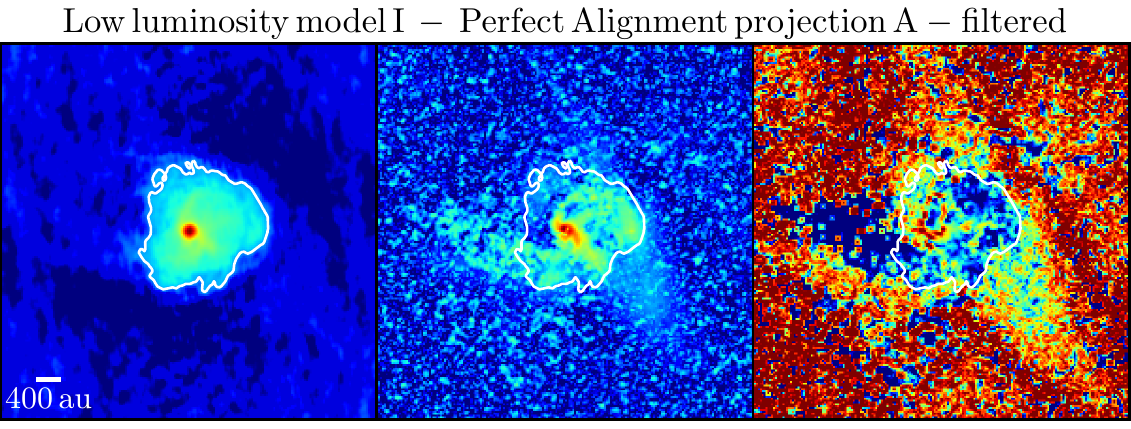}}
\vspace{-0.2cm}
\subfigure{\includegraphics[scale=\scaleSP,clip,trim= 1.9cm 3.1cm 0.2cm 2.65cm]{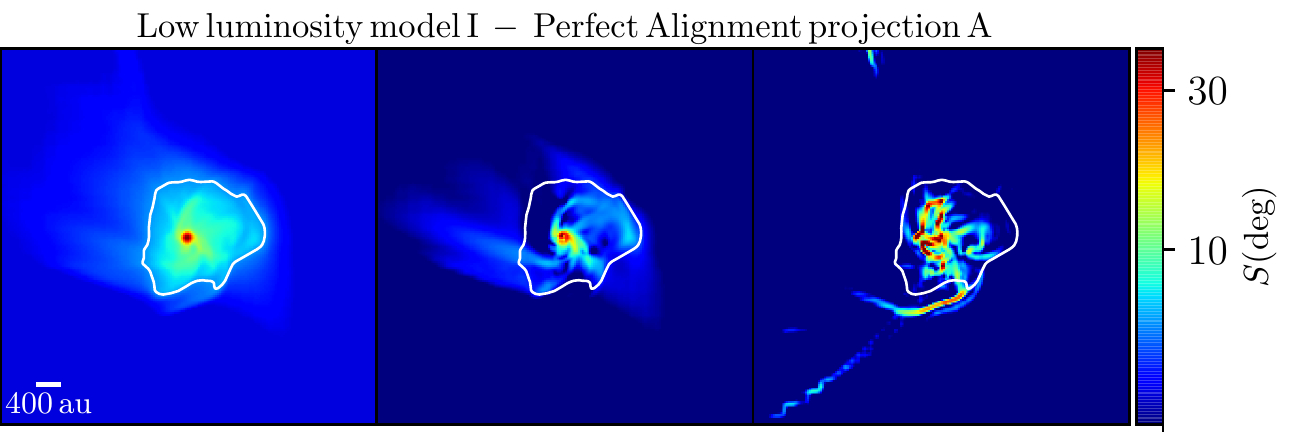}}
\subfigure{\includegraphics[scale=\scaleSP,clip,trim= 1.9cm 3.1cm 1.85cm 2.65cm]{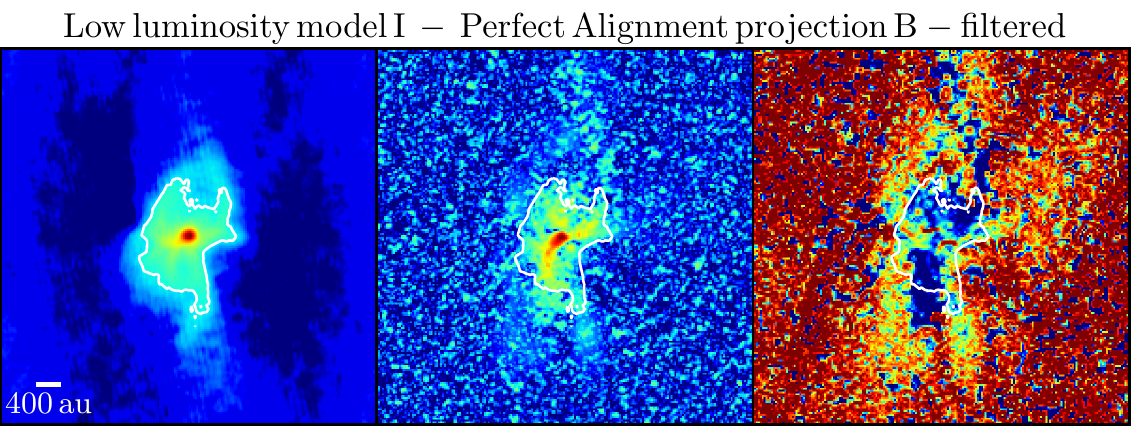}}
\vspace{-0.2cm}
\subfigure{\includegraphics[scale=\scaleSP,clip,trim= 1.9cm 3.1cm 0.2cm 2.65cm]{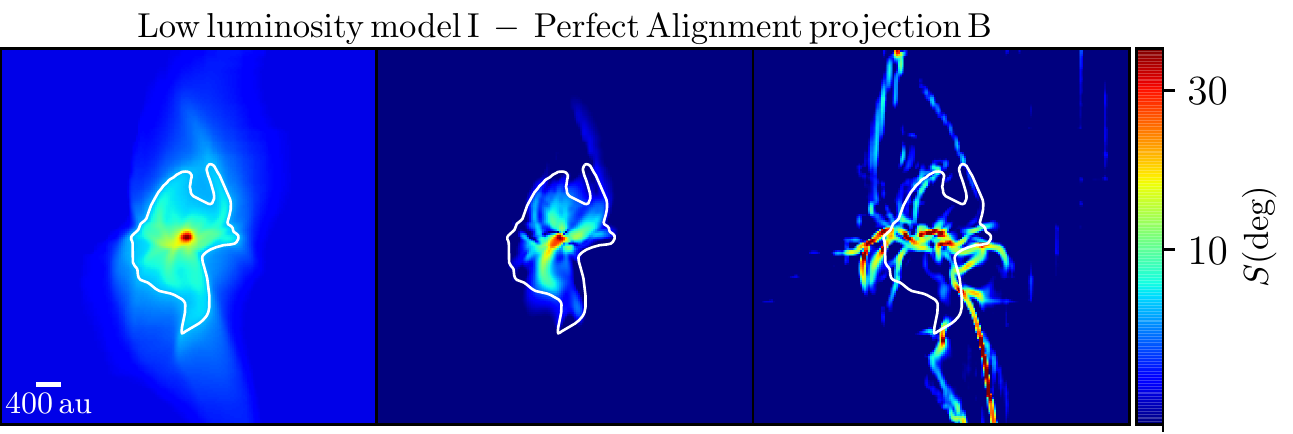}}
\subfigure{\includegraphics[scale=\scaleSP,clip,trim= 1.9cm 3.1cm 1.85cm 2.65cm]{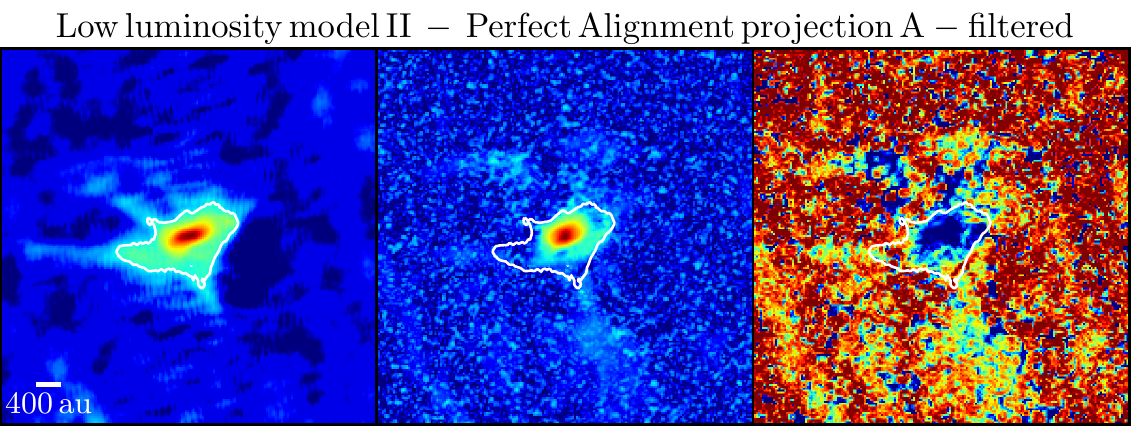}}
\vspace{-0.2cm}
\subfigure{\includegraphics[scale=\scaleSP,clip,trim= 1.9cm 3.1cm 0.2cm 2.65cm]{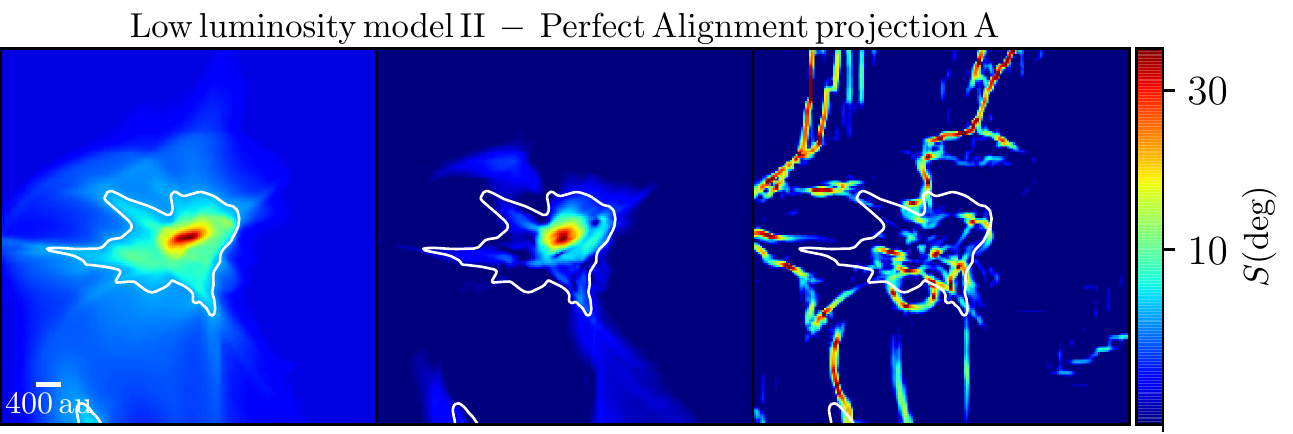}}
\subfigure{\includegraphics[scale=\scaleSP,clip,trim= 1.9cm 3.1cm 1.85cm 2.65cm]{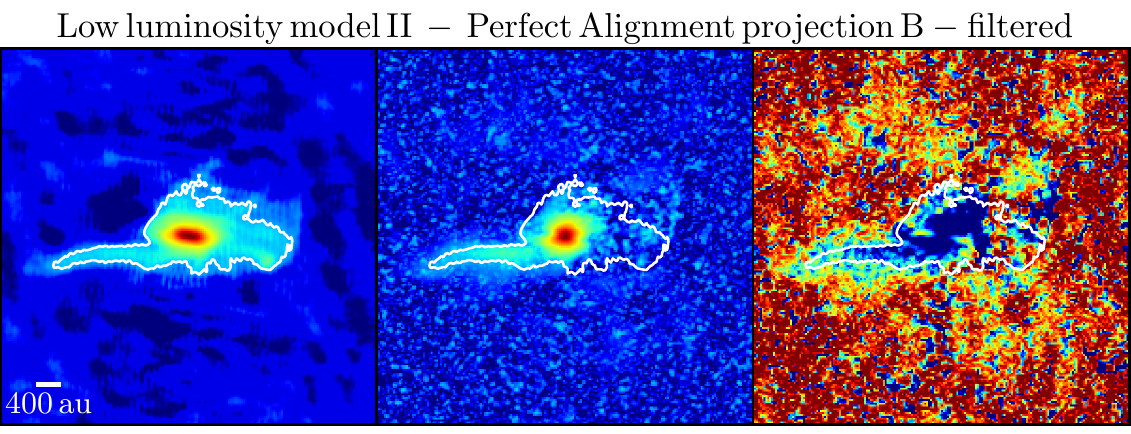}}
\vspace{-0.2cm}
\subfigure{\includegraphics[scale=\scaleSP,clip,trim= 1.9cm 3.1cm 0.2cm 2.65cm]{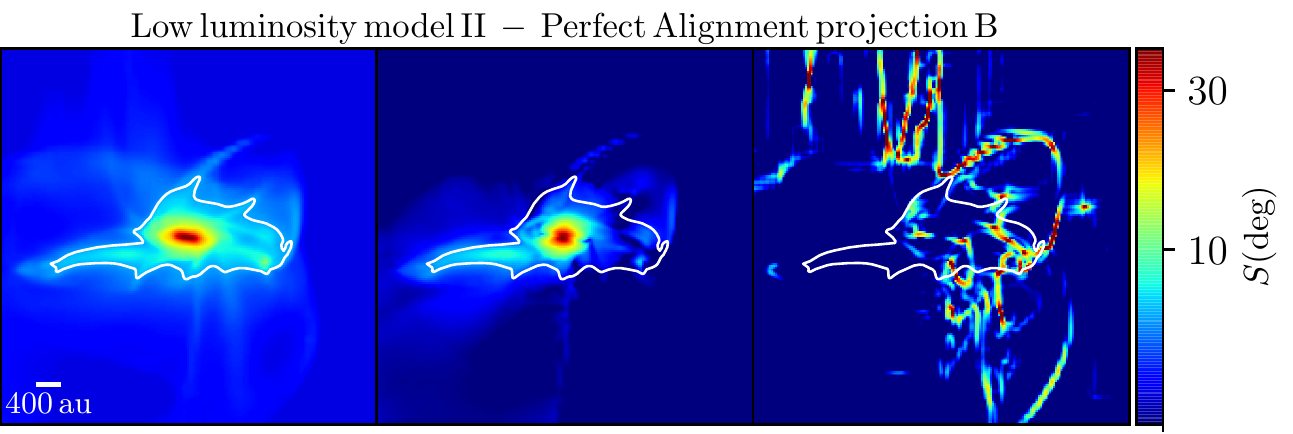}}
\subfigure{\includegraphics[scale=\scaleSP,clip,trim= 1.9cm 3.1cm 1.85cm 2.65cm]{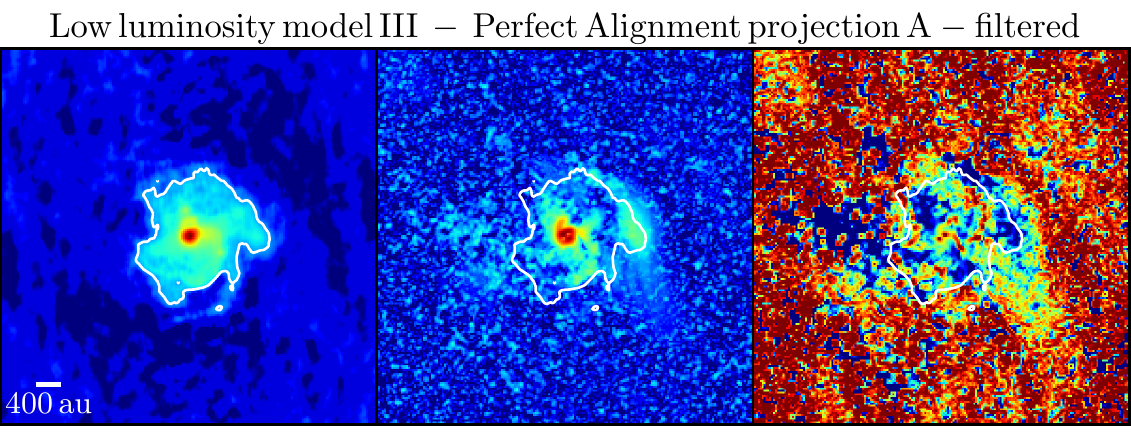}}
\vspace{-0.2cm}
\subfigure{\includegraphics[scale=\scaleSP,clip,trim= 1.9cm 3.1cm 0.2cm 2.65cm]{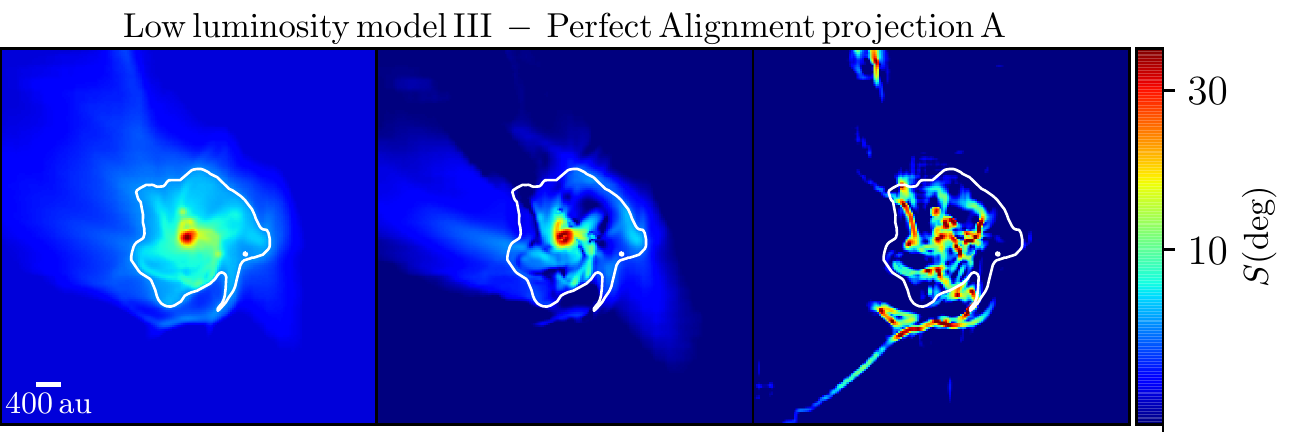}}
\subfigure{\includegraphics[scale=\scaleSP,clip,trim= 1.9cm 3.1cm 1.85cm 2.65cm]{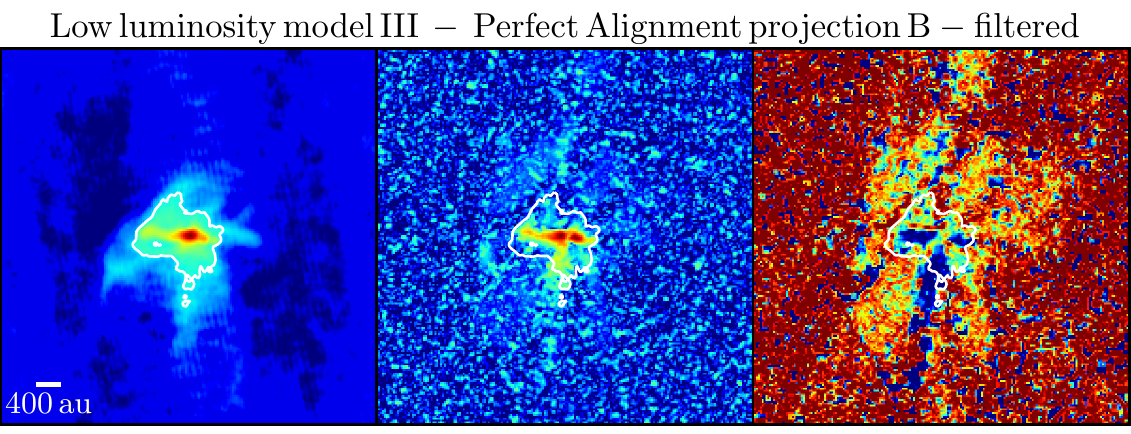}}
\vspace{-0.2cm}
\subfigure{\includegraphics[scale=\scaleSP,clip,trim= 1.9cm 3.1cm 0.2cm 2.65cm]{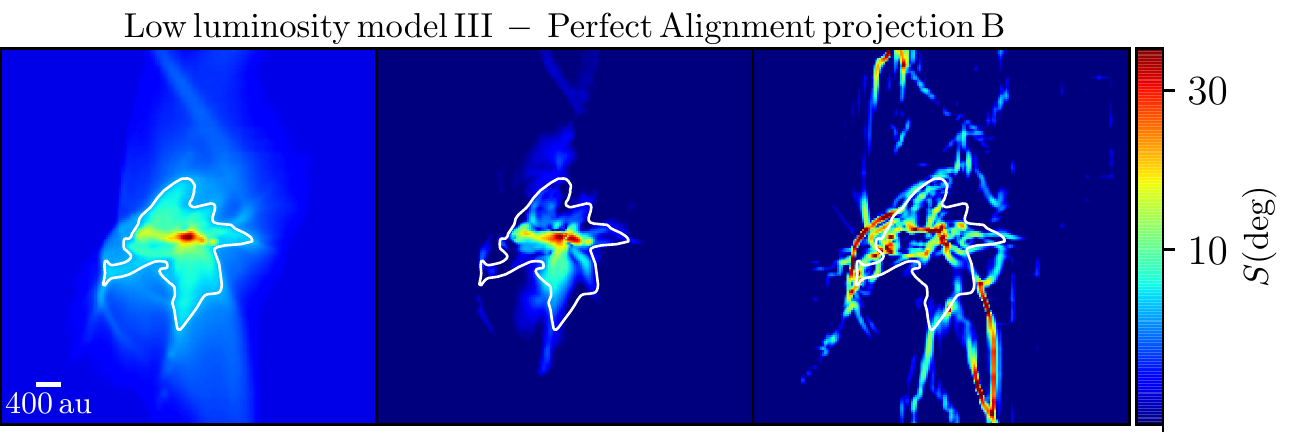}}
\caption[]{\small Same as Figure \ref{fig:S_I_pol_maps_sources}, but for the simulations. On the left the cores are filtered, on the right they are not.}
\label{fig:S_I_pol_maps_simu1}
\vspace{0.2cm}
\end{figure*}

\def\extansion{}

\begin{figure*}[!tbph]
\centering
\vspace{-0.1cm}
\subfigure{\includegraphics[scale=\scaleSP,clip,trim= 1.9cm 3.1cm 1.85cm 2.65cm]{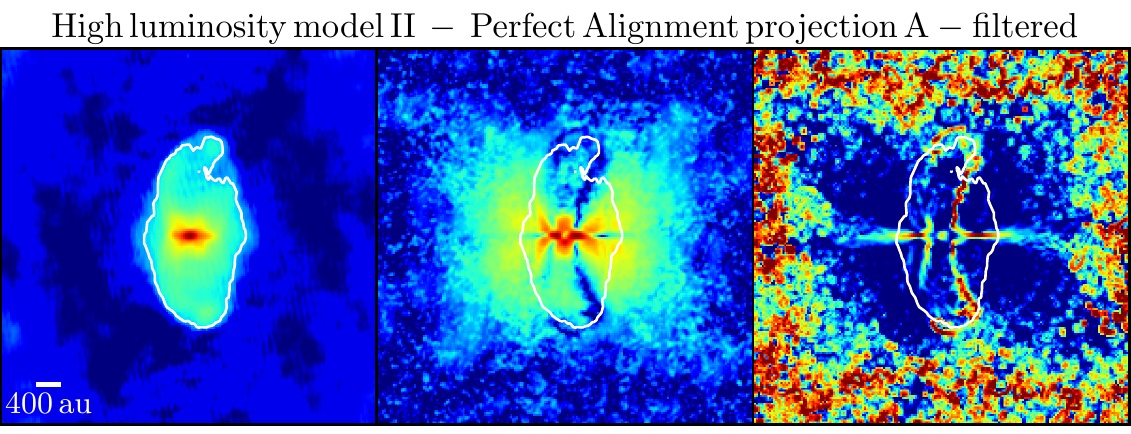}}
\vspace{-0.2cm}
\subfigure{\includegraphics[scale=\scaleSP,clip,trim= 1.9cm 3.1cm 0.2cm 2.65cm]{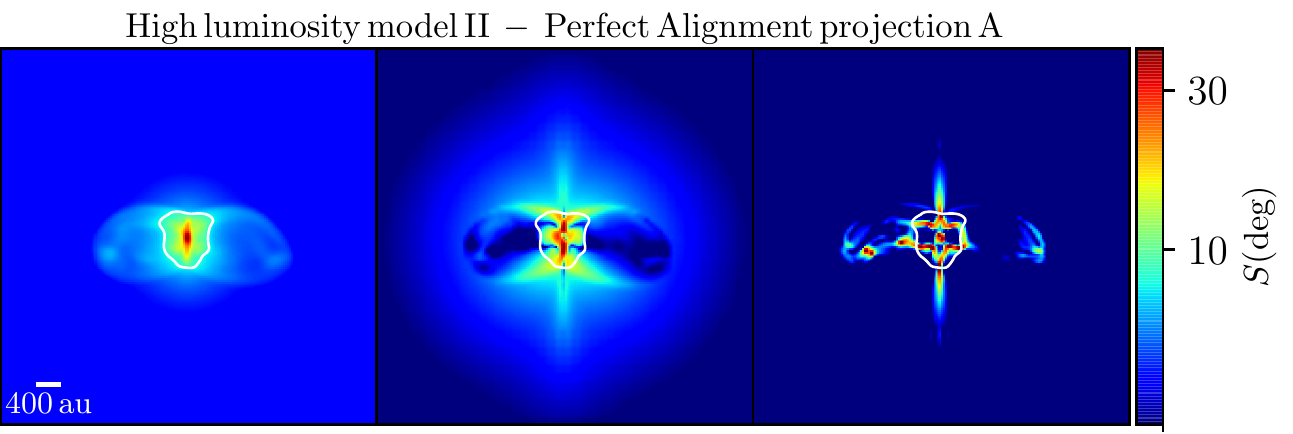}}
\subfigure{\includegraphics[scale=\scaleSP,clip,trim= 1.9cm 3.1cm 1.85cm 2.65cm]{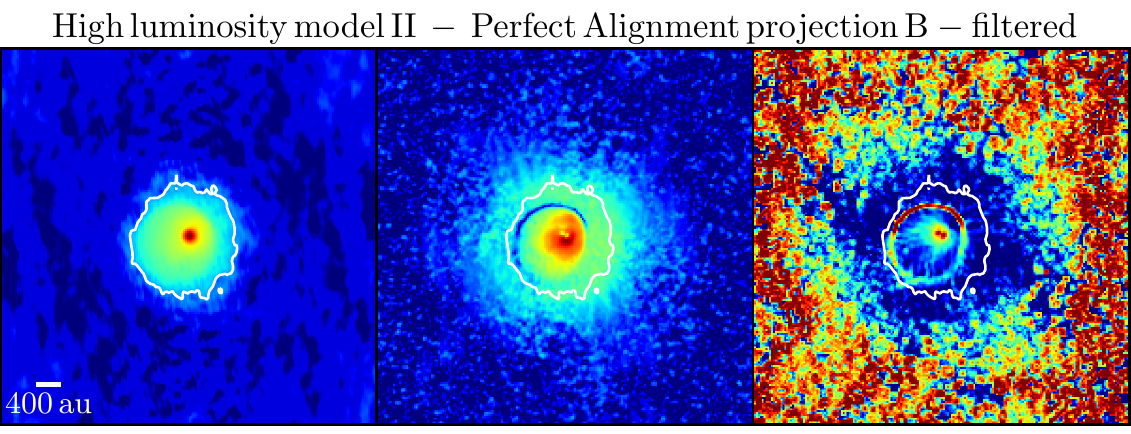}}
\vspace{-0.2cm}
\subfigure{\includegraphics[scale=\scaleSP,clip,trim= 1.9cm 3.1cm 0.2cm 2.65cm]{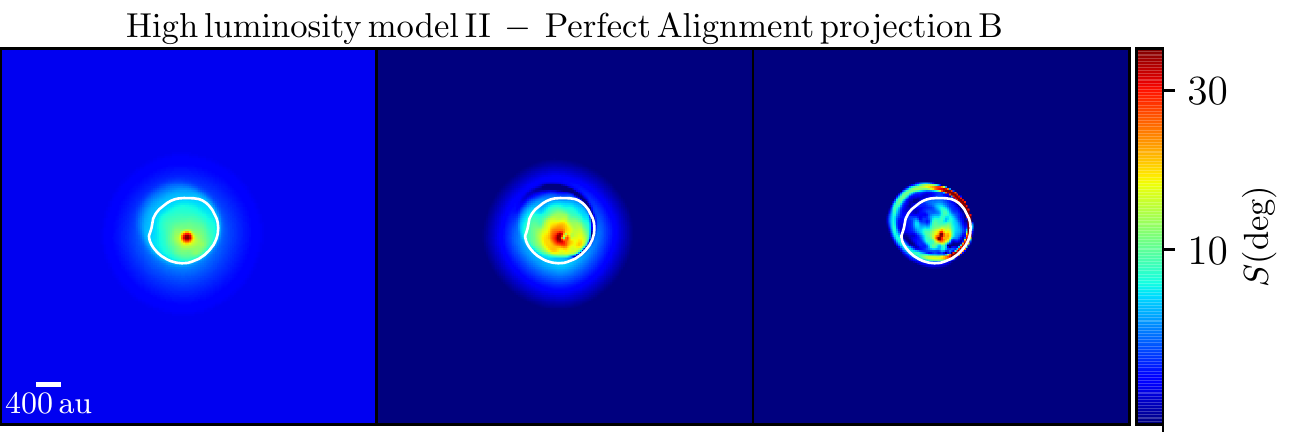}}
\subfigure{\includegraphics[scale=\scaleSP,clip,trim= 1.9cm 3.1cm 1.85cm 2.65cm]{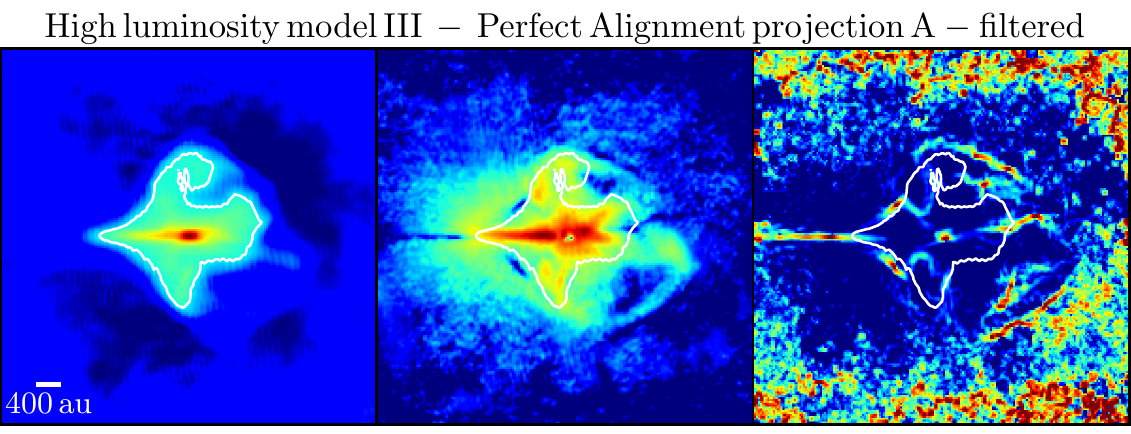}}
\vspace{-0.2cm}
\subfigure{\includegraphics[scale=\scaleSP,clip,trim= 1.9cm 3.1cm 0.2cm 2.65cm]{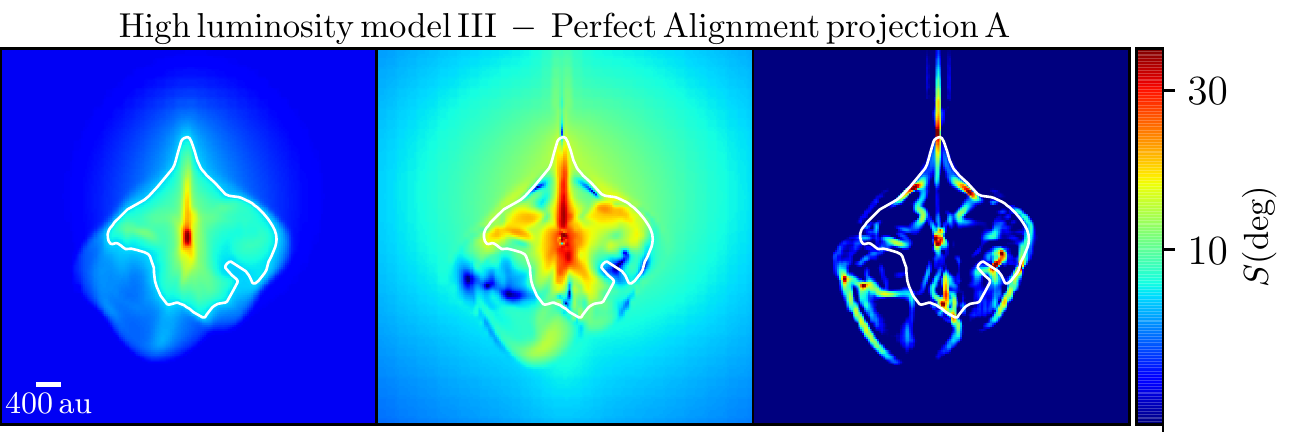}}
\subfigure{\includegraphics[scale=\scaleSP,clip,trim= 1.9cm 3.1cm 1.85cm 2.65cm]{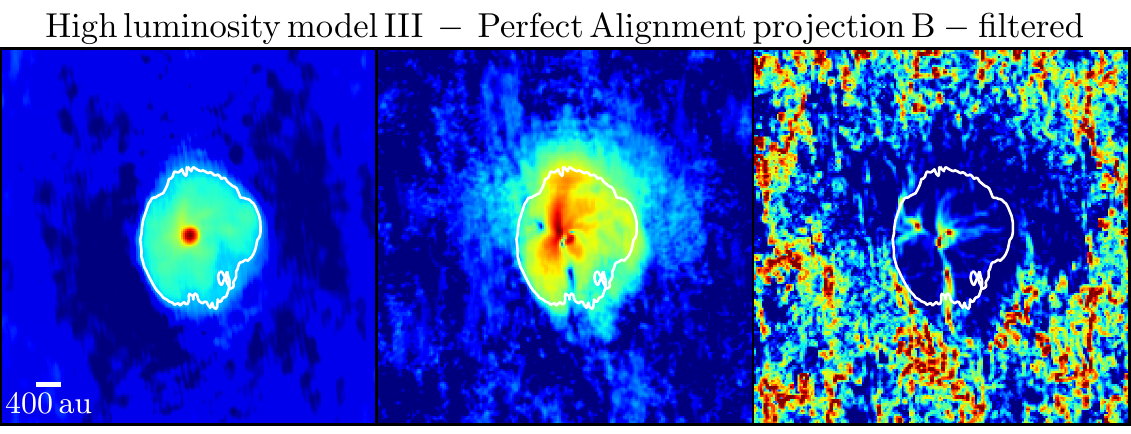}}
\vspace{-0.2cm}
\subfigure{\includegraphics[scale=\scaleSP,clip,trim= 1.9cm 3.1cm 0.2cm 2.65cm]{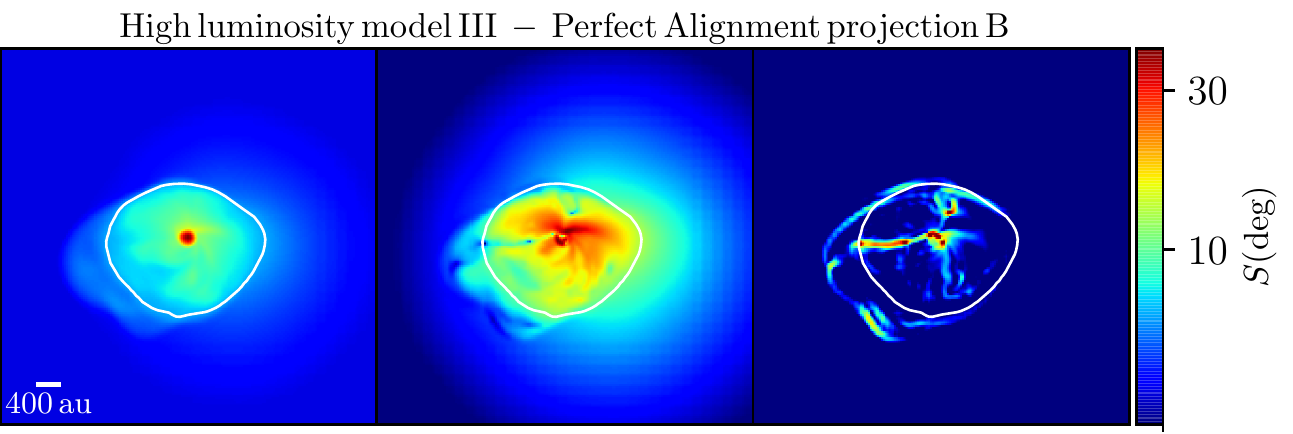}}
\caption[]{\small Same as Figure \ref{fig:S_I_pol_maps_sources}, but for the simulations. On the left the cores are filtered, on the right they are not.}
\label{fig:S_I_pol_maps_simu2}
\vspace{0.2cm}
\end{figure*}

We present in Figures \ref{fig:scatter_S_Pfrac_simu_all_wturbu} and \ref{fig:scatter_S_Pfrac_simu_all_woturbu} the distribution of the polarization fraction $\Pf$ as a function of the dispersion of polarization angles $\S$ in the synthetically observed maps, separating the three simulations that implement initial turbulence and have lower total mass from the three others that do not implement turbulence and have a much higher total mass. In each Figure, we plot $\Pf$ versus $\S$ using PA and RATs, before and after spatial filtering.

In Figure \ref{fig:scatter_S_Pfrac_simu_all_wturbu}, we see that both $\Pf$ and $\S$ are higher in the case of perfect alignment. In the case where we use the perfect alignment hypothesis, the detected polarized emission covers larger regions of the core than with RATs. Indeed, fewer grains are aligned if we assume RATs; this explains the distribution of polarization fraction, which is directly sensitive to the dust grain alignment efficiency, and is lower in the case of RATs. It also explains the lack of detection when assuming RATs and filtering the maps, as we add atmospheric noise to non-filtered maps that are only marginally polarized. The correlation between $\S$ and $\Pf$ seems to be the closest to the observational correlation presented in Figure \ref{fig:scatter_S_Pfrac_Alma_merged} when we consider the lower-mass cases where the simulations have initial turbulence, have perfect grain alignment, and are spatially filtered.

The results from the statistics using the other set of three simulations (see Figure \ref{fig:scatter_S_Pfrac_simu_all_woturbu}), which have higher mass and no turbulence, behave differently. The observed $\S$ versus $\Pf$ correlations do not vary significantly whether we filter the maps or not, or whether we use RATs or perfect alignment. This can be explained by the fact that the central heating source is much hotter than in the lower-mass simulations: the simulations used in Figure \ref{fig:scatter_S_Pfrac_simu_all_woturbu} have larger initial masses, and are synthetically observed at later times in terms of core evolution, which means that their sinks are more massive, being on the order of a few solar masses. In consequence, RATs appear to be so efficient that the statistics obtained from these simulations are very close to those obtained when we assume perfect alignment.
The correlations fitted to the distributions do not vary significantly within these four sub-cases; however, we still notice that on average, the distributions from the perfect alignment cases tend to have larger values of $\S$ and $\Pf$, which is the expected behavior.

\def\scaleSP{0.34}
\begin{figure*}[!tbph]
\centering
\vspace{-0.1cm}
\subfigure{\includegraphics[scale=\scaleSP,clip,trim= 1.6cm 1.4cm 3cm 2.9cm]{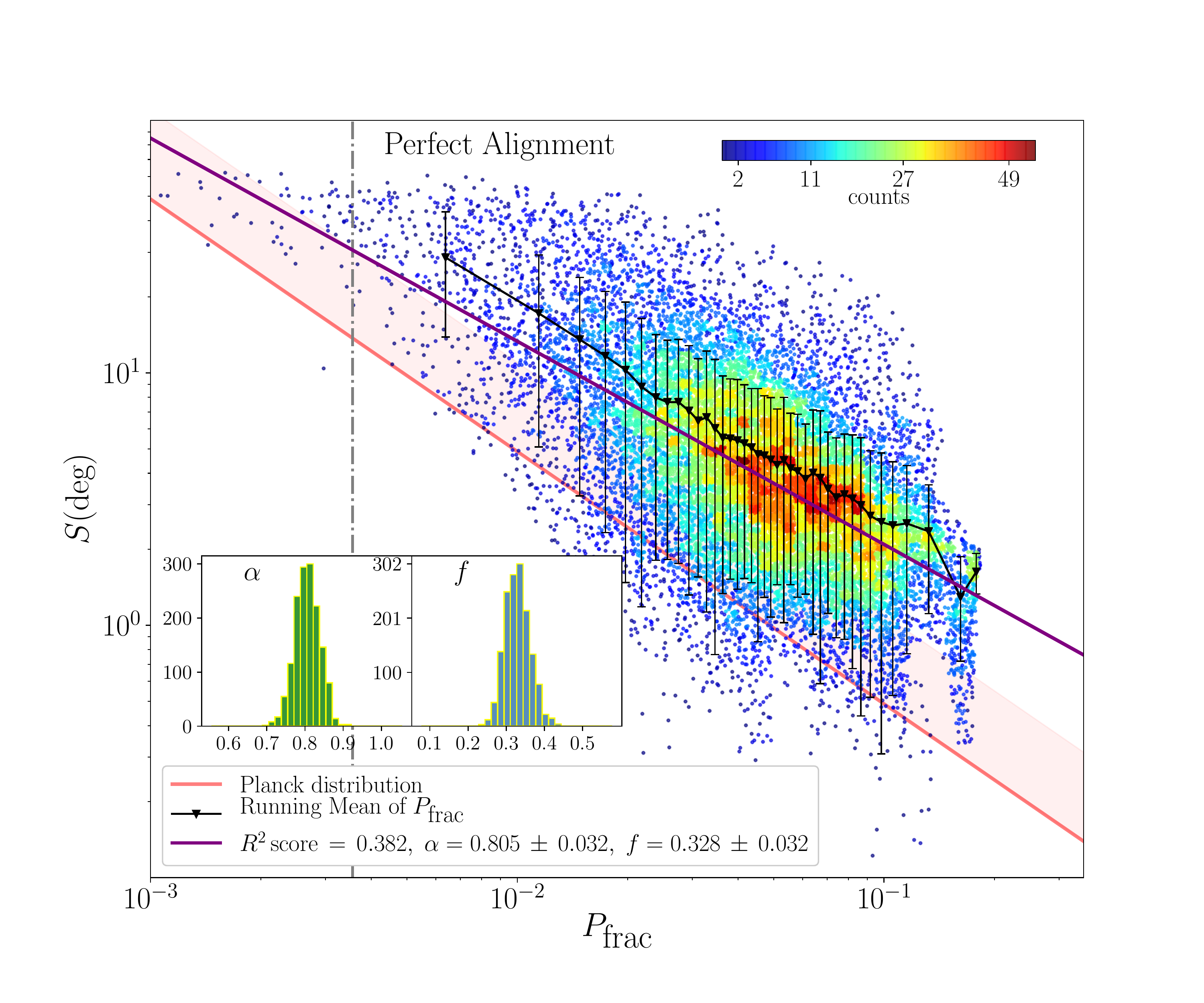}}
\subfigure{\includegraphics[scale=\scaleSP,clip,trim= 1.6cm 1.4cm 3cm 2.9cm]{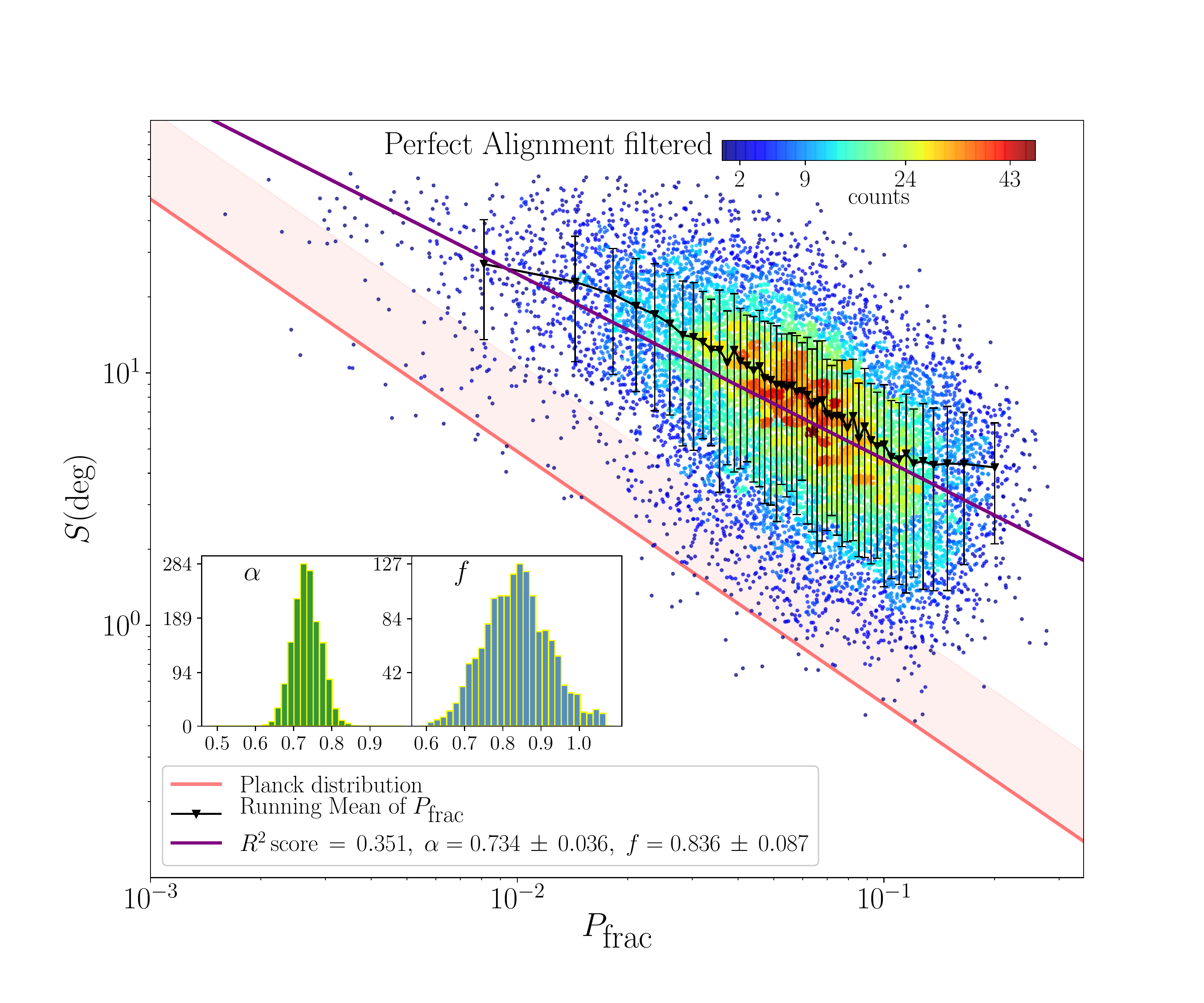}}
\subfigure{\includegraphics[scale=\scaleSP,clip,trim= 1.6cm 1.4cm 3cm 2.9cm]{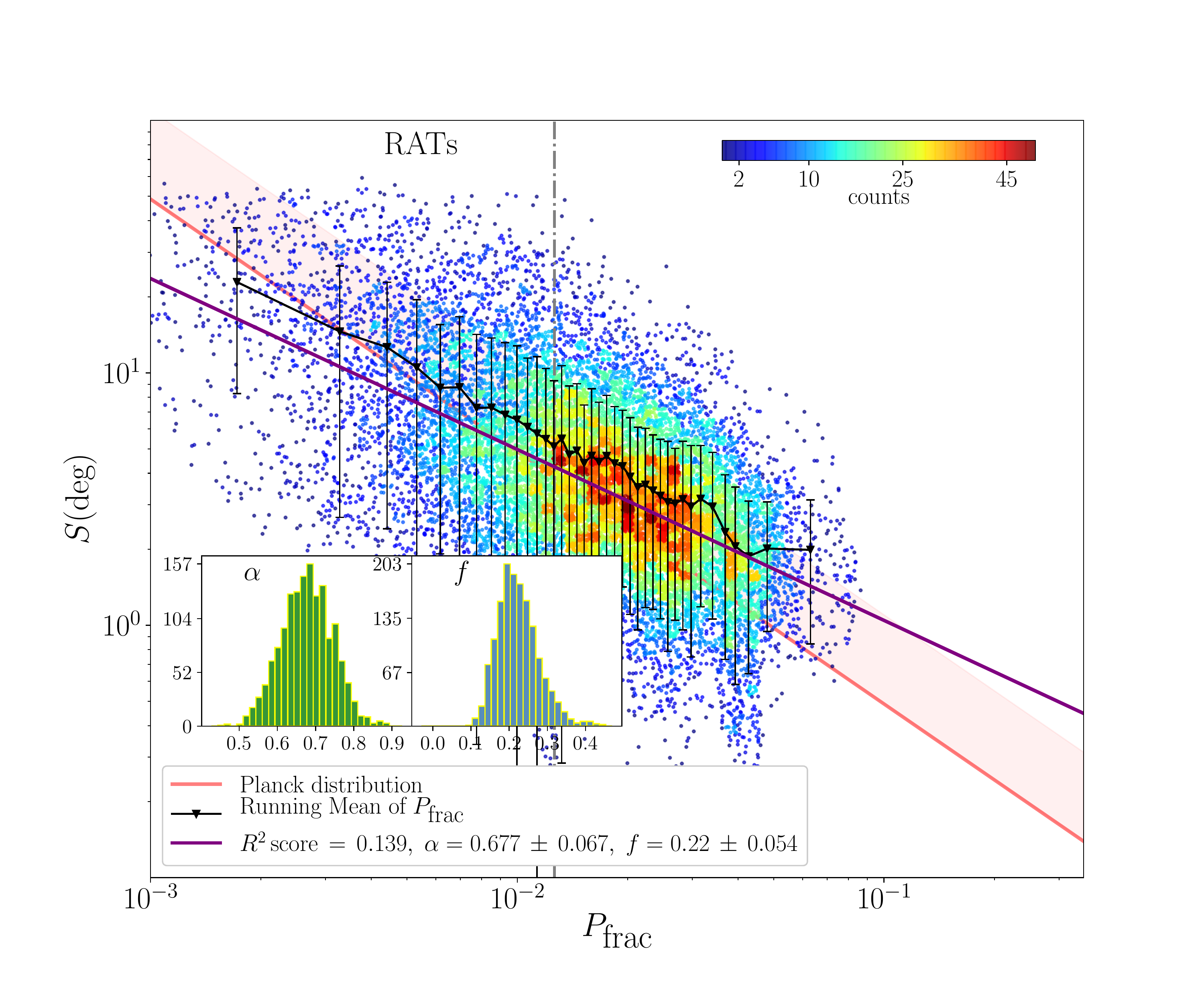}}
\subfigure{\includegraphics[scale=\scaleSP,clip,trim= 1.6cm 1.4cm 3cm 2.9cm]{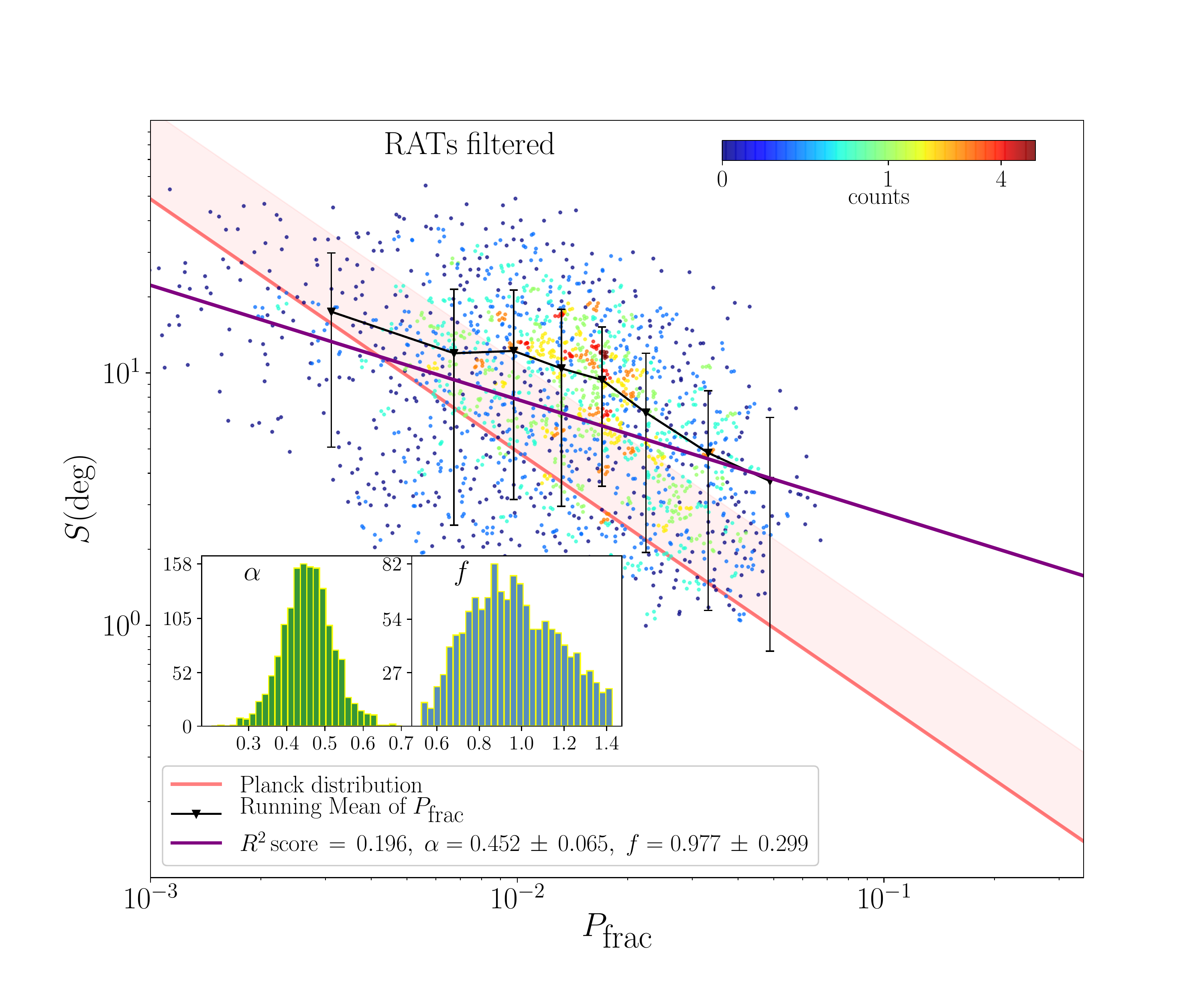}}
\caption[]{\small Statistics toward all the low-luminosity simulations. Same as Figure \ref{fig:scatter_S_Pfrac_Alma_merged}. \textit{Top-left}: perfect alignment before filtering. \textit{Top-right}: perfect alignment after filtering. \textit{Bottom-left}: RATs before filtering. \textit{Bottom-right}: RATs after filtering.}
\vspace{0.2cm}
\label{fig:scatter_S_Pfrac_simu_all_wturbu}
\end{figure*}

\begin{figure*}[!tbph]
\centering
\vspace{-0.1cm}
\subfigure{\includegraphics[scale=\scaleSP,clip,trim= 1.6cm 1.4cm 3cm 2.9cm]{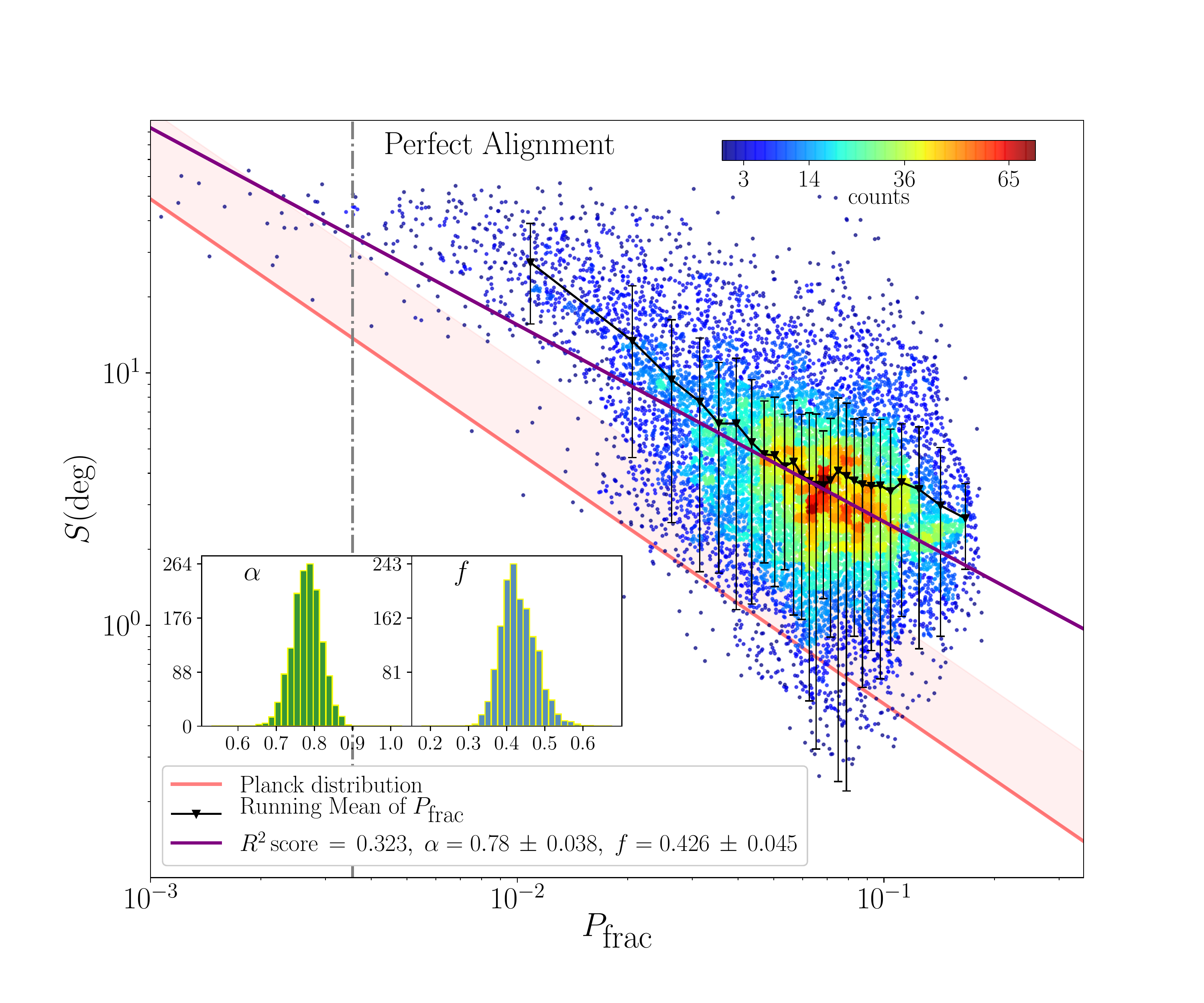}}
\subfigure{\includegraphics[scale=\scaleSP,clip,trim= 1.6cm 1.4cm 3cm 2.9cm]{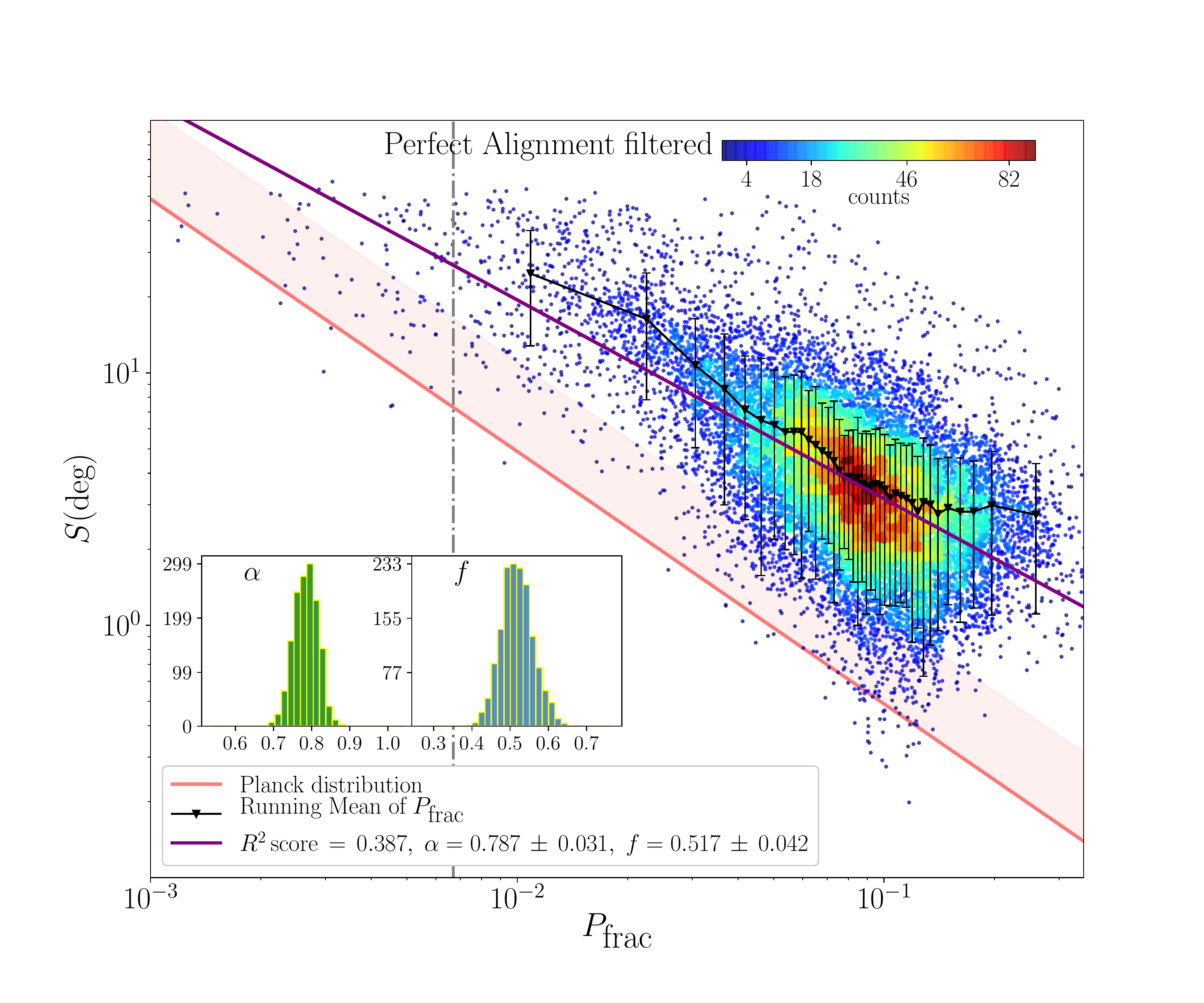}}
\subfigure{\includegraphics[scale=\scaleSP,clip,trim= 1.6cm 1.4cm 3cm 2.9cm]{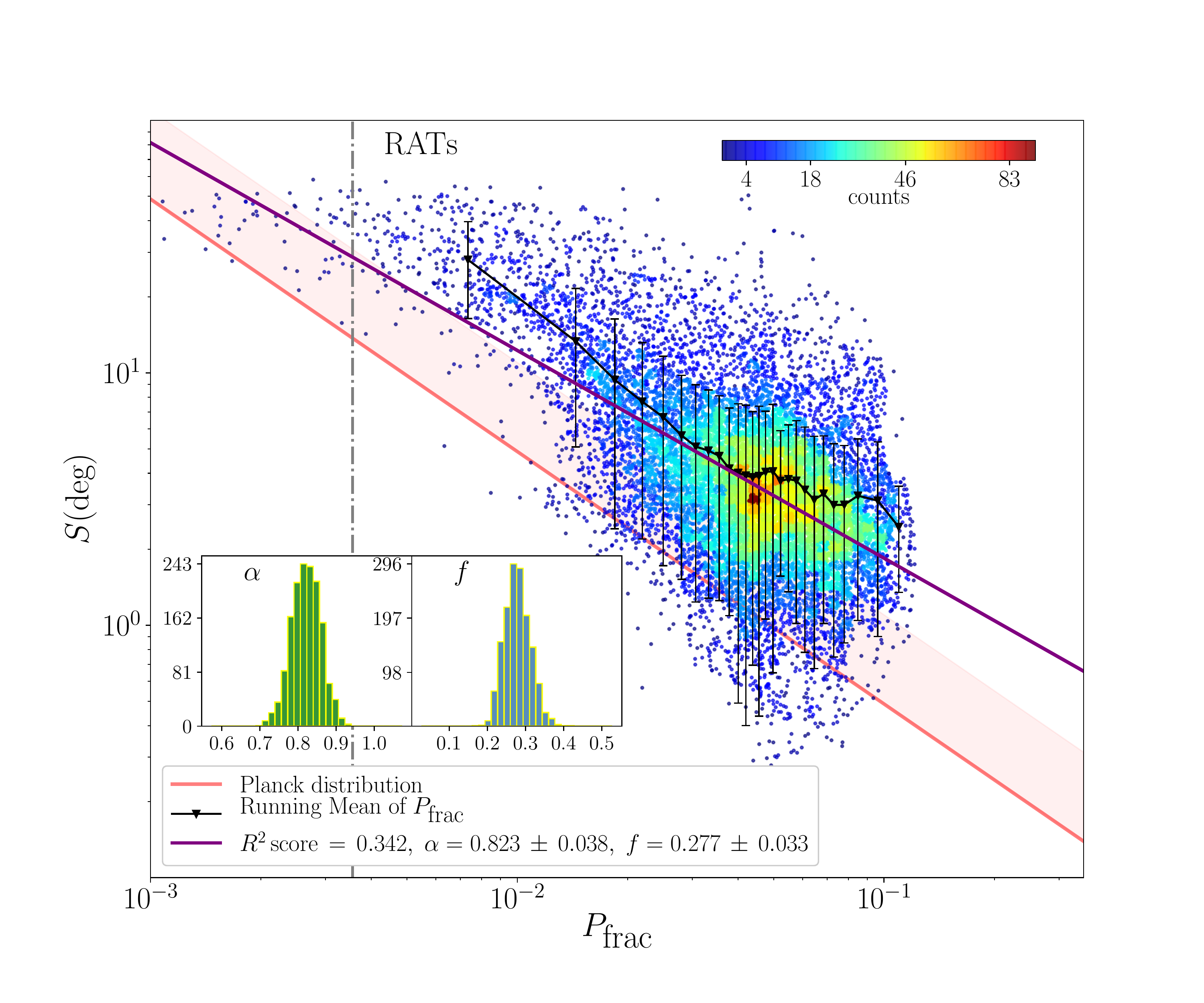}}
\subfigure{\includegraphics[scale=\scaleSP,clip,trim= 1.6cm 1.4cm 3cm 2.9cm]{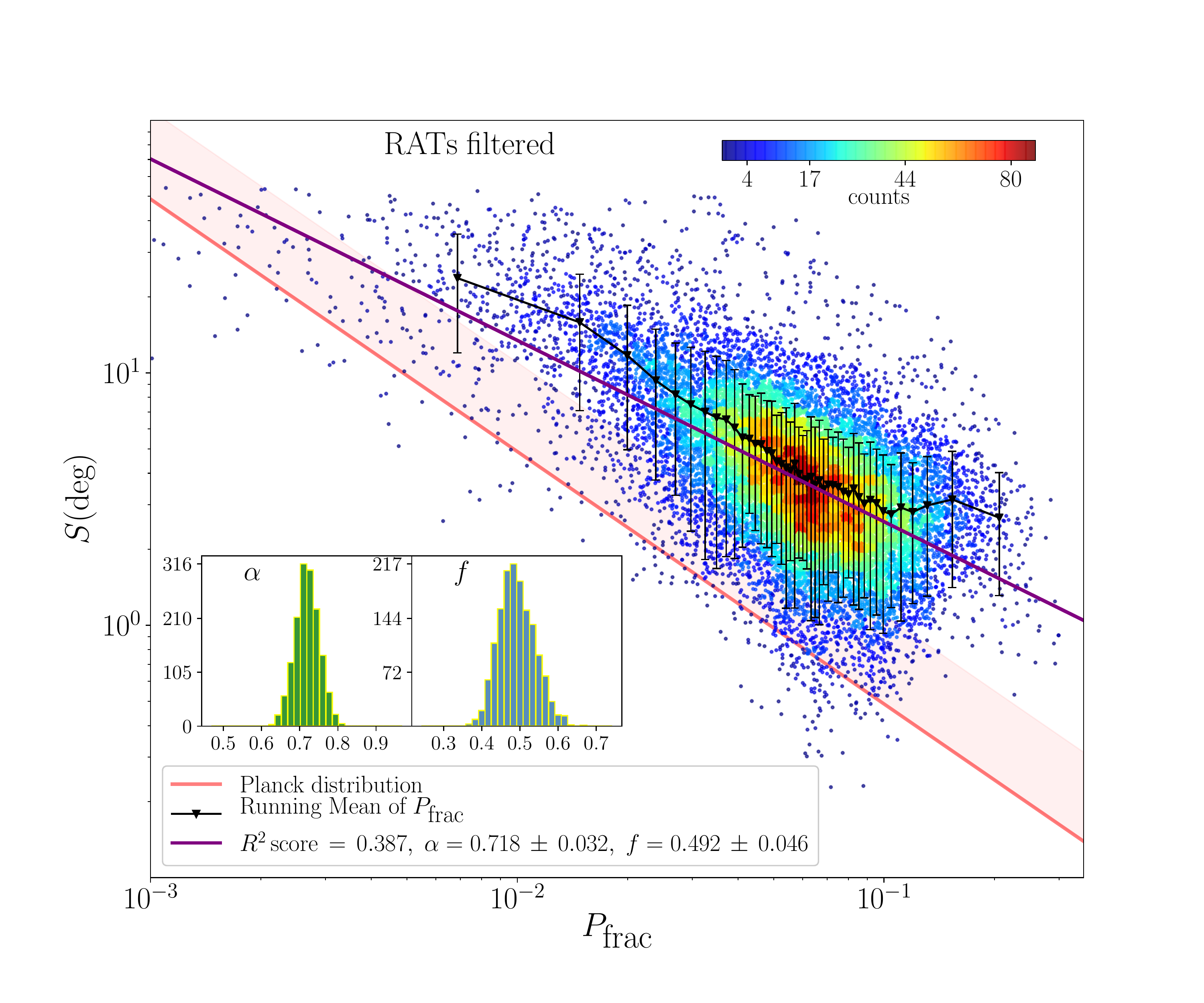}}
\caption[]{\small Same as Figure \ref{fig:scatter_S_Pfrac_simu_all_wturbu}, for all of the high-luminosity simulations.}
\vspace{0.2cm}
\label{fig:scatter_S_Pfrac_simu_all_woturbu}
\end{figure*}

\clearpage
\section{$\Pi$ Investigations}
\label{app:PI_attempt}

In the analytical model of \citet{Planck2018XII}, they demonstrate the $\StimesP$ estimator can trace the $\mathcal{P}_\textrm{frac,max}$ parameter, given some assumptions such as that the intensity maps should not vary strongly and there should be only small differences of polarization position angles $\Delta\phi$ between adjacent cells, implying that $\tan{\Delta\phi}\approx\Delta\phi$ and $Q_jQ-U_jU\,\backsimeq\,P^2$. Finally, the assumption that $\langle S \rangle$ and $\sqrt{\langle S^2 \rangle}$ behave the same is also made in the analytical model.
These assumptions may not be valid in the emission maps of Class 0 protostellar cores, as we observe, for example, strong gradients in Stokes $I$ maps. We present here a new estimator of $\mathcal{P}_\textrm{frac,max}$, called $\Pi$, the derivation of which does not require these assumptions.  This new estimator will be investigated in detail by Guillet et al., in preparation.  In order to derive the relation between $\Pi$, $\mathcal{P}_\textrm{frac,max}$, and $f_\textrm{m}(\delta)$,  we follow the same method presented in Appendix \ref{app:Planck_model} and Appendix E of \citet{Planck2018XII}:

\begin{equation}
\Pi\,=\,\frac{1}{2P}\sqrt{\frac{1}{N}\sum_j{\left( \frac{Q_jU-QU_j}{I_j}\right)^{2}}}
\end{equation}

Therefore, we have:
\begin{equation}
\Pi\,=\,\mathcal{P}_\textrm{frac,max}\frac{f_\textrm{m}(\delta)}{\sqrt{6N}}
\end{equation}

\begin{figure*}[!tbph]
\subfigure{\includegraphics[scale=0.34
,clip,trim= 1.5cm 1.2cm 2.5cm 1.9cm]
{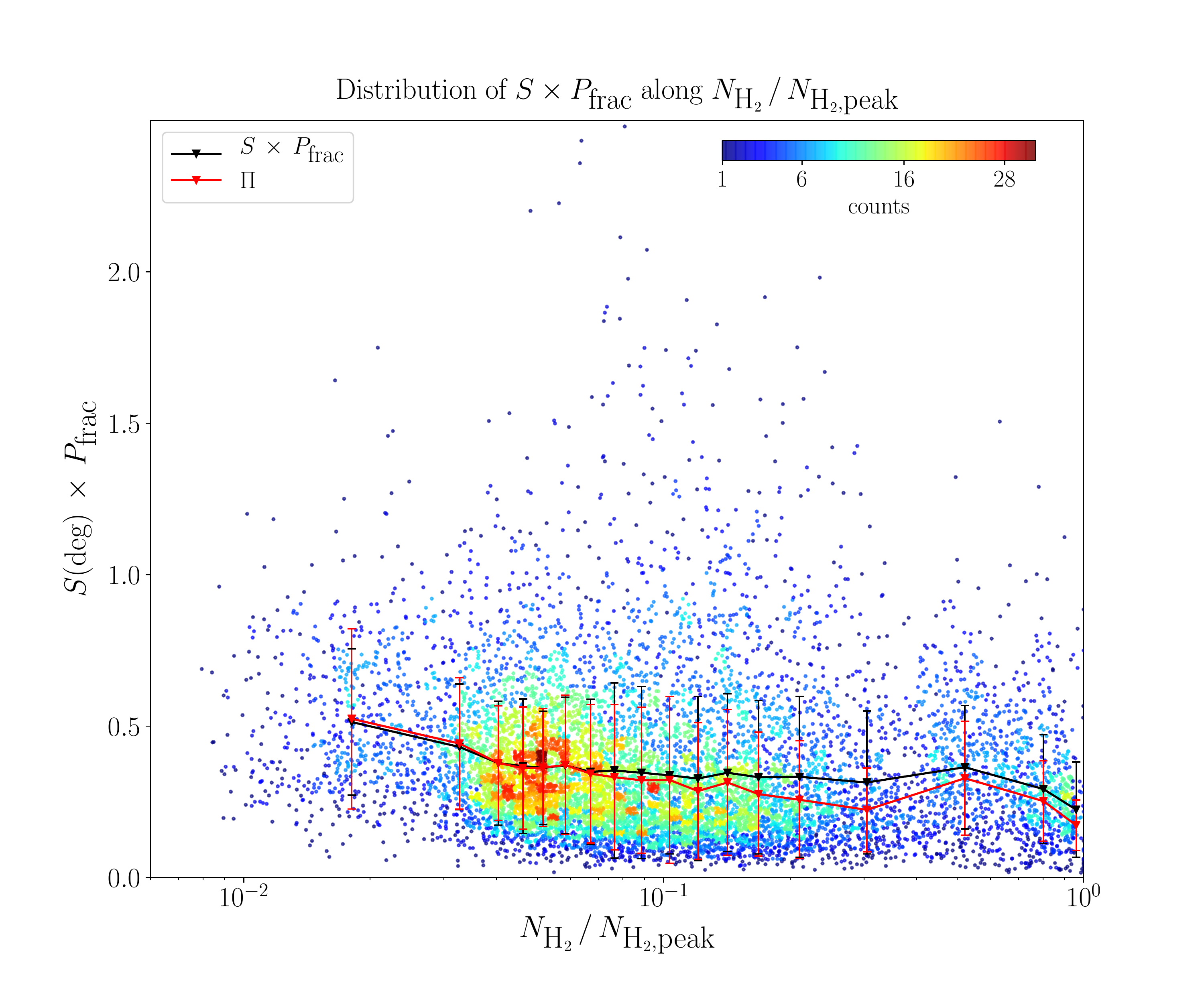}}
\subfigure{\includegraphics[scale=0.34
,clip,trim= 1.5cm 1.2cm 2.5cm 1.9cm]
{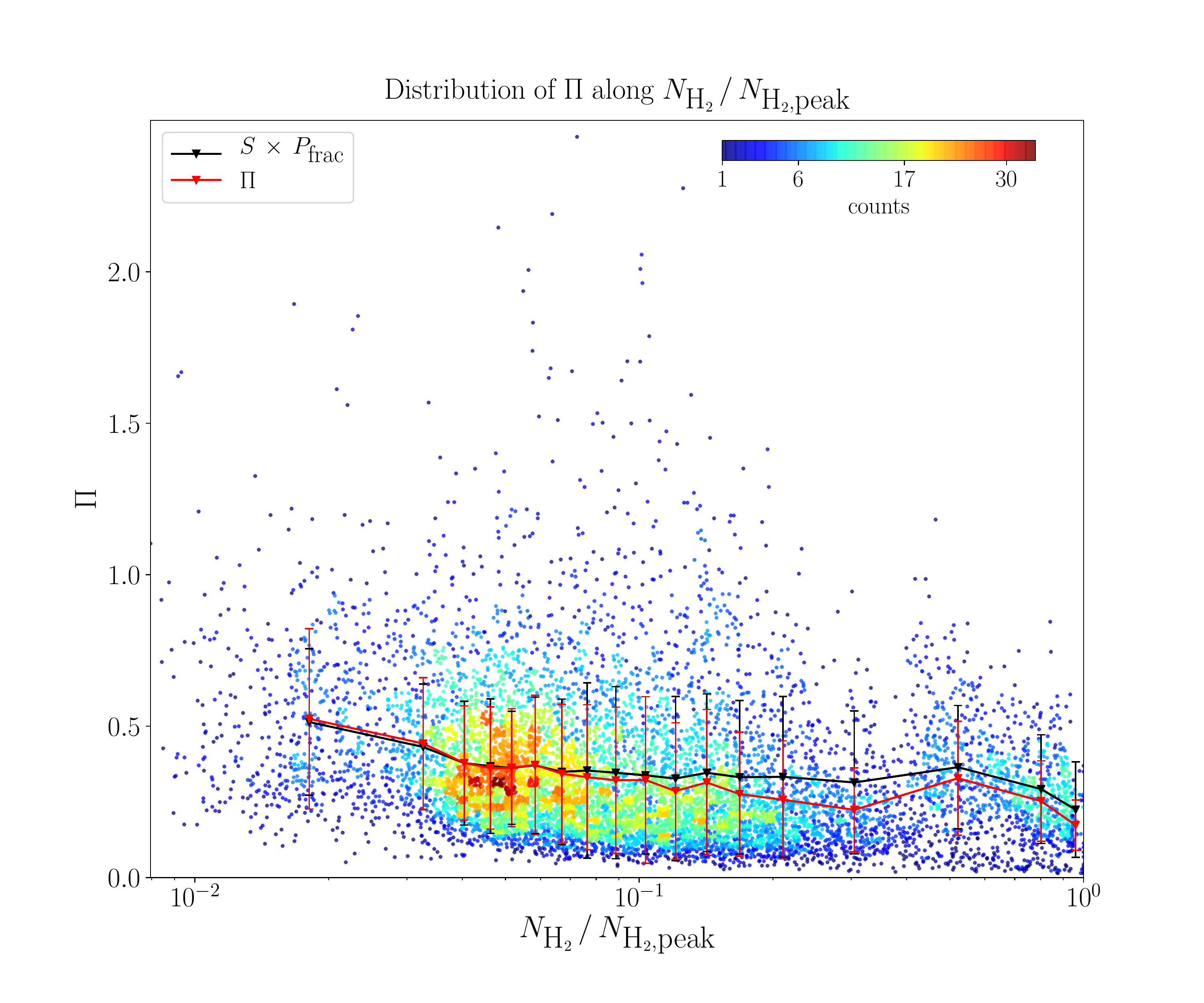}}
\caption[]{\small Distributions of $\StimesP$ (\textit{left}) and $\Pi$ (\textit{right}) as a function of the column density $N_{\textrm{H}_2}$, normalized in each core by its maximum value $N_{\textrm{H}_2\textrm{,peak}}$. The color scale represents number density of points in the plots. The solid black (red) lines and black (red) points represent the running mean of $\StimesP$ ($\Pi$); the associated error bars are $\pm$ the standard deviation of each bin. To facilitate the visual comparison, the running means of both $\StimesP$ and $\Pi$ a re plotted in both panels.}
\label{fig:scatter_coldens_norma_PiS}
\end{figure*}

\begin{figure*}[!tbph]
\subfigure{\includegraphics[scale=0.34
,clip,trim=  1.5cm 1.2cm 2.5cm 1.9cm]
{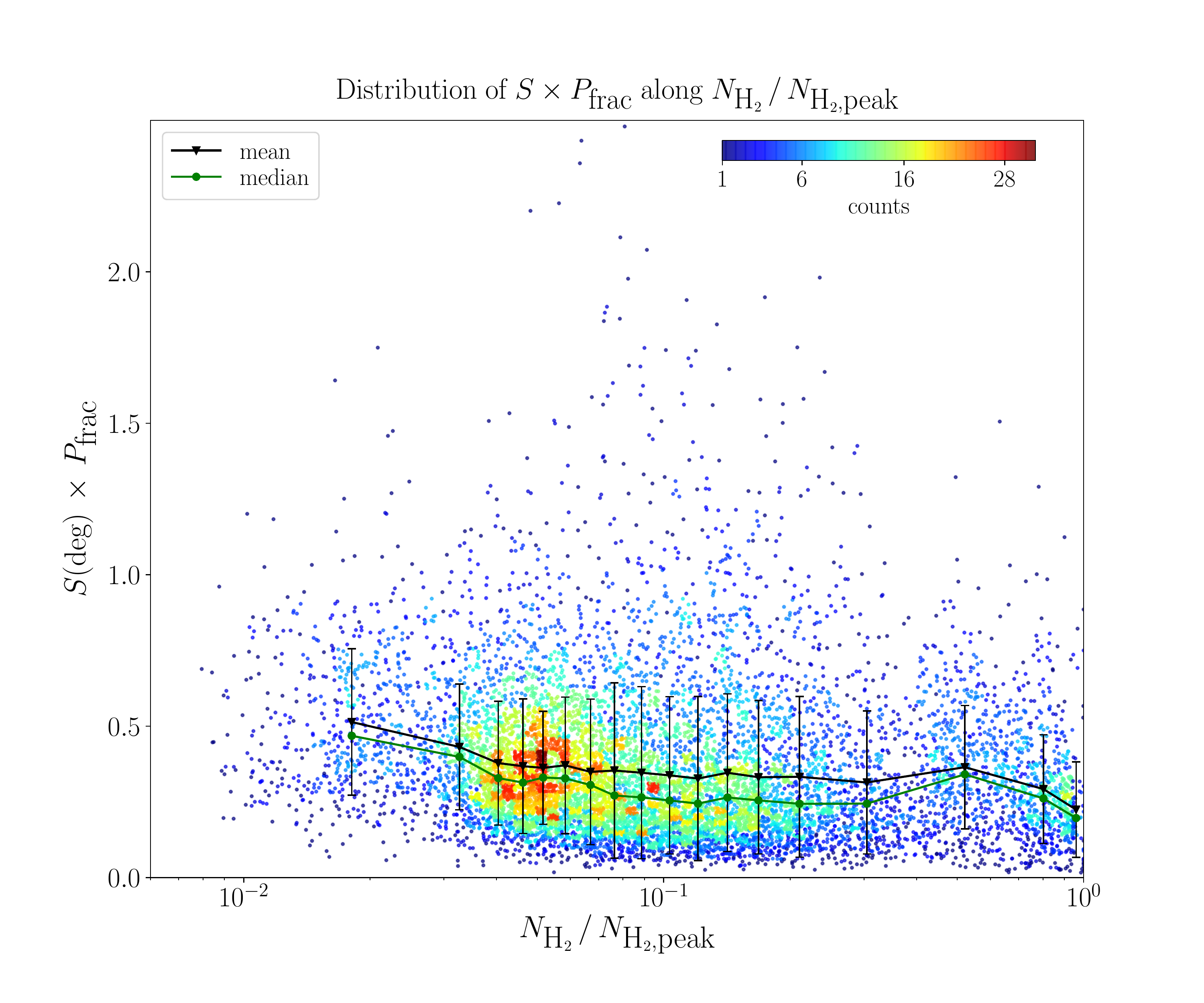}}
\subfigure{\includegraphics[scale=0.34
,clip,trim=  1.5cm 1.2cm 2.5cm 1.9cm]
{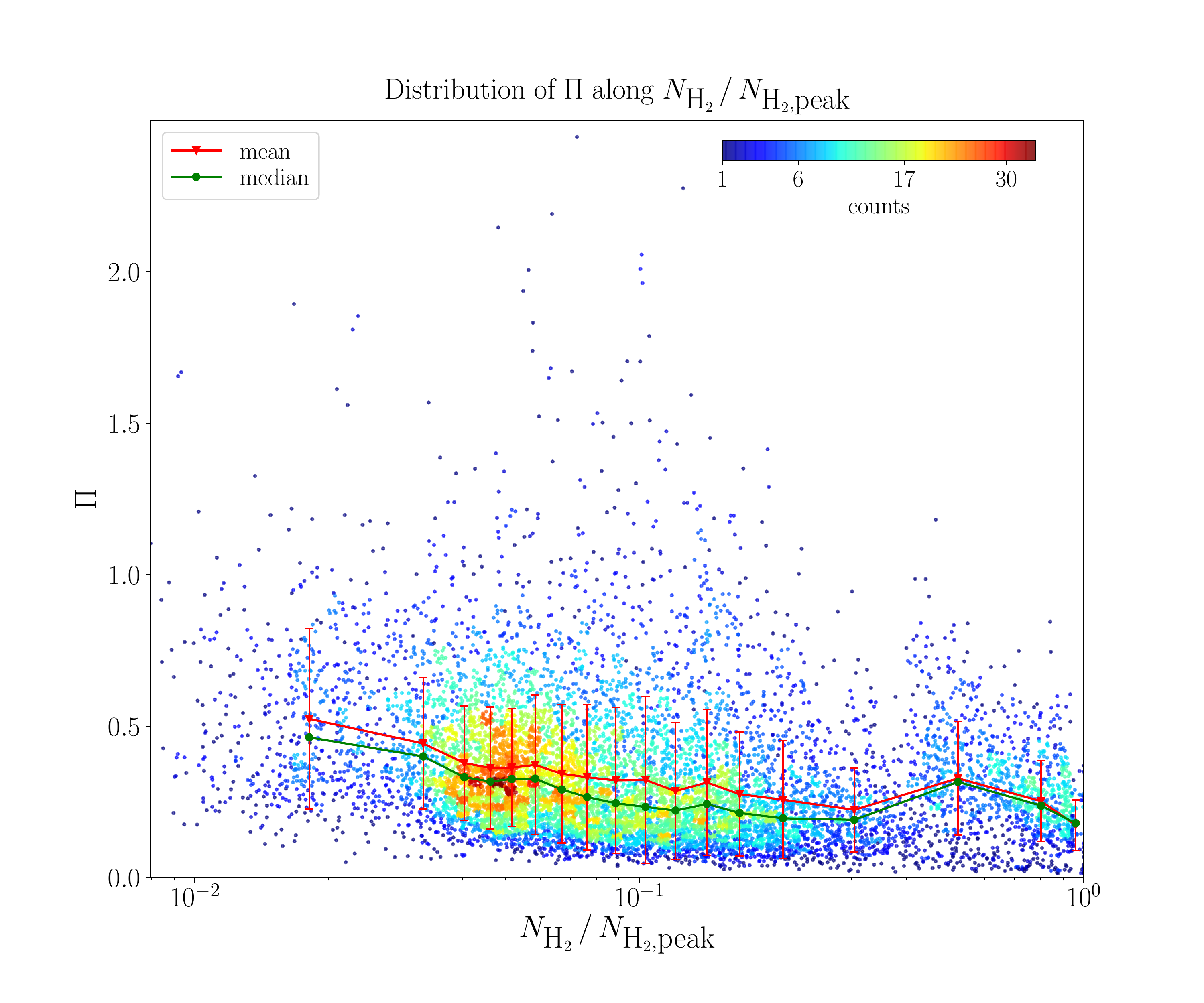}}
\caption[]{\small Same as Figure \ref{fig:scatter_coldens_norma_PiS}, but comparing the mean (shown in both panels of Figure \ref{fig:scatter_coldens_norma_PiS}) with the median.}
\label{fig:scatter_coldens_norma_median}
\end{figure*}

In Figures \ref{fig:scatter_coldens_norma_PiS}, \ref{fig:scatter_coldens_norma_median}, and \ref{fig:Sp_PI_pol_maps_sources} we show the comparisons between the results provided by $\Pi$ and $\StimesP$. These plots show that the results are only marginally different, and thus we do not recompute our results using this new estimator.

\def\scaleSP{0.67}

\begin{figure*}[!tbph]
\centering
\vspace{-0.2cm}
\subfigure{\includegraphics[scale=\scaleSP,clip,trim= 1.9cm 3.0cm 1.85cm 2.6cm]{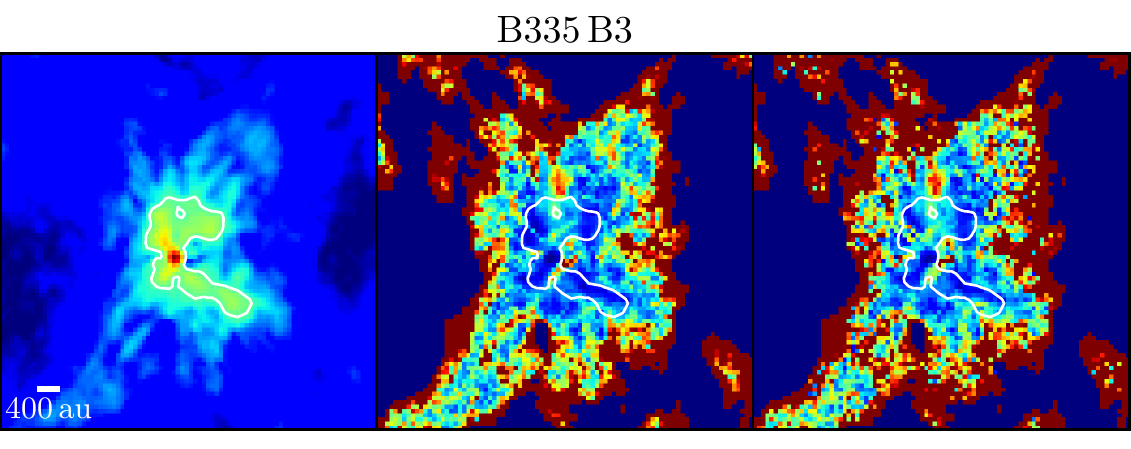}}
\vspace{-0.4cm}
\subfigure{\includegraphics[scale=\scaleSP,clip,trim= 1.9cm 3.0cm 0.2cm 2.6cm]{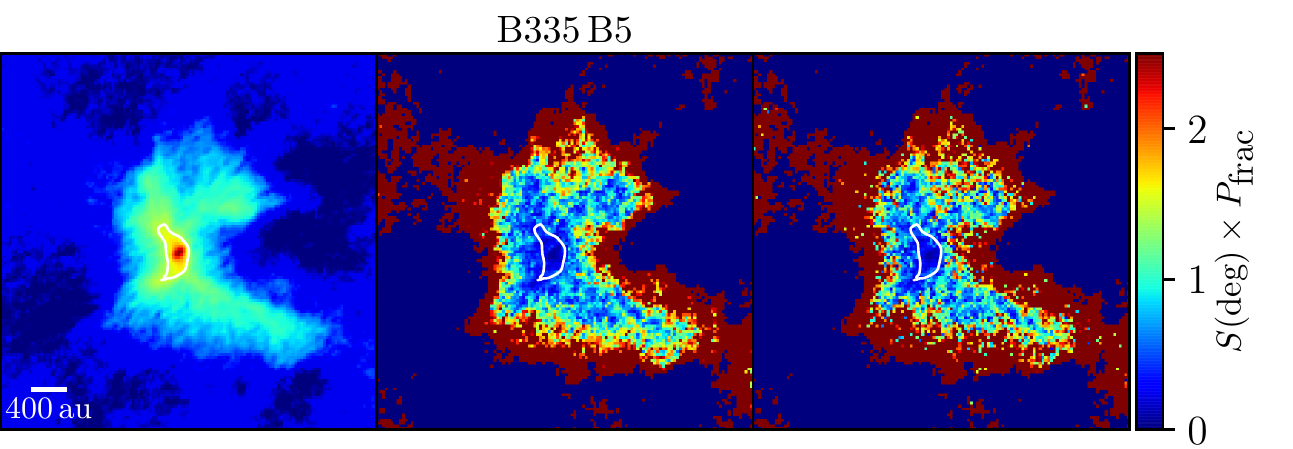}}
\vspace{-0.2cm}
\subfigure{\includegraphics[scale=\scaleSP,clip,trim= 1.9cm 3.0cm 1.85cm 2.6cm]{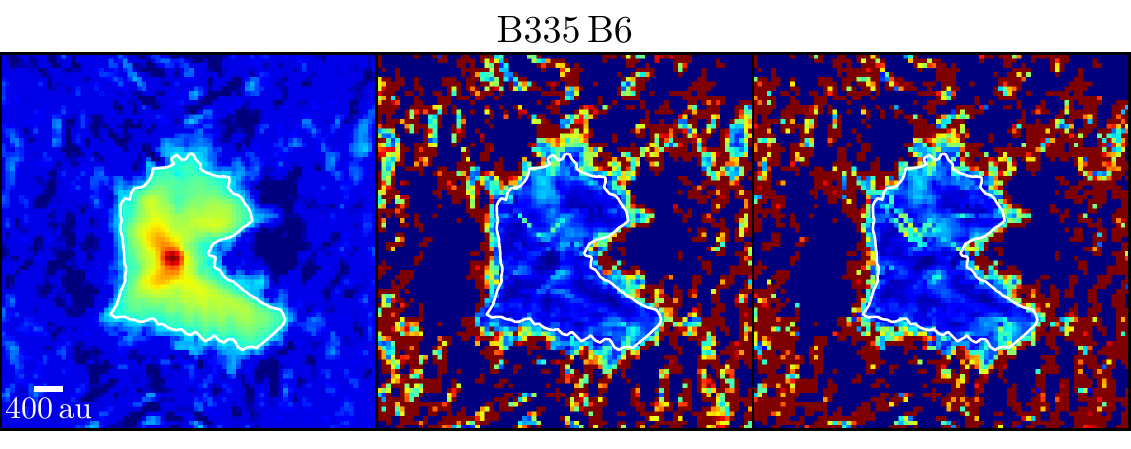}}
\vspace{-0.2cm}
\subfigure{\includegraphics[scale=\scaleSP,clip,trim= 1.9cm 3.0cm 0.2cm 2.6cm]{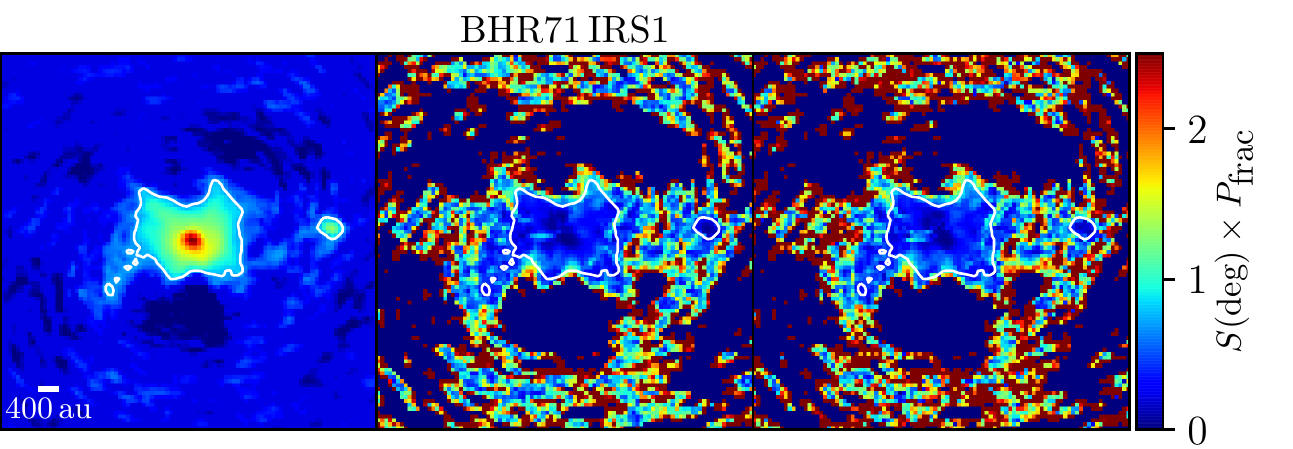}}
\vspace{-0.2cm}
\subfigure{\includegraphics[scale=\scaleSP,clip,trim= 1.9cm 3.0cm 1.85cm 2.6cm]{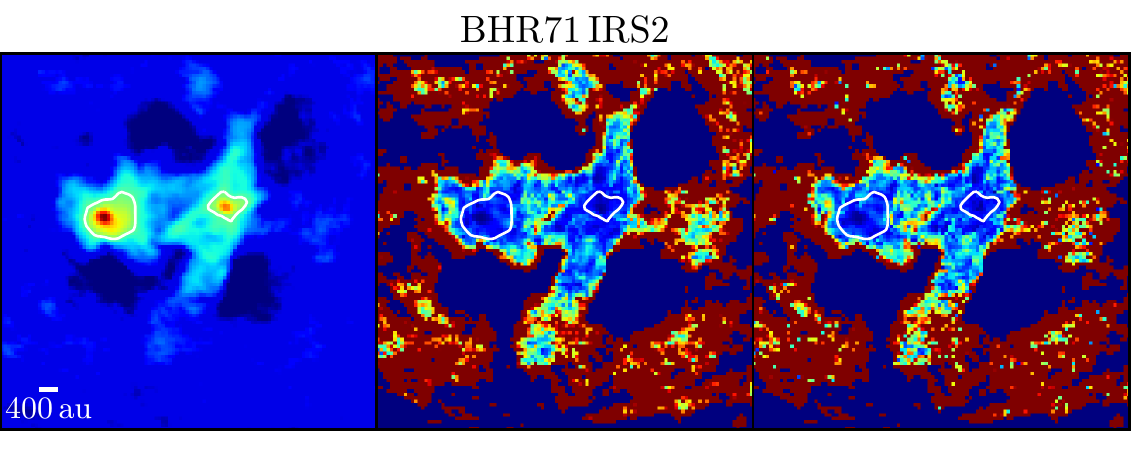}}
\vspace{-0.2cm}
\subfigure{\includegraphics[scale=\scaleSP,clip,trim= 1.9cm 3.0cm 0.2cm 2.6cm]{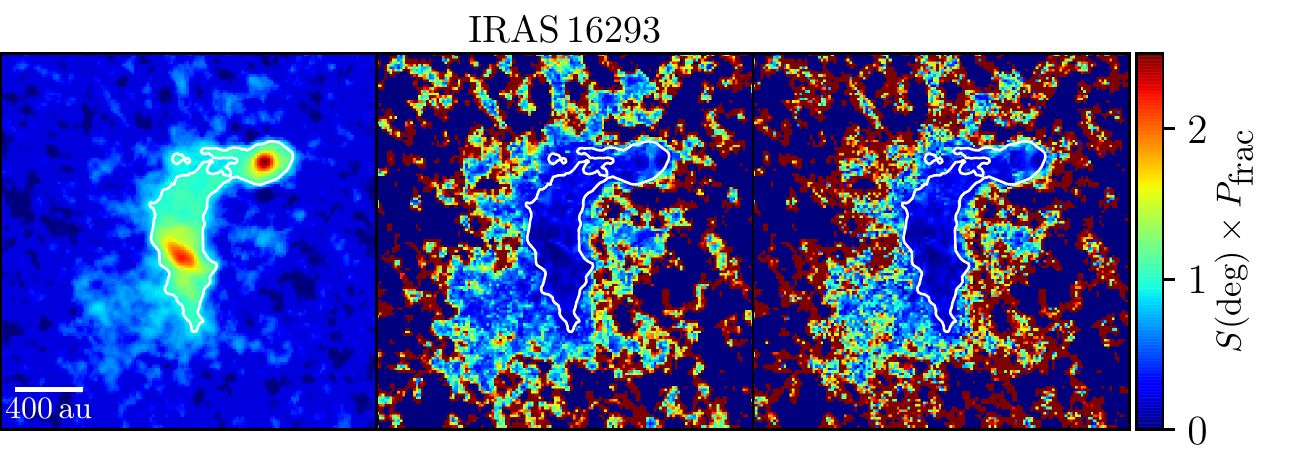}}
\vspace{-0.2cm}
\subfigure{\includegraphics[scale=\scaleSP,clip,trim= 1.9cm 3.0cm 1.85cm 2.6cm]{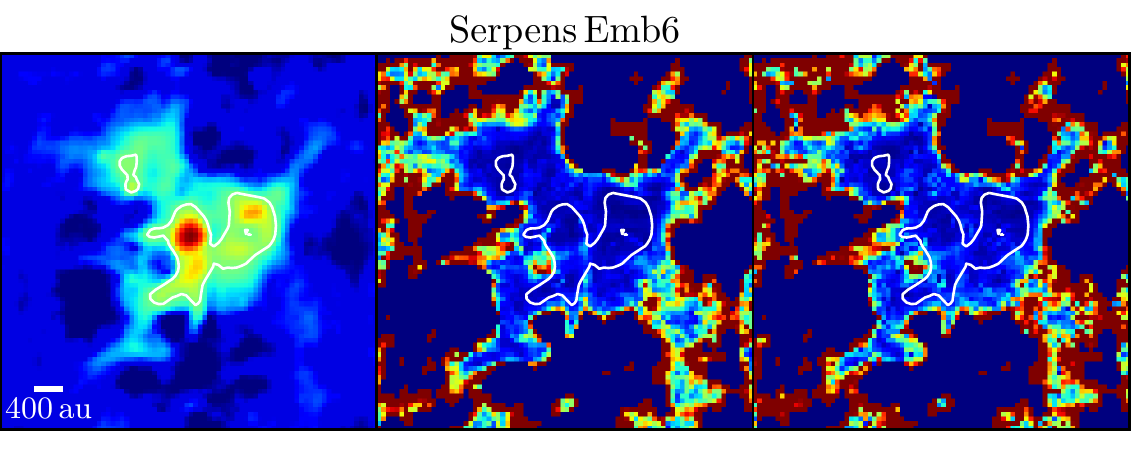}}
\vspace{-0.2cm}
\subfigure{\includegraphics[scale=\scaleSP,clip,trim= 1.9cm 3.0cm 0.2cm 2.6cm]{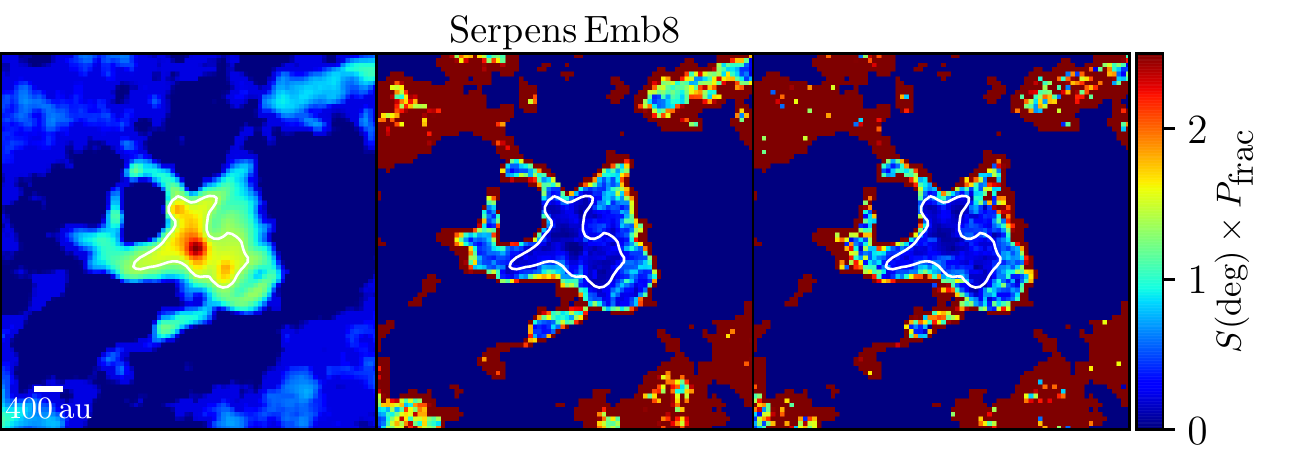}}
\vspace{-0.2cm}
\subfigure{\includegraphics[scale=\scaleSP,clip,trim= 1.9cm 3.0cm 1.85cm 2.6cm]{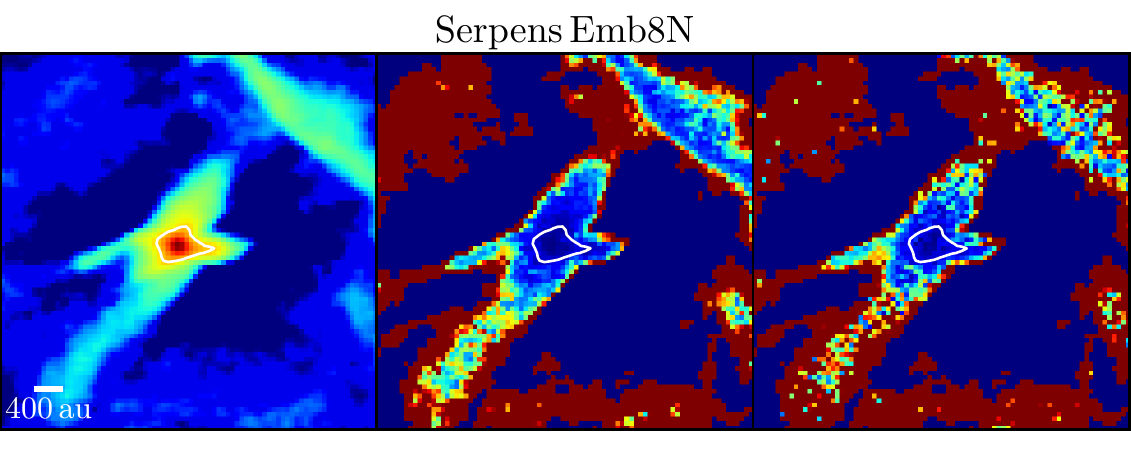}}
\vspace{-0.2cm}
\subfigure{\includegraphics[scale=\scaleSP,clip,trim= 1.9cm 3.0cm 0.2cm 2.6cm]{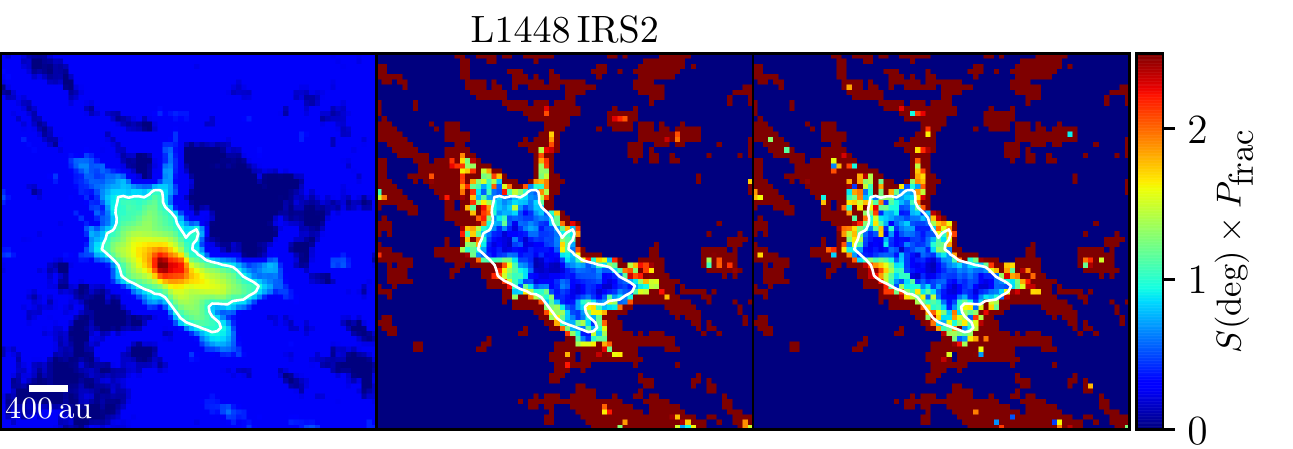}}
\vspace{-0.2cm}
\subfigure{\includegraphics[scale=\scaleSP,clip,trim= 1.9cm 3.0cm 1.85cm 2.6cm]{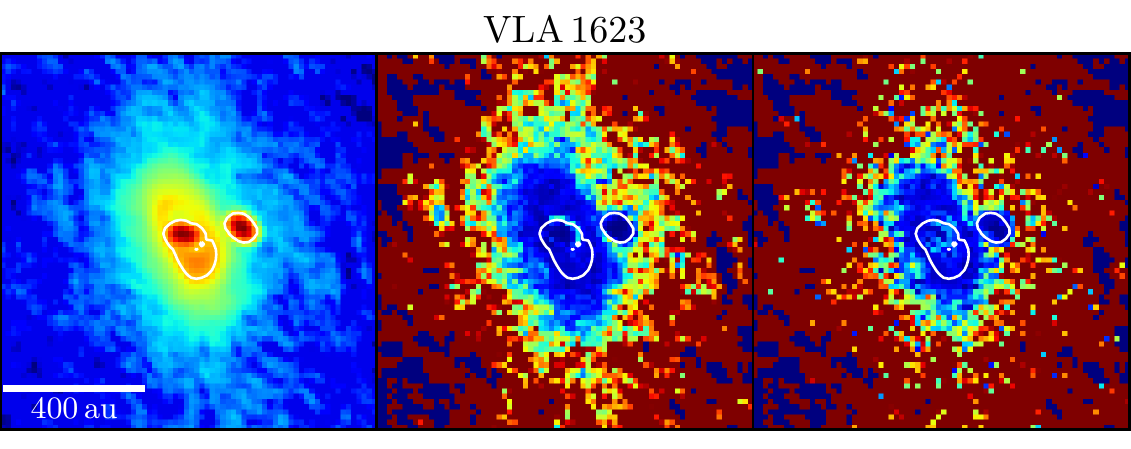}}
\vspace{-0.2cm}
\subfigure{\includegraphics[scale=\scaleSP,clip,trim= 1.9cm 3.0cm 0.2cm 2.6cm]{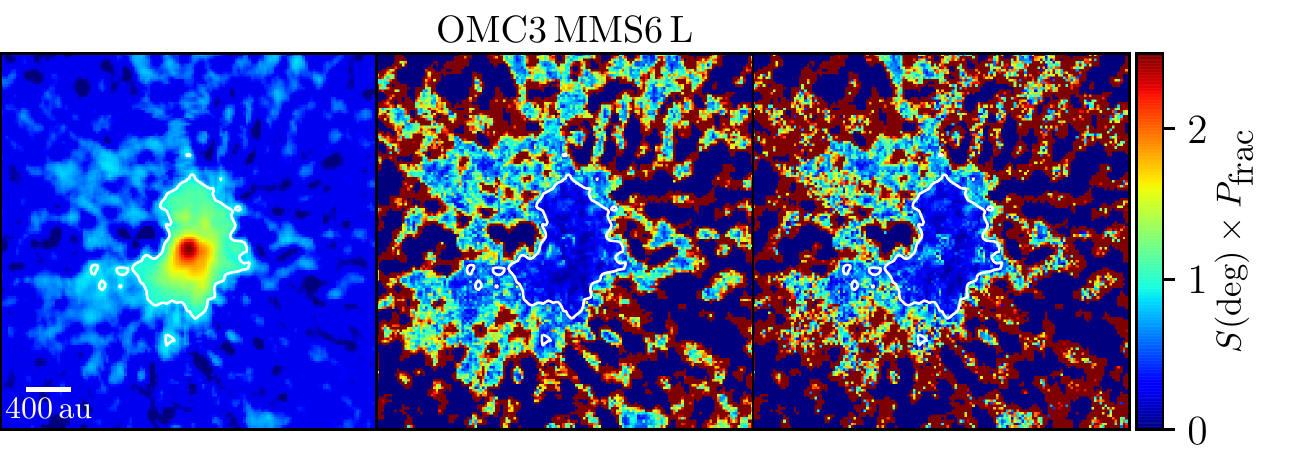}}
\vspace{-0.2cm}
\subfigure{\includegraphics[scale=\scaleSP,clip,trim= 1.9cm 3.0cm 1.85cm 2.6cm]{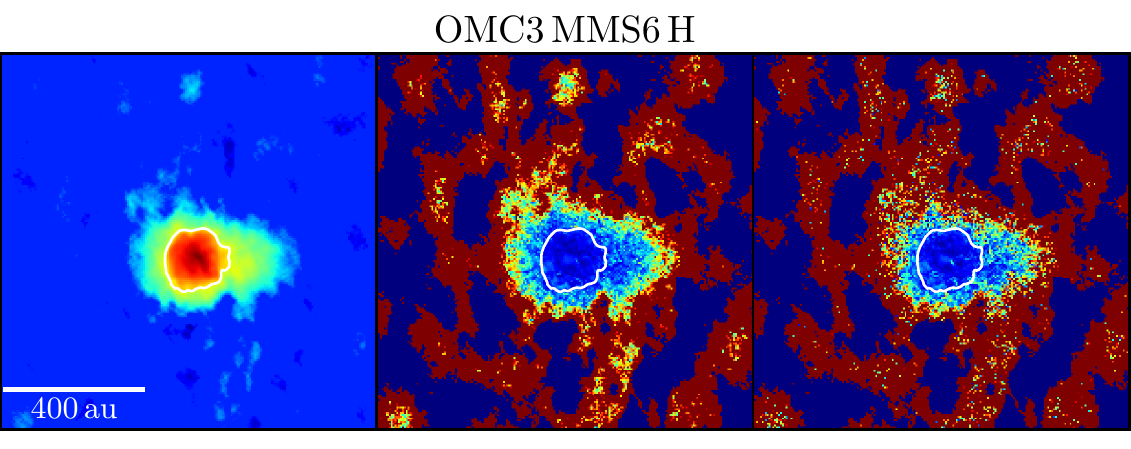}}
\vspace{-0.2cm}
\subfigure{\includegraphics[scale=\scaleSP,clip,trim= 1.9cm 3.0cm 0.2cm 2.6cm]{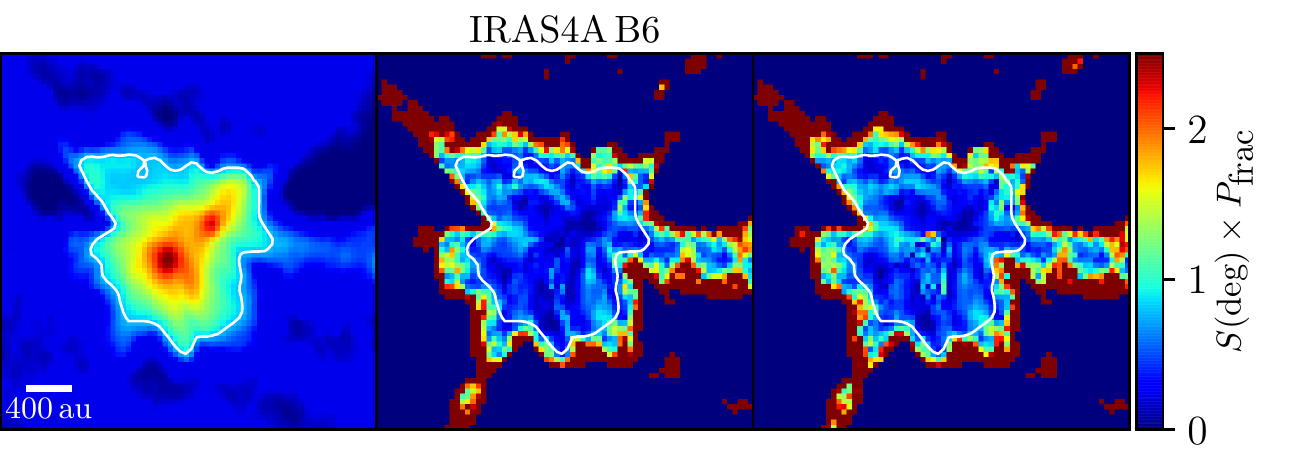}}
\vspace{-0.2cm}
\subfigure{\includegraphics[scale=\scaleSP,clip,trim= 1.9cm 3.0cm 0.2cm 2.6cm]{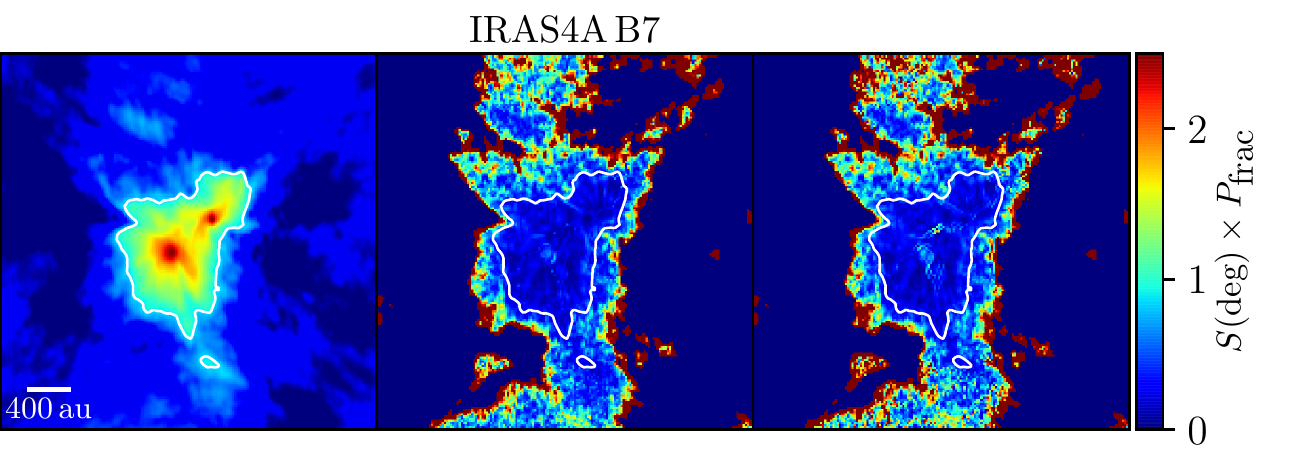}}
\caption[]{\small \textit{Left:} Stokes $I$; \textit{center:} $\Pi$; \textit{right:} $\StimesP$. The white contour represents the threshold in Stokes $I$ used to select the data.}
\label{fig:Sp_PI_pol_maps_sources}
\vspace{0.2cm}
\end{figure*}


\end{appendix}

\clearpage
\bibliography{ms}
\bibliographystyle{apj}

\end{document}